\documentclass[preprint,12pt]{elsarticle}

\UseRawInputEncoding

\usepackage{amssymb}
\usepackage{tikz}
\usepackage{psfrag}
\usepackage{pstricks}
\usepackage{array}
\csname PSTplotLoaded\endcsname
\let\PSTplotLoaded 
\ifx\PSTricksLoaded \else\input pstricks.tex\fi
\ifx\PSTXKeyLoaded \else \input pst-xkey.tex\fi
\ifx\PSTtoolsloaded \else\input pst-tools.tex\fi
\ifx\PSTtoolsloaded \else\input pstricks-add.tex\fi
\ifx\PSTFPloaded \else   \input pst-fp.tex  \fi
\ifx\MultidoLoaded \else \input multido.tex \fi
\def\fileversion{1.92}
\def\filedate{2019/05/16}
\edef\TheAtCode{\the\catcode`\@}
\catcode`\@=11
\pst@addfams{pst-plot}
\def\pst@linetype{2}%
\SpecialCoor

\newdimen\pstRadUnit
\newdimen\pstRadUnitInv
\begingroup
\catcode`\{=13
\catcode`\}=13
\catcode`\(=13
\catcode`\)=13
\catcode`\,=13
\catcode`\!=1
\catcode`\*=2
\catcode`\ =13
\catcode`\_=13
\catcode`\^^M=13
\gdef\pst@datadelimiters!% Begin def
\catcode`\{=13%
\catcode`\}=13%
\catcode`\(=13%
\catcode`\)=13%
\catcode`\,=13%
\catcode`\ =13%
\catcode`\^^M=13%
\def,##1!%
\ifcat\noexpand,\noexpand##1%
\expandafter##1%
\else\space%
D\space##1%
\fi*%
\let(,\let),\let{,\let},\let ,\let^^M,\let_\@empty*% End def
\endgroup
\define@key[psset]{pst-plot}{ignoreLines}[0]{\def\psk@ignoreLines{#1}}
\psset[pst-plot]{ignoreLines=0}
\newcount\pst@linecnt
\begingroup
\catcode`\,=13
\catcode`\_=13
\gdef\savedata@#1[#2]{%
  \xdef\pst@tempg{#2_}%
  \endgroup
  \let#1\pst@tempg
  \global\let\pst@tempg\relax
  \ignorespaces}
\gdef\readdata@{%
  \read1 to \pst@tempA
  \ifnum\pst@linecnt=\psk@nStep
    \global\pst@linecnt=0
    \expandafter\readdata@@\pst@tempA_\@nil
  \fi
  \global\advance\pst@linecnt by 1
  \ifeof1\else\expandafter\readdata@\fi}
\gdef\pst@@readfile#1#2\@nil{\addto@pscode{,#1#2}}%
\gdef\readdata@@#1#2\@nil{\xdef\pst@tempg{\pst@tempg,#1#2}}%
\endgroup
\def\readdata{\@ifnextchar[{\readdata@i}{\readdata@i[]}}
\def\readdata@i[#1]#2#3{%
  \openin1=#3
  \begingroup
  \ifx#1\relax#1\else\psset{#1}\fi
  \def\pst@tempg{}%
  \ifeof1
    \@pstrickserr{Data file `#3' not found.}\@ehpa
  \else
    \pst@datadelimiters
    \catcode`\[=1
    \catcode`\]=2
    \pst@cnta=0
    \loop \ifnum\the\pst@cnta<\psk@ignoreLines
      \advance\pst@cnta by 1\relax
      \read1 to \pst@tempA
    \repeat
    \psDEBUG[pst-plot]{>>> ignored \the\pst@cnta\space data lines}%
    \global\pst@linecnt=\psk@nStep
    \readdata@
  \fi
  \endgroup
  \global\let#2\pst@tempg%
  \global\let\pst@tempg\relax%
\ignorespaces}
\def\pst@readfile#1{{\let\readdata@@\pst@@readfile\readdata\pst@tempg{#1}}}
\def\pst@altreadfile#1{%
  \openin1=#1
  \ifeof1
    \@pstrickserr{Data file `#1' not found.}\@ehpa
  \else
    \catcode`\{=10
    \catcode`\}=10
    \catcode`\(=10
    \catcode`\)=10
    \catcode`\,=10
    \catcode`\^^M=10
    \catcode`\[=1
    \catcode`\]=2
    \pst@@altreadfile
  \fi}
\def\pst@@altreadfile{%
  \read1 to \pst@tempg
  \expandafter\pst@@@altreadfile\pst@tempg\@empty\@nil
  \ifeof1\else\expandafter\pst@@@altreadfile\fi}
\def\pst@@@altreadfile#1#2\@nil{\addto@pscode{#1#2}}%
\def\savedata#1{\begingroup\pst@datadelimiters\savedata@{#1}}
\newread\RCD@file
\def\psreadDataColumn{\@ifnextchar[\psreadDataColumn@i{\psreadDataColumn@i[]}}
\def\psreadDataColumn@i[#1]{%
  \psset{#1}%
  \psreadDataColumn@ii
}
\def\psreadDataColumn@ii#1#2#3#4{%    #1:column #2:delimiter #3:\result #4:data file
  \immediate\openin\RCD@file=#4\relax
  \global\let#3=\@empty
  \pst@cnta=0
  \loop \ifnum\the\pst@cnta<\psk@ignoreLines
      \advance\pst@cnta by 1\relax
      \read\RCD@file to \@tempa
  \repeat
  \loop
    \read\RCD@file to \@tempa
    \ifeof\RCD@file\else
      \edef\@tempa{\@tempa#2}%
      \def\reserved@b{}%
      \@tempswafalse
      \@tempcnta=#1\relax
    \expandafter\@tfor\expandafter\reserved@a
      \expandafter:\expandafter=\@tempa\do{% loop over every token
      \if\reserved@a#2\relax%                delimiter?
        \advance\@tempcnta \m@ne
        \ifnum \@tempcnta=\z@
          \expandafter\g@addto@macro\expandafter#3\expandafter{\reserved@b\space}%
          \@tempswatrue
        \fi
        \def\reserved@b{}% ???
      \else
        \edef\reserved@b{\reserved@b\reserved@a}
      \fi
      \if@tempswa\@break@tfor\fi
    }%
  \repeat
  \immediate\closein\RCD@file
}

\def\beginplot@line{\begin@OpenObj}
\def\endplot@line{\psline@ii}
\def\beginplot@polygon{\begin@ClosedObj}
\def\endplot@polygon{\pspolygon@ii}
\def\beginplot@curve{\begin@OpenObj}
\def\endplot@curve{\pscurve@ii}
\def\beginplot@ecurve{\begin@OpenObj}
\def\endplot@ecurve{\psecurve@ii}
\def\beginplot@ccurve{\begin@ClosedObj}
\def\endplot@ccurve{\psccurve@ii}
\def\beginplot@dots{\begin@SpecialObj}
\def\endplot@dots{\psdots@ii}
\define@key[psset]{pst-plot}{Hue}[180]{\pst@getint{#1}\pst@HueAngle}
\psset[pst-plot]{Hue=180}
\def\beginplot@colordots{\begin@SpecialObj}
\def\endplot@colordots{%
  \addto@pscode{%
    \psk@dotsize
    \@nameuse{psds@\psk@dotstyle}
    newpath
    /MaxValue 0 def
    /m n 2 mul def
    n { 
      dup MaxValue gt { dup /MaxValue ED } if
      m 2 roll
    } repeat
    n { dup MaxValue div  % y y
      \pst@number\psyunit div abs % to orig y value
      \pst@HueAngle\space 360 div exch dup sethsbcolor % 180 Y Y hsb color
      transform floor .5 add exch floor
      .5 add exch itransform Dot stroke } repeat }%
  \end@SpecialObj%
}
\def\beginplot@bubble{\begin@SpecialObj}
\def\endplot@bubble{%
  \addto@pscode{%
    newpath
    n { dup % x y y
      \pst@number\psyunit div abs % to orig y value
      transform floor .5 add exch floor
      .5 add exch itransform 
      0 360 arc \psk@fill 
      stroke } repeat }%
  \end@SpecialObj%
}
\def\beginplot@bezier{\begin@OpenObj}
\def\endplot@bezier{\psbezier@ii}
\def\beginplot@cbezier{\begin@ClosedObj}
\def\endplot@cbezier{\pscbezier@ii}
\def\beginplot@cspline{\begin@OpenObj}%% Christoph Bersch
\def\endplot@cspline{\pscspline@ii}
\let\beginplot@LineToYAxis\beginplot@line  % all from pst-plot added 2007-06-26 (hv)
\def\endplot@LineToYAxis{\psLineToYAxis@ii}
\let\beginqp@LineToYAxis\beginqp@line
\let\doqp@LineToYAxis\doqp@line
\let\endqp@LineToYAxis\endqp@line
\let\testqp@LineToYAxis\testqp@line
\let\beginplot@LineToXAxis\beginplot@line
\def\endplot@LineToXAxis{\psLineToXAxis@ii}
\let\beginqp@LineToXAxis\beginqp@line
\let\doqp@LineToXAxis\doqp@line
\let\endqp@LineToXAxis\endqp@line
\let\testqp@LineToXAxis\testqp@line
\newif\ifPst@interrupt \Pst@interruptfalse
\define@key[psset]{pst-plot}{barwidth}[0.25cm]{\pst@getlength{#1}\Add@barwidth}
\psset[pst-plot]{barwidth=0.25cm}
\define@key[psset]{pst-plot}{interrupt}[]{\expandafter\pst@interrupt#1,,,\@nil}
\def\pst@interrupt#1,#2,#3,#4\@nil{%
  \ifx\relax#1\relax \Pst@interruptfalse
  \else
    \Pst@interrupttrue
    \def\pst@interrupt@YMax{#1 }%
    \def\pst@interrupt@YMaxSep{#2 }%
    \def\pst@interrupt@YMaxDiff{#3 }%
  \fi
}
\def\psbar@ii{\addto@pscode{false \tx@NArray \psbar@iii}}

\def\psbar@iii{%
  \ifPst@interrupt
    /YMax \pst@interrupt@YMax \strip@pt\psyunit\space mul def
    /YMaxSep \pst@interrupt@YMaxSep \strip@pt\psyunit\space mul def
    /YMaxDiff \pst@interrupt@YMaxDiff \strip@pt\psyunit\space mul def
    /Tilde { % on stack DX
      /Op ED % add or sub
      /DX ED
      currentpoint 2 copy
      /Y ED /X ED   % x y  
      X DX add Y YMaxSep 2 div Op   
      X DX dup add add Y           
      curveto
      currentpoint 2 copy pop /X ED 
      X DX add Y YMaxSep 2 div neg Op  
      X DX dup add add Y    
      curveto      
    } def  
    newpath
    n { 
      /Yval exch def /Xval exch def 
      Xval \number\Add@barwidth 0.5 mul sub 0 moveto 
      Yval YMax le {  
        0 Yval rlineto \number\Add@barwidth 0 rlineto 
        0 Yval neg rlineto \number\Add@barwidth neg 0 rlineto
      }{
        0 YMax rlineto 
        \number\Add@barwidth 4 div 
        { add } Tilde
        0 YMax neg rlineto 
        \number\Add@barwidth neg 0 rlineto
        closepath
        Xval \number\Add@barwidth 0.5 mul sub YMax YMaxSep add moveto 
        0 Yval YMax sub YMaxSep sub YMaxDiff sub rlineto 
        \number\Add@barwidth 0 rlineto 
        0 Yval YMax YMaxSep add sub YMaxDiff sub neg rlineto 
        \number\Add@barwidth 4 div neg 
        { sub } Tilde
      } ifelse
    } repeat
  \else
    newpath
    n { 
      /Yval exch def /Xval exch def 
      Xval \number\Add@barwidth 0.5 mul sub 0 moveto 
      0 Yval rlineto \number\Add@barwidth 0 rlineto 
      0 Yval neg rlineto \number\Add@barwidth neg 0 rlineto
    } repeat
  \fi
}%
\def\beginplot@bar{\begin@SpecialObj}
\def\endplot@bar{%
  \psbar@ii\psk@fillstyle\ifpsshadow\pst@closedshadow\fi%
  \pst@stroke
  \end@SpecialObj}
\def\psybar@ii{\addto@pscode{false \tx@NArray \psybar@iii}}
\def\psybar@iii{%
  newpath
  n { 
    /Yval exch def /Xval exch def 
    0 Yval \number\Add@barwidth 0.5 mul sub moveto 
    Xval 0 rlineto 0 \number\Add@barwidth rlineto 
    Xval neg 0 rlineto 0 \number\Add@barwidth neg rlineto
  } repeat
}%
\def\beginplot@ybar{\begin@SpecialObj}
\def\endplot@ybar{%
  \psybar@ii\psk@fillstyle\ifpsshadow\pst@closedshadow\fi%
  \pst@stroke
  \end@SpecialObj}
\def\psLSM@ii{\addto@pscode{ false \tx@NArray \psLSM@iii }}
\def\psLSM@iii{%
  /xiSquare 0 def				% xi*xi
  /xi 0 def					% xi
  /fi 0 def					% f(xi)
  /xifi 0 def					% xi*f(xi)
  exch dup dup /xEnd ED /xStart ED exch
  n { 						% number of data pairs
    /Yval ED /Xval ED 				% save x y values
    /xi xi Xval add def				% sum xi
    /xiSquare xiSquare Xval dup mul add def	% sum xi*xi
    /xifi xifi Xval Yval mul add def		% sum xi*yi, same as xi*f(xi)
    /fi fi Yval add def				% sum yi, same as f(xi)
    Xval xStart lt { /xStart Xval def } if	% find the lowest xi
    Xval xEnd gt { /xEnd Xval def } if		% find the largest xi
  } repeat
  /u xiSquare fi mul xi xifi mul sub n xiSquare mul xi dup mul sub div def
  /v n xifi mul xi fi mul sub n xiSquare mul xi dup mul sub div def
  \Pst@Debug\space 0 gt { 			% print the equation
    /NimbusSanL-Regu findfont 12 scalefont setfont	
    0 -50 moveto (y=) show 			% print y=
    v \pst@number\psyunit \pst@number\psxunit div div 20 string cvs show ( x+) show		% m*x+
    u \pst@number\psyunit div 20 string cvs show } if
  newpath
  (\psk@xStart) length 0 gt 			% special start value?
    { \psk@xStart\space \pst@number\psxunit mul }
    { xStart } ifelse 
  dup v mul u add 				% xStart f(xStart)  
  moveto		 			% goto first point x1 y(x1)
  (\psk@xEnd) length 0 gt 			% special end value?
    { \psk@xEnd\space \pst@number\psxunit mul }
    { xEnd } ifelse 
  dup v mul u add 				% xEnd f(xEnd)	
  lineto					% line to second point x2 y(x2)
}%
\def\beginplot@LSM{\begin@SpecialObj}
\def\endplot@LSM{%
  \psLSM@ii\psk@fillstyle\ifpsshadow\pst@closedshadow\fi%
  \pst@stroke
  \end@SpecialObj}
\define@key[psset]{pst-plot}{IQLfactor}{\pst@checknum{#1}\pst@IQLfactor}
\psset[pst-plot]{IQLfactor=1.5}
\define@key[psset]{pst-plot}{postAction}[]{\def\psk@postAction{%
  \ifx\relax#1\relax\else\pst@number\psyunit div #1 \pst@number\psyunit mul \fi }}
\psset[pst-plot]{postAction=}
\define@key[psset]{pst-plot}{mediancolor}[black]{\pst@getcolor{#1}\median@linecolor}
\psset[pst-plot]{mediancolor=black}
\define@boolkey[psset]{pst-plot}[Pst@]{markMedian}[true]{}
\psset[pst-plot]{markMedian=false}

\def\psBoxplot@ii{%
  \addto@pscode{
    /Barwidth \number\Add@barwidth 2 div def  
    /Endwidth Barwidth \psk@arrowlength\space mul def  
   NArray bubblesort
   /NArray ED 				% save sorted array
   [ NArray { yUnit mul } forall ] /NArray ED % multiply with y unit
   NArray 0 get /MinVal ED		% save minimum
   NArray m 1 sub get /MaxVal ED	% maximum
   m 2 div cvi /M ED 			% the middle
   NArray length 2 mod 0 eq {		% even numbers of entries
     M 1 sub NArray exch get 		% even number of values
     NArray M get          		% and the upper one
     add 2 div /Median ED  		% the median
   }{
     NArray M get /Median ED  		% odd numbers of values
   } ifelse
   m 4 mod 0 eq {	  		% get the lower Quartil even/odd
     m 4 div cvi dup 1 sub NArray exch get
     exch NArray exch get
     add 2 div floor /LowerQuartil ED
   }{ 
     NArray M 2 div cvi get /LowerQuartil ED 
   } ifelse				% end even/odd 
   m 0.75 mul dup dup cvi sub 0 eq {	% get the upper Quartil
     cvi dup 1 sub NArray exch get exch NArray exch get
     add 2 div floor /UpperQuartil ED
   }{					% upper quartil
     NArray m 0.75 mul floor cvi get /UpperQuartil ED
   } ifelse 
   /IQL UpperQuartil LowerQuartil sub \pst@IQLfactor\space mul def
   0 1 m 1 sub { % Index on stack
     dup /Index ED
     NArray exch get LowerQuartil sub abs IQL sub 0 gt { 
       \psk@dotsize
       \@nameuse{psds@\psk@dotstyle}
       0 NArray Index get \psk@postAction
       Dot
       NArray Index LowerQuartil UpperQuartil LowerQuartil sub \pst@IQLfactor\space mul sub 
       dup /MinVal ED put % replace with 1.5 IQL
       NArray Index 1 add get /MinVal ED 
    } { exit } ifelse
   } for
   m 1 sub -1 0 {	% Index on stack
     dup /Index ED
     NArray exch get UpperQuartil sub abs IQL sub 0 gt { 
       \psk@dotsize
       \@nameuse{psds@\psk@dotstyle}
       0 NArray Index get \psk@postAction\space
       Dot
       NArray Index UpperQuartil LowerQuartil sub \pst@IQLfactor\space mul UpperQuartil add 
       dup /MaxVal ED put % replace with 1.5 IQL
       NArray Index 1 sub get /MaxVal ED 
     }{ exit } ifelse
   } for
   Endwidth neg MaxVal \psk@postAction moveto			% we are on top / lower whisker
   Endwidth dup add 0 rlineto 
   0 MaxVal \psk@postAction moveto 
   0 UpperQuartil \psk@postAction lineto			% upper quartil
   MinVal \psk@postAction MaxVal \psk@postAction lt {
     0 LowerQuartil \psk@postAction moveto			% line to lower whisker
     0 MinVal \psk@postAction lineto 
     Endwidth neg MinVal \psk@postAction moveto 
     Endwidth dup add 0 rlineto 
   } if
   gsave
   \pst@number\pslinewidth SLW
   \pst@usecolor\pslinecolor
   \tx@setStrokeTransparency 
   \@nameuse{psls@\pslinestyle}
   stroke
   grestore
   newpath
   Barwidth neg LowerQuartil \psk@postAction moveto	% lower quartil
   Barwidth neg UpperQuartil \psk@postAction lineto
   Barwidth dup add 0 rlineto
   Barwidth LowerQuartil \psk@postAction lineto
   closepath
   \pst@usecolor\psfillcolor
   gsave \pst@usecolor\psfillcolor \tx@setTransparency fill grestore
   \@nameuse{psls@solid}
   \ifPst@markMedian
     \pst@number\pslabelsep neg Median moveto currentpoint 
     /YMedian ED /XMedian ED 
      Barwidth neg Median \psk@postAction lineto  % median
   \else
      Barwidth neg Median \psk@postAction moveto  % median
   \fi
   Barwidth dup add 0 rlineto 
   \pst@number\pslinewidth SLW
   \pst@usecolor\median@linecolor
   \tx@setStrokeTransparency
   stroke
  }
}% 
\def\beginplot@Boxplot{\init@pscode}
\def\endplot@Boxplot{%
  \psBoxplot@ii\psk@fillstyle\ifpsshadow\pst@closedshadow\fi%
  \pst@stroke
  \end@SpecialObj}
\def\psBoxplot{\def\pst@par{}\pst@object{psBoxplot}}
\def\psBoxplot@i#1{%
  \leavevmode
  \pst@killglue
  \begingroup
  \addbefore@par{barwidth=40pt,arrowlength=0.75}%
  \addto@par{plotstyle=Boxplot}%
  \use@par
  \@nameuse{beginplot@\psplotstyle}%
  \addto@pscode{
    /D {} def
    [ #1 ] /NArray ED 
    NArray aload length /m ED
    /xUnit \pst@number\psxunit def
    /yUnit \pst@number\psyunit def
  }%
  \@nameuse{endplot@\psplotstyle}%
  \ignorespaces%
}
\define@key[psset]{pst-plot}{plotstyle}[line]{%
  \@ifundefined{beginplot@#1}%
    {\@pstrickserr{Plot style `#1' not defined}\@eha}%
    {\def\psplotstyle{#1}}}
\psset[pst-plot]{plotstyle=line}
\define@key[psset]{pst-plot}{plotpoints}[50]{%
  \pst@cntg=#1\relax
  \ifnum\pst@cntg<2
    \@pstrickserr{plotpoints parameter must be at least 2}\@ehpa
  \else
    \advance\pst@cntg-1
    \edef\psk@plotpoints{\the\pst@cntg\space}%
  \fi}
\psset[pst-plot]{plotpoints=50}
\define@key[psset]{pst-plot}{yMaxValue}[1.e30]{\def\psk@yMaxValue{#1 }\def\psk@yMinValue{#1 neg }}
\psset{yMaxValue=1.e30}
\define@key[psset]{pst-plot}{yMinValue}[-1.e30]{\def\psk@yMinValue{#1 }}
\psset{yMinValue=-1.e30}
\def\beginqp@line{\pst@oplineto}
\def\doqp@line{ 
  dup
  \psk@yMaxValue \pst@number\psyunit mul gt 
    { moveto }
    { dup \psk@yMinValue \pst@number\psyunit mul lt 
      { moveto }
      { L } ifelse 
    } ifelse
}
\def\endqp@line{%
  \ifPst@variableLW \addto@pscode{ \pst@flattenpath }\fi%
  \end@OpenObj}%

\def\testqp@line{%
  \ifdim\pslinearc>\z@\else
    \ifshowpoints\else
      \ifx\psk@arrowA\@empty
        \ifx\psk@arrowB\@empty
          \@psttrue
        \fi
      \fi
    \fi
  \fi}
\def\beginqp@polygon{moveto }
\def\doqp@polygon{ 
      dup
      \psk@yMaxValue \pst@number\psyunit mul gt 
      { moveto }{ 
          dup
          \psk@yMinValue \pst@number\psyunit mul lt 
          { moveto }{ L } ifelse 
      } ifelse
}
\def\endqp@polygon{%
  \addto@pscode{closepath}%
  \end@ClosedObj}
\def\testqp@polygon{%
  \ifdim\pslinearc>\z@\else
    \ifshowpoints\else
      \@psttrue
    \fi
  \fi}
\def\beginqp@dots{%
  \psk@dotsize
  \@nameuse{psds@\psk@dotstyle}
  Dot }
\def\doqp@dots{Dot }
\def\endqp@dots{\end@SpecialObj}
\def\testqp@dots{\@psttrue}
\def\beginqp@bezier{/n 0 def \pst@oplineto}
\def\doqp@bezier{/n n 1 add def n 3 mod 0 eq { % we need 3 points   
    dup \psk@yMaxValue\space \pst@number\psyunit mul gt 
    { moveto pop pop pop pop}
    { dup \psk@yMinValue\space \pst@number\psyunit mul lt 
      { moveto pop pop pop pop}{ curveto } ifelse 
    } ifelse 
  } if
}
\def\endqp@bezier{%
  \addto@pscode{n 3 mod { pop pop } repeat}
  \end@OpenObj}%
\def\testqp@bezier{%
  \ifshowpoints\else
    \ifx\psk@arrowA\@empty
      \ifx\psk@arrowB\@empty
        \@psttrue
      \fi
    \fi
  \fi}
\def\beginqp@cbezier{/n 0 def moveto }
\def\doqp@cbezier{\doqp@bezier}
\def\endqp@cbezier{%
  \addto@pscode{n 3 mod { pop pop } repeat closepath}
  \end@ClosedObj}%
\def\testqp@cbezier{\ifshowpoints\else\@psttrue\fi}
\def\tx@LineToYAxis{LineToYAxis }
\def\psLineToYAxis@ii{%
\addto@pscode{\pst@cp \psline@iii \psk@Ox\space \pst@number\psxunit mul \tx@LineToYAxis}%
\end@OpenObj}
\def\tx@LineToXAxis{LineToXAxis }
\def\psLineToXAxis@ii{%
\addto@pscode{\pst@cp \psline@iii \psk@Oy\space \pst@number\psyunit mul \tx@LineToXAxis}%
\end@OpenObj}
\define@key[psset]{pst-plot}{PSfont}[NimbusRomNo9L-Regu]{\def\psk@PSfont{/#1 }}
\define@key[psset]{pst-plot}{valuewidth}[10]{\pst@getint{#1}\psk@valuewidth }
\define@key[psset]{pst-plot}{fontscale}[10]{\pst@checknum{#1}\psk@fontscale }
\define@key[psset]{pst-plot}{decimals}[-1]{\pst@getint{#1}\psk@decimals }
\psset[pst-plot]{PSfont=NimbusRomNo9L-Regu,fontscale=10,valuewidth=10,decimals=-1}
\newdimen\psxlabelsep
\newdimen\psylabelsep
\define@key[psset]{pst-plot}{xlabelsep}[5pt]{\pssetlength\psxlabelsep{#1}}
\define@key[psset]{pst-plot}{ylabelsep}[5pt]{\pssetlength\psylabelsep{#1}}
\psset[pst-plot]{xlabelsep=5pt,ylabelsep=5pt}

\newif\ifPst@valuesStar\Pst@valuesStarfalse
\newif\ifPst@xvalues\Pst@xvaluesfalse
\def\psvalues@ii{\addto@pscode{ false \tx@NArray \psvalues@iii }}
\def\psvalues@iii{
  \psk@PSfont findfont \psk@fontscale scalefont setfont 
  newpath 
  n { /yO ED /xO ED
      gsave
      \ifPst@xvalues
        xO \pst@number\psxunit div
      \else
        yO \pst@number\psyunit div
      \fi
      \psk@decimals 0 eq { cvi } if
      \psk@decimals 0 gt { 10 \psk@decimals exp dup 3 1 roll mul cvi exch div } if
      \psk@valuewidth string cvs /Str ED
      \ifPst@valuesStar
      Str stringwidth pop /yS \psk@fontscale def /xS ED 
      gsave newpath 
        xO \ifPst@xvalues \pst@number\pslabelsep add \fi 
        yO \ifPst@xvalues \psk@fontscale 4 div sub \else \pst@number\pslabelsep add \fi 
        moveto \ifx\psk@rot\@empty\else\psk@rot rotate \fi
        xS 0 rlineto 0 yS rlineto xS neg 0 rlineto 0 yS neg rlineto 
        closepath  1 setgray fill stroke 
      grestore 
      \fi
      xO \ifPst@xvalues \pst@number\pslabelsep add \fi
      yO \ifPst@xvalues \psk@fontscale 4 div sub \else \pst@number\pslabelsep add \fi 
      moveto \ifx\psk@rot\@empty\else\psk@rot rotate \fi 
      Str show 
      grestore } repeat 
}%
\def\beginplot@values{\Pst@valuesStarfalse\begin@SpecialObj}
\expandafter\def\csname beginplot@values*\endcsname{\Pst@valuesStartrue\begin@SpecialObj}
\def\beginplot@xvalues{\Pst@valuesStarfalse\begin@SpecialObj}
\expandafter\def\csname beginplot@xvalues*\endcsname{\Pst@valuesStartrue\begin@SpecialObj}
\def\endplot@values{%
  \Pst@xvaluesfalse%  
  \psvalues@ii%
  \pst@stroke
  \end@SpecialObj}
\@namedef{endplot@values*}{\endplot@values}
\def\endplot@xvalues{%
  \Pst@xvaluestrue%  
  \psvalues@ii%
  \pst@stroke
  \end@SpecialObj}
\@namedef{endplot@xvalues*}{\endplot@xvalues}
\def\psdataplot{\def\pst@par{}\pst@object{dataplot}}
\def\dataplot{\def\pst@par{}\pst@object{dataplot}}
\def\dataplot@i#1{%
  \pst@killglue
  \begingroup
    \use@par
    \@pstfalse
    \@nameuse{testqp@\psplotstyle}%
    \if@pst
      \dataplot@ii{\addto@pscode{#1}}%
    \else
      \listplot@ii{\addto@pscode{#1}}%
    \fi
  \endgroup
  \ignorespaces}
\def\dataplot@ii#1{%
  \@nameuse{beginplot@\psplotstyle}%
    \addto@pscode{%
      /Dx { \pst@number\psxunit mul /D { Dy } def } def
      /Dy { \pst@number\psyunit mul Do /D { Dx } def } def
      /D { /D { Dx } def } def
      /Do {
        \@nameuse{beginqp@\psplotstyle}%
        /Do { \@nameuse{doqp@\psplotstyle}} def
      } def}%
    #1%			% this is \pst@readfile{#1} for fileplot
    \addto@pscode{ D }%
  \@nameuse{endqp@\psplotstyle}}
\def\psfileplot{\def\pst@par{}\pst@object{fileplot}}
\def\fileplot{\def\pst@par{}\pst@object{fileplot}}
\def\fileplot@i#1{%
  \pst@killglue%
  \begingroup%
    \use@par%
    \@pstfalse%
    \@nameuse{testqp@\psplotstyle}%
    \if@pst\dataplot@ii{\pst@readfile{#1}}\else\listplot@ii{\pst@altreadfile{#1}}\fi%
  \endgroup%
  \ignorespaces}
\def\pslistplot{\pst@object{listplot}}
\def\listplot{\pst@object{listplot}}
\def\listplot@i#1{\listplot@ii{\addto@pscode{#1}}}
\def\listplot@ii#1{%
  \@nameuse{beginplot@\psplotstyle}%
  \addto@pscode{/D {} def mark}%
  #1%
  \addto@pscode{
    \tx@PreparePoints
    \pst@number\psxunit
    \pst@number\psyunit
    \tx@ScalePoints
  }%
  \@nameuse{endplot@\psplotstyle}%
}
\define@boolkey[psset]{pst-plot}[Psk@]{xyValues}[true]{}
\define@boolkey[psset]{pst-plot}[Pst@]{ChangeOrder}[true]{}
\psset[pst-plot]{xyValues,ChangeOrder=false}
\pst@def{PreparePoints}<{%
  counttomark /m exch def
  /maxYValues \psk@plotNoMax\space def
  /YValuePos \psk@plotNo\space def
  /XValuePos \psk@plotNoX\space def
  /maxYvalue \ifx\empty\psk@plotYMax false \else true \fi def
  maxYvalue { /YMaxValue \psk@plotYMax\space def } if
  \ifPsk@xyValues\else % we have only y values
    /mm m def
    /M m 1 add def
    m { mm exch M 2 roll /M M 1 add def /mm mm 1 sub def } repeat
    /m m dup add def
  \fi
  \ifPst@ChangeOrder
    /m0 m def
    m maxYValues 1 add div 1 sub cvi {
      m0 maxYValues 1 add roll /m0 m0 maxYValues 1 add sub def
    } repeat
  \fi
  /n m maxYValues 1 add div cvi def
  XValuePos 1 gt {% multiple x values?            x y y xNo y
    n {
      maxYValues 1 add XValuePos neg roll    % y x y y xNo
      dup /XValue ED
      maxYValues 1 add XValuePos 1 sub  roll % y y xNo y x
      pop XValue                             % y y xNo y xNo
      maxYValues 1 add 1 roll                % xNo y y xNo y 
      m maxYValues 1 add  roll               % next values
    } repeat
  } if % no multiple data files
  maxYValues 1 gt {% multiple data files?           x y y yNo y 
    n {
      maxYValues YValuePos 1 sub neg roll % x yNo y y y ...
      maxYValues 1 sub { pop } repeat     % x yNo
      /m m maxYValues 1 sub sub def
      m 2 roll
    } repeat
  } if % no multiple data files
  /xMax -99999 def /yMax -99999 def
  /xP 0 def /yP 0 def
  m copy
  n {
    /y exch def /x exch def
    xMax x lt { /xMax x def } if
    yMax y lt {/yMax y def } if
    xP x gt { /xP x def } if
    yP y gt { /yP y def } if
  } repeat
  \psk@xStep\space 0 gt \psk@yStep\space 0 gt or (\psk@xStart) length 0 gt or
  (\psk@yStart) length 0 gt or (\psk@xEnd) length 0 gt or (\psk@yEnd) length 0 gt or {
    (\psk@xStart) length 0 gt {\psk@xStart\space }{ xP } ifelse /xStart exch def
    (\psk@yStart) length 0 gt {\psk@yStart\space }{ yP } ifelse /yStart exch def
    (\psk@xEnd) length 0 gt { \psk@xEnd\space }{ xMax } ifelse /xEnd exch def
    (\psk@yEnd) length 0 gt { \psk@yEnd\space }{ yMax } ifelse /yEnd exch def
    n {
      m -2 roll
      2 copy /yVal exch def /xVal exch def
      xVal xP ge
      yVal yP ge and
      xVal xEnd le and
      yVal yEnd le and
      xVal xStart ge and
      yVal yStart ge and {
        /xP xP \psk@xStep\space add def
        /yP yP \psk@yStep\space add def
      }{%
        pop pop
        /m m 2 sub def
      } ifelse
    } repeat
  }{%
    /ncount 1 def
    (\psk@nEnd) length 0 gt { \psk@nEnd\space }{ m } ifelse 
    /nEnd exch def
    n {
      m -2 roll
      \psk@nStep\space 1 gt { ncount \psk@nStart\space sub \psk@nStep\space mod 0 eq }{ true } ifelse
        ncount nEnd le and
        ncount \psk@nStart\space ge and not {
          pop pop
          /m m 2 sub def
        } if
        /ncount ncount 1 add def
      } repeat
  } ifelse
  maxYvalue { % delete values gt YMaxValue
    /m0 m def
    n {
      dup YMaxValue gt { pop pop /m m 2 sub def } if
      m -2 roll 
    } repeat
  } if
}>
\define@boolkey[psset]{pst-plot}[Pst@]{polarplot}[true]{}
\define@boolkey[psset]{pst-plot}[Pst@]{VarStep}[true]{}
\define@key[psset]{pst-plot}{PlotDerivative}{\def\psk@PlotDerivative{#1}}%
\define@key[psset]{pst-plot}{VarStepEpsilon}{\def\psk@VarStepEpsilon{#1}}%
\define@key[psset]{pst-plot}{method}{\def\psk@method{#1}}%     	   adams - rk4
\def\@rkiv{rk4}%		Runge-Kutta 4  method
\def\@varrkiv{varrkiv}%		Runge-Kutta 4 with an adaptive step method
\def\@adams{adams}%		Adams method
\def\@default{default}%		Adams method
\psset[pst-plot]{VarStep=false,PlotDerivative=none,VarStepEpsilon=default,polarplot=false,method={}}
\def\psplotinit#1{\xdef\psplot@init{#1 }}
\def\psplot@init{}
\def\psplot{\def\pst@par{}\pst@object{psplot}}
\def\psplot@i#1#2{\@ifnextchar[{\psplot@x{#1}{#2}}{\psplot@x{#1}{#2}[]}}
\def\psplot@x#1#2[#3]#4{%
  \pst@killglue
  \begingroup
    \use@par
    \@nameuse{beginplot@\psplotstyle}%
    \ifPst@polarplot
      \addto@pscode{
        \psplot@init
        #3 
        /x #1 def
        /x1 #2 def
        /dx x1 x sub \psk@plotpoints div def
        /F@pstplot \ifPst@algebraic (#4)
                    \ifx\psk@PlotDerivative\@none\else
                      \psk@PlotDerivative\space { (x) tx@Derive begin Derive end } repeat
                    \fi\space
                    tx@AlgToPs begin AlgToPs end cvx
                 \else { #4 } \fi  def
        \ifPst@VarStep
          /StillZero 0 def /LastNonZeroStep dx def
          /F2@pstplot tx@Derive begin (#4) (x) Derive (x) Derive end
                     \ifx\psk@PlotDerivative\@none\else
                       \psk@PlotDerivative\space { (x) tx@Derive begin Derive end } repeat
                     \fi\space
                    tx@AlgToPs begin AlgToPs end cvx def
          /epsilon12 \ifx\psk@VarStepEpsilon\@default tx@Derive begin F2@pstplot end dx 3 exp abs mul abs
                    \else\psk@VarStepEpsilon\space 12 mul \fi def
          /ComputeStep {
            dup 1e-4 lt
            { pop StillZero 2 ge { LastNonZeroStep 2 mul } { LastNonZeroStep } ifelse /StillZero StillZero 1 add def }
            { epsilon12 exch div 1 3 div exp /StillZero 0 def }
            ifelse } bind def
        \fi
        /xy {% Adapted from \parametricplot@i
          F@pstplot x \ifPst@algebraic RadtoDeg \fi PtoC
          \pst@number\psyunit mul exch
          \pst@number\psxunit mul exch
        } def}%
    \else% polarplot
    \addto@pscode{
      \psplot@init
      #3 
      /x #1 def
      /x1 #2 def
      /dx x1 x sub \psk@plotpoints div def
      /F@pstplot \ifPst@algebraic (#4)
                    \ifx\psk@PlotDerivative\@none\else
                      \psk@PlotDerivative\space { (x) tx@Derive begin Derive end } repeat
                    \fi\space
                    tx@AlgToPs begin AlgToPs end cvx
                 \else { #4 } \fi  def
      \ifPst@VarStep
         /StillZero 0 def /LastNonZeroStep dx def
         /F2@pstplot tx@Derive begin (#4) (x) Derive (x) Derive end
                     \ifx\psk@PlotDerivative\@none\else
                       \psk@PlotDerivative\space { (x) tx@Derive begin Derive end } repeat
                     \fi\space
                    tx@AlgToPs begin AlgToPs end cvx def
         /epsilon12 \ifx\psk@VarStepEpsilon\@default tx@Derive begin F2@pstplot end dx 3 exp abs mul abs
                    \else\psk@VarStepEpsilon\space 12 mul \fi def
         /ComputeStep {
           dup 1e-4 lt
           { pop StillZero 2 ge { LastNonZeroStep 2 mul } { LastNonZeroStep } ifelse /StillZero StillZero 1 add def }
           { epsilon12 exch div 1 3 div exp /StillZero 0 def }
           ifelse } bind def
      \fi
      /xy { x \pst@number\psxunit mul F@pstplot \pst@number\psyunit mul
      } def}%
    \fi
    \gdef\psplot@init{}%
    \ifx\pslinestyle\psls@@symbol
      \psplot@iii
    \else
      \@pstfalse
      \@nameuse{testqp@\psplotstyle}%
      \if@pst\psplot@ii\else\psplot@iii\fi
    \fi
  \endgroup
  \ignorespaces}
\def\psplot@ii{%
  \ifPst@VarStep%
    \addto@pscode{%
      mark xy \@nameuse{beginqp@\psplotstyle}
      { F2@pstplot abs ComputeStep
        x 2 copy add dup x1 gt {pop x1} if /x exch def F2@pstplot abs ComputeStep
        /x 3 -1 roll def 2 copy gt { exch } if pop
        /x x 3 -1 roll add dup x1 gt {pop x1} if def
        xy \@nameuse{doqp@\psplotstyle}
        x x1 eq { exit } if} loop}%
  \else
    \pst@killglue%
    \addto@pscode{
      /ps@Exit false def
      xy \@nameuse{beginqp@\psplotstyle}
      \ifx\psk@method\@varrkiv\else\psk@plotpoints 1 sub \fi {
        /x x dx add \ifx\psk@method\@varrkiv  dup x1 gt { pop x1 } if \fi def
        xy \@nameuse{doqp@\psplotstyle}
        \ifx\psk@method\@varrkiv  x x1 eq { exit } if \fi
      } 
      ps@Exit { exit } if
      \ifx\psk@method\@varrkiv loop \else repeat \fi
      ps@Exit not {
        /x x1 def
        xy \@nameuse{doqp@\psplotstyle}
      } if }%
  \fi%
  \@nameuse{endqp@\psplotstyle}}
\def\psplot@iii{%
  \ifPst@VarStep%
    \addto@pscode{
      /n 2 def
      mark
      { xy n 2 roll F2@pstplot abs
        ComputeStep x 2 copy add dup x1 gt {pop x1} if
        /x exch def F2@pstplot abs ComputeStep
        /x 3 -1 roll def 2 copy gt { exch } if pop
        /x x 3 -1 roll dup /LastNonZeroStep exch def add dup x1 gt {pop x1} if def /n n 2 add def
        x x1 eq { exit } if } loop
      xy 
      n 2 roll}%
  \else\pst@killglue%
    \addto@pscode{
      mark
      /n 2 def
      \ifx\psk@method\@varrkiv\else\psk@plotpoints\fi {
        xy
        n 2 roll
        /n n 2 add def
        /x x dx add \ifx\psk@method\@varrkiv  dup x1 gt { pop x1 } if \fi def
        \ifx\psk@method\@varrkiv  x x1 eq { exit } if \fi
      } \ifx\psk@method\@varrkiv loop\else repeat \fi \space
      /x x1 def
      xy 
      2 copy \tx@UserCoor 2 array astore /FinalState ED
      n 2 roll}%
  \fi%
  \@nameuse{endplot@\psplotstyle}}
\def\psparametricplot{\pst@object{parametricplot}}% 	hv 2008-11-22
\def\parametricplot{\pst@object{parametricplot}}
\def\parametricplot@i#1#2{\@ifnextchar[{\parametricplot@x{#1}{#2}}{\parametricplot@x{#1}{#2}[]}}
\def\parametricplot@x#1#2[#3]{\@ifnextchar[{\parametricplot@xi{#1}{#2}[#3]}{\parametricplot@xi{#1}{#2}[#3][]}}
\def\parametricplot@xi#1#2[#3][#4]#5{%
  \pst@killglue%
  \begingroup%
    \use@par%
    \@nameuse{beginplot@\psplotstyle}%
    \addto@pscode{%
      #3 %prefix PS code
      \psplot@init
      /t #1 def
      /t1 #2 def
      /dt t1 t sub \psk@plotpoints div def
      /F@pstplot \ifPst@algebraic (#5)
                    \ifx\psk@PlotDerivative\@none\else
                      \psk@PlotDerivative\space { (t) tx@Derive begin Derive end } repeat
                    \fi\space
                    tx@AlgToPs begin AlgToPs end cvx
                 \else { #5 } \fi  def
      \ifPst@VarStep
         /StillZero 0 def /LastNonZeroStep dt def
         /F2@pstplot tx@Derive begin (#5) (t) Derive (t) Derive end
                     \ifx\psk@PlotDerivative\@none\else
                       \psk@PlotDerivative\space { (t) tx@Derive begin Derive end } repeat
                     \fi\space
                    tx@AlgToPs begin AlgToPs end cvx def
         /epsilon12 \ifx\psk@VarStepEpsilon\@default
                       tx@Derive begin F2@pstplot end Pyth
                       dt 3 exp abs mul
                    \else\psk@VarStepEpsilon\space 12 mul \fi def
         /ComputeStep {
           dup 1e-4 lt
           { pop StillZero 2 ge { LastNonZeroStep 2 mul } { LastNonZeroStep } ifelse /StillZero StillZero 1 add def }
           { epsilon12 exch div 1 3 div exp /StillZero 0 def }
           ifelse } bind def
      \fi
      /xy {
        \ifPst@algebraic F@pstplot \else #5 \fi
        \pst@number\psyunit mul exch
        \pst@number\psxunit mul exch
      } def
      }%
    \gdef\psplot@init{}%
    \@pstfalse
    \@nameuse{testqp@\psplotstyle}%
    \if@pst\parametricplot@ii{#4}\else\parametricplot@iii{#4}\fi
  \endgroup%
  \ignorespaces}
\def\parametricplot@ii#1{% para is the post code
  \ifPst@VarStep%
    \addto@pscode{%
      mark xy \@nameuse{beginqp@\psplotstyle}
      { F2@pstplot Pyth ComputeStep
        t 2 copy add dup t1 gt {pop t1} if /t exch def F2@pstplot Pyth ComputeStep
        /t 3 -1 roll def 2 copy gt { exch } if pop
        /t t 3 -1 roll add dup t1 gt {pop t1} if def
        xy \@nameuse{doqp@\psplotstyle}
        t t1 eq { exit } if } loop}%
  \else\pst@killglue%
    \addto@pscode{%
      /ps@Exit false def
      xy \@nameuse{beginqp@\psplotstyle}
      \psk@plotpoints 1 sub {
        /t t dt add def
        xy \@nameuse{doqp@\psplotstyle}
        ps@Exit { exit } if 
      } repeat
      ps@Exit not {
        /t t1 def
        xy \@nameuse{doqp@\psplotstyle}
      } if 
    }%
  \fi%
  \addto@pscode{ #1 }%
  \@nameuse{endqp@\psplotstyle}}
\def\parametricplot@iii#1{%
  \ifPst@VarStep%
    \addto@pscode{%
      /n 2 def
      mark
      { xy n 2 roll F2@pstplot Pyth
        ComputeStep t 2 copy add dup t1 gt {pop t1} if
        /t exch def F2@pstplot Pyth ComputeStep
        /t 3 -1 roll def 2 copy gt { exch } if pop
        /t t 3 -1 roll dup /LastNonZeroStep exch def add dup t1 gt {pop t1} if def /n n 2 add def
        t t1 eq { exit } if } loop
      xy 
      2 copy \tx@UserCoor 2 array astore /FinalState ED
      n 2 roll}%
  \else\pst@killglue%
    \addto@pscode{
      mark
      /n 2 def
      \psk@plotpoints {
        xy
        n 2 roll
        /n n 2 add def
        /t t dt add def
      } repeat
      /t t1 def
      xy
      n 2 roll}%
  \fi%
  \addto@pscode{ #1 }%
  \@nameuse{endplot@\psplotstyle}}
\newdimen\psk@subticksize\psk@subticksize=\z@
\newdimen\pst@xticksizeA
\newdimen\pst@xticksizeB
\newdimen\pst@xticksizeC
\newdimen\pst@yticksizeA
\newdimen\pst@yticksizeB
\newdimen\pst@yticksizeC
\define@key[psset]{pst-plot}{ticks}[all]{\pst@expandafter\psset@@ticks{#1}\@nil\psk@ticks}
\def\psset@@ticks#1#2\@nil#3{%
  \ifx#1a\let#3\z@\else%				0=a)ll
    \ifx#1x\let#3\@ne\else%				1=x
      \ifx#1y\let#3\tw@\else%				2=y
        \ifx#1n\let#3\thr@@\else%			3=n)one
          \@pstrickserr{Bad argument: `#1#2'}\@ehpa
  \fi\fi\fi\fi}
\psset[pst-plot]{ticks=all}
\define@key[psset]{pst-plot}{labels}[all]{\pst@expandafter\psset@@ticks{#1}\@nil\psk@labels}% same as ticks
\psset[pst-plot]{labels=all}
\define@key[psset]{pst-plot}{Ox}[0]{\def\psk@Ox{#1}}
\psset[pst-plot]{Ox=0}
\define@key[psset]{pst-plot}{Dx}[1]{\def\psk@Dx{#1}}
\psset[pst-plot]{Dx=1}
\define@key[psset]{pst-plot}{dx}[0]{%
  \pssetxlength\pst@dimg{#1}%
  \edef\psk@dx{\number\pst@dimg}}
\psset[pst-plot]{dx=0}
\define@key[psset]{pst-plot}{Oy}[0]{\def\psk@Oy{#1}}
\psset[pst-plot]{Oy=0}
\define@key[psset]{pst-plot}{Dy}[1]{\def\psk@Dy{#1}}
\psset[pst-plot]{Dy=1}
\define@key[psset]{pst-plot}{dy}[0]{%
  \pssetylength\pst@dimg{#1}%
  \edef\psk@dy{\number\pst@dimg}}
\psset[pst-plot]{dy=0}
\define@boolkey[psset]{pst-plot}[]{showXorigin}[true]{}
\define@boolkey[psset]{pst-plot}[]{showYorigin}[true]{}
\define@boolkey[psset]{pst-plot}[]{showorigin}[true]{%
  \ifshoworigin
    \showXorigintrue\showYorigintrue
  \else
    \showXoriginfalse\showYoriginfalse
  \fi
}
\psset[pst-plot]{showorigin=true}
\long\def\psrotatebox#1#2{%
  \leavevmode
  \Grot@setangle{#1}%
  \setbox\z@\hbox{{#2}}%
  \Grot@x\z@
  \Grot@y\z@
  \Grot@box}
\def\Grot@setangle#1{\edef\Grot@angle{#1}}
\def\Grot@Px#1#2#3{%
        #1\Grot@cos#2%
        \advance#1-\Grot@sin#3}
\def\Grot@Py#1#2#3{%
        #1\Grot@sin#2%
        \advance#1\Grot@cos#3}
\def\Grot@box{%
  \begingroup
  \CalculateSin\Grot@angle
  \CalculateCos\Grot@angle
  \edef\Grot@sin{\UseSin\Grot@angle}%
  \edef\Grot@cos{\UseCos\Grot@angle}%
  \Grot@r\wd\z@  \advance\Grot@r-\Grot@x
  \Grot@l\z@     \advance\Grot@l-\Grot@x
  \Grot@h\ht\z@  \advance\Grot@h-\Grot@y
  \Grot@d-\dp\z@ \advance\Grot@d-\Grot@y
  \ifdim\Grot@sin\p@>\z@
    \ifdim\Grot@cos\p@>\z@
      \Grot@Py\Grot@height \Grot@r\Grot@h%B
      \Grot@Px\Grot@right  \Grot@r\Grot@d%E
      \Grot@Px\Grot@left   \Grot@l\Grot@h%C
      \Grot@Py\Grot@depth  \Grot@l\Grot@d%D
    \else
      \Grot@Py\Grot@height \Grot@r\Grot@d%E
      \Grot@Px\Grot@right  \Grot@l\Grot@d%D
      \Grot@Px\Grot@left   \Grot@r\Grot@h%B
      \Grot@Py\Grot@depth  \Grot@l\Grot@h%C
    \fi
  \else
    \ifdim\Grot@cos\p@<\z@
      \Grot@Py\Grot@height \Grot@l\Grot@d%D
      \Grot@Px\Grot@right  \Grot@l\Grot@h%C
      \Grot@Px\Grot@left   \Grot@r\Grot@d%E
      \Grot@Py\Grot@depth  \Grot@r\Grot@h%B
    \else
      \Grot@Py\Grot@height \Grot@l\Grot@h%C
      \Grot@Px\Grot@right  \Grot@r\Grot@h%B
      \Grot@Px\Grot@left   \Grot@l\Grot@d%D
      \Grot@Py\Grot@depth  \Grot@r\Grot@d%E
    \fi
  \fi
  \advance\Grot@height\Grot@y
  \advance\Grot@depth\Grot@y
  \Grot@Px\dimen@  \Grot@x\Grot@y
  \Grot@Py\dimen@ii \Grot@x\Grot@y
  \dimen@-\dimen@     \advance\dimen@-\Grot@left
  \dimen@ii-\dimen@ii \advance\dimen@ii\Grot@y
  \setbox\z@\hbox{%
    \kern\dimen@
    \raise\dimen@ii\hbox{\Grot@start\box\z@\Grot@end}}%
  \ht\z@\Grot@height
  \dp\z@-\Grot@depth
  \advance\Grot@right-\Grot@left\wd\z@\Grot@right
  \leavevmode\box\z@
  \endgroup}
\define@key[psset]{pst-plot}{labelFontSize}[{}]{\def\psk@xlabelFontSize{#1}\def\psk@ylabelFontSize{#1}}%
\define@key[psset]{pst-plot}{xlabelFontSize}[{}]{\def\psk@xlabelFontSize{#1}}%
\define@key[psset]{pst-plot}{ylabelFontSize}[{}]{\def\psk@ylabelFontSize{#1}}%
\define@boolkey[psset]{pst-plot}[Pst@]{mathLabel}[true]{%
  \ifPst@mathLabel%
    \Pst@xmathLabeltrue \Pst@ymathLabeltrue
    \def\pshlabel##1{$\psk@xlabelFontSize##1$}%
    \def\psvlabel##1{$\psk@ylabelFontSize##1$}%
  \else%
    \Pst@xmathLabelfalse \Pst@ymathLabelfalse
    \def\pshlabel##1{\psk@xlabelFontSize##1}%
    \def\psvlabel##1{\psk@ylabelFontSize##1}%
  \fi}
\define@boolkey[psset]{pst-plot}[Pst@]{xmathLabel}[true]{%
  \ifPst@xmathLabel%
    \def\pshlabel##1{$\psk@xlabelFontSize##1$}\else\def\pshlabel##1{\psk@xlabelFontSize##1}\fi}
\define@boolkey[psset]{pst-plot}[Pst@]{ymathLabel}[true]{%
  \ifPst@ymathLabel%
    \def\psvlabel##1{$\psk@ylabelFontSize##1$}\else\def\psvlabel##1{\psk@ylabelFontSize##1}\fi}
\psset[pst-plot]{labelFontSize={},mathLabel}
\define@boolkey[psset]{pst-plot}[Pst@]{xAxis}[true]{}
\define@boolkey[psset]{pst-plot}[Pst@]{yAxis}[true]{}
\define@boolkey[psset]{pst-plot}[Pst@]{xyAxes}[true]{%
    \@nameuse{Pst@xAxis#1}\@nameuse{Pst@yAxis#1}}%
\psset[pst-plot]{xAxis,yAxis}%
\define@key[psset]{pst-plot}{xlabelPos}[b]{\pst@expandafter\psset@@xlabelPos#1\@nil}
\define@key[psset]{pst-plot}{ylabelPos}[l]{\pst@expandafter\psset@@ylabelPos#1\@nil}
\def\psset@@xlabelPos#1#2\@nil{%
  \ifx#1t\relax
    \let\psk@xlabelPos\tw@%		2=top
    \pst@xticksizeC=\pst@xticksizeB
  \else
    \ifx#1a\relax
      \let\psk@xlabelPos\@ne %	 	1=axis
      \pst@xticksizeC=\z@
    \else
      \def\psk@xlabelPos{\z@}%		0=bottom	
      \pst@xticksizeC=\pst@xticksizeA
  \fi\fi
}
\def\psset@@ylabelPos#1#2\@nil{%
  \ifx#1r\relax
    \def\psk@ylabelPos{\tw@}%		2=right
    \pst@yticksizeC=\pst@yticksizeB
  \else
    \ifx#1a\relax
      \def\psk@ylabelPos{\@ne}% 	1=axis
      \pst@yticksizeC=\z@
    \else 
      \def\psk@ylabelPos{\z@}%		0=left	
      \pst@yticksizeC=\pst@yticksizeA
  \fi\fi
}
\psset[pst-plot]{xlabelPos=b, ylabelPos=l}%
\define@key[psset]{pst-plot}{xyDecimals}[{}]{\def\psk@xDecimals{#1}\def\psk@yDecimals{#1}}
\define@key[psset]{pst-plot}{xDecimals}[{}]{\def\psk@xDecimals{#1}}
\define@key[psset]{pst-plot}{yDecimals}[{}]{\def\psk@yDecimals{#1}}
\psset[pst-plot]{xyDecimals={}}%
\define@key[psset]{pst-plot}{xlogBase}[{}]{\def\psk@xlogBase{#1}}
\define@key[psset]{pst-plot}{ylogBase}[{}]{\def\psk@ylogBase{#1}}
\define@key[psset]{pst-plot}{xylogBase}[{}]{\def\psk@xlogBase{#1}\def\psk@ylogBase{#1}}%
\psset[pst-plot]{xylogBase={}}%
\define@key[psset]{pst-plot}{trigLabelBase}[0]{\pst@getint{#1}{\psk@xtrigLabelBase}\let\psk@ytrigLabelBase\psk@xtrigLabelBase}
\define@key[psset]{pst-plot}{xtrigLabelBase}[0]{%
  \pst@getint{#1}{\psk@xtrigLabelBase}\psset{xtrigLabels}}
\define@key[psset]{pst-plot}{ytrigLabelBase}[0]{%
  \pst@getint{#1}{\psk@ytrigLabelBase}\psset{ytrigLabels}}
\psset[pst-plot]{trigLabelBase=0}
\define@key[psset]{pst-plot}{fractionLabelBase}[0]{\pst@getint{#1}{\psk@xfractionLabelBase}\let\psk@yfractionLabelBase\psk@xfractionLabelBase}
\define@key[psset]{pst-plot}{xfractionLabelBase}[0]{%
  \pst@getint{#1}{\psk@xfractionLabelBase}\psset{xfractionLabels}}
\define@key[psset]{pst-plot}{yfractionLabelBase}[0]{%
  \pst@getint{#1}{\psk@yfractionLabelBase}\psset{yfractionLabels}}
\psset[pst-plot]{fractionLabelBase=0}
\def\setDefaulthLabels{%
  \ifPst@xmathLabel\def\pshlabel##1{$\psk@xlabelFontSize##1$}\else\def\pshlabel##1{\psk@xlabelFontSize##1}\fi
  \def\pst@@@hlabel##1{%
      \edef\@xyDecimals{\psk@xDecimals}%
      \ifnum\psk@labels<\tw@\relax% labels=all|x
        \ifx\psk@xlogBase\@empty
          \pshlabel{\psk@xlabelFontSize\expandafter\@LabelComma##1..\@nil\psk@xlabelFactor}%
        \else
          \ifPst@xmathLabel
            \pshlabel{\psk@xlabelFontSize\psk@xlogBase^{\expandafter\@stripDecimals##1..\@nil}}%
          \else
            \pshlabel{\psk@xlabelFontSize\psk@xlogBase\textsuperscript{\expandafter\@stripDecimals##1..\@nil}}%
          \fi
        \fi
      \fi
    }%
    \ifPst@xmathLabel\def\pshlabel##1{$\psk@xlabelFontSize##1$}\else\def\pshlabel##1{\psk@xlabelFontSize##1}\fi
}
\def\setTrighLabels{%
    \def\pst@@@hlabel##1{\pshlabel{##1}}%
    \def\pshlabel##1{%
      \ifnum\psk@xtrigLabelBase<2
        \def\de@nominator{\@ne}\else\def\de@nominator{\psk@xtrigLabelBase}\fi
      \def\pst@tempA{##1}% 
      \pst@abs{\pst@tempA}\pst@cntm 
      \pst@mod{\pst@cntm}{\de@nominator}\pst@cntp % cntb=##1 modulo trigLabelBase
      \ifnum\@ne>\pst@cntp                  % 1 > modulo -> then we have pi/x
        \pst@cnto=\pst@cntm \divide\pst@cnto by \de@nominator  
	\ifPst@xmathLabel
          $\psk@xlabelFontSize
  	  \ifnum\pst@tempA<0 -\fi
          \ifnum\pst@cnto=\@ne                % #1 = trigLabelBase
            \pi                 	      % print pi
          \else
            \ifnum\pst@cnto=\z@ 0\else
            \the\pst@cnto\pi 	              % print \pst@cnto/\de@nominator pi
          \fi\fi$%   
	\else%
          \psk@xlabelFontSize
  	  \ifnum\pst@tempA<0 -\fi
          \ifnum\pst@cnto=\@ne%                % #1 = trigLabelBase
            $\pi$%                             % print pi
          \else%
            \the\pst@cnto$\pi$%                % print \pst@cnto/\de@nominator pi
          \fi%
	\fi%
      \else%
	\ifPst@xmathLabel%
          $\psk@xlabelFontSize%
          \ifnum\pst@cntp=\@ne%                % < 1 pi?
            \if\pst@cntm=\@ne%
              \frac{\pi}{\de@nominator}%   % pi/x
            \else\ifnum\pst@tempA=-1 \frac{-\pi}{\de@nominator}%
              \else \ifnum\pst@tempA=1 \frac{\pi}{\de@nominator}%
                \else\frac{\pst@tempA\pi}{\de@nominator}% (x pi)/y
            \fi\fi\fi%
          \else%
            \ifnum\pst@tempA=1 \frac{\pi}{\de@nominator}%
            \else\ifnum\pst@tempA=\de@nominator \pi%
              \else\frac{\pst@tempA\pi}{\de@nominator}% 
          \fi\fi\fi$%
	\else%
          \psk@xlabelFontSize%
          \ifnum\pst@cntp=\@ne%                % < 1 pi?
            \if\pst@cntm=\@ne%
              $\frac{\pi}{\de@nominator}$%   % pi/x
            \else\ifnum\pst@tempA=-1 $\frac{-\pi}{\de@nominator}$%
              \else \ifnum\pst@tempA=1 $\frac{\pi}{\de@nominator}$%
                \else$\frac{\pst@tempA\pi}{\de@nominator}$% (x pi)/y
            \fi\fi\fi
          \else
            \ifnum\pst@tempA=1 $\frac{\pi}{\de@nominator}$%
            \else\ifnum\pst@tempA=\de@nominator $\pi$%
              \else$\frac{\pst@tempA\pi}{\de@nominator}$% 
          \fi\fi\fi
	\fi
      \fi
    }%
}
\def\setDefaultvLabels{%
  \ifPst@ymathLabel\def\psvlabel##1{$\psk@ylabelFontSize##1$}\else\def\psvlabel##1{\psk@ylabelFontSize##1}\fi
    \def\pst@@@vlabel##1{%
      \edef\@xyDecimals{\psk@yDecimals}%
      \ifodd\psk@labels % labelss=all||y (0,2)
      \else%
        \ifx\psk@ylogBase\@empty
          \psvlabel{\expandafter\@LabelComma##1..\@nil\psk@ylabelFactor}%
        \else%
          \ifPst@ymathLabel%
            \psvlabel{\psk@ylogBase^{\expandafter\@stripDecimals##1..\@nil }}%
	  \else
            \psvlabel{\psk@ylogBase\textsuperscript{\expandafter\@stripDecimals##1..\@nil }}%
          \fi%
        \fi%
      \fi%
    }%
}%
\def\setTrigvLabels{%
  \def\pst@@@vlabel##1{\psvlabel{##1}}%
    \def\psvlabel##1{%
      \ifnum\psk@ytrigLabelBase<2 \def\de@nominator{\@ne}\else\def\de@nominator{\psk@ytrigLabelBase}\fi
      \def\pst@tempA{##1} 
      \pst@abs{\pst@tempA}\pst@cntm 
      \pst@mod{\pst@cntm}{\de@nominator}\pst@cntp % cntb=##1 modulo trigLabelBase
      \ifnum\@ne>\pst@cntp                  % 1 > modulo -> then we have pi/x
        \pst@cnto=\pst@cntm \divide\pst@cnto by \de@nominator  
	\ifPst@ymathLabel%
          $\psk@ylabelFontSize
  	  \ifnum\pst@tempA<0 -\fi
          \ifnum\pst@cnto=\@ne                % #1 = trigLabelBase
            \pi                 	      % print pi
          \else
            \the\pst@cnto\pi 	              % print \pst@cnto/\de@nominator pi
          \fi$%   
	\else%
          \psk@ylabelFontSize%
  	  \ifnum\pst@tempA<0 -\fi
          \ifnum\pst@cnto=\@ne%                % #1 = trigLabelBase
            $\pi$%                             % print pi
          \else
            \the\pst@cnto$\pi$%                % print \pst@cnto/\de@nominator pi
          \fi
	\fi
      \else
	\ifPst@ymathLabel%
          $\psk@ylabelFontSize
          \ifnum\pst@cntp=\@ne%                % < 1 pi?    $
            \if\pst@cntm=\@ne%
              \frac{\pi}{\de@nominator}%   % pi/x
            \else\ifnum\pst@tempA=-1 \frac{-\pi}{\de@nominator}%
              \else \ifnum\pst@tempA=1 \frac{\pi}{\de@nominator}%
                \else\frac{\pst@tempA\pi}{\de@nominator}% (x pi)/y
            \fi\fi\fi%
          \else%
            \ifnum\pst@tempA=1 \frac{\pi}{\de@nominator}%
            \else\ifnum\pst@tempA=\de@nominator \pi%
              \else\frac{\pst@tempA\pi}{\de@nominator}% 
          \fi\fi\fi$%
	\else
          \psk@ylabelFontSize
          \ifnum\pst@cntp=\@ne%                % < 1 pi?
            \if\pst@cntm=\@ne
              $\frac{\pi}{\de@nominator}$%   % pi/x
            \else\ifnum\pst@tempA=-1 $\frac{-\pi}{\de@nominator}$%
              \else \ifnum\pst@tempA=1 $\frac{\pi}{\de@nominator}$%
                \else$\frac{\pst@tempA\pi}{\de@nominator}$% (x pi)/y
            \fi\fi\fi
          \else
            \ifnum\pst@tempA=1 $\frac{\pi}{\de@nominator}$%
            \else\ifnum\pst@tempA=\de@nominator $\pi$%
              \else$\frac{\pst@tempA\pi}{\de@nominator}$% 
          \fi\fi\fi
	\fi
      \fi
    }%
}%$
\def\setFractionvLabels{%
  \def\pst@@@vlabel##1{\psvlabel{##1}}%
  \def\psvlabel##1{%
      \ifnum\psk@yfractionLabelBase<2 \def\de@nominator{\@ne}\else\def\de@nominator{\psk@yfractionLabelBase}\fi
      \def\pst@tempA{##1}% 
      \pst@abs{\pst@tempA}\pst@cntm 
      \pst@mod{\pst@cntm}{\de@nominator}\pst@cntp % cntb=##1 modulo trigLabelBase
      \ifnum\@ne>\pst@cntp                  % 1 > modulo -> then we have pi/x
        \pst@cnto=\pst@cntm \divide\pst@cnto by \de@nominator  
	\ifPst@ymathLabel$\psk@ylabelFontSize\ifnum\pst@tempA<0 -\fi\the\pst@cnto\psk@ylabelFactor$%
	\else             \psk@ylabelFontSize\ifnum\pst@tempA<0 -\fi\the\pst@cnto\psk@ylabelFactor
	\fi
      \else
	\ifPst@ymathLabel
          $\psk@ylabelFontSize
          \ifnum\pst@cntp=\@ne                % < 1?    $
            \if\pst@cntm=\@ne
              \frac{1}{\de@nominator}\psk@ylabelFactor%   % 1/x
            \else\ifnum\pst@tempA=-1 \frac{-1}{\de@nominator}\psk@ylabelFactor%
              \else \ifnum\pst@tempA=1 \frac{1}{\de@nominator}\psk@ylabelFactor%
                \else\frac{\pst@tempA}{\de@nominator}\psk@ylabelFactor% x/y
            \fi\fi\fi
          \else
            \ifnum\pst@tempA=1 \frac{1}{\de@nominator}\psk@ylabelFactor%
            \else\ifnum\pst@tempA=\de@nominator 1\psk@xlabelFactor \else\frac{\pst@tempA}{\de@nominator}\psk@ylabelFactor%
          \fi\fi\fi$
	\else
          \psk@ylabelFontSize
          \ifnum\pst@cntp=\@ne%                % < 1?
            \if\pst@cntm=\@ne
              $\frac{1}{\de@nominator}\psk@ylabelFactor$%   % 1/x
            \else\ifnum\pst@tempA=-1 $\frac{-1}{\de@nominator}\psk@ylabelFactor$%
              \else \ifnum\pst@tempA=1 $\frac{1}{\de@nominator}\psk@ylabelFactor$%
                \else$\frac{\pst@tempA}{\de@nominator}\psk@ylabelFactor$% x/y
            \fi\fi\fi%
          \else%
            \ifnum\pst@tempA=1 $\frac{1}{\de@nominator}\psk@ylabelFactor$%
            \else\ifnum\pst@tempA=\de@nominator 1\psk@ylabelFactor
              \else$\frac{\pst@tempA}{\de@nominator}\psk@ylabelFactor$%   %$
          \fi\fi\fi
	\fi
      \fi
    }%
}%$
\def\setFractionhLabels{%
  \def\pst@@@hlabel##1{\pshlabel{##1}}%
  \def\pshlabel##1{%
      \ifnum\psk@xfractionLabelBase<2 \def\de@nominator{\@ne}\else\def\de@nominator{\psk@xfractionLabelBase}\fi
      \def\pst@tempA{##1}% 
      \pst@abs{\pst@tempA}\pst@cntm 
      \pst@mod{\pst@cntm}{\de@nominator}\pst@cntp% cntb=##1 modulo trigLabelBase
      \ifnum\@ne>\pst@cntp                  % 1 > modulo -> then we have 1/x
        \pst@cnto=\pst@cntm \divide\pst@cnto by \de@nominator  
	\ifPst@xmathLabel$\psk@xlabelFontSize\ifnum\pst@tempA<0 -\fi\the\pst@cnto\psk@xlabelFactor$%
	\else             \psk@xlabelFontSize\ifnum\pst@tempA<0 -\fi\the\pst@cnto\psk@xlabelFactor
	\fi
      \else
	\ifPst@xmathLabel
          $\psk@xlabelFontSize% $
          \ifnum\pst@cntp=\@ne
            \if\pst@cntm=\@ne \frac{1}{\de@nominator}\psk@xlabelFactor%   % 1/x
            \else\ifnum\pst@tempA=-1 \frac{-1}{\de@nominator}\psk@xlabelFactor%
              \else\ifnum\pst@tempA=1 \frac{1}{\de@nominator}\psk@xlabelFactor%
                \else\frac{\pst@tempA}{\de@nominator}\psk@xlabelFactor% x/y
            \fi\fi\fi%
          \else%
            \ifnum\pst@tempA=1 \frac{1}{\de@nominator}\psk@xlabelFactor%
            \else\ifnum\pst@tempA=\de@nominator 1\psk@xlabelFactor\else\frac{\pst@tempA}{\de@nominator}\psk@xlabelFactor%
          \fi\fi\fi$
	\else
          \psk@xlabelFontSize
          \ifnum\pst@cntp=\@ne
            \if\pst@cntm=\@ne $\frac{1}{\de@nominator}\psk@xlabelFactor$%            % 1/x
            \else\ifnum\pst@tempA=-1 $\frac{-1}{\de@nominator}\psk@xlabelFactor$%
              \else \ifnum\pst@tempA=1 $\frac{1}{\de@nominator}\psk@xlabelFactor$%
                \else$\frac{\pst@tempA}{\de@nominator}\psk@xlabelFactor$% x/y
            \fi\fi\fi
          \else
            \ifnum\pst@tempA=1 $\frac{1}{\de@nominator}\psk@xlabelFactor$%
            \else\ifnum\pst@tempA=\de@nominator 1\psk@xlabelFactor%
              \else$\frac{\pst@tempA}{\de@nominator}\psk@xlabelFactor$%   %$
          \fi\fi\fi
	\fi
      \fi
    }%
}%$
\define@boolkey[psset]{pst-plot}[Pst@]{xtrigLabels}[true]{%
  \ifPst@xtrigLabels \setTrighLabels \else \setDefaulthLabels\fi}
\define@boolkey[psset]{pst-plot}[Pst@]{ytrigLabels}[true]{%
  \ifPst@ytrigLabels \setTrigvLabels \else \setDefaultvLabels \fi}
\define@boolkey[psset]{pst-plot}[Pst@]{trigLabels}[true]{%
  \ifPst@trigLabels\psset[pst-plot]{xtrigLabels,ytrigLabels=false}
  \else            \psset[pst-plot]{xtrigLabels=false,ytrigLabels=false}%
  \fi}
\psset[pst-plot]{trigLabels=false}
\define@boolkey[psset]{pst-plot}[Pst@]{xfractionLabels}[true]{%
  \ifPst@xfractionLabels \setFractionhLabels \else \setDefaulthLabels \fi}
\define@boolkey[psset]{pst-plot}[Pst@]{yfractionLabels}[true]{%
  \ifPst@yfractionLabels \setFractionvLabels \else \setDefaultvLabels \fi}
\define@boolkey[psset]{pst-plot}[Pst@]{fractionLabels}[true]{%
  \ifPst@fractionLabels \setFractionhLabels\setFractionvLabels\Pst@xfractionLabelstrue\Pst@yfractionLabelstrue\fi}
\psset[pst-plot]{fractionLabels=false}
\def\psk@logLines{3}
\define@key[psset]{pst-plot}{logLines}[none]{\pst@expandafter\psset@@logLines#1\@nil\psk@logLines}
\def\psset@@logLines#1#2\@nil#3{%
  \ifx#1a\relax
    \let#3\z@
    \Pst@maxxTickstrue\Pst@maxyTickstrue
    \set@xticksize{0 4pt}\set@yticksize{0 4pt}%
    \def\psk@xsubticksize{1}\def\psk@ysubticksize{1}%
  \else
    \ifx#1x\relax
      \let#3\@ne
      \Pst@maxxTickstrue\Pst@maxyTicksfalse
      \set@xticksize{0 4pt}\def\psk@xsubticksize{1}%
    \else
      \ifx#1y\relax
        \let#3\tw@
	\Pst@maxyTickstrue\Pst@maxxTicksfalse
	\set@yticksize{0 4pt}\def\psk@ysubticksize{1}%
      \else
        \ifx#1n\let#3\thr@@\else
          \@pstrickserr{Bad argument: `#1#2'}\@ehpa
  \fi\fi\fi\fi}
\psset[pst-plot]{logLines=none}%
\define@key[psset]{pst-plot}{ylabelFactor}[\relax]{\def\psk@ylabelFactor{#1}}
\define@key[psset]{pst-plot}{xlabelFactor}[\relax]{\def\psk@xlabelFactor{#1}}
\define@boolkey[psset]{pst-plot}[Pst@]{showOriginTick}[true]{}%
\psset[pst-plot]{xlabelFactor=\relax,ylabelFactor=\relax,showOriginTick}%

\def\psxTick{\pst@object{psxTick}}% idea by Martin Chicoine
\def\psxTick@i{\@ifnextchar({\psxTick@ii{0}}\psxTick@ii}
\def\psxTick@ii#1(#2)#3{{%
  \pst@killglue
  \addbefore@par{arrows=-,linewidth=\psk@xtickwidth\pslinewidth}
  \ifPst@xtrigLabels\addto@par{xtrigLabels=false}\fi 
  \use@par
  \edef\temp@coor{(!#2 \pst@number\pst@xticksizeB \pst@number\psyunit div)(!#2 \pst@number\pst@xticksizeA \pst@number\psyunit div)}%
  \expandafter\psline\temp@coor
  \rput[t]{#1}(! \psk@origin 
                 #2 \pst@number\psxlabelsep \pst@number\pst@xticksizeB add
                 \pst@number\psyunit div neg ){\pshlabel{#3\vphantom{1}}}%
  }\ignorespaces}
\def\psyTick{\pst@object{psyTick}}% idea by Martin Chicoine
\def\psyTick@i{\@ifnextchar({\psyTick@ii{0}}\psyTick@ii}
\def\psyTick@ii#1(#2)#3{{%
  \pst@killglue
  \addbefore@par{arrows=-,linewidth=\psk@ytickwidth\pslinewidth}
  \ifPst@ytrigLabels \setDefaultvLabels \fi
  \use@par
  \edef\temp@coor{(!\pst@number\pst@yticksizeB \pst@number\psxunit div #2)(!\pst@number\pst@yticksizeA \pst@number\psxunit div #2)}%
  \expandafter\psline\temp@coor
    \rput[r]{#1}(!\psk@origin
                  \pst@number\pst@yticksizeB \pst@number\psylabelsep add
                  \pst@number\psxunit div neg #2){\psvlabel{#3}}}\ignorespaces}
\define@boolkey[psset]{pst-plot}[Pst@]{markPoint}[true]{}%
\psset[pst-plot]{markPoint}
\def\psCoordinates{\pst@object{psCoordinates}}
\def\psCoordinates@i(#1){%
  \pst@killglue%
  \begingroup
  \addbefore@par{showpoints=false,markPoint}
  \use@par
  \psline(#1|0,0)(#1)% single lines to allow arrows
  \psline(#1)(0,0|#1)%
  \ifPst@markPoint\psdot(#1)\fi%
  \endgroup
  \ignorespaces
}
\def\stripDecimals#1{\expandafter\@stripDecimals#1..\@nil}
\def\@stripDecimals#1.#2.#3\@nil{%
  \def\pst@dummy{#1}%
  \ifx\pst@dummy\@empty\the\@zero\else#1\fi% the integer part
}
\newcount\@digitcounter\@digitcounter=0\relax
\def\@inc@digitcounter{\global\advance\@digitcounter by 1\relax}
\def\@get@digitcounter{\the\@digitcounter\relax}
\def\@Reset@digitcounter{\global\@digitcounter=0\relax}
\def\@zeroFill{%
  \ifnum \@xyDecimals>\@get@digitcounter
    \bgroup
      0\@inc@digitcounter\@zeroFill
    \egroup
  \fi
}
\def\@process@digits#1#2;{%
  \ifx *#1\@zeroFill\else#1\@inc@digitcounter 
  \ifnum\@xyDecimals>\@get@digitcounter\expandafter\@process@digits#2;\fi\fi%
}
\def\@writeDecimals#1{%
  \ifx\@xyDecimals\@empty% take value as is
    \def\@tempa{#1}% write only if not empty
    \ifx\@tempa\@empty% write nothing
    \else\ifmmode\expandafter\mathord\expandafter{\psk@decimalSeparator}\else\psk@decimalSeparator\fi#1\fi%
  \else% write only \xy@decimals
    \ifnum\@xyDecimals>\@zero
      \ifmmode\expandafter\mathord\expandafter{\psk@decimalSeparator}\else\psk@decimalSeparator\fi%
        \@Reset@digitcounter
        \expandafter\@process@digits#1*;%
      \fi%
  \fi%
}
\def\@LabelComma#1.#2.#3\@nil{%
  \def\pst@tempA{#1}%
  \ifx\pst@tempA\@empty\the\@zero\else#1\fi% the integer part
  \def\pst@tempA{#2}%
  \ifx\pst@tempA\@empty\@writeDecimals{}\else\@writeDecimals{#2}\fi
}
\def\set@xticksize#1{%
  \pst@expandafter\pst@getydimdim{#1} {} {}\@nil% y-unit!! 
  \ifdim\pst@dimm>\pst@dimn% 		%	first > second value
    \pst@xticksizeA=\the\pst@dimn%
    \pst@xticksizeB=\the\pst@dimm%
  \else%
    \pst@xticksizeA=\the\pst@dimm%
    \pst@xticksizeB=\the\pst@dimn%	first > second value
  \fi%
  \edef\psk@xticksize{\pst@number\pst@xticksizeA \pst@number\pst@xticksizeB}%
  \ifnum\psk@xlabelPos<\z@\relax% top
    \pst@xticksizeC=\pst@dimn
  \else
    \pst@xticksizeC=\pst@dimm%	bottom	
  \fi
}
\def\set@yticksize#1{%
  \pst@expandafter\pst@getxdimdim{#1} {} {}\@nil% x-unit!
  \ifdim\pst@dimm>\pst@dimn\relax%   		%	first > second value
    \pst@yticksizeA=\the\pst@dimn%
    \pst@yticksizeB=\the\pst@dimm%
  \else%
    \pst@yticksizeA=\the\pst@dimm%
    \pst@yticksizeB=\the\pst@dimn%	first > second value
  \fi%
  \edef\psk@yticksize{\pst@number\pst@yticksizeA \pst@number\pst@yticksizeB}%
  \ifnum\psk@ylabelPos<\z@	% right	
    \pst@yticksizeC=\pst@dimn%
  \else%
      \pst@yticksizeC=\pst@dimo%  left
  \fi%
}
\newif\ifPst@maxxTicks
\newif\ifPst@maxyTicks
\define@key[psset]{pst-plot}{ticksize}[-4pt 4pt]{%
  \def\pst@tempA{max}%
  \def\pst@tempB{#1}%
  \ifx\pst@tempA\pst@tempB%
    \Pst@maxxTickstrue\Pst@maxyTickstrue%
    \set@xticksize{0 4pt}\set@yticksize{0 4pt}%
  \else%
    \Pst@maxxTicksfalse\Pst@maxyTicksfalse%
    \set@xticksize{#1}\set@yticksize{#1}%
  \fi}
\define@key[psset]{pst-plot}{xticksize}{%
  \def\pst@tempA{max}%
  \def\pst@tempB{#1}%
  \ifx\pst@tempA\pst@tempB
    \Pst@maxxTickstrue\set@xticksize{0 4pt}%
  \else\set@xticksize{#1}\Pst@maxxTicksfalse\fi}
\define@key[psset]{pst-plot}{yticksize}{%
  \def\pst@tempA{max}%
  \def\pst@tempB{#1}%
  \ifx\pst@tempA\pst@tempB%
    \Pst@maxyTickstrue\set@yticksize{0 4pt}%
  \else\set@yticksize{#1}\Pst@maxyTicksfalse\fi}%
\psset[pst-plot]{ticksize=-4pt 4pt}
\define@key[psset]{pst-plot}{tickstyle}[full]{\pst@expandafter\psset@@tickstyle{#1}\@nil}
\def\psset@@tickstyle#1#2\@nil{%
  \ifx#1f\let\psk@tickstyle\z@\else			% 0=f)ull
    \ifx#1t\let\psk@tickstyle\@ne			% 1=t)op
      \edef\psk@xticksize{0 \pst@number\pst@xticksizeB}%
      \edef\psk@yticksize{0 \pst@number\pst@yticksizeB}%
    \else\ifx#1b\let\psk@tickstyle\m@ne			% -1=b)ottom
      \edef\psk@xticksize{\pst@number\pst@xticksizeA 0}%
      \edef\psk@yticksize{\pst@number\pst@yticksizeA 0}%
      \else\ifx#1i\let\psk@tickstyle\tw@%		% 2=i)nner (for frame)
        \else\@pstrickserr{Bad tick style: `#1#2'}\@ehpa
  \fi\fi\fi\fi}
\psset[pst-plot]{tickstyle=full}
\define@key[psset]{pst-plot}{subticks}[1]{\def\psk@xsubticks{#1}\def\psk@ysubticks{#1}}
\define@key[psset]{pst-plot}{xsubticks}[1]{\def\psk@xsubticks{#1}}
\define@key[psset]{pst-plot}{ysubticks}[1]{\def\psk@ysubticks{#1}}
\define@key[psset]{pst-plot}{subticksize}[0.75]{\def\psk@xsubticksize{#1}\def\psk@ysubticksize{#1}}
\define@key[psset]{pst-plot}{xsubticksize}[0.75]{\def\psk@xsubticksize{#1}}
\define@key[psset]{pst-plot}{ysubticksize}[0.75]{\def\psk@ysubticksize{#1}}
\define@key[psset]{pst-plot}{tickwidth}[0.5\pslinewidth]{%
  \pst@getlength{#1}\psk@xtickwidth%
  \pst@getlength{#1}\psk@ytickwidth}
\define@key[psset]{pst-plot}{xtickwidth}[0.5\pslinewidth]{\pst@getlength{#1}\psk@xtickwidth}
\define@key[psset]{pst-plot}{ytickwidth}[0.5\pslinewidth]{\pst@getlength{#1}\psk@ytickwidth}
\define@key[psset]{pst-plot}{subtickwidth}[0.25\pslinewidth]{%
  \pst@getlength{#1}\psk@xsubtickwidth%
  \pst@getlength{#1}\psk@ysubtickwidth}
\define@key[psset]{pst-plot}{xsubtickwidth}[0.25\pslinewidth]{\pst@getlength{#1}\psk@xsubtickwidth}
\define@key[psset]{pst-plot}{ysubtickwidth}[0.25\pslinewidth]{\pst@getlength{#1}\psk@ysubtickwidth}
\define@key[psset]{pst-plot}{labelOffset}[0pt]{%
  \pst@getlength{#1}\psk@xlabelOffset%
  \pst@getlength{#1}\psk@ylabelOffset}
\define@key[psset]{pst-plot}{xlabelOffset}[0pt]{\pst@getlength{#1}\psk@xlabelOffset}
\define@key[psset]{pst-plot}{ylabelOffset}[0pt]{\pst@getlength{#1}\psk@ylabelOffset}
\define@key[psset]{pst-plot}{frameOffset}[0pt]{\pst@getlength{#1}\psk@frameOffset}
\define@key[psset]{pst-plot}{tickcolor}[black]{%
    \pst@getcolor{#1}\psk@xtickcolor%
    \pst@getcolor{#1}\psk@ytickcolor}
\define@key[psset]{pst-plot}{xtickcolor}[black]{\pst@getcolor{#1}\psk@xtickcolor}
\define@key[psset]{pst-plot}{ytickcolor}[black]{\pst@getcolor{#1}\psk@ytickcolor}
\define@key[psset]{pst-plot}{subtickcolor}[gray]{%
  \pst@getcolor{#1}\psk@xsubtickcolor%
  \pst@getcolor{#1}\psk@ysubtickcolor}
\define@key[psset]{pst-plot}{xsubtickcolor}[gray]{\pst@getcolor{#1}\psk@xsubtickcolor}
\define@key[psset]{pst-plot}{ysubtickcolor}[gray]{\pst@getcolor{#1}\psk@ysubtickcolor}
\define@key[psset]{pst-plot}{xticklinestyle}[solid]{%
  \@ifundefined{psls@#1}%
    {\@pstrickserr{Line style `#1' not defined}\@eha}%
    {\def\psxticklinestyle{#1}}}
\define@key[psset]{pst-plot}{xsubticklinestyle}[solid]{%
  \@ifundefined{psls@#1}%
    {\@pstrickserr{Line style `#1' not defined}\@eha}%
    {\def\psxsubticklinestyle{#1}}}
\define@key[psset]{pst-plot}{yticklinestyle}[solid]{%
  \@ifundefined{psls@#1}%
    {\@pstrickserr{Line style `#1' not defined}\@eha}%
    {\def\psyticklinestyle{#1}}}
\define@key[psset]{pst-plot}{ysubticklinestyle}[solid]{%
  \@ifundefined{psls@#1}%
    {\@pstrickserr{Line style `#1' not defined}\@eha}%
    {\def\psysubticklinestyle{#1}}}
\define@key[psset]{pst-plot}{ticklinestyle}[solid]{%
  \@ifundefined{psls@#1}%
    {\@pstrickserr{Line style `#1' not defined}\@eha}%
    {\def\psxticklinestyle{#1}\def\psyticklinestyle{#1}}}
\define@key[psset]{pst-plot}{subticklinestyle}[solid]{%
  \@ifundefined{psls@#1}%
    {\@pstrickserr{Line style `#1' not defined}\@eha}%
    {\def\psxsubticklinestyle{#1}\def\psysubticklinestyle{#1}}}
\psset[pst-plot]{subticksize=0.75,subticks=1,tickcolor=black,ticklinestyle=solid,
  subticklinestyle=solid,subtickcolor=gray,tickwidth=0.5\pslinewidth,
  subtickwidth=0.25\pslinewidth,labelOffset=0pt,frameOffset=0pt}
\define@key[psset]{pst-plot}{nStep}[1]{\def\psk@nStep{#1}}
\define@key[psset]{pst-plot}{nStart}[0]{\def\psk@nStart{#1}}
\define@key[psset]{pst-plot}{nEnd}[{}]{\def\psk@nEnd{#1}}
\define@key[psset]{pst-plot}{xStep}[0]{\def\psk@xStep{#1}}
\define@key[psset]{pst-plot}{yStep}[0]{\def\psk@yStep{#1}}
\define@key[psset]{pst-plot}{xStart}[{}]{\def\psk@xStart{#1}}
\define@key[psset]{pst-plot}{xEnd}[{}]{\def\psk@xEnd{#1}}
\define@key[psset]{pst-plot}{yStart}[{}]{\def\psk@yStart{#1}}
\define@key[psset]{pst-plot}{yEnd}[{}]{\def\psk@yEnd{#1}}
\define@key[psset]{pst-plot}{plotNoX}[1]{\def\psk@plotNoX{#1}}
\define@key[psset]{pst-plot}{plotNo}[1]{\def\psk@plotNo{#1}}
\define@key[psset]{pst-plot}{plotNoMax}[1]{\def\psk@plotNoMax{#1}}
\define@key[psset]{pst-plot}{plotYMax}[1]{\def\psk@plotYMax{#1}}
\psset[pst-plot]{nStep=1, nStart=0, nEnd={},%
  xStep=0, yStep=0, xStart={}, xEnd={},  yStart={}, yEnd={}, 
  plotNo=1,plotNoMax=1,plotNoX=1,plotYMax={}}%
\def\pstScalePoints(#1,#2)#3#4{%
  \def\pstXScale{#1 }%
  \def\pstYScale{#2 }%
  \def\pstXPSScale{#3 }%
  \def\pstYPSScale{#4 }%
  \pst@def{ScalePoints}<%
    /yVal ED /xVal ED
    /yPSOp { #4 yVal mul #2 mul } def
    /xPSOp { #3 xVal mul #1 mul } def
    counttomark dup dup cvi eq not { exch pop } if
    /m exch def /n m 2 div cvi def
    n {
      \ifPst@polarplot exch cvi 360 mod PtoC \fi  % x cvi 360 mod PtoC
      yPSOp m 1 roll xPSOp m 1 roll 
      /m m 2 sub
      def } repeat>%
}
\pstScalePoints(1,1){}{}% the default -> no special operators
\def\psxs@none{\let\psk@arrowA\@empty\let\psk@arrowB\@empty\psxs@axes}
\def\psxs@axes{{%
  \ifPst@xAxis\psxs@@axes\pst@dima\pst@dimb\pst@dimc\pst@dimd{}{x}\fi%
  \ifPst@yAxis\psxs@@axes\pst@dima\pst@dimb\pst@dimc\pst@dimd{exch}{y}\fi%
}}
\newif\ifSpecialLabelsDone
\def\psaxes{\pst@object{psaxes}}
\def\psaxes@i{%
  \let\pst@par@save\pst@par
  \pst@getarrows\psaxes@ii}
\def\psaxes@ii(#1){\@ifnextchar({\psaxes@iii(#1)}{\psaxes@iv(0,0)(0,0)(#1)}}
\def\psaxes@iii(#1)(#2){\@ifnextchar({\psaxes@iv(#1)(#2)}{\psaxes@iv(#1)(#1)(#2)}}
\def\psaxes@iv(#1)(#2)(#3){\@ifnextchar[{\psaxes@v(#1)(#2)(#3)}{\psaxes@vii(#1)(#2)(#3)}}%
\def\psaxes@v(#1)(#2)(#3)[#4]{\@ifnextchar[{\psaxes@vi(#1)(#2)(#3)[#4]}{\psaxes@vi(#1)(#2)(#3)[#4][]}}%
\def\psaxes@vi(#1)(#2)(#3)[#4,#5][#6,#7]{%
  \psaxes@vii(#1)(#2)(#3)%
  \let\pst@par\pst@par@save
  \begingroup
  \SpecialCoor
  \use@par
  \ifshowgrid\psgrid[style=gridstyleA]\fi
  \uput{\psxlabelsep}[#5](#3|#1){#4}\uput{\psylabelsep}[#7](#1|#3){#6}%
  \endgroup
  \ignorespaces
}
\def\psaxes@vii(#1,#2)(#3,#4)(#5,#6){%
  \pst@killglue
  \begingroup
  \ifdim\pst@dimc<\z@\relax 
    \ifdim\pst@dimd<\z@\relax % axes show to left and down
      \addbefore@par{xlabelPos=t,ylabelPos=r}%
  \fi\fi
  \use@par%	now the same with an optional unit=... in par
  \pssetxlength\pst@dimc{#5}% ur-x
  \pssetylength\pst@dimd{#6}% ur-y
    \pssetxlength\pst@dimg{#1}% o-x
    \pssetylength\pst@dimh{#2}% o-y
    \pssetxlength\pst@dima{#3}% ll-x
    \pssetylength\pst@dimb{#4}% ll-y
    \pst@dima=\dimexpr\pst@dima-\pst@dimg\relax
    \pst@dimb=\dimexpr\pst@dimb-\pst@dimh\relax
    \pst@dimc=\dimexpr\pst@dimc-\pst@dimg\relax
    \pst@dimd=\dimexpr\pst@dimd-\pst@dimh\relax
   \setbox\pst@hbox=\hbox\bgroup
    \ifshowgrid\psgrid[style=gridstyleA]\fi
    \@nameuse{psxs@\psk@axesstyle}%  \psxs@axes or \psxs@frame or \psxs@polar
    \ifPst@xAxis
      \SpecialLabelsDonefalse
      \begingroup
      \ifnum\psk@dx=\z@
        \pst@dimg=\psk@Dx\psxunit
        \ifdim\pst@dimg<\p@ 
          \pst@cnta=\psk@Dx
          \edef\psk@Dx{\the\numexpr-1*\pst@cnta}%
        \fi% v.1.21
        \edef\psk@dx{\number\pst@dimg}%
      \fi
      \pst@hlabels{\pst@dimc}{\psk@arrowB}{#3}{#5}% Right
      \ifPst@yAxis\showoriginfalse\fi
      \pst@hlabels{\pst@dima}{\psk@arrowA}{#3}{#5}% Left
      \endgroup
    \fi
    \ifPst@yAxis
      \SpecialLabelsDonefalse
      \begingroup
      \ifdim\pst@dima=\z@ \else\ifPst@xtrigLabels\showoriginfalse\fi\fi
      \ifnum\psk@dy=\z@
        \pst@dimg=\psk@Dy\psyunit
        \ifdim\pst@dimg<\p@ 
          \pst@cnta=\psk@Dy
          \edef\psk@Dy{\the\numexpr-1*\pst@cnta}%
        \fi% v.1.21
        \edef\psk@dy{\number\pst@dimg}%
      \fi
      \pst@vlabels{\pst@dimb}{\psk@arrowA}{#4}{#6}%
      \ifPst@xAxis\ifdim\pst@dima<\z@ \showoriginfalse\fi\fi % no 0 when x- axis is crossing
      \pst@vlabels{\pst@dimd}{\psk@arrowB}{#4}{#6}%
      \endgroup
    \fi
  \egroup%
  \pssetxlength\pst@dimg{#1}%
  \pssetylength\pst@dimh{#2}%
  \leavevmode
  \psput@cartesian\pst@hbox
  \endgroup
  \ignorespaces
}
\newif\ifis@yAxis%
\def\psxs@@axes#1#2#3#4#5#6{% llx,lly,urx,ury,exch,x|y,arrowA,arrowB
  \pst@killglue
  \begin@SpecialObj
    \ifx#6x\relax%				% x-axis?
      \is@yAxisfalse
      \ifnum\psk@dx=\z@
        \pst@dimg=\psk@Dx\psxunit
        \def\psk@dx{\number\pst@dimg}%
      \fi
    \else
      \is@yAxistrue
      \ifnum\psk@dy=\z@
        \pst@dimg=\psk@Dy\psyunit
        \def\psk@dy{\number\pst@dimg}%
      \fi
    \fi
    \let\pst@linetype\pst@arrowtype
    \def\pst@axes{axes}%
    \pst@addarrowdef
    \addto@pscode{
      /showOrigin \ifPst@showOriginTick true \else false \fi def 	% ticks for 0/0 ?
      \ifis@yAxis 0 \pst@number#4 \else \pst@number#3 0 \fi
      \ifis@yAxis 0 \pst@number#2 \else \pst@number#1 0 \fi
      ArrowA
      CP 4 2 roll
      ArrowB 
      2 copy
      /yEnd exch def /xEnd exch def
      \ifx\psk@axesstyle\@none   
        pop pop % axesstyle = none (only ticks) or frame (already drawn)
      \else
        gsave                              		% save current state
        L                                  		% the line with arrows 
        \@nameuse{psls@\pslinestyle}                 	% linestyle for the axes
        stroke                                       	% draw the main line
        grestore
      \fi
      /yStart exch def
      /xStart exch def
      \number\psk@ticks\space dup 2 mod 0 eq \ifis@yAxis true \else false \fi and 
      exch 2 lt \ifis@yAxis false \else true \fi and or {
      /viceversa 
        \ifis@yAxis\pst@number#2 \pst@number#4 \else\pst@number#1 \pst@number#3 \fi
         gt { true }{ false } ifelse def           % other way round
      /epsilon 0.01 def                            % rounding errors
      /minTickline \ifis@yAxis \pst@number#1 \else \pst@number#2 \fi def
      /maxTickline \ifis@yAxis \pst@number#3 \else \pst@number#4 \fi def
      /dT \ifis@yAxis \psk@dy \else \psk@dx \fi\space abs  % added abs 2006-07-07
        65536 div viceversa { neg } if def                 % div to get pt instead of sp
      /DT \ifis@yAxis \psk@Dy \else \psk@Dx \fi\space abs viceversa { neg } if def  
      /subTNo \ifis@yAxis\psk@ysubticks\else\psk@xsubticks\fi \space def
      subTNo 0 gt { /dsubT dT subTNo div def}{ /dsubT 0 def } ifelse  % deltaSubTick
      \ifis@yAxis \psk@yticksize \else \psk@xticksize \fi
      /tickend exch def /tickstart exch def
      /Twidth \ifis@yAxis \psk@ytickwidth \else \psk@xtickwidth \fi\space def
      /subTwidth \ifis@yAxis \psk@ysubtickwidth \else \psk@xsubtickwidth \fi\space def
      /STsize \ifis@yAxis \psk@ysubticksize \else \psk@xsubticksize \fi\space def
      /TColor {
        \ifis@yAxis\pst@usecolor\psk@ytickcolor
        \else\pst@usecolor\psk@xtickcolor\fi\space } def
      /subTColor {
        \ifis@yAxis\pst@usecolor\psk@ysubtickcolor
        \else\pst@usecolor\psk@xsubtickcolor\fi\space } def
      /MinValue { \ifis@yAxis yStart \else xStart \fi
        \ifx\psk@arrowA\@empty\else 
          \psk@arrowsize\space CLW mul add \psk@arrowlength\space mul 
           viceversa { sub epsilon add }{ add epsilon sub } ifelse \fi } def
      /MaxValue { \ifis@yAxis yEnd \else xEnd \fi 
        \ifx\psk@arrowB\@empty\else
          \psk@arrowsize\space CLW mul add \psk@arrowlength\space mul 
           viceversa { add epsilon sub }{ sub epsilon add } ifelse \fi } def
      /logLines {
        \ifnum\psk@logLines=\z@ true \else         % all axes
          \ifnum\psk@logLines<\tw@                 % x axis
            \ifis@yAxis false \else true \fi       % do we have x or y axis
          \else
            \ifnum\psk@logLines<\thr@@             % y axis
              \ifis@yAxis true \else false \fi     % do we have x or y axis
            \else 
              false                                % no one
            \fi
          \fi
        \fi
      } def
      /LSstroke {                                  % set linestyle and stroke
        \ifis@yAxis \@nameuse{psls@\psyticklinestyle}
        \else       \@nameuse{psls@\psxticklinestyle}\fi 
        stroke} def
      /subLSstroke {                               % set sublinestyle and stroke
        \ifis@yAxis \@nameuse{psls@\psysubticklinestyle}
        \else       \@nameuse{psls@\psxsubticklinestyle}\fi 
        stroke} def
      0 dT MaxValue 1 add {                        % the positive part of the axes, step unit is pt
        /cntTick exch def                          % the index
        logLines {                                 % log lines?
          gsave
          1 1 DT {
           1 sub /OffSet exch def
          -10 subTNo 1 add div dup 10 add exch dup -0.1 mul 1 add {                   % do not write a line for 10 and 1, trace lines between 10 and 1 by steps of 10/subTno
            /dx exch def                           % save index
            /x dx log OffSet add \ifis@yAxis\pst@number\psyunit\else\pst@number\psxunit\fi\space mul cntTick add def       %
            x abs MaxValue abs le {                % out of range?
	      \ifis@yAxis
	        \ifPst@maxyTicks true \else false \fi
	      \else
	        \ifPst@maxxTicks true \else false \fi
	      \fi
                { x minTickline #5 moveto
                  x maxTickline #5 lineto }
                { x tickstart STsize mul #5 moveto
                  x tickend STsize mul #5 lineto } ifelse
            } if
          } for } for
          subTwidth SLW subTColor                  % set line width and subtick color
          subLSstroke
          grestore                                 % restore main tick status
          stroke
          /dsubT 0 def                             % no other subticks
        } if 					   % end logLines
        dsubT abs 0 gt {                           % du we have subticks?
          gsave                                    % save graphic state
          /cntsubTick cntTick dsubT add def
          subTNo 1 sub {
            cntsubTick abs MaxValue abs le {       % out of range?
    	    \ifis@yAxis
              \ifPst@maxyTicks true \else false \fi
    	    \else
              \ifPst@maxxTicks true \else false \fi
    	    \fi
              { cntsubTick minTickline STsize mul #5 moveto
                cntsubTick maxTickline STsize mul #5 lineto }
              { cntsubTick tickstart STsize mul #5 moveto
                cntsubTick tickend STsize mul #5 lineto } ifelse
            }{ exit }  ifelse
            /cntsubTick cntsubTick dsubT add def
          } repeat 
          subTwidth SLW subTColor               % set line width and subtick color
          subLSstroke
          grestore                              % restore tick status
        } if
        showOrigin {
          gsave
          \ifis@yAxis
            \ifPst@maxyTicks true \else false \fi
          \else
            \ifPst@maxxTicks true \else false \fi
          \fi
            { cntTick minTickline #5 moveto
              cntTick maxTickline #5 lineto }
            { cntTick tickstart #5 moveto        % line begin main Tick
              cntTick tickend #5 lineto } ifelse % lineto tick end
          Twidth SLW TColor                      % set line width and tick color
          LSstroke
          grestore
        }{ /showOrigin true def } ifelse         % only for the very first tick valid
      } for
      /showOrigin \ifPst@showOriginTick true \else false \fi def % ticks for 0/0 ?
      /dT dT neg def                               % the other side of the axis
      /dsubT dsubT neg def
      0 dT MinValue epsilon viceversa { add }{ sub } ifelse {
        /cntTick exch def
        logLines {                                 % log lines?
          gsave
          1 1 DT cvi {
            1 sub /OffSet exch def
          -10 subTNo 1 add div dup 10 add exch dup -0.1 mul 1 add {                   % do not write a line for 10 and 1, trace lines between 10 and 1 by steps of 10/subTno
            /dx exch def                           % save index
            /x dx log OffSet add \ifis@yAxis\pst@number\psyunit\else\pst@number\psxunit\fi\space mul cntTick add def
            x abs MinValue abs le {                % out of range?
	      \ifis@yAxis
	        \ifPst@maxyTicks true \else false \fi
	      \else
	        \ifPst@maxxTicks true \else false \fi
	      \fi
                { x minTickline #5 moveto
                  x maxTickline #5 lineto }
                { x tickstart STsize mul #5 moveto
                  x tickend STsize mul #5 lineto } ifelse
            } if
          } for } for
          /dsubT 0 def 
          subTwidth SLW subTColor                  % set line width and subtick color
          subLSstroke
          grestore
        }                                          % end loglines
        dsubT abs 0 gt {                           % do we have subticks?
          gsave                                    % save main state
          /cntsubTick cntTick dsubT add def
          subTNo 1 sub {
            cntsubTick abs MinValue abs le {       % out of range?
              cntsubTick tickstart STsize mul #5 moveto
              cntsubTick tickend STsize mul #5 lineto
            }{ exit } ifelse
            /cntsubTick cntsubTick dsubT add def
          } repeat % for
          subTwidth SLW subTColor                  % set line width and subtick color
          subLSstroke
          grestore                                 % restore main state
        } if
        showOrigin {
          gsave
          cntTick tickstart #5 moveto         	% line begin main Tick
          cntTick tickend #5 lineto    	       	% lineto tick end
          Twidth SLW TColor                         % set line width and tick color
          LSstroke
          grestore
        }{ /showOrigin true def } ifelse         % only for the very first tick valid
      } for
    } if
   }%	end of \pscode
  \end@SpecialObj%
  \ifx\psk@axesstyle\@none\else
    \ifPst@yAxis\psline[linecolor=\pslinecolor](0,#2)(0,#4)\fi
    \ifPst@xAxis\psline[linecolor=\pslinecolor](#1,0)(#3,0)\fi
  \fi
  \ignorespaces
}%
\def\psxs@frame{%
  \psset{axesstyle=none}%
  \begin@SpecialObj%
    \addto@pscode{					% the frame
      \pst@number\pst@dima \psk@frameOffset sub \pst@number\pst@dimb \psk@frameOffset sub moveto 	% lower left
      \pst@number\pst@dimc \psk@frameOffset add \pst@number\pst@dimb \psk@frameOffset sub L	% upper left
      \pst@number\pst@dimc \psk@frameOffset add \pst@number\pst@dimd \psk@frameOffset add L 	% upper right
      \pst@number\pst@dima \psk@frameOffset sub \pst@number\pst@dimd \psk@frameOffset add L 	% lower right
      closepath 
      }%
    \pst@stroke%
    \psk@fillstyle%
  \end@SpecialObj%
  \let\psk@arrowA\@empty%
  \let\psk@arrowB\@empty%
  \pst@xticksizeC=\z@\pst@yticksizeC=\z@  
  \ifPst@xAxis\psxs@@axes\pst@dima\pst@dimb\pst@dimc\pst@dimd{}{x}\fi%		x axis
  \ifPst@yAxis\psxs@@axes\pst@dima\pst@dimb\pst@dimc\pst@dimd{ exch }{y}\fi%	y axis
  \ifnum\psk@tickstyle=\tw@	% llx,lly,urx,ury,exch,x|y,arrowA,arrowB	
    \psDEBUG[psxs@frame]{psk@tickstyle=2 (inner)}%
    \psDEBUG[psxs@frame]{pst@dima=\pst@number\pst@dima}%
    \psDEBUG[psxs@frame]{pst@dimb=\pst@number\pst@dimb}%
    \psDEBUG[psxs@frame]{pst@dimc=\pst@number\pst@dimc}%
    \psDEBUG[psxs@frame]{pst@dimd=\pst@number\pst@dimd}%
    \ifPst@xAxis\psxs@@axes\pst@dima\pst@dimb\pst@dimc\pst@dimd{ neg \pst@number\pst@dimd add }{x}\fi%	% upper x axis
    \ifPst@yAxis\psxs@@axes\pst@dima\pst@dimb\pst@dimc\pst@dimd{ neg \pst@number\pst@dimc add exch }{y}\fi%  right y axis
  \fi%
}
\def\psxs@polar{% (rx,ry) % all other values are ignored
  \pst@killglue
  \begingroup
  \edef\pst@dimC{\strip@pt\pst@dimc}% 			RadiusX
  \pstFPDiv\pstR@dius{\pst@dimC}{\strip@pt\psxunit}%	in cm and as integer
  \edef\pst@dimD{\strip@pt\pst@dimd}% 			RadiusX
  \pstFPDiv\psk@EndAngle{\pst@dimD}{\strip@pt\psyunit}%	in cm and as integer
  \ifnum\psk@EndAngle=0 \def\psk@EndAngle{360}\fi
  \use@keep@par
  \pstFPDiv\pstN@lpha{\psk@EndAngle}{\psk@Dy}% 			No. of (int) main lines
  \pstFPdiv\pstd@lpha{\psk@Dy}{\psk@ysubticks}% 	sub dAlpha
  \pstFPdiv\pstdR@dius{1}{\psk@xsubticks}%		sub dRadius
  \pst@cntm=\psk@xsubticks\advance\pst@cntm by \m@ne
  \multido{\iA=\psk@Dx+\psk@Dx,\rB=\pstdR@dius+\psk@Dx,\iB=0+1}{\pstR@dius}{%
    \multido{\rA=\rB+\pstdR@dius}{\the\pst@cntm}{%
      \psarc[linestyle=\psxsubticklinestyle,
         linecolor=\psk@xsubtickcolor,linewidth=\psk@xsubtickwidth pt](0,0){\rA}{0}{\psk@EndAngle}}    
    \psarc[linestyle=\psxticklinestyle,linecolor=\psk@xtickcolor,
		linewidth=\psk@xtickwidth pt](0,0){\iA}{0}{\psk@EndAngle}%
    \ifnum\psk@labels<2\relax% is all or x (0,1)
      \uput[-45](\iB,0){\pshlabel{\iB}}\uput[45](0,\iB){\pshlabel{\iB}}%
    \fi%
  }%
  \pst@cntm=\psk@ysubticks\advance\pst@cntm by \m@ne
  \multido{\iA=\psk@Oy+\psk@Dy,\rB=\pstd@lpha+\psk@Dy}{\pstN@lpha}{%
    \multido{\rA=\rB+\pstd@lpha}{\the\pst@cntm}{\psline[linestyle=\psysubticklinestyle,
      linecolor=\psk@ysubtickcolor,linewidth=\psk@ysubtickwidth pt](\pstR@dius;\rA)} 
    \psline[linestyle=\psyticklinestyle,
      linecolor=\psk@ytickcolor,linewidth=\psk@ytickwidth pt](\pstR@dius;\iA)%
    \ifodd\psk@labels\else% is all or y (0,3)
      \uput[\iA](\pstR@dius;\iA){\psvlabel{\iA\psk@ylabelFactor}}%
    \fi%
  }%
  \ifnum\psk@EndAngle<360 \psline[linestyle=\psyticklinestyle,
      linecolor=\psk@ytickcolor,linewidth=\psk@ytickwidth pt](\pstR@dius;0)\fi
  \endgroup\ignorespaces%
  \Pst@xAxisfalse\Pst@yAxisfalse%
}
\def\@polar{polar}
\define@key[psset]{pst-plot}{axesstyle}[axes]{%
  \@ifundefined{psxs@#1}%
    {\@pstrickserr{Axes style `#1' not defined}\@eha}%
    {\def\psk@axesstyle{#1}%
     \ifx\psk@axesstyle\@polar\psset{Dy=30}\fi}}
\psset[pst-plot]{axesstyle=axes}
\define@key[psset]{pst-plot}{xLabels}[]{\def\psk@xLabels{#1}}
\define@key[psset]{pst-plot}{xLabelsRot}[0]{\pst@getangle{#1}\pst@xLabelsRot}
\psset[pst-plot]{xLabels=,xLabelsRot=0}
\define@key[psset]{pst-plot}{yLabels}[]{\def\psk@yLabels{#1}}
\define@key[psset]{pst-plot}{yLabelsRot}[0]{\pst@getangle{#1}\pst@yLabelsRot}
\psset[pst-plot]{yLabels=,yLabelsRot=0}
\def\pst@hlabels#1#2#3#4{%
  \ifSpecialLabelsDone
  \else
    \kern\psk@xlabelOffset pt            % set the x offset?
    \ifx\empty\psk@xLabels
      \ifdim#1=\z@
      \else                   % start from 0 ?
        \ifx#2\empty
        \else
          \advance#1\ifdim#1>\z@-\fi7\pslinewidth
        \fi
        \pst@cnta=#1\relax                % Distance (in sp) to end.
        \divide\pst@cnta\psk@dx\relax     % Number of ticks/labels
        \ifnum\pst@cnta=\z@
        \else
          \pst@dimb=\psk@dx sp            % Space between ticks.
          \ifnum\psk@labels<\tw@ \ifPst@xAxis\pst@@hlabels\fi\fi
          \showoriginfalse
        \fi
      \fi
   \else
     \ifnum\psk@xlabelPos=\tw@ \def\pst@tempC{90}\else\def\pst@tempC{-90}\fi
       \pstFPsub\pst@pmtempa{#4}{#3}%
       \pstFPDiv\pst@pmtempb{\pst@pmtempa}{\psk@Dx}%
       \pstFPadd\pst@pmtempc{\pst@pmtempb}{-1}%
       \pstFPadd\pst@pmtempd{\pst@pmtempb}{1}%
       \ifdim\pst@pmtempb pt < \z@ 
         \def\pst@pmtempe{\pst@int{\pst@pmtempc}}%
       \else
         \def\pst@pmtempe{\pst@int{\pst@pmtempd}}%
       \fi
       \multido{\nA=0+1,\rA=#3+\psk@Dx}{\pst@pmtempe}{%
         \ifdim \nA pt < \z@ \def\nB{-\nA} \else \def\nB{\nA} \fi
         \uput{\psxlabelsep}[\pst@tempC]{\pst@xLabelsRot}(\rA,0){%
              \strut\expandafter\pshlabel\expandafter{\psPutXLabel{\nB}}}}%
       \SpecialLabelsDonetrue
    \fi
  \fi
}
\def\pst@@hlabels{%
  \setbox\z@=\vbox{%			save all in a box
    \ifcase\psk@xlabelPos
      \vskip-\pst@xticksizeA\vskip\psxlabelsep\or% 1
      \vskip-1ex\vskip-\pslabelsep\or% 2
      \vskip-\pst@xticksizeB\vskip-\psxlabelsep\vskip-1ex% 3
    \fi
    \ifnum\pst@cnta<\z@ \pst@dimb=-\pst@dimb\fi
    \hbox to \z@{%
      \ifshoworigin\hbox to \z@{\hss\pst@@@hlabel{\psk@Ox}\hss}\fi
      \mmultido{\nA=\psk@Ox+\psk@Dx}{\pst@cnta}{%
        \hskip\pst@dimb \hbox to \z@{\hss
          \ifdim\nA pt=\z@\relax\ifshoworigin\pst@@@hlabel{0}\fi
          \else\expandafter\pst@@@hlabel{\nA}%
          \fi% prevent -0, doesn't work with \ifnum
        \hss}%
      }\hss%    1.85
    }%
  }\ht\z@\z@ \dp\z@\z@ \box\z@}% set all values to zero
\def\pst@vlabels#1#2#3#4{%
  \ifSpecialLabelsDone\else
      \ifx\empty\psk@yLabels
        \ifdim#1=\z@\else
          \ifx#2\empty\else\ifdim#1>\z@ \advance#1 by -7\pslinewidth\else\advance#1 by 7\pslinewidth\fi\fi
          \pst@cnta=#1\relax           %      % Distance (in sp) to end.
          \divide\pst@cnta\psk@dy\relax%   % Number of ticks/labels
          \ifnum\pst@cnta=\z@\else
            \pst@dima=\psk@dy sp%            % Space between ticks.
            \ifodd\number\psk@labels\else\ifPst@yAxis\pst@@vlabels\fi
          \fi
          \showoriginfalse
        \fi
      \fi
    \else
	\pstFPsub\pst@pmtempa{#4}{#3}%
	\pstFPDiv\pst@pmtempb{\pst@pmtempa}{\psk@Dy}%
	\pstFPadd\pst@pmtempc{\pst@pmtempb}{-1}%
	\pstFPadd\pst@pmtempd{\pst@pmtempb}{1}%
	\ifdim\pst@pmtempb pt < \z@ \def\pst@pmtempe{\pst@int{\pst@pmtempc}}\else\def\pst@pmtempe{\pst@int{\pst@pmtempd}}\fi
	\multido{\nA=0+1,\rA=#3+\psk@Dy}{\pst@pmtempe}{%
	  \ifdim \nA pt < \z@ \def\nB{-\nA}\else \def\nB{\nA}\fi
	  \ifnum\psk@ylabelPos=0
            \uput{\psylabelsep}[180]{\pst@yLabelsRot}(0,\rA){%
              \strut\expandafter\psvlabel\expandafter{\psPutYLabel{\nB}}}%
          \else
            \uput{\psylabelsep}[0]{\pst@yLabelsRot}(0,\rA){%
              \strut\expandafter\psvlabel\expandafter{\psPutYLabel{\nB}}}%
          \fi
        }%  
      \SpecialLabelsDonetrue
    \fi
  \fi
}
\def\pst@@vlabels{%
  \vbox to\z@{%
   \vbox to -\psk@ylabelOffset pt{}% the y label offset
    \ifnum\pst@cnta>\z@ \pst@dima=-\pst@dima\fi%  up or down label positions
    \offinterlineskip
    \ifshoworigin
      \vbox to \z@{\vss\hbox to\z@{%
        \ifcase\psk@ylabelPos
	  \hss\pst@@@vlabel{\psk@Oy}\hskip\psylabelsep\hskip-\pst@yticksizeA\or%
	  \hskip\pslabelsep\hss\pst@@@vlabel{\psk@Oy}\hss\or%		% right labels
	  \hskip\pst@yticksizeB\hskip\psylabelsep\pst@@@vlabel{\psk@Oy}%
	\fi}\vss}%
    \fi
    \mmultido{\nA=\psk@Oy+\psk@Dy}{\pst@cnta}{%
      \vbox to\pst@dima{\vss}%
      \vbox to \z@{%
        \vss\hbox to\z@{%
        \ifcase\psk@ylabelPos % and also check for -0
	  \hss\ifdim\nA pt=\z@ \ifshoworigin\pst@@@vlabel{0}\fi\else\pst@@@vlabel{\nA}\fi
	    \hskip\psylabelsep\hskip-\pst@yticksizeA\or% top = 1
	  \hss\ifdim\nA pt=\z@ \ifshoworigin\pst@@@vlabel{0}\fi\else\pst@@@vlabel{\nA}\fi
	  \ifdim\psylabelsep=\z@\hss\else\kern-\psylabelsep\fi\or% right=2
	  \hskip\pst@yticksizeB\hskip\psylabelsep
	  \ifdim\nA pt=\z@ \ifshoworigin\pst@@@vlabel{0}\fi\else\pst@@@vlabel{\nA}\fi% bottom
	\fi}\vss}%
    }\vss}%
}
\define@key[psset]{pst-plot}{xAxisLabel}[x]{\def\psk@xAxisLabel{#1}}
\define@key[psset]{pst-plot}{yAxisLabel}[y]{\def\psk@yAxisLabel{#1}}
\psset[pst-plot]{xAxisLabel=x,yAxisLabel=y}
\define@key[psset]{pst-plot}{xAxisLabelPos}[{}]{\def\psk@xAxisLabelPos{#1}}
\define@key[psset]{pst-plot}{yAxisLabelPos}[{}]{\def\psk@yAxisLabelPos{#1}}
\psset[pst-plot]{yAxisLabelPos={},xAxisLabelPos={}}
\newdimen\psk@llx
\newdimen\psk@lly
\newdimen\psk@urx
\newdimen\psk@ury
\define@key[psset]{pst-plot}{llx}[\z@]{\pssetxlength\psk@llx{#1}}
\define@key[psset]{pst-plot}{lly}[\z@]{\pssetylength\psk@lly{#1}}
\define@key[psset]{pst-plot}{urx}[\z@]{\pssetxlength\psk@urx{#1}}
\define@key[psset]{pst-plot}{ury}[\z@]{\pssetylength\psk@ury{#1}}
\psset[pst-plot]{llx=\z@, lly=\z@, urx=\z@, ury=\z@}% prevents rounding errors 
\define@boolkey[psset]{pst-plot}[Pst@]{psgrid}[true]{}
\define@key[psset]{pst-plot}{gridpara}[{}]{\def\psk@gridpara{#1}}
\define@key[psset]{pst-plot}{gridcoor}[\relax]{\def\psk@gridcoor{#1}}
\psset[pst-plot]{psgrid=false,gridpara={gridlabels=0pt,gridcolor=red!30,subgridcolor=green!30,subgridwidth=0.5\pslinewidth,
  subgriddiv=5},gridcoor=\relax}
\define@key[psset]{pst-plot}{axespos}[bottom]{\pst@expandafter\psset@@axespos{#1}\@nil}
\def\psset@@axespos#1#2\@nil{%
  \ifx#1b\let\psk@axespos\z@\else		% 0=b)bottom
    \ifx#1t\let\psk@axespos\@ne			% 1=t)op
      \else\@pstrickserr{Bad axes position: `#1#2'}\@ehpa
  \fi\fi}
\psset[pst-plot]{axespos=b}
\newdimen\pst@xunit
\newdimen\pst@yunit
\def\pslegend{\@ifnextchar[\pslegend@i{\pslegend@i[rt]}}
\def\pslegend@i[#1]{\@ifnextchar({\pslegend@ii[#1]}{\pslegend@ii[#1](\pst@number\pslabelsep,\pst@number\pslabelsep)}}
\def\pslegend@ii[#1](#2,#3)#4{%
  \gdef\pslegend@ref{#1}%
  \xdef\pslegend@sepx{#2 }%
  \xdef\pslegend@sepy{#3 }%
  \gdef\pslegend@text{#4}}
\newpsstyle{legendstyle}{fillstyle=solid,fillcolor=white,linewidth=0.5pt}
\def\pslegend@iii[#1](#2){\rput[#1](#2){\psframebox[style=legendstyle]{%
  \footnotesize\tabcolsep=2pt%
  \tabular[t]{@{}ll@{}}\pslegend@text\endtabular}}\global\let\pslegend@text\relax}
\let\pslegend@text\relax% define it as empty
\def\psgraph{\pst@object{psgraph}}
\def\psgraph@i{%
  \let\psgraph@para\pst@par
  \let\psk@save@arrowA\psk@arrowA
  \let\psk@save@arrowB\psk@arrowB
  \pst@getarrows\psgraph@ii}
\def\psgraph@ii(#1,#2){\catcode`\!=12\relax
  \@ifnextchar({\psgraph@iii(#1,#2)}{\psgraph@iv(0,0)(#1,#2)}}
\def\psgraph@iii(#1,#2)(#3,#4){\@ifnextchar({\psgraph@v(#1,#2)(#3,#4)}{\psgraph@iv(#1,#2)(#3,#4)}}
\def\psgraph@iv(#1,#2)(#3,#4)#5#6{%  no special origin defined
  \pst@killglue%
  \begingroup
  \use@keep@par
  \pstFPsub\pst@tempA{#3}{#1}%
  \pst@dimm=#5
  \pst@dimo=\pst@tempA pt
  \pstFPdiv\pst@@dx{\strip@pt\pst@dimm}{\pst@tempA}%
  \pst@xunit=\pst@@dx\p@
  \ifx!#6\let\pst@yunit=\pst@xunit\else
    \pst@dimm=#6
    \pstFPsub\pst@tempA{#4}{#2}%
    \pstFPdiv\pst@@dy{\strip@pt\pst@dimm}{\pst@tempA}%
    \pst@yunit=\pst@@dy\p@
  \fi
  \pst@dimm=#1\pst@xunit\advance\pst@dimm by \psk@llx
  \pst@dimn=#2\pst@yunit\advance\pst@dimn by \psk@lly
  \pst@dimo=#3\pst@xunit\advance\pst@dimo by \psk@urx
  \pst@dimp=#4\pst@yunit\advance\pst@dimp by \psk@ury
  \if@star\pspicture*(\pst@dimm,\pst@dimn)(\pst@dimo,\pst@dimp)\else
  \pspicture(\pst@dimm,\pst@dimn)(\pst@dimo,\pst@dimp)\fi
  \let\psxunit\pst@xunit \let\psyunit\pst@yunit
  \ifdim\pst@xunit=\pst@yunit\relax\psset{runit=\pst@xunit}\fi%
  \bgroup
    \use@par
  \ifPst@psgrid
     \expandafter\psset\expandafter{\psk@gridpara}%
      \rput[lb](0,0){\expandafter\psgrid\psk@gridcoor}  
  \fi
    \ifnum\psk@axespos=0
      \expandafter\psaxes\expandafter[\psgraph@para](#1,#2)(#3,#4)%
    \else
      \xdef\psgraph@coor{(#1,#2)(#3,#4)(#5,#6)}%
    \fi
  \egroup
  \psgraph@vi(#1,#2)(#1,#2)(#3,#4)%
}
\def\psgraph@v(#1,#2)(#3,#4)(#5,#6)#7#8{%  with special origin
  \pst@killglue%
  \let\psgraph@para\pst@par
  \begingroup%
  \use@keep@par
  \pstFPsub\pst@tempA{#5}{#3}%
  \pst@dimm=#7%
  \pst@dimo=\pst@tempA pt%
  \pstFPdiv\pst@@dx{\strip@pt\pst@dimm}\pst@tempA%
  \pst@xunit=\pst@@dx\p@%
  \ifx!#8\let\pst@yunit=\pst@xunit\else
    \pst@dimm=#8%
    \pstFPsub\pst@tempA{#6}{#4}%
    \pstFPdiv\pst@@dy{\strip@pt\pst@dimm}\pst@tempA%
    \pst@yunit=\pst@@dy\p@%
  \fi%
  \pst@dima=#3\pst@xunit \advance\pst@dima by \psk@llx%
  \pst@dimb=#4\pst@yunit \advance\pst@dimb by \psk@lly%
  \pst@dimc=#5\pst@xunit \advance\pst@dimc by \psk@urx%
  \pst@dimd=#6\pst@yunit \advance\pst@dimd by \psk@ury%
  \if@star\pspicture*(\pst@dima,\pst@dimb)(\pst@dimc,\pst@dimd)\else%
          \pspicture(\pst@dima,\pst@dimb)(\pst@dimc,\pst@dimd)\fi%
  \psset{xunit=\pst@xunit,yunit=\pst@yunit}
  \ifdim\pst@xunit=\pst@yunit \psset{runit=\pst@xunit}\fi%
  \bgroup%
    \use@par%
  \ifPst@psgrid
     \expandafter\psset\expandafter{\psk@gridpara}%
      \rput[lb](0,0){\expandafter\psgrid\psk@gridcoor}
  \fi%
    \ifnum\psk@axespos=0
      \psaxes(#1,#2)(#3,#4)(#5,#6)%
    \else
      \xdef\psgraph@coor{(#1,#2)(#3,#4)(#5,#6)}%
    \fi
  \egroup
  \psgraph@vi(#1,#2)(#3,#4)(#5,#6)%
}
\def\setxLabelC@@r#1,#2(#3,#4)(#5){%
  \pst@getcoor{#5}\pst@tempB%
  \ifx c#1 
    \pssetylength\pst@dimm{#2}%
    \rput(! #4 #3 add 2 div \pst@number\pst@dimm \pst@tempB\space exch pop add 
      \pst@number\psyunit div ){\psk@xAxisLabel}%
  \else%
    \pst@getcoor{\psk@xAxisLabelPos}\pst@tempA%
    \rput(! \pst@tempA\space \pst@tempB\space exch pop add \tx@UserCoor ){\psk@xAxisLabel}%
  \fi}
\def\setyLabelC@@r#1,#2(#3,#4)(#5){%
  \pst@getcoor{#5}\pst@tempB%
  \ifx c#2
    \pssetxlength\pst@dimm{#1}%
    \rput{90}(! \pst@number\pst@dimm \pst@tempB\space pop add \pst@number\psxunit div #4 #3 add 2 div ){\psk@yAxisLabel}%
  \else%
    \pst@getcoor{\psk@yAxisLabelPos}\pst@tempA%
    \rput{90}(! \pst@tempB\space pop \pst@tempA\space 3 1 roll add exch \tx@UserCoor ){\psk@yAxisLabel}%
  \fi}
\def\psgraph@vi(#1,#2)(#3,#4)(#5,#6){%
  \ifx\psk@xAxisLabel\@empty\else%
    \ifx\psk@xAxisLabelPos\@empty\uput[0](#5,#2){\psk@xAxisLabel}%
    \else\expandafter\setxLabelC@@r\psk@xAxisLabelPos(#3,#5)(#1,#2)\fi%
  \fi%
  \ifx\psk@yAxisLabel\@empty\else%
    \ifx\psk@yAxisLabelPos\@empty\uput[90](#1,#6){\psk@yAxisLabel}%
    \else\expandafter\setyLabelC@@r\psk@yAxisLabelPos(#4,#6)(#1,#2)\fi%
  \fi%
  \def\lt@@{lt}\def\lb@@{lb}\def\rb@@{rb}%
  \ifx\pslegend@ref\lb@@    \gdef\pslegend@coor{#3 \pslegend@sepx \pst@number\psxunit div add 
                                                   \pslegend@sepy \pst@number\psyunit div}%
  \else%
    \ifx\pslegend@ref\lt@@  \gdef\pslegend@coor{#3 \pslegend@sepx \pst@number\psxunit div add 
                                                #6 \pslegend@sepy \pst@number\psyunit div sub}%
    \else%
      \ifx\pslegend@ref\rb@@\gdef\pslegend@coor{#5 \pslegend@sepx \pst@number\psxunit div sub 
                                                   \pslegend@sepy \pst@number\psyunit div}%
      \else                 \gdef\pslegend@coor{#5 \pslegend@sepx \pst@number\psxunit div sub 
                                                #6 \pslegend@sepy \pst@number\psyunit div sub}%
      \fi%
    \fi%
  \fi%
  \xdef\psgraphLLx{#3}\xdef\psgraphLLy{#4}\xdef\psgraphURx{#5}\xdef\psgraphURy{#6}%
  \global\let\psk@arrowA\psk@save@arrowA
  \global\let\psk@arrowB\psk@save@arrowB
  \ignorespaces
}
\def\endpsgraph{%
  \ifx\relax\pslegend@text\relax \else\pslegend@iii[\pslegend@ref](!\pslegend@coor)\fi
  \expandafter\psset\expandafter{\psgraph@para}%
  \ifnum\psk@axespos>0
    \expandafter\psaxes\psgraph@coor
  \fi
  \endpspicture
  \endgroup\ignorespaces}
\@namedef{psgraph*}{\psgraph*}
\@namedef{endpsgraph*}{\endpsgraph}
\def\psPutXLabel#1{%
  \global\pst@cnto=0\relax
  \global\pst@cntp=#1\relax
  \expandafter\get@Label\psk@xLabels,,\@nil%
}
\def\psPutYLabel#1{%        
  \global\pst@cnto=0\relax
  \global\pst@cntp=#1\relax
  \expandafter\get@Label\psk@yLabels,,\@nil%
}
\def\get@Label#1,#2,#3\@nil{%
    \ifnum\the\pst@cnto<\the\pst@cntp
      \global\advance\pst@cnto by \@ne 
      \ifx\relax#3\relax\else\expandafter\get@Label#2,#3\@nil\fi%
    \else #1\fi%
}
\def\psVectorfield{\pst@object{psVectorfield}}
\def\psVectorfield@i(#1,#2)(#3,#4)#5{{%
  \addbefore@par{Dx=0.1,Dy=0.1,Ox=3,arrows=->,linewidth=0.2pt}%
  \begin@SpecialObj
  \SpecialCoor
  \pstFPsub\pst@tempA{#3}{#1}%
  \pstFPsub\pst@tempB{#4}{#2}%
  \pstFPDiv{\pst@tempC}{\pst@tempA}{\psk@Dx}%
  \pstFPDiv{\pst@tempD}{\pst@tempB}{\psk@Dy}%
  \pstVerb{ /yStrich \ifPst@algebraic (#5) tx@AlgToPs begin AlgToPs end cvx
                \else { #5 } \fi def }%
  \multido{\rX=#1+\psk@Dx}{\numexpr\pst@tempC+1}{%
    \multido{\rY=#2+\psk@Dy}{\numexpr\pst@tempD+1}{%
       \psline%
         (! /x \rX\space def 
            /y \rY\space def 
            /yTemp yStrich \psk@Dx\space \psk@Ox\space div mul def 
            \rX\space \psk@Dx\space \psk@Ox\space div sub \rY\space yTemp sub)%
         (! /x \rX\space def 
            /y \rY\space def 
            /yTemp yStrich \psk@Dx\space \psk@Ox\space div mul def 
            \rX\space \psk@Dx\space \psk@Ox\space div add \rY\space yTemp add)%
   }}%
  \end@SpecialObj
}\ignorespaces}  
\def\psFixpoint{\pst@object{psFixpoint}}
\def\psFixpoint@i#1#2#3{% #1: xStart #2: f(x) #3: number of iterations
  \pst@killglue%
  \begingroup%
  \use@par%
  \@nameuse{beginplot@\psplotstyle}%
  \addto@pscode{
    \psplot@init
      /x #1 def
      /F@pstplot \ifPst@algebraic (#2) tx@AlgToPs begin AlgToPs end cvx
                 \else { #2 } \fi  def
      /xy { x \pst@number\psxunit mul F@pstplot dup /x ED \pst@number\psyunit mul } def 
  }%
  \gdef\psplot@init{}%
  \@pstfalse%
  \@nameuse{testqp@\psplotstyle}%
  \addto@pscode{
      mark
      x \pst@number\psxunit mul 0
      /n 2 def
      #3 {
        xy 
        dup dup 
        /n n 4 add def
      } repeat 
  }%
  \@nameuse{endplot@\psplotstyle}%
  \endgroup%
  \ignorespaces}
\define@boolkey[psset]{pst-plot}[Pst@]{showDerivation}[true]{}
\psset{showDerivation}
\def\psNewton{\pst@object{psNewton}}
\def\psNewton@i#1#2{\@ifnextchar[{\psNewton@ii{#1}{#2}}{\psNewton@iii{#1}{#2}}}
\def\psNewton@ii#1#2[#3]#4{% #1:xStart #2:f(x) #3:f'(x) #4:number of iterations
  \pst@killglue%
  \begingroup%
  \addbefore@par{showDerivation}%
  \use@par%
  \@nameuse{beginplot@\psplotstyle}%
  \addto@pscode{
    \psplot@init
      /x #1 def
      /F@pstplot \ifPst@algebraic (#2) tx@AlgToPs begin AlgToPs end cvx \else { #2 } \fi  def
      /F@pstplotDerive \ifPst@algebraic (#3) tx@AlgToPs begin AlgToPs end cvx \else { #3 } \fi  def
      /newxVal { % y on stack
        F@pstplotDerive % we have m
        div neg %\pst@number\psxunit div % new x val = -y0/m
      } def
  }%
  \gdef\psplot@init{}%
  \@pstfalse%
  \@nameuse{testqp@\psplotstyle}%
  \addto@pscode{
      mark
      x 0 \tx@ScreenCoor % start point
      /n 2 def
      #4 {
        F@pstplot /yVal ED
        x yVal \tx@ScreenCoor
        /n n 2 add def
        yVal newxVal x add /x ED
        x 0 \tx@ScreenCoor 
        \ifPst@showDerivation /n n 4 add def \else moveto /n n 2 add def\fi
      } repeat 
      pstack
  }%
  \@nameuse{endplot@\psplotstyle}%
  \endgroup%
  \ignorespaces}
\def\psNewton@iii#1#2#3{% #1:xStart #2:f(x) #3:number of iterations
  \pst@killglue%
  \begingroup%
  \addbefore@par{VarStepEpsilon=0.01,showDerivation}%
  \use@par%
  \@nameuse{beginplot@\psplotstyle}%
  \addto@pscode{
    \psplot@init
      /epsilon \psk@VarStepEpsilon\space def
      /x #1 def
      /F@pstplot \ifPst@algebraic (#2) tx@AlgToPs begin AlgToPs end cvx \else { #2 } \fi  def
      /newxVal { % y on stack
        /saveX x def
        saveX epsilon add /x ED F@pstplot saveX epsilon sub /x ED F@pstplot sub epsilon dup add div % we have m
        div neg % new x val = -y0/m
        /x saveX def
      } def
  }%
  \gdef\psplot@init{}%
  \@pstfalse%
  \@nameuse{testqp@\psplotstyle}%
  \addto@pscode{
      mark
      x 0 \tx@ScreenCoor % start point
      /n 2 def
      #3 {
        F@pstplot /yVal ED
        x yVal \tx@ScreenCoor
        yVal newxVal x add /x ED
        x 0 \tx@ScreenCoor 
        \ifPst@showDerivation /n n 4 add def \else moveto /n n 2 add def\fi
      } repeat 
  }%
  \@nameuse{endplot@\psplotstyle}%
  \endgroup%
  \ignorespaces}
\def\psResetPlotValues{%
  \psset{method={}}%
}%
\catcode`\@=\TheAtCode\relax
 
\csname PSTnodesLoaded\endcsname
\let\PSTnodesLoaded 
\ifx\PSTricksLoaded \else\input pstricks.tex \fi\relax
\ifx\PSTXKeyLoaded \else \input pst-xkey \fi
\def\fileversion{1.42}
\def\filedate{2019/03/03}
\edef\TheAtCode{\the\catcode`\@}
\catcode`\@=11
\pstheader{pst-node.pro}
\pst@addfams{pst-node}
\SpecialCoor
\define@boolkey[psset]{pst-node}[Pst@]{trueAngle}[true]{}
\psset[pst-node]{trueAngle=false}
\define@boolkey[psset]{pst-node}[Pst@]{storeNodeInfo}[true]{}
\psset[pst-node]{storeNodeInfo=false}
\def\pst@nodedict{tx@NodeDict begin }
\def\pst@zapspace#1 #2{%
#1%
\ifx#2\@empty\else\expandafter\pst@zapspace\fi
#2}
\def\pst@getnode#1#2{\pst@expandafter\pst@@getnode{#1},,\@nil#2}
\def\pst@@getnode#1,#2,#3\@nil#4{%
  \ifx\@empty#3\@empty
    \edef#4{/N@\pst@zapspace#1 \@empty\space}%
  \else
    \pst@cntg=#1\relax
    \pst@cnth=#2\relax
    \edef#4{/N@M-\ifnum\psmatrixcnt=\z@ 1\else\the\psmatrixcnt\fi
    -\the\pst@cntg-\the\pst@cnth\space}%
  \fi}
\def\tx@NewNode{/NodeScale {\ifx\pstnodescale\@undefined  \else\pstnodescale \fi} def NewNode }
\def\psopenNodeFile{%
  \pst@Verb{ %globaldict begin
    (\jobname.nodes) (w) file /NodeFile exch def 
  }}
\def\pscloseNodeFile{\pstVerb{ tx@NodeDict begin NodeFile closefile end }}
\define@boolkey[psset]{pst-node}[Pst@]{showNode}[true]{\ifPst@showNode\psopenNodeFile\fi}% write coors into the logfile
\define@boolkey[psset]{pst-node}[Pst@]{markNode}[true]{}% makr the node with its name
\define@boolkey[psset]{pst-node}[Pst@]{saveNodeCoors}[true]{}
\define@key[psset]{pst-node}{NodeCoorPrefix}[]{\def\psk@NodeCoorPrefix{#1}}% if empty it is N-<Name>.x|y
\psset[pst-node]{saveNodeCoors=false,showNode=false,markNode=false,NodeCoorPrefix=}% <NodeCoorPrefix<Name>x|y>
\def\pst@newnode#1#2#3#4{%
\pst@killglue
\leavevmode
\pst@getnode{#1}\pst@thenode
\pst@Verb{
  \ifPst@saveNodeCoors
    \ifx\relax#3\relax 0 0 \else gsave \pst@dict STV CP T end #3 \tx@UserCoor grestore \fi 
    \if$\psk@NodeCoorPrefix$
      /N-#1.y exch def
      /N-#1.x exch def
    \else
      /\psk@NodeCoorPrefix#1y exch def
      /\psk@NodeCoorPrefix#1x exch def
    \fi
  \fi
  \pst@nodedict
  {#3}
  \ifx\psk@name\relax false \else \psk@name true \fi
  \pst@thenode
  #2
  {#4}
  \ifPst@showNode 
  exch dup /NodeType ED 
  exch
   NodeType 10 eq {  % pnode type
    5 copy 
    cvlit aload pop
    20 string cvs (; )   6 2 roll % InitPnode
    20 string cvs (; )   7 2 roll % type
    20 string cvs (; )   8 2 roll %/N@Name
    20 string cvs (; )   9 2 roll % true/false
    cvlit dup length 2 eq 
      { aload pop exch 
        20 string cvs (; ) 11 2 roll 
        20 string cvs (, ) 12 2 roll  % x,y
        (\string\n)                   % add newline
        13 array astore concatstringarray 
      }
      { 255 string cvs (; ) 10 2 roll 
        (\string\n)                   % add newline
        11 array astore concatstringarray 
      } ifelse 
    NodeFile exch writestring 
  } if
  NodeType 14 eq {  % dotnode
    5 copy 
    /@@temp ED 
    @@temp  % to get X Y
    4 -1 roll cvlit pop
    ( OvalNodePos ) (; )  5 2 roll
    20 string cvs (; )   6 2 roll % type
    20 string cvs (; )   7 2 roll %/N@Name
    20 string cvs (; )   8 2 roll % true/false
    Y 20 string cvs (; ) 10 2 roll
    X 20 string cvs (, ) 12 2 roll
    (\string\n)                   % add newline
    13 array astore concatstringarray 
    tx@NodeDict begin NodeFile exch writestring end
  } if
  \fi
  \tx@NewNode
  end 
}%
\global\let\psk@name\relax%
\pstree@nodehook%
\global\let\pstree@nodehook\relax}
\let\pstree@nodehook\relax
\define@boolkey[psset]{pst-node}[Pst@]{nodealign}[true]{}
\psset[pst-node]{nodealign=false}

\def\pst@nodealign{%
\pst@dimg=\ht\pst@hbox
\advance\pst@dimg by -\dp\pst@hbox
\divide\pst@dimg by \tw@
\lower\pst@dimg}
\def\tx@InitPnode{InitPnode }
\def\pnode{\@ifnextchar[{\pnode@i}{\pnode@iii}}
\def\pnode@i[#1]{\@ifnextchar({\pnode@ii[#1]}{\pnode@ii[#1](0,0)}}
\def\pnode@ii[#1](#2)#3{%
  \pst@getcoor{#1}\pst@tempA%
  \pst@getcoor{#2}\pst@tempB%
  \pst@newnode{#3}{10}{\pst@tempA \pst@tempB 3 -1 roll add 3 1 roll add exch }{\tx@InitPnode}%
  \ifPst@showNode\psdot(#3)\uput[\ifx\psk@rot\@empty0\else\psk@rot\fi]{0}(#3){#3}\fi
  \ignorespaces}
\def\pnode@iii{\@ifnextchar({\pnode@}{\pnode@(0,0)}}
\def\pnode@(#1)#2{%
  \pst@@getcoor{#1}%
  \pst@newnode{#2}{10}{\pst@coor}{\tx@InitPnode}%
  \ifPst@showNode\psdot(#2)\uput[\ifx\psk@rot\@empty0\else\psk@rot\fi]{0}(#2){#2}\fi
  \ignorespaces}
\def\pnodes{\@ifnextchar[{\pnodes@i}{\pnodes@i[0,0]}}
\def\pnodes@i[#1]{\@ifnextchar({\psnodes@ii[#1]}{\pnodes@ii}}
\def\psnodes@ii[#1](#2)#3{%
  \pnode[#1](#2){#3}%
  \@ifnextchar({\psnodes@ii[#1]}{}%
}
\def\tx@InitCnode{InitCnode }
\def\cnode{\pst@object{cnode}}
\def\cnode@i{\@ifnextchar({\cnode@ii}{\cnode@ii(0,0)}}
\def\cnode@ii(#1)#2#3{%
  \leavevmode
  \hbox{%
    \use@par
    \pst@@getcoor{#1}%
    \pssetlength\pst@dimc{#2}%
    \pst@dimg=\psk@dimen\pslinewidth
    \advance\pst@dimc-\pst@dimg
    \advance\pst@dimc.5\pslinewidth
    \ifPst@nodealign
      \kern\pst@dimc
      \vrule width\z@ height \pst@dimc depth \pst@dimc
    \fi
    \pscircle@do(#1){#2}%
    \pst@newnode{#3}{11}{\pst@coor \pst@number\pst@dimc}{\tx@InitCnode}%
    \ifPst@nodealign\kern\pst@dimc\fi%
  }%
  \ignorespaces}
\def\Cnode{\pst@object{Cnode}}
\def\Cnode@i{\@ifnextchar({\Cnode@ii}{\Cnode@ii(0,0)}}
\def\Cnode@ii(#1)#2{\cnode@ii(#1){\psk@radius}{#2}}%
\def\cnodeput{\pst@object{cnodeput}}
\def\cnodeput@i{\@ifnextchar({\cnodeput@iii}{\cnodeput@ii}}
\def\cnodeput@ii#1{%
  \addto@par{rot={#1}}%
  \@ifnextchar({\cnodeput@iii}{\cnodeput@iii(\z@,\z@)}%
}
\def\cnodeput@iii(#1)#2{%
  \pst@killglue
  \@fixedradiusfalse
  \def\pst@nodehook{\cnodeput@iv{#2}}%
  \pst@makebox{\cput@v{#1}}%
}
\def\cnodeput@iv#1{%
  \pst@newnode{#1}{11}{\pscirclebox@iv \pst@number\pslinewidth add}{\tx@InitCnode}%
  \global\let\pst@nodehook\relax
  \ignorespaces
}
\def\Cnodeput{\pst@object{Cnodeput}}
\def\Cnodeput@i{\@ifnextchar({\Cnodeput@iii}{\Cnodeput@ii}}
\def\Cnodeput@ii#1{%
  \addto@par{rot={#1}}%
  \@ifnextchar({\Cnodeput@iii}{\Cnodeput@iii(\z@,\z@)}}
\def\Cnodeput@iii(#1)#2{%
  \pst@killglue
  \@fixedradiustrue
  \def\pst@nodehook{\Cnodeput@iv{#2}}%
  \pst@makebox{\cput@v{#1}}%
}
\def\Cnodeput@iv#1{%
  \pst@newnode{#1}{11}{%
    \pst@number{\wd\pst@hbox} 2 div \pst@number\pst@dima % x y
    \pst@number\pst@dimb \pst@number\pslinewidth \psk@dimen .5 sub mul sub }% r
       {\tx@InitCnode}%
  \global\let\pst@nodehook\relax}
\def\circlenode{\pst@object{circlenode}}
\def\circlenode@i#1{\pst@makebox{\circlenode@ii{#1}}}
\def\circlenode@ii#1{%
  \begingroup
  \pst@useboxpar
  \setbox\pst@hbox=\hbox{%
    \cnodeput@iv{#1}%
    \pscirclebox@iii
    \box\pst@hbox}%
  \ifPst@nodealign \psboxseptrue \fi
  \ifpsboxsep \pscirclebox@sep \fi
  \leavevmode
  \ifPst@nodealign\pst@nodealign\fi
  \box\pst@hbox
  \endgroup}
\def\Circlenode{\pst@object{Circlenode}}
\def\Circlenode@i#1{\pst@makebox{\Circlenode@ii{#1}}}
\def\Circlenode@ii#1{%
\begingroup
  \pst@useboxpar
  \pst@dima=\ht\pst@hbox
  \advance\pst@dima by -\dp\pst@hbox
  \divide\pst@dima by \tw@
  \pssetlength\pst@dimb\psk@radius
  \setbox\pst@hbox=\hbox{%
  \Cnodeput@iv{#1}%
  \pscircle(.5\wd\pst@hbox,\pst@dima){\pst@dimb}%
  \box\pst@hbox}%
  \ifPst@nodealign \psboxseptrue \fi
  \ifpsboxsep \psCirclebox@sep \fi
  \leavevmode
  \ifPst@nodealign\pst@nodealign\fi
  \box\pst@hbox
  \endgroup}
\def\tx@GetRnodePos{GetRnodePos }
\def\tx@InitRnode{InitRnode }
\def\psnode{\pst@object{psnode}}
\def\psnode@i{\@ifnextchar(\psnode@ii{\psnode@ii(0,0)}}
\def\psnode@ii(#1)#2#3{%    #1: coordinates, #2: node name,  #3 contents
  \rput(#1){\rnode{#2}{#3}}}
\def\rnode{\@ifnextchar[{\rnode@i}{\def\pst@par{}\rnode@ii}}
\def\rnode@i[#1]{\def\pst@par{ref=#1}\rnode@ii}
\def\rnode@ii#1{\pst@makebox{\rnode@iii\rnode@iv{#1}}}
\def\rnode@iii#1#2{%
\leavevmode
\begingroup
\pst@useboxpar
#1%
\ifPst@nodealign\lower\pst@dimb\fi
\hbox{%
\pst@newnode{#2}{16}{%
\pst@number{\ht\pst@hbox}%
\pst@number{\dp\pst@hbox}%
\pst@number{\wd\pst@hbox}%
\pst@number\pst@dima%
\pst@number\pst@dimb}%
{\tx@InitRnode}%
\box\pst@hbox}%
\endgroup}
\def\rnode@iv{%
\pst@dima=\psk@xref\wd\pst@hbox
\ifx\psk@yref\relax
\pst@dimb=\z@
\else
\pst@dimb=\ht\pst@hbox
\advance\pst@dimb\dp\pst@hbox
\pst@dimb=\psk@yref\pst@dimb
\advance\pst@dimb-\dp\pst@hbox
\fi}
\define@key[psset]{pst-node}{href}[0]{\pst@checknum{#1}\psk@href}
\psset[pst-node]{href=0}
\define@key[psset]{pst-node}{vref}[0.7ex]{\def\psk@vref{#1}}
\psset[pst-node]{vref=0.7ex}
\def\Rnode{\pst@object{Rnode}}
\def\Rnode@i#1{\pst@makebox{\rnode@iii\Rnode@ii{#1}}}
\def\Rnode@ii{%
\use@par
\pst@dima=\psk@href\wd\pst@hbox
\advance\pst@dima\wd\pst@hbox
\divide\pst@dima 2
\pssetlength\pst@dimb{\psk@vref}}
\def\tx@DiaNodePos{DiaNodePos }
\def\dianode{\pst@object{dianode}}
\def\dianode@i#1{\pst@makebox{\dianode@ii{#1}}}
\def\dianode@ii#1{%
\begingroup
\pst@useboxpar
\psdiabox@iii
\setbox\pst@hbox=\hbox{%
\pst@newnode{#1}{14}{}{%
/X \pst@number\pst@dima def
/Y \pst@number\pst@dimb def
/w \pst@number\pst@dimc 2 mul def
/h \pst@number\pst@dimd 2 mul def
/NodePos { \tx@DiaNodePos } def}%
\box\pst@hbox}%
\ifPst@nodealign\psboxseptrue\fi
\ifpsboxsep\psdiabox@sep\fi
\leavevmode
\ifPst@nodealign\lower\pst@dimb\fi
\box\pst@hbox
\endgroup}
\def\tx@TriNodePos{TriNodePos }
\def\tx@InitTriNode{InitTriNode }
\def\trinode{\pst@object{trinode}}
\def\trinode@i#1{\pst@makebox{\trinode@ii{#1}}}
\def\trinode@ii#1{%
  \begingroup%
  \pst@useboxpar%
  \pstribox@iii
  \setbox\pst@hbox=\hbox{%
    \pst@newnode{#1}{14}{}{
      \pst@number\pst@dimc
      \pst@number\pst@dimd
      \ifodd\psk@trimode
        exch
        \pst@number\pst@dima
      \else
        \pst@number\pst@dimb
      \fi
      \psk@trimode
      \pst@number{\wd\pst@hbox}
      \pst@number{\ht\pst@hbox}
      \pst@number{\dp\pst@hbox}
      \tx@InitTriNode
    }%
    \box\pst@hbox%
  }%
  \ifPst@nodealign\psboxseptrue\fi
  \ifpsboxsep\pstribox@sep\fi
  \leavevmode
  \ifPst@nodealign\lower\pst@tempa\fi
  \box\pst@hbox%
  \endgroup}
\def\tx@OvalNodePos{OvalNodePos }
\def\ovalnode{\pst@object{ovalnode}}
\def\ovalnode@i#1{\pst@makebox{\ovalnode@ii{#1}}}
\def\ovalnode@ii#1{%
\begingroup
\pst@useboxpar
\psovalbox@iii
\setbox\pst@hbox=\hbox{%
\pst@newnode{#1}{14}{}{%
/X \pst@number\pst@dima def
/Y \pst@number\pst@dimb def
/w \pst@number\pst@dimc def
/h \pst@number\pst@dimd def
/NodePos { \tx@OvalNodePos } def}%
\unhbox\pst@hbox}%
\ifPst@nodealign\psboxseptrue\fi
\ifpsboxsep\psovalbox@sep\fi
\leavevmode
\ifPst@nodealign\lower\pst@dimb\fi
\box\pst@hbox
\endgroup}
\def\dotnode{\pst@object{dotnode}}
\def\dotnode@i{\@ifnextchar({\dotnode@ii}{\dotnode@ii(\z@,\z@)}}
\def\dotnode@ii(#1)#2{%
  \leavevmode
  \hbox{%
    \use@par
    \pst@@getcoor{#1}%
    \pst@getdotsize
    \pstree@nodehook
    \ifPst@nodealign
      \pst@dima=\pst@dimg
      \kern\pst@dima
      \vrule width\z@ height \pst@dimh depth \pst@dimh
    \fi
    \pst@newnode{#2}{14}{}{
      \pst@coor
      /Y exch def /X exch def
      /w \pst@number\pst@dimg def
      /h \pst@number\pst@dimh def
      /NodePos { \tx@OvalNodePos } def}%
    \psdot@ii(#1)%
    \ifPst@nodealign\kern\pst@dima\fi}%
  \ifPst@markNode\uput[\ifx\psk@rot\@empty0\else\psk@rot\fi]{0}(#2){#2}\fi
  \ignorespaces}
\def\dotnodes{\pst@object{dotnodes}}
\def\dotnodes@i{\use@par\dotnodes@ii}
\def\dotnodes@ii(#1)#2{%
  \dotnode(#1){#2}%
  \@ifnextchar(\dotnodes@ii{\def\pst@par{}}}
\define@key[psset]{pst-node}{framesize}{\pst@expandafter\psset@@framesize{#1} \@nil}
\def\psset@@framesize#1 #2\@nil{%
  \pssetlength\pst@dimg{#1}%
  \divide\pst@dimg2
  \edef\psk@framewidth{\pst@number\pst@dimg}%
  \ifx\@empty#2\@empty
    \let\psk@frameheight\psk@framewidth
  \else
    \pssetlength\pst@dimg{#2}%
    \divide\pst@dimg2
    \edef\psk@frameheight{\pst@number\pst@dimg}%
  \fi}
\psset[pst-node]{framesize=10pt}
\def\fnode{\pst@object{fnode}}
\def\fnode@i{\@ifnextchar({\fnode@ii}{\fnode@ii(\z@,\z@)}}
\def\fnode@ii(#1)#2{%
  \leavevmode
  \pst@killglue
  \hbox{%
    \use@par%
    \begin@ClosedObj%
    \ifPst@nodealign
      \kern\psk@framewidth\p@
      \vrule width\z@ height \psk@frameheight\p@ depth \psk@frameheight\p@
      \edef\pst@coor{0 0 }%
    \else\pst@@getcoor{#1}\fi
    \pst@newnode{#2}{14}{}{
      \pst@coor
      /Y exch def /X exch def
      /d \psk@dimen .5 sub CLW mul neg def
      /r \psk@framewidth d add def
      /l r neg def
      /u \psk@frameheight d add def
      /d u neg def
      /NodePos { \tx@GetRnodePos } def}%
    \addto@pscode{
      /x2 \psk@framewidth CLW \psk@dimen mul sub def
      /y2 \psk@frameheight CLW \psk@dimen mul sub def
      \pst@coor 2 copy
      y2 sub /y1 ED
      x2 sub /x1 exch def
      y2 add /y2 exch def
      x2 add /x2 exch def
      \psk@cornersize
      1 index 0 eq { pop pop \tx@Rect } { \tx@OvalFrame } ifelse}%
    \def\pst@linetype{2}%
    \showpointsfalse%
    \end@ClosedObj%
    \ifPst@nodealign\kern\psk@framewidth\p@\fi}% end of \hbox
  \ignorespaces}
\define@key[psset]{}{XnodesepA}{\pst@getlength{#1}\psk@nodesepA\def\psk@nodeseptypeA{2 }}
\define@key[psset]{}{XnodesepB}{\pst@getlength{#1}\psk@nodesepB\def\psk@nodeseptypeB{2 }}
\define@key[psset]{}{Xnodesep}{%
    \pst@getlength{#1}\psk@nodesepA
    \let\psk@nodesepB\psk@nodesepA
    \def\psk@nodeseptypeA{2 }%
    \def\psk@nodeseptypeB{2 }}
\define@key[psset]{}{YnodesepA}{\pst@getlength{#1}\psk@nodesepA\def\psk@nodeseptypeA{1 }}
\define@key[psset]{}{YnodesepB}{\pst@getlength{#1}\psk@nodesepB\def\psk@nodeseptypeB{1 }}
\define@key[psset]{}{Ynodesep}{%
    \pst@getlength{#1}\psk@nodesepA
    \let\psk@nodesepB\psk@nodesepA
    \def\psk@nodeseptypeA{1 }%
    \def\psk@nodeseptypeB{1 }}
\define@key[psset]{pst-node}{nodesepA}[0pt]{\pst@getlength{#1}\psk@nodesepA \def\psk@nodeseptypeA{0 }}
\define@key[psset]{pst-node}{nodesepB}[0pt]{\pst@getlength{#1}\psk@nodesepB \def\psk@nodeseptypeB{0 }}
\define@key[psset]{pst-node}{nodesep}[0pt]{%
  \pst@getlength{#1}\psk@nodesepA
  \let\psk@nodesepB\psk@nodesepA
  \def\psk@nodeseptypeA{0 }%
  \def\psk@nodeseptypeB{0 }}
\psset[pst-node]{nodesep=0pt}
\define@key[psset]{pst-node}{armA}[10pt]{\pst@getlength{#1}\psk@armA \def\psk@armtypeA{0 }}
\define@key[psset]{pst-node}{armB}[10pt]{\pst@getlength{#1}\psk@armB \def\psk@armtypeB{0 }}
\define@key[psset]{pst-node}{arm}[10pt]{%
  \pst@getlength{#1}\psk@armA
  \let\psk@armB\psk@armA
  \def\psk@armtypeA{0 }%
  \def\psk@armtypeB{0 }}
\psset[pst-node]{arm=10pt}
\define@key[psset]{pst-node}{XarmA}[]{\pst@getlength{#1}\psk@armA \def\psk@armtypeA{1 }}
\define@key[psset]{pst-node}{XarmB}[]{\pst@getlength{#1}\psk@armB \def\psk@armtypeB{1 }}
\define@key[psset]{pst-node}{Xarm}{%
  \pst@getlength{#1}\psk@armA
  \let\psk@armB\psk@armA
  \def\psk@armtypeA{1 }%
  \def\psk@armtypeB{1 }}
\define@key[psset]{pst-node}{YarmA}[]{\pst@getlength{#1}\psk@armA \def\psk@armtypeA{2 }}
\define@key[psset]{pst-node}{YarmB}[]{\pst@getlength{#1}\psk@armB \def\psk@armtypeB{2 }}
\define@key[psset]{pst-node}{Yarm}[]{%
  \pst@getlength{#1}\psk@armA
  \let\psk@armB\psk@armA
  \def\psk@armtypeA{2 }%
  \def\psk@armtypeB{2 }}
\define@key[psset]{pst-node}{offsetA}[0pt]{\pst@getlength{#1}\psk@offsetA}
\define@key[psset]{pst-node}{offsetB}[0pt]{\pst@getlength{#1}\psk@offsetB}
\define@key[psset]{pst-node}{offset}[0pt]{\pst@getlength{#1}\psk@offsetA\let\psk@offsetB\psk@offsetA}
\psset[pst-node]{offset=0pt}
\define@key[psset]{pst-node}{angleA}[0]{\pst@getangle{#1}\psk@angleA}
\define@key[psset]{pst-node}{angleB}[0]{\pst@getangle{#1}\psk@angleB}%
\define@key[psset]{pst-node}{angle}[0]{%
  \pst@getangle{#1}\psk@angleA
  \let\psk@angleB\psk@angleA}
\psset[pst-node]{angle=0}
\define@key[psset]{pst-node}{arcangleA}[8]{\pst@getangle{#1}\psk@arcangleA}
\define@key[psset]{pst-node}{arcangleB}[8]{\pst@getangle{#1}\psk@arcangleB}%
\define@key[psset]{pst-node}{arcangle}[8]{%
  \pst@getangle{#1}\psk@arcangleA
  \let\psk@arcangleB\psk@arcangleA}
\psset[pst-node]{arcangle=8}
\define@key[psset]{pst-node}{ncurvA}[0.67]{\pst@checknum{#1}\psk@ncurvA}
\define@key[psset]{pst-node}{ncurvB}[0.67]{\pst@checknum{#1}\psk@ncurvB}%
\define@key[psset]{pst-node}{ncurv}[0.67]{\pst@checknum{#1}\psk@ncurvA\let\psk@ncurvB\psk@ncurvA}
\psset[pst-node]{ncurv=0.67}
\def\tx@GetCenter{GetCenter }
\def\tx@XYPos{XYPos }
\def\tx@GetEdge{GetEdge }
\def\tx@AddOffset{AddOffset }
\def\tx@GetEdgeA{GetEdgeA }
\def\tx@GetEdgeB{GetEdgeB }
\def\tx@GetArmA{GetArmA }
\def\tx@GetArmB{GetArmB }
\def\check@arrow#1#2{%
  \check@@arrow#2-\@nil
  \if@pst\addto@par{arrows=#2}\def\next{#1}%
  \else\def\next{#1{#2}}\fi
  \next}
\def\check@@arrow#1-#2\@nil{%
\ifx\@nil#2\@nil\@pstfalse\else\@psttrue\fi}
\def\tx@InitNC{InitNC }
\def\nc@object#1#2#3#4#5{%
  \csname begin@#1Obj\endcsname
  \showpointsfalse
  \pst@getnode{#2}\pst@tempa
  \pst@getnode{#3}\pst@tempb
  \gdef\npos@default{#4 }%
  \addto@pscode{%
    /NCLW CLW def
    \pst@nodedict
    \psk@offsetA
    \psk@offsetB neg
    \psk@nodesepA
    \psk@nodesepB
    \psk@nodeseptypeA
    \psk@nodeseptypeB
    \pst@tempa
    \pst@tempb
    \tx@InitNC { #5 } if
    end }%
  \def\use@pscode{%
    \pst@Verb{gsave \tx@STV newpath \pst@code\space grestore}%
    \gdef\pst@code{}}%
  \csname end@#1Obj\endcsname
  \pst@shortput}
\def\npos@default{.5 }
\def\pc@object#1{%
  \@ifnextchar({\pc@@object#1}{\pst@getarrows{\pc@@object#1}}}
\def\pc@@object#1(#2)(#3){%
  \pnode(#2){@@A}\pnode(#3){@@B}%
  #1{@@A}{@@B}}
\def\tx@LPutLine{LPutLine }
\def\tx@LPutLines{LPutLines }
\def\tx@BezierMidpoint{BezierMidpoint }
\def\tx@HPosBegin{HPosBegin }
\def\tx@HPosEnd{HPosEnd }
\def\tx@HPutLine{HPutLine }
\def\tx@HPutLines{HPutLines }
\def\tx@VPosBegin{VPosBegin }
\def\tx@VPosEnd{VPosEnd }
\def\tx@VPutLine{VPutLine }
\def\tx@VPutLines{VPutLines }
\def\tx@HPutCurve{HPutCurve }
\def\tx@NCCoor{NCCoor }
\def\tx@NCLine{NCLine }
\def\ncline{\pst@object{ncline}}
\def\ncline@i{\check@arrow{\ncline@ii}}
\def\ncline@ii#1#2{\nc@object{Open}{#1}{#2}{.5}{\tx@NCLine}}
\def\pcline{\pst@object{pcline}}
\def\pcline@i{\pc@object\ncline@ii}
\def\ncLine{\pst@object{ncLine}}
\def\ncLine@i{\check@arrow{\ncLine@ii}}
\def\ncLine@ii#1#2{\nc@object{Open}{#1}{#2}{.5}%
{\tx@NCLine /LPutPos { xB yB xA yA \tx@LPutLine } def}}
\def\tx@NCLines{NCLines }
\def\nclines{\pst@object{nclines}}
\def\nclines@i{\check@arrow\nclines@ii}
\def\nclines@ii#1#2{%
\begingroup
\use@par
\def\pst@aftercoors{\nclines@iii{#1}{#2}}%
\def\pst@coors{}%
\pst@@getcoors}
\def\nclines@iii#1#2{%
\nc@object{Open}{#1}{#2}{.5}{%
tx@Dict begin \psline@iii pop end
mark \pst@coors \tx@NCLines}%
\endgroup
\ignorespaces}
\def\tx@NCCurve{NCCurve }
\def\nccurve{\pst@object{nccurve}}
\def\nccurve@i{\check@arrow{\nccurve@ii}}
\def\nccurve@ii#1#2{\nc@object{Open}{#1}{#2}{.5}{%
  /AngleA \psk@angleA\space def /AngleB \psk@angleB\space def
  \psk@ncurvB\space \psk@ncurvA\space
  \tx@NCCurve}}
\def\pccurve{\pst@object{pccurve}}
\def\pccurve@i{\pc@object\nccurve@ii}
\def\ncarc{\pst@object{ncarc}}
\def\ncarc@i{\check@arrow{\ncarc@ii}}
\def\ncarc@ii#1#2{\nc@object{Open}{#1}{#2}{.5}{%
  yB yA sub xB xA sub \tx@Atan dup
  \psk@arcangleA\space add /AngleA exch def
  \psk@arcangleB\space sub 180 add /AngleB exch def
  \psk@ncurvB\space \psk@ncurvA\space
  \tx@NCCurve}}
\def\pcarc{\pst@object{pcarc}}
\def\pcarc@i{\pc@object\ncarc@ii}
\def\tx@NCAngles{NCAngles }
\define@boolkey[psset]{pst-node}[Pst@]{pcRef}[true]{}
\psset[pst-node]{pcRef=false}
\def\ncangles{\pst@object{ncangles}}
\def\ncangles@i{\check@arrow{\ncangles@ii}}
\def\ncangles@ii#1#2{%
  \nc@object{Open}{#1}{#2}{1.5}{\ncangles@iii \tx@NCAngles}}
\def\ncangles@iii{
  tx@Dict begin \psline@iii pop end
  /AngleA \psk@angleA def
  /AngleB \psk@angleB def
  /ArmA \psk@armA \ifPst@pcRef 
    GetEdgeA yA yA1 sub dup mul xA xA1 sub dup mul add sqrt sub \fi def
  /ArmB \psk@armB def
  /ArmTypeA \psk@armtypeA def
  /ArmTypeB \psk@armtypeB def }
\def\pcangles{\pst@object{pcangles}}
\def\pcangles@i{\pc@object\ncangles@ii}
\def\tx@NCAngle{NCAngle }
\def\ncangle{\pst@object{ncangle}}
\def\ncangle@i{\check@arrow{\ncangle@ii}}
\def\ncangle@ii#1#2{%
\nc@object{Open}{#1}{#2}{1.5}{\ncangles@iii \tx@NCAngle}}
\def\pcangle{\pst@object{pcangle}}
\def\pcangle@i{\pc@object\ncangle@ii}
\def\tx@NCBar{NCBar }
\def\ncbar{\pst@object{ncbar}}
\def\ncbar@i{\check@arrow{\ncbar@ii}}
\def\ncbar@ii#1#2{\nc@object{Open}{#1}{#2}{1.5}{%
\ncangles@iii /AngleB \psk@angleA def \tx@NCBar}}
\def\pcbar{\pst@object{pcbar}}
\def\pcbar@i{\pc@object\ncbar@ii}
\define@key[psset]{pst-node}{lineAngle}[0]{%
  \ifdim#1pt=\z@\else\psset{armB=0.5}\fi
  \def\psk@lineAngle{#1}}%
\psset[pst-node]{lineAngle=0}%
\def\tx@NCDiag{NCDiag }
\def\ncdiag{\pst@object{ncdiag}}
\def\ncdiag@i{\check@arrow{\ncdiag@ii}}
\def\ncdiag@ii#1#2{%
  \nc@object{Open}{#1}{#2}{1.5}{\ncangles@iii \psk@lineAngle\space \tx@NCDiag}}
\def\pcdiag{\pst@object{pcdiag}}
\def\pcdiag@i{\pc@object\ncdiag@ii}
\def\tx@NCDiagg{NCDiagg }
\def\ncdiagg{\pst@object{ncdiagg}}
\def\ncdiagg@i{\check@arrow{\ncdiagg@ii}}
\def\ncdiagg@ii#1#2{%
  \nc@object{Open}{#1}{#2}{.5}{\ncangles@iii \psk@lineAngle\space \tx@NCDiagg}}
\def\pcdiagg{\pst@object{pcdiagg}}
\def\pcdiagg@i{\pc@object\ncdiagg@ii}
\def\tx@NCLoop{NCLoop }
\define@key[psset]{pst-node}{loopsize}{\pst@getlength{#1}\psk@loopsize}
\psset[pst-node]{loopsize=1cm}
\def\ncloop{\pst@object{ncloop}}
\def\ncloop@i{\check@arrow{\ncloop@ii}}
\def\ncloop@ii#1#2{%
\nc@object{Open}{#1}{#2}{2.5}%
{\ncangles@iii /loopsize \psk@loopsize def \tx@NCLoop}}
\def\pcloop{\pst@object{pcloop}}
\def\pcloop@i{\pc@object\ncloop@ii}
\def\tx@NCCircle{NCCircle }
\def\nccircle{\pst@object{nccircle}}
\def\nccircle@i{\check@arrow{\nccircle@ii}}
\def\nccircle@ii#1#2{%
\pssetlength\pst@dima{#2}%
\nc@object{Open}{#1}{#1}{.5}{%
/AngleA \psk@angleA def
/r \pst@number\pst@dima def
\tx@NCCircle \psarc@v end}}
\def\tx@NCBox{NCBox }
\def\ncbox{\pst@object{ncbox}}
\def\ncbox@i{\check@arrow{\ncbox@ii}}
\def\ncbox@ii#1#2{%
\def\pst@linetype{2}%
\nc@object{Closed}{#1}{#2}{.5}{%
tx@Dict begin \psline@iii pop end
\psk@boxheight \psk@boxdepth
\tx@NCBox}}
\def\pcbox{\pst@object{pcbox}}
\def\pcbox@i{\pc@object\ncbox@ii}
\def\tx@NCArcBox{NCArcBox }
\define@key[psset]{pst-node}{boxheight}[0.4cm]{\pst@getlength{#1}\psk@boxheight}
\define@key[psset]{pst-node}{boxdepth}[0.4cm]{\pst@getlength{#1}\psk@boxdepth}
\define@key[psset]{pst-node}{boxsize}[0.4cm]{%
  \pst@getlength{#1}\psk@boxheight%  
  \let\psk@boxdepth\psk@boxheight}
\psset[pst-node]{boxsize=0.4cm}
\def\ncarcbox{\pst@object{ncarcbox}}
\def\ncarcbox@i{\check@arrow{\ncarcbox@ii}}
\def\ncarcbox@ii#1#2{%
\def\pst@linetype{1}%
\nc@object{Closed}{#1}{#2}{.5}{%
\psk@arcangleA \psk@boxheight \psk@boxdepth \pst@number\pslinearc
\tx@NCArcBox}}
\def\pcarcbox{\pst@object{pcarcbox}}
\def\pcarcbox@i{\pc@object\ncarcbox@ii}
\def\tx@Tfan{Tfan }
\begingroup
\catcode`\:=13
\gdef\pst@activerot{\def:{\string:}}
\endgroup
\define@key[psset]{pst-node}{nrot}[0]{%
  \begingroup
  \pst@activerot
  \pst@expandafter{\@ifnextchar:{\psset@@nrot}{\psset@@rot}}{#1}\@nil
  \global\let\pst@tempg\psk@rot
  \endgroup
  \let\psk@nrot\pst@tempg}
\def\psset@@nrot:#1\@nil{%
  \psset@@rot#1\@nil
  \edef\psk@rot{NAngle \ifx\psk@rot\@empty\else\psk@rot add \fi}}
\psset[pst-node]{nrot=0}
\def\tx@LPutCoor{LPutCoor }
\def\tx@LPut{LPut }
\define@key[psset]{pst-node}{npos}[{}]{%
  \def\pst@tempa{#1}%
  \ifx\pst@tempa\@empty\def\psk@npos{\npos@default}\else\pst@checknum{#1}\psk@npos\fi}
\psset[pst-node]{npos=}
\def\ncput{\pst@object{ncput}}
\def\ncput@i{\pst@killglue\pst@makebox{\ncput@ii}}
\def\ncput@ii{%
  \begingroup%
  \use@par%
  \if@star\pst@starbox\fi%
  \pst@makesmall\pst@hbox%
  \pst@rotate\psk@nrot\pst@hbox%
  \ncput@iii%
  \endgroup%
  \pst@shortput}
\def\ncput@iii{%
  \leavevmode%
  \hbox{%
    \pst@Verb{
      \pst@nodedict
      /t \psk@npos def
      \tx@LPut
      end
      \tx@PutBegin}%
    \box\pst@hbox%
    \pst@Verb{\tx@PutEnd}}}
\def\naput{\pst@object{naput}}
\def\naput@i{\pst@killglue\pst@makebox{\naput@ii{NAngle 90 add}}}
\def\naput@ii#1{%
  \begingroup
  \use@par
  \if@star\pst@starbox\fi
  \def\psk@refangle{#1 }%
  \let\psk@rot\psk@nrot
  \pst@Verb{ 
    gsave  STV CP T /ps@refangle {#1 } def 
    /ps@rot { \psk@rot } def grestore }%ADDED (MJS)
  \uput@vii
  {exch pop add a \tx@PtoC h1 add exch w1 add exch }%
  {tx@Dict /NCLW known { NCLW add } if }%
  \ncput@iii
  \endgroup
  \pst@shortput}
\def\nbput{\pst@object{nbput}}
\def\nbput@i{\pst@killglue\pst@makebox{\naput@ii{NAngle 90 sub}}}
\define@key[psset]{pst-node}{tpos}[0.5]{%
  \pst@checknum{#1}\psk@tpos
  \ifdim\psk@tpos \p@<\z@
    \def\psk@tpos{.5}%
    \@pstrickserr{Bad `tpos' value: `#1'. Must be 0<tpos<1}\@ehpa
  \else
    \ifdim\psk@tpos \p@>\p@
      \def\psk@tpos{.5}%
      \@pstrickserr{Bad `tpos' value: `#1'. Must be 0<tpos<1}\@ehpa%
    \fi%
  \fi}
\psset[pst-node]{tpos=0.5}
\def\nlput{\pst@object{nlput}}
\def\nlput@i(#1)(#2)#3#4{%
  \begin@SpecialObj
  \psLDNode(#1)(#2){#3}{temp@lnput}
  \pcline[linestyle=none](#1)(temp@lnput)%
  \ncput[npos=1]{#4}%
  \end@SpecialObj}
\def\tvput{\pst@object{tvput}}
\def\tvput@i{\pst@makebox{\psput@tput{H}{1}}}
\def\tlput{\pst@object{tlput}}
\def\tlput@i{\pst@makebox{\psput@tput{H}{true}}}
\def\trput{\pst@object{trput}}
\def\trput@i{\pst@makebox{\psput@tput{H}{false}}}
\def\thput{\pst@object{thput}}
\def\thput@i{\pst@makebox{\psput@tput{V}{1}}}
\def\taput{\pst@object{taput}}
\def\taput@i{\pst@makebox{\psput@tput{V}{true}}}
\def\tbput{\pst@object{tbput}}
\def\tbput@i{\pst@makebox{\psput@tput{V}{false}}}
\def\tx@HPutAdjust{HPutAdjust }
\def\tx@VPutAdjust{VPutAdjust }
\def\psput@tput#1#2{%
  \begingroup
  \use@par
  \pst@tputmakesmall
  \leavevmode
  \hbox{%
    \pst@Verb{%
      \pst@nodedict
      /t \psk@tpos \pst@tposflip def
      tx@NodeDict /HPutPos known
        { #1PutPos }
        { CP /Y exch def /X exch def /NAngle 0 def /NCLW 0 def }
      ifelse
      /Sin NAngle sin def
      /Cos NAngle cos def
      /s \pst@number\pslabelsep NCLW add def
      /l \pst@number\pst@dima def
      /r \pst@number\pst@dimb def
      /h \pst@number\pst@dimc def
      /d \pst@number\pst@dimd def
      \ifnum1=0#2 \else
        /flag #2 def
        \csname tx@#1PutAdjust\endcsname
      \fi
      \tx@LPutCoor
      end
      \tx@PutBegin}%
    \box\pst@hbox
    \pst@Verb{\tx@PutEnd}}%
  \endgroup
  \pst@shortput}
\def\pst@tposflip{}
\def\pst@tputmakesmall{%
\pst@dima=\wd\pst@hbox
\divide\pst@dima 2
\pst@dimg=\psk@href\pst@dimg
\pst@dimb\pst@dima
\advance\pst@dima\pst@dimg % leftsize
\advance\pst@dimb-\pst@dimg % rightsize
\pst@dimd=\psk@vref\relax
\pst@dimc=\ht\pst@hbox
\advance\pst@dimc-\pst@dimd % height
\advance\pst@dimd\dp\pst@hbox % depth
\setbox\pst@hbox=\hbox to\z@{%
\kern-\pst@dima\vbox to\z@{\vss\box\pst@hbox\vskip-\pst@dimd}\hss}}
\def\MakeShortNab#1#2{%
  \def\pst@shortput@nab{%
    \def\pst@tempg{\next}%
    \ifx#1\next
      \let\pst@tempg\naput
    \else
      \ifx#2\next
        \let\pst@tempg\nbput
      \else
        \ifx\@sptoken\next
          \let\pst@tempg\pst@shortput
        \fi
      \fi
    \fi
    \pst@tempg}}
\MakeShortNab{^}{_}
\def\MakeShortTablr#1#2#3#4{%
  \def\pst@shortput@tablr{%
    \def\pst@tempg{\next}%
    \ifx#1\next
      \let\pst@tempg\taput
    \else
      \ifx#2\next
        \let\pst@tempg\tbput
      \else
        \ifx#3\next
          \let\pst@tempg\tlput
        \else
          \ifx#4\next
            \let\pst@tempg\trput
          \else
            \ifx\@sptoken\next
              \let\pst@tempg\pst@shortput
            \fi
          \fi
        \fi
      \fi
    \fi
    \pst@tempg}}
\MakeShortTablr{^}{_}{<}{>}
\def\MakeShortTab#1#2{%
  \def\pst@shortput@tab{%
    \def\pst@tempg{\next}%
    \ifx#1\next
      \def\pst@tempg{%
        \@nameuse{%
          t\ifodd\psk@treemode\ifpstreeflip b\else a\fi
          \else\ifpstreeflip r\else l\fi\fi put}}%
    \else
      \ifx#2\next
        \def\pst@tempg{%
          \@nameuse{%
            t\ifodd\psk@treemode\ifpstreeflip a\else b\fi
            \else\ifpstreeflip l\else r\fi\fi put}}%
      \else
        \ifx\@sptoken\next
          \let\pst@tempg\pst@shortput
        \fi
      \fi
    \fi
    \pst@tempg}}
\MakeShortTab{^}{_}
\define@key[psset]{pst-node}{shortput}[none]{%
  \def\pst@tempg{#1}%
  \ifx\pst@tempg\@none
    \let\pst@shortput\ignorespaces
  \else
    \@ifundefined{pst@shortput@#1}%
     {\@pstrickserr{Bad short put: `#1'}\@ehpa}%
     {\edef\pst@shortput{\noexpand\afterassignment\expandafter\noexpand
      \csname pst@shortput@#1\endcsname\noexpand\let\noexpand\next}}%
  \fi}
\psset[pst-node]{shortput=none}
\def\lput{\def\pst@par{}\pst@ifstar{\@ifnextchar[{\lput@i}{\lput@ii}}}
\def\lput@i[#1]{\addto@par{ref=#1}\lput@ii}
\def\lput@ii{\@ifnextchar({\lput@iv}{\lput@iii}}
\def\lput@iii#1{\addto@par{nrot=#1}\@ifnextchar({\lput@iv}{\ncput@i}}
\def\lput@iv(#1){\addto@par{npos=#1}\ncput@i}
\def\mput{\def\pst@par{}\pst@ifstar{\@ifnextchar[{\mput@i}{\ncput@i}}}
\def\mput@i[#1]{\addto@par{ref=#1}\ncput@i}
\def\Lput{\def\pst@par{}\pst@ifstar{\@ifnextchar[{\Lput@ii}{\Lput@i}}}
\def\Lput@i#1{\addto@par{labelsep=#1}\Lput@ii}
\def\Lput@ii[#1]{\addto@par{ref={#1}}\@ifnextchar({\Lput@iv}{\Lput@iii}}
\def\Lput@iii#1{\addto@par{nrot={#1}}\@ifnextchar({\Lput@iv}{\Lput@v}}
\def\Lput@iv(#1){\addto@par{npos=#1}\Lput@v}
\def\Lput@v{\pst@killglue\pst@makebox{\Lput@vi}}
\def\Lput@vi{%
\begingroup
\use@par
\if@star\pst@starbox\fi
\Rput@vi
\pst@makesmall\pst@hbox
\ifx\psk@rot\@empty\else\pst@rotate{ps@rot }\pst@hbox\fi% (MJS)
\ncput@iii
\endgroup
\pst@shortput}
\def\Mput{\def\pst@par{}\pst@ifstar{\@ifnextchar[{\Mput@ii}{\Mput@i}}}
\def\Mput@i#1{\addto@par{labelsep=#1}\Mput@ii}
\def\Mput@ii[#1]{\addto@par{ref={#1}}\Lput@v}
\def\aput@#1{\def\pst@par{}\pst@ifstar{\@ifnextchar[{\aput@i#1}{\aput@ii#1}}}
\def\aput@i#1[#2]{\addto@par{labelsep=#2}\aput@ii#1}
\def\aput@ii#1{\@ifnextchar({\aput@iv#1}{\aput@iii#1}}
\def\aput@iii#1#2{\addto@par{nrot=#2}\@ifnextchar({\aput@iv#1}{#1}}
\def\aput@iv#1(#2){\addto@par{npos=#2}#1}
\def\aput{\aput@\naput@i}
\def\bput{\aput@\nbput@i}
\def\Aput{\def\pst@par{}\pst@ifstar{\@ifnextchar[{\Aput@i}{\naput@i}}}
\def\Aput@i[#1]{\addto@par{labelsep=#1}\naput@i}
\def\Bput{\def\pst@par{}\pst@ifstar{\@ifnextchar[{\Bput@i}{\nbput@i}}}
\def\Bput@i[#1]{\addto@par{labelsep=#1}\nbput@i}
\def\node@coor#1;#2\@nil{%  for normal nodes (name)
  \pst@getnode{#1}\pst@tempg
  \edef\pst@coor{%
    \pst@nodedict
    tx@NodeDict \pst@tempg known
    \pslbrace \pst@tempg load \tx@GetCenter \psrbrace
    \pslbrace 0 0 \psrbrace ifelse
    end }}
\def\Node@coor[#1]#2;#3\@nil{%  for special nodes ([...]{node}node)
\begingroup
\psset{angle=0,#1}%   angle=0, to prevent problems if angle is set globally
\@ifnextchar\bgroup{\Node@@@coor}% we have an additional node [...]{node}
                   {\Node@@coor}#2\@nil% we have  [...]node
\endgroup
\let\pst@coor\pst@tempg}
\def\Node@@coor#1\@nil{%
\pst@getnode{#1}\pst@tempg
\xdef\pst@tempg{%
\pst@nodedict
tx@NodeDict \pst@tempg known
  { \psk@nodesepA \psk@angleA 
    \pst@tempg load \psk@nodeseptypeA \tx@GetEdge
    \psk@offsetA \psk@angleA \tx@AddOffset
    \pst@tempg load \tx@GetCenter
    3 -1 roll add 3 1 roll add exch }
  { CP } ifelse end }}
\def\Node@@@coor#1{%   [...]{#1}node
\pst@@getcoor{#1}%
\def\psk@angleA{%
  \pst@tempg load \tx@GetCenter \pst@coor
  3 -1 roll sub 3 1 roll sub neg \tx@Atan \psk@angleB add
  }%
\Node@@coor}
\def\nput{\pst@object{nput}}
\def\nput@i#1#2{\pst@killglue\pst@makebox{\nput@ii{#1}{#2}}}
\def\nput@ii#1#2{%
  \begingroup
  \use@par
  \if@star\pst@starbox\fi%
  \psset[pstricks]{refangle=#1}%
  \let\psk@angleA\psk@refangle
  \edef\psk@nodesepA{\pst@number\pslabelsep}%
  \def\psk@nodeseptypeA{0 }%
  \pslabelsep\z@
  \uput@vi
  \Node@@coor#2\@nil
  \let\pst@coor\pst@tempg
  \leavevmode
  \psput@special\pst@hbox
  \endgroup
  \ignorespaces}
\newcount\psrow
\newcount\pscol
\newcount\psmatrixcnt
\newskip\psrowsep
\newskip\pscolsep
\define@key[psset]{pst-node}{colsep}[1.5cm]{\pssetlength\pscolsep{#1}}
\define@key[psset]{pst-node}{rowsep}[1.5cm]{\pssetlength\psrowsep{#1}}
\psset[pst-node]{colsep=1.5cm}
\psset[pst-node]{rowsep=1.5cm}
\newif\ifpsmatrix
\ifx\mscount\@undefined\let\mscount\@multicnt\fi
\def\psmatrix{\begingroup{\ifnum0=`}\fi % Don't want to expand any &.
  \@ifnextchar[{\psmatrix@i}{\ifnum0=`{\fi}{}\psmatrix@ii}}
\def\psmatrix@i[#1]{%
  \ifnum0=`{\fi}{}%
  \psset{#1}%
  \psmatrix@ii}
\def\psmatrix@ii{%
  \KillGlue
  \edef\psm@beginmath{%
    \ifmmode$\m@th\ifinner\textstyle\else\displaystyle\fi\fi}%
  \edef\psm@endmath{\ifmmode$\fi}%
  \let\\\psm@cr
  \advance\psmatrixcnt by \@ne
  \def\psm@thenode{M-\the\psmatrixcnt-\the\psrow-\the\pscol}%
  \tabskip\z@
  \psrow=\@ne
  \pscol\z@
  \psset{shortput=tablr}%
  \leavevmode
  \vbox\bgroup\halign\bgroup&%
  \begingroup
  \global\advance\pscol by \@ne
  \csname psrowhook\romannumeral\psrow\endcsname
  \csname pscolhook\romannumeral\pscol\endcsname
  \psm@beginnode##\psm@endnode\endgroup
  \cr}
\def\endpsmatrix{%
  \crcr\egroup\unskip\egroup
  \endgroup}
\def\psm@cr{{\ifnum0=`}\fi\ps@ifnextchar[{\psm@@cr}{\psm@@@cr{}}}
\def\psm@@cr[#1]{\psm@@@cr{\vskip#1\relax}}
\def\psm@@@cr#1{%
  \ifnum0=`{\fi}{}\cr
  \noalign{%
  \global\advance\psrow 1
  \global\pscol\z@
  \vskip\psrowsep
  #1}}
\def\psm@beginnode{%
  \@ifnextchar\psm@endnode
    {\let\psm@endnode@i\relax\setbox\pst@hbox=\hbox{}}%
    {\pst@object{psm@beginnode}}}
\def\psm@beginnode@i{%
  \setbox\pst@hbox=\hbox\bgroup
  \psm@beginmath
  \begingroup
  \ignorespaces}
\def\psm@endnode@i{%
  \unskip
  \endgroup
  \psm@endmath
  \egroup
  \use@par
  \@psttrue}
\def\psm@endnode{%
  \@pstfalse
  \psm@endnode@i
  \ifnum\pscol>1\relax \pshskip\pscolsep \fi
  \psk@mnodesize
  \hfil
  \Pst@nodealigntrue
  \if@pst\csname mnode@\psk@mnode\endcsname
  \else\csname mnode@\psk@emnode\endcsname\fi
  \psk@mcol
  \psk@@mnodesize}
\def\psspan#1{\global\mscount#1\relax\pstloop\ifnum\mscount>\@ne\sp@n\repeat}
\def\pstloop#1\repeat{\gdef\pstiterate{#1\relax\expandafter\pstiterate\fi}%
  \pstiterate
  \let\pstiterate\relax}
\define@key[psset]{pst-node}{name}[\relax]{\pst@getnode{#1}\psk@name}
\let\psk@name\relax
\define@key[psset]{pst-node}{mcol}[c]{%
  \ifx r#1\relax\let\psk@mcol\relax\else
    \ifx l#1\relax\let\psk@mcol\hfill\else
    \let\psk@mcol\hfil\fi\fi}
\psset[pst-node]{mcol=c}
\define@key[psset]{pst-node}{mnodesize}[-1pt]{%
  \pssetlength\pst@dimg{#1}%
  \ifdim\pst@dimg<\z@
    \let\psk@mnodesize\relax
    \let\psk@@mnodesize\relax
  \else
    \edef\psk@mnodesize{\noexpand\hbox to\number\pst@dimg sp\noexpand\bgroup}%
    \let\psk@@mnodesize\egroup
  \fi}
\psset[pst-node]{mnodesize=-1pt}
\def\mnode@R{\rnode@iii\Rnode@ii{\psm@thenode}}
\def\mnode@r{\rnode@iii\rnode@iv{\psm@thenode}}
\def\mnode@oval{\ovalnode@ii{\psm@thenode}}
\def\mnode@tri{\trinode@ii{\psm@thenode}}
\def\mnode@dia{\dianode@ii{\psm@thenode}}
\def\mnode@C{{\Pst@nodealigntrue\cnode@ii(\z@,\z@){\psk@radius}{\psm@thenode}}}
\def\mnode@f{{\Pst@nodealigntrue\fnode@ii(\z@,\z@){\psm@thenode}}}
\def\mnode@circle{\circlenode@ii{\psm@thenode}}
\def\mnode@Circle{\Circlenode@ii{\psm@thenode}}
\def\mnode@p{\pnode(\z@,\z@){\psm@thenode}}
\def\mnode@dot{\dotnode@ii(\z@,\z@){\psm@thenode}}
\def\mnode@none{\box\pst@hbox}
\define@key[psset]{pst-node}{mnode}[R]{%
  \@ifundefined{mnode@#1}%
    {\@pstrickserr{\string\psmatrix\space node `#1' not defined.}\@ehpa}%
    {\edef\psk@mnode{#1}}}
\define@key[psset]{pst-node}{emnode}[none]{%
  \@ifundefined{mnode@#1}%
    {\@pstrickserr{\string\psmatrix\space node `#1' not defined.}\@ehpa}%
    {\edef\psk@emnode{#1}}}
\psset[pst-node]{mnode=R,emnode=none}
\def\nccoil{\pst@object{nccoil}}
\def\nccoil@i{\check@arrow{\nccoil@ii}}
\def\nccoil@ii#1#2{\nc@object{Open}{#1}{#2}{.5}{
  \tx@NCCoor
  tx@Dict begin
  4 2 roll
  \psk@coilwidth \pscoilheight
  \psk@coilarmA \psk@coilarmB
  \psk@coilaspect \psk@coilinc
  \pst@coildict \tx@Coil end
  end}%
}
\def\nczigzag{\pst@object{nczigzag}}
\def\nczigzag@i{\check@arrow{\nczigzag@ii}}
\def\nczigzag@ii#1#2{\nc@object{Open}{#1}{#2}{.5}{
  \tx@NCCoor
  tx@Dict begin
  4 2 roll
  \pscoilheight
  \psk@coilwidth
  \psk@coilarmA
  \psk@coilarmB
  \pst@coildict \tx@ZigZag end
  \psline@iii
  \tx@Line
  end}%
}
\def\psGetNodeCenter#1{ tx@NodeDict begin /N@#1 load GetCenter end % x y on stack in system coor
  \pst@number\psyunit div /#1.y exch def 	% /#1.y in user coor
  \pst@number\psxunit div /#1.x exch def }	% /#1.x in user coor
\def\psGetEdgeA#1#2{
  tx@NodeDict begin \psk@offsetA \psk@offsetB neg 
    \psk@nodesepA \psk@nodesepB 0 0 
    /N@#1 /N@#2 InitNC { NCCoor } if pop pop \tx@UserCoor end}
\def\psGetEdgeB#1#2{
  tx@NodeDict begin \psk@offsetA \psk@offsetB neg 
    \psk@nodesepA \psk@nodesepB 0 0 
    /N@#1 /N@#2 InitNC { NCCoor } if 4 2 roll pop pop \tx@UserCoor end}
\def\ncbarr{\pst@object{ncbarr}}
\def\ncbarr@i#1#2{%
  \begingroup
  \use@par%
  \psLNode(#1)(#2){0.5}{barr@tempNode}%
  \pst@dimc=\psk@angleA pt
  \pst@dimd=180pt
  \ifdim\pst@dimc=\z@\else\ifdim\pst@dimc=\pst@dimd\else\psset{angleA=0}\fi\fi
  \ncbar[arrows=-]{#1}{barr@tempNode}
  \ifdim\psk@angleA pt=\z@\relax
    \ncbar[angleA=180,angleB=180]{barr@tempNode}{#2}
  \else\ncbar[angleA=0,angleB=0]{barr@tempNode}{#2}\fi%
  \endgroup%
}
\def\psLNode(#1)(#2)#3#4{%
  \pst@getcoor{#1}\pst@tempA%
  \pst@getcoor{#2}\pst@tempB%
  \pnode(!
    \pst@tempA /YA exch \pst@number\psyunit div def
    /XA exch \pst@number\psxunit div def
    \pst@tempB /YB exch \pst@number\psyunit div def
    /XB exch \pst@number\psxunit div def
    /dx XB XA sub def
    /dy YB YA sub def
    XA dx #3\space mul add YA dy #3\space mul add){#4}}
\def\psLCNode(#1)#2(#3)#4#5{%
  \pst@getcoor{#1}\pst@tempA%
  \pst@getcoor{#3}\pst@tempB%
  \pnode(!
    \pst@tempA /YA exch \pst@number\psyunit div def
    /XA exch \pst@number\psxunit div def
    \pst@tempB /YB exch \pst@number\psyunit div def
    /XB exch \pst@number\psxunit div def
    XA #2\space mul XB #4\space mul add
    YA #2\space mul YB #4\space mul add){#5}}
\def\psLDNode(#1)(#2)#3#4{%  
  \pst@getcoor{#1}\pst@tempA%
  \pst@getcoor{#2}\pst@tempB%
  \pssetlength\pst@dimb{#3}%
  \pnode(!%
    \pst@tempA /YA exch \pst@number\psyunit div def
    /XA exch \pst@number\psxunit div def
    \pst@tempB /YB exch \pst@number\psyunit div def
    /XB exch \pst@number\psxunit div def
    /dx XB XA sub def
    /dy YB YA sub def
    /angle dy dx Atan def
    /linelength \pst@number\pst@dimb \pst@number\psunit div def
    XA linelength angle cos mul add YA linelength angle sin mul add ){#4}%
}
\def\psRelNode{\pst@object{psRelNode}}
\def\psRelNode@i(#1)(#2)#3#4{{% A - B - factor - node name
  \use@par
  \pst@getcoor{#1}\pst@tempA%
  \pst@getcoor{#2}\pst@tempB%
  \pnode(!
    \pst@tempA /YA exch \pst@number\psyunit div def
    /XA exch \pst@number\psxunit div def
    \pst@tempB /YB exch \pst@number\psyunit div def
    /XB exch \pst@number\psxunit div def
    /AlphaStrich \psk@angleA\space def
    /unit \pst@number\psyunit \pst@number\psxunit div def % yunit/xunit
    /dx XB XA sub  def
    /dy YB YA sub \ifPst@trueAngle\space unit mul \fi\space def
    /laenge dy dup mul dx dup mul add sqrt #3 mul def
    /Alpha dy dx atan def 
    /beta Alpha AlphaStrich add def
    laenge beta cos mul XA add
    laenge beta sin mul \ifPst@trueAngle\space unit div \fi\space YA add ){#4}%
}\ignorespaces}
\define@cmdkeys[psset]{pstricks-add}[PSTPSPNk@]{% Christophe Jorssen 2007
  blName,bcName,brName,
  clName,ccName,crName,
  tlName,tcName,trName}[]{}%
\psset[pstricks-add]{%
  blName=PSPbl,bcName=PSPbc,brName=PSPbr,
  clName=PSPcl,ccName=PSPcc,crName=PSPcr,
  tlName=PSPtl,tcName=PSPtc,trName=PSPtr}
\def\psDefPSPNodes{\def\pst@par{}\pst@object{psDefPSPNodes}}
\def\psDefPSPNodes@i{%
  \pst@killglue
  \begingroup
  \use@par
  \expandafter\psDefPSPNodes@ii\pic@coor}
\def\psDefPSPNodes@ii(#1)(#2)(#3){%
    \pnode(#1){PSPN@temp}\pnode([angle=45]PSPN@temp){\PSTPSPNk@blName}
    \pnode(#3){PSPN@temp}\pnode([angle=-135]PSPN@temp){\PSTPSPNk@trName}
    \pnode(\PSTPSPNk@blName|\PSTPSPNk@trName){\PSTPSPNk@tlName}
    \pnode(\PSTPSPNk@trName|\PSTPSPNk@blName){\PSTPSPNk@brName}
    \ncline[linestyle=none]{\PSTPSPNk@blName}{\PSTPSPNk@tlName}
    \ncput[npos=.5]{\pnode{\PSTPSPNk@clName}}
    \ncline[linestyle=none]{\PSTPSPNk@blName}{\PSTPSPNk@brName}
    \ncput[npos=.5]{\pnode{\PSTPSPNk@bcName}}
    \pnode(\PSTPSPNk@brName|\PSTPSPNk@clName){\PSTPSPNk@crName}
    \pnode(\PSTPSPNk@bcName|\PSTPSPNk@trName){\PSTPSPNk@tcName}
    \pnode(\PSTPSPNk@bcName|\PSTPSPNk@clName){\PSTPSPNk@ccName}
  \endgroup
  \ignorespaces}
\def\psDefBoxNodes#1#2{\rnode[tl]{#1:tl}{\rnode[Bl]{#1:Bl}{\rnode[tr]{#1:tr}{%
\rnode[bl]{#1:bl}{\rnode[Br]{#1:Br}{\rnode[br]{#1:br}{#2}}}}}}%
\pnode(!\psGetNodeCenter{#1:bl}
          \psGetNodeCenter{#1:tl} 
          #1:bl.x #1:tl.x add 2 div #1:bl.y #1:tl.y add 2 div ){#1:Cl}%
\pnode(!\psGetNodeCenter{#1:tr}
          \psGetNodeCenter{#1:br} 
          #1:tr.x #1:br.x add 2 div #1:tr.y #1:br.y add 2 div ){#1:Cr}%
\pnode(!\psGetNodeCenter{#1:Cl}
          \psGetNodeCenter{#1:Cr} 
          #1:Cl.x #1:Cr.x add 2 div #1:Cl.y #1:Cr.y add 2 div ){#1:C}%
\pnode(!\psGetNodeCenter{#1:Br}
          \psGetNodeCenter{#1:Bl} 
          #1:Br.x #1:Bl.x add 2 div #1:Br.y #1:Bl.y add 2 div ){#1:BC}%
\pnode(!\psGetNodeCenter{#1:tr}
          \psGetNodeCenter{#1:tl} 
          #1:tr.x #1:tl.x add 2 div #1:tr.y #1:tl.y add 2 div ){#1:tC}%
\pnode(!\psGetNodeCenter{#1:br}
          \psGetNodeCenter{#1:bl} 
          #1:br.x #1:bl.x add 2 div #1:br.y #1:bl.y add 2 div ){#1:bC}}%
\newcount\pst@args%used in several macros
\newcount\num@pts
\newcount\pst@argcnt% for use in parsing node expressions
\def\PST@root{}
\let\pst@next\relax
\def\my@tempA{}
\def\my@tempB{}
\def\my@tempC{}
\def\my@tempD{}
\def\my@next{}
\newif\if@paren%
\newif\if@equal%
\newif\if@colon%
\newif\ifshow
\def\plussign{+}\def\minussign{-}
\def\defaultvalue#1#2{%#1 is a command, #2 is a value, possibly a command
  \ifdefined#1\ifx#1\@empty\xdef#1{#2}\fi\else\xdef#1{#2}\fi}%
\def\testAlg#1|#2\@nil{%
\ifx\relax#2\relax%
   \let\my@next\psparnode\xdef\my@tempD{}%
\else%
   \let\my@next\algparnode\xdef\my@tempD{A}% algebraic
\fi}%
\def\trim #1{\expandafter\trim@\expandafter{#1 }#1}%
\def\trim@ #1{\trim@@ @#1 @ #1 @ @@}%
\def\trim@@ #1@ #2@ #3@@{\trim@@@\empty #2 @}%
\def\unbrace#1{#1}%
\unbrace{\def\trim@@@ #1 } #2@#3{\expandafter\def%
  \expandafter #3\expandafter {#1}}%
\def\hasparen#1(#2\@nil{%check if expression contains a (--call with \hasparen#1(\@nil
  \ifx\relax#2\relax \@parenfalse \else \@parentrue\fi}%
\def\hasequal#1=#2\@nil{%check if expression contains a =--call with \hasequal#1=\@nil
  \ifx\relax#2\relax \@equalfalse \else \@equaltrue\fi
  \hascolon#2:\@nil}%
\def\hascolon#1:#2\@nil{%check if expression contains a :--call with \hascolon#1:\@nil
\ifx\relax#2\relax \@colonfalse \else \@colontrue\fi}%
\def\equalwhat#1=#2:#3\@nil{{#2}{#3}}%
\def\parsenodexn#1(#2)#3\@nil{%
  \def\coeffA{#1}\edef\nodeA{#2}%
  \trim\coeffA%
  \ifx\nodeA\@empty\else%
    \pnode(#2){@@TMP}%
    \ifx\coeffA\@empty\def\coeffA{1}\else%
      \ifx\coeffA\plussign\def\coeffA{1}\else\ifx\coeffA\minussign\def\coeffA{-1}\fi\fi\fi% 
  \edef\cmd{\noexpand\psLCNode(@TMP\the\pst@argcnt){1}(@@TMP){\coeffA}{@TMP}}%
  \cmd%
  \advance\pst@argcnt by \@ne%
  \pnode(@TMP){@TMP\the\pst@argcnt}%
  \parsenodexn#3\@nil%
  \fi}%
\def\normalvec(#1)#2{%
  \psRelNodeVar(0,0)(#1)(0,1){#2}}%
\def\curvepnode#1#2#3{%
  \edef\my@tempA{#2}% x(t) y(t) expanded
  \expandafter\testAlg\my@tempA|\@nil\my@next {#1}{#2}{#3}}
\def\psparnode#1#2#3{%
  \pnode(!/t #1 def #2){#3}%
  \pnode(!/t #1 .001 sub def #2 
          /t #1 .001 add def 
           #2 3 -1 roll sub 3 1 roll sub neg 
           2 copy Pyth dup 3 1 roll div 3 1 roll div ){#3tang}}%unit tangent vector at t
\def\algparnode#1#2#3{%
  \pstVerb{tx@Dict begin /Func (#2) AlgParser cvx def end }
  \pnode(!/t #1 def Func){#3}
  \pnode(!/t #1 .001 sub def Func 
          /t #1 .001 add def 
          Func 3 -1 roll sub 3 1 roll sub neg 
          2 copy Pyth dup 3 1 roll div 3 1 roll div ){#3tang}%unit tangent vector at t
}%
\def\nodex#1{%
\expandafter\hasparen#1(\@nil%
\if@paren%it's an expression
  \pnode(0,0){@TMP0}%
  \pst@argcnt=0%
  \expandafter\parsenodexn#1()\@nil%
\else%
  \def\my@tempC{#1}%
  \ifx\my@tempC\@empty\pnode(0,0){@TMP}\else\pnode(#1){@TMP}\fi%
\fi}%\if@paren
\def\nodexn#1#2{%
\expandafter\hasparen#1(\@nil%%%      hv 20130917 use \expandafter
\if@paren%it's an expression
  \pnode(0,0){@TMP0}%
  \pst@argcnt=0%
  \parsenodexn#1()\@nil%
  \pnode(@TMP){#2}%
\else%
  \def\my@tempC{#1}%
  \ifx\my@tempC\@empty\pnode(0,0){#2}\else\pnode(#1){#2}\fi%
\fi}%\if@paren
\def\psxline{\pst@object{psxline}}%
\def\psxline@i{\@ifnextchar({\psxline@iii}{\psxline@ii}}%
\def\psxline@ii#1{%
\addto@par{arrows=#1}%
\psxline@iii}%
\def\psxline@iii(#1)#2#3{{%#1=basepoint, #2,#3 node expressions
\pst@killglue%
\use@par%
\nodexn{#2}{@TMP@a}%
\AplusB(#1)(@TMP@a){@TMP@A}%
\nodexn{#3}{@TMP@a}%
\AplusB(#1)(@TMP@a){@TMP@B}%
\psline(@TMP@A)(@TMP@B)%
}%
\ignorespaces}%
\def\curvepnodes{\pst@object{curvepnodes}}
\def\curvepnodes@i#1#2#3#4{{%optional [plotpoints=xx]
  \pst@killglue
  \use@par
  \edef\my@tempA{#3}% x(t) y(t) expanded
  \expandafter\testAlg\my@tempA|\@nil %
  \pstVerb{% 
	tx@Dict begin % so we can use definitions from tx@Dict
	/t0 #1 def
	/t1 #2 def  
	 t1 t0 sub end \psk@plotpoints div /dt exch def }%
  \pst@cntc=\psk@plotpoints\relax%\psk@plotpoints=plotpoints-1
  \advance\pst@cntc by \@ne\relax %=plotpoints
  \ifx\my@tempD\@empty\pstVerb{tx@Dict begin /Func (#3) cvx def end }%add tx@Dict
  \else\pstVerb{tx@Dict begin /Func (#3 ) AlgParser cvx def end }%
  \fi%
    \multido{\i=0+1}{\pst@cntc}{%
      \pnode(! /t #1 dt \i\space mul add def Func ){#4\i}}% remove t before Func
    \expandafter\xdef \csname #4nodecount\endcsname {\psk@plotpoints}%
    \ifnum\Pst@Debug>0 \typeout{Created nodes #40 .. #4\psk@plotpoints}\fi%
}\ignorespaces}%
\def\fnpnode{\pst@object{fnpnode}}
\def\fnpnode@i#1#2#3{{%optional [algebraic]
  \pst@killglue
  \use@par
  \ifPst@algebraic\pnode(*#1 {#2}){#3}\else\pnode(! /x #1 def x #2){#3}\fi
}\ignorespaces}%
\def\fnpnodes{\pst@object{fnpnodes}}
\def\fnpnodes@i#1#2#3#4{{%optional [algebraic]
\pst@killglue
\use@par
\pst@dima=#1pt \pst@dimb=#2pt \advance\pst@dimb -\pst@dima%
\pst@cnta=\psk@plotpoints \relax %=plotpoints-1
\def\PST@root{#4}
\divide\pst@dimb by \pst@cnta%plotpoint-1 intervals
\pst@cntc=\pst@cnta %
\advance\pst@cntc by 1 \relax %=plotpoints
\ifPst@algebraic 
  \multido{\i=0+1}{\pst@cntc}{\pnode(*{\pst@number\pst@dima} {#3}){#4\i}%  hv 20130713
  \advance\pst@dima \pst@dimb}%
\else
    \multido{\i=0+1}{\pst@cntc}{\pnode(!/x \pst@number\pst@dima\space def x #3){#4\i}%
  \advance\pst@dima \pst@dimb}%
\fi%
  \expandafter\xdef \csname \PST@root nodecount\endcsname {\the\pst@cnta}%
  \ifnum\Pst@Debug>0 \typeout{Created nodes #40 .. #4\the\pst@cnta}\fi%
}\ignorespaces}%
\def\AtoB(#1)(#2)#3{\psLCNodeVar(#1)(#2)(-1,1){#3}}
\def\AplusB(#1)(#2)#3{\psLCNodeVar(#1)(#2)(1,1){#3}}
\def\midAB(#1)(#2)#3{\psLCNodeVar(#1)(#2)(.5,.5){#3}}
\def\psnline{\pst@object{psnline}}%line of nodes
\def\psnline@i{\pst@getarrows{\psnline@ii}}
\def\psnline@ii(#1,#2)#3{{%
\pst@killglue%
\use@par%
\pst@cnta=#2 \relax\advance\pst@cnta by 1
\edef\@tmp{}%
\multido{\i=#1+1}{\pst@cnta}{\xdef\@tmp{\@tmp(#3\i)}}%
\expandafter\psline\@tmp}%
\ignorespaces}%
\def\psnpolygon{\pst@object{psnpolygon}}%polygon of nodes
\def\psnpolygon@i{\pst@getarrows{\psnpolygon@ii}}
\def\psnpolygon@ii(#1,#2)#3{{%
\pst@killglue%
\use@par%
\pst@cnta=#2 \relax\advance\pst@cnta by 1
\edef\@tmp{}%
\multido{\i=#1+1}{\pst@cnta}{\xdef\@tmp{\@tmp(#3\i)}}%
\expandafter\pspolygon\@tmp}%
\ignorespaces}%
\def\psncurve{\pst@object{psncurve}}%line of nodes
\def\psncurve@i{\pst@getarrows{\psncurve@ii}}
\def\psncurve@ii(#1,#2)#3{{%
\pst@killglue%
\use@par%
\pst@cnta=#2 \relax\advance\pst@cnta by 1
\edef\@tmp{}%
\multido{\i=#1+1}{\pst@cnta}{\xdef\@tmp{\@tmp(#3\i)}}%
\expandafter\pscurve\@tmp}%
\ignorespaces}%
\def\psnccurve{\pst@object{psnccurve}}%line of nodes
\def\psnccurve@i{\pst@getarrows{\psnccurve@ii}}
\def\psnccurve@ii(#1,#2)#3{{%
\pst@killglue%
\use@par%
\pst@cnta=#2 \relax\advance\pst@cnta by 1
\xdef\@tmp{}%
\multido{\i=#1+1}{\pst@cnta}{\xdef\@tmp{\@tmp(#3\i)}}%
\expandafter\psccurve\@tmp}%
\ignorespaces}%
\def\shownode(#1){%display node user coords in console
  \pst@killglue%
  \pstVerb{% 
    gsave tx@Dict begin %
    tx@NodeDict /N@#1 known { % known node
      /tmpar [(Node #1: ) <28> () (, ) () <29>] def %
      /str 12 string def 
      STV CP T \psGetNodeCenter{#1}\space 
      tmpar 2 #1.x str cvs put 
      /str 12 string def 
      tmpar 4 #1.y str cvs put 
      tmpar concatstringarray = }%
    {% not known
      (Node #1: (NOT KNOWN)) = %
    } ifelse %
    end grestore }%
  \ignorespaces}%
\def\pnodes@ii#1{\getnodelist{#1}{}}
\def\getnodelist#1#2{%
\pst@args=0 \relax%
\def\PST@root{#1}%
\def\pst@next{#2}% command to perform after reading list
\getnext@Node}%
\def\getnext@Node{\@ifnextchar({\getnext@Node@i}%
  {\advance\pst@args by \m@ne \expandafter\xdef \csname \PST@root nodecount\endcsname {\the\pst@args}
  \ifnum\Pst@Debug>0 \typeout{Created nodes \PST@root0 .. \PST@root\the\pst@args}\fi% 
  \pst@next}%
}%
\def\getnext@Node@i(#1){%
\pnode(#1){\PST@root\the\pst@args}%
\advance\pst@args by \@ne\relax%
\getnext@Node}%
\def\psLCNodeVar(#1)(#2)(#3)#4{%
\pst@getcoor{#1}\my@tempA%
\pst@getcoor{#2}\my@tempB%
\pnode(#3){tmpLCn@de}%
\pnode(!%
  \my@tempA /YA exch \pst@number\psyunit div def
  /XA exch \pst@number\psxunit div def
  \my@tempB /YB exch \pst@number\psyunit div def
  /XB exch \pst@number\psxunit div def %stack now empty
  \psGetNodeCenter{tmpLCn@de}\space
  XA tmpLCn@de.x mul XB tmpLCn@de.y mul add
  YA tmpLCn@de.x mul YB tmpLCn@de.y mul add){tmpLCn@deA}%
\pnode(tmpLCn@deA){#4}%
}%
\def\psRelNodeVar{\pst@object{psRelNodeVar}}
\def\psRelNodeVar@i(#1)(#2)(#3)#4{{% A - B - factor;angle - node name
  \use@par
  \pst@getcoor{#1}\my@tempA%
  \pst@getcoor{#2}\my@tempB%
   \pnode(#3){tmpn@de}%
\pnode(!
  /unit \pst@number\psyunit \pst@number\psxunit div def % yunit/xunit
    \my@tempA /YA exch \pst@number\psyunit div def
    /XA exch \pst@number\psxunit div def
    \my@tempB /YB exch \pst@number\psyunit div YA sub 
    \ifPst@trueAngle\space unit mul \fi\space def
    /XB exch \pst@number\psxunit div XA sub def
    \psGetNodeCenter{tmpn@de}
    XB tmpn@de.x mul YB tmpn@de.y mul sub
    YB tmpn@de.x mul XB tmpn@de.y mul add
    \ifPst@trueAngle\space unit div \fi\space 
   YA add exch XA add exch %x, y coords on stack
    ){#4}%
}}
\def\psRelLineVar{\pst@object{psRelLineVar}}
\def\psRelLineVar@i{\@ifnextchar({\psRelLineVar@iii}{\psRelLineVar@ii}}
\def\psRelLineVar@ii#1{%
  \addto@par{arrows=#1}%
  \psRelLineVar@iii}
\def\psRelLineVar@iii(#1)(#2)(#3)#4{{%
  \pst@killglue
  \use@par
  \psRelNodeVar(#1)(#2)(#3){#4}%
  \psline(#1)(#4)%
}\ignorespaces}
\def\rhombus#1(#2)(#3)#4#5{% \rhombus{m}(B)(D){A}{C} 
\AtoB(#2)(#3){node@P}% P=BD
\pnode(! %compute angle and scale in PS
/tmp \psGetNodeCenter{node@P} node@P.x node@P.y 
Pyth 2 div def %tmp=half-length of BD
/ang tmp #1\space div Acos def %ang=angle from BD to BC & BA
#1\space tmp 2 mul div %scale factor s=m/BD
dup ang cos mul exch ang sin mul ){node@A1}% s cos(ang), s sin(ang)
\pnode(! \psGetNodeCenter{node@A1} node@A1.x node@A1.y neg ){node@A2}%reflect in x axis
\psRelNodeVar(#2)(#3)(node@A1){#4}%
\psRelNodeVar(#2)(#3)(node@A2){#5}%
}%
\def\psrline{\pst@object{psrline}}% relative lines
\def\psrline@i{\@ifnextchar({\psrline@iii}{\psrline@ii}}%
\def\psrline@ii#1{%
\addto@par{arrows=#1}%
\psrline@iii}%
\def\psrline@iii{%
\getnodelist{@tmpnode}{\psrline@iv}%
}%
\def\psrline@iv{%
   \ifnum\pst@args<0\else%do nothing
      \pnode(@tmpnode0){@tmpnodeB0}%
      \multido{\iA=1+1,\iB=0+1}{\pst@args}{%
      \AplusB(@tmpnodeB\iB)(@tmpnode\iA){@tmpnodeB\iA}}%
      \psrline@v%
   \fi%
}%
\def\psrline@v{{%finish up
  \pst@killglue%
  \use@par%
  \xdef\tmp{(@tmpnodeB0)}%
  \multido{\i=1+1}{\pst@args}%
{\xdef\tmp{\tmp(@tmpnodeB\i)}}%
\expandafter\psline\tmp%
}\ignorespaces}%
\def\polyIntersections#1#2(#3)(#4){%
\def\nodenameA{#1}\def\nodenameB{#2}%
\pnode(#3){P@A}\pnode(#4){P@B}%
\@ifnextchar({\polyIntersections@next}{\polyIntersections@ii}%
}%
\def\polyIntersections@ii#1#2{%
\def\root@node{#1}\num@pts=#2 \relax%
\polyIntersections@iii}% 
\def\polyIntersections@next{%read as many points as exist
\def\root@node{P@}\getnodelist{P@}{\num@pts=\pst@args \relax\polyIntersections@iii}%
}%
\def\polyIntersections@iii{%nodes are now XXX0....XXXn, n=num@pts
\pst@cnta=\num@pts \relax\advance\pst@cnta by 1 \relax%
\pstVerb{%
 /xarray \the\pst@cnta\space array def
 /yarray \the\pst@cnta\space array def  tx@Dict begin }%
\multido{\i=0+1}{\the\pst@cnta}{\pstVerb{ \psGetNodeCenter{\root@node\i} xarray \i\space \root@node\i.x put yarray \i\space \root@node\i.y put }}%
\pstVerb{ /tposmin 100 def /tnegmax -100 def %/argposmin 0 def /argposmax 0 def 
\psGetNodeCenter{P@B} \psGetNodeCenter{P@A} 
/dx P@B.x P@A.x sub def 
/dy P@B.y P@A.y sub def 
/lenAB dx dy Pyth def
/oldx xarray 0 get def /oldy yarray 0 get def 
1 1 \the\num@pts\space {/k exch def /newx xarray k get def /newy yarray k get def 
/ddx newx oldx sub def /ddy newy oldy sub def 
/det ddy dx mul ddx dy mul sub def
det abs lenAB ddx ddy Pyth mul .001 mul gt 
{/ac oldx P@A.x sub def /bd oldy P@A.y sub def 
 /tt  ac ddy mul bd ddx mul sub det div def %solve for t value at intersection
 /ss ac  dy mul bd dx mul sub det div def % solve for s value at intersection
ss 0 ge 
   {ss 1 le 
        {tt 0 lt {tt tnegmax gt {/tnegmax tt def} if } {tt tposmin lt {/tposmin tt def} if } ifelse }
    if } % ss 1 le
if }%ss 0 ge
 if %det>
 /oldx newx def /oldy newy def} for end }%
\pnode(! \psGetNodeCenter{P@A} \psGetNodeCenter{P@B} P@B.x P@A.x sub  tposmin mul P@A.x add  P@B.y P@A.y sub tposmin  mul P@A.y add ){\nodenameA}%
\pnode(! \psGetNodeCenter{P@A} \psGetNodeCenter{P@B} P@B.x P@A.x sub tnegmax mul P@A.x add P@B.y P@A.y sub tnegmax mul P@A.y add){\nodenameB}%
}%
\def\actualscale#1 #2 scale{% extract x-scale from, eg,  {2. 2. scale}
#1}
\def\psGetCenter#1{ tx@NodeDict begin /N@#1 load GetCenter end }% x y on stack in system coor
\def\ArrowNotch{\pst@object{ArrowNotch}}
\def\ArrowNotch@i#1#2#3#4{{%
\pst@killglue%
\use@par%
\def\inc{-1}%
\ifx#3<\def\inc{1}\fi% -1 means notch to left of arrowhead
\pstVerb{ 
    1 \psk@arrowinset\space sub \psk@arrowlength\space \psk@arrowsize\space  
    \pst@number\pslinewidth \space mul add  mul mul 
    \expandafter\actualscale\psk@arrowscale \space  mul 
    /hh exch def /hh1 hh .05 sub def }% PS variable hh contains dist from tip to notch of arrow, in pts
\def\root@node{#1}\num@pts=\csname\root@node nodecount\endcsname %
\pst@cntb=\num@pts \advance\pst@cntb by \@ne%actual node count
\pst@cnta=\num@pts \advance\pst@cnta by \thr@@%size of PS array
\pst@cntc=#2 \relax% index of center of circle
\ifnum\pst@cntc>\num@pts \pnode(0,0){#4}\else
\pstVerb{%
/PythSq { dup mul exch dup mul add } def
/PtSub {					%  xA yA xB yB
  3 -1 roll 		% xA xB yB yA
  sub neg		% xA xB yA-yB
  3 1 roll 		% yB-yA xA xB
  sub			% yB-yA xA-xB
  exch                     % xB-xA yA-yB
} def
  /xarray \the\pst@cnta\space array def
  /yarray \the\pst@cnta\space array def  
  tx@Dict begin }% end pstVerb
\multido{\i=0+1,\ib=1+1}{\the\pst@cntb}{\pnode(! \psGetCenter{\root@node\i}\space  % center on stack in system coords, not user coords
yarray \ib\space 3 -1 roll put xarray \ib\space 3 -1 roll put 0 0 ){@tmp}}% end multido
\pnode(! xarray 1 get dup yarray 1 get dup 3 1 roll % x1 y1 x1 y1
xarray 2 get yarray 2 get PtSub  % x1 y1 x1-x2 y1-y2
2 copy Pyth hh div 2 div dup % x1 y1 x1-x2 y1-y2 d d ,d->d/2*hh
3 1 roll % x1 y1 x1-x2 d y1-y2 d
div 3 1 roll div %x1 y1  (y1-y2)/d (x1-x2)/d
3 1 roll %x1  (x1-x2)/d y1  (y1-y2)/d
add 3 1 roll add %  y1-(y2-y1)/d x1-(x2-x1)/d
 xarray 0 3 -1 roll put yarray 0 3 -1 roll put %stack empty
 xarray length 2 sub /topnum exch def 
 xarray topnum get dup yarray topnum get dup 3 1 roll %xn yn xn yn
topnum 1 sub /topnum exch def xarray topnum get yarray topnum get % xn yn xn yn x(n-1) y(n-1)
3 -1 roll sub  neg 3 1 roll sub exch % xn yn (xn-x(n-1)) (yn-y(n-1))
2 copy Pyth hh div 2 div dup % xn yn (xn-x(n-1)) (yn-y(n-1)) d d (d->d/(2*h))
3 1 roll div 3 1 roll div %xn yn (xn-x(n-1))/d (yn-y(n-1))/d
3 -1 roll add 3 1 roll % y(n+1) x(n+1)
topnum 2 add /topnum exch def xarray topnum 3 -1 roll put yarray topnum 3 -1 roll put % empty
 /oldcindex \the\pst@cntc\space 1 add def %position in array
 xarray oldcindex get /xc exch def yarray oldcindex get /yc exch def
/inc \inc\space def %+1 for left facing arrow, else -1 
/cindex oldcindex def 
{cindex inc add /cindex exch def xarray cindex get xc sub yarray cindex get yc sub Pyth dup hh1 gt 
{ exit } if } loop % exit from loop with cindex the first index of an external point--dist on stack
 hh1 .1 add lt { xarray cindex get yarray cindex get } %else within segment
{ xarray cindex inc sub get dup yarray cindex inc sub get dup 4 -1 roll exch 
xarray cindex get yarray cindex get PtSub /dy1 exch def /dx1 exch def dx1 dy1 PythSq /Aterm exch def 
 2 copy xc yc PtSub % x(n-inc) y(n-inc) (x(n-inc)-xc) (y(n-inc)-yc)
 2 copy 2 copy 3 -1 roll mul 3 1 roll mul add hh dup mul sub % x(n-inc) y(n-inc) (y(n-inc)-yc) (x(n-inc)-xc) (x(n-inc)-xc)^2+(y(n-inc)-yc)^2-hh^2
 Aterm div /Cterm exch def  % x(n-inc) y(n-inc)  (y(n-inc)-yc) (x(n-inc)-xc) , Cterm=((x(n-inc)-xc)^2+(y(n-inc)-yc)^2-hh^2)/(Aterm) (<0)
 dx1 dy1 %  x(n-inc) y(n-inc) (y(n-inc)-yc)  (x(n-inc)-xc)  dx1 dy1
 4 1 roll mul 3 1 roll mul add Aterm div /Bterm exch def %   x(n-inc) y(n-inc) , Bterm=( (x(n-inc)-xc)*dx1+(y(n-inc)-yc)*dy1)/Aterm
 Bterm abs neg dup dup mul Cterm sub sqrt add dup /tval exch def
 dup dx1 dy1 4 1 roll mul 3 1 roll mul  % x(n-inc) y(n-inc) t*dx1 t*dy1
 PtSub } ifelse % x y screen coords of arrow notch now on stack---convert to user x y
 \pst@number\psyunit div exch \pst@number\psxunit div exch  %use coords now on stack
){#4}\fi%
\pstVerb { end } %tx@Dict
}\ignorespaces}%
\def\saveDataAsNodes#1#2{%  Filename NodePrefix
  \psLoopIndex=0\relax
  \typeout{Open file #1}%
  \openin7=#1
  \loop
    \read7 to \@Data
    \ifeof7\else
      \ifx\@Data\@empty
      \else
        \pnode(!\@Data){#2\the\psLoopIndex}%
        \typeout{#2\the\psLoopIndex -> \@Data}%
	\advance\psLoopIndex by 1
        \let\@oldData\@Data
      \fi
  \repeat
  \closein7
  \advance\psLoopIndex by -1
  \pnode(!\@oldData){#2Last}%  
}
\catcode`\@=\TheAtCode\relax
 
\csname PSTcoilsLoaded\endcsname
\let\PSTcoilsLoaded 
\ifx\PSTricksLoaded   \else\input pstricks.tex\fi
\ifx\PSTnodeLoaded    \else\fi
\ifx\PSTXKeyLoaded    \else\input pst-xkey \fi
\def\fileversion{1.07}
\def\filedate{2015/05/13}
\edef\TheAtCode{\the\catcode`\@}
\catcode`\@=11
\pst@addfams{pst-coil}
\pstheader{pst-coil.pro}
\edef\pst@theheaders{\pst@theheaders,pst-coil.pro}
\def\pst@CoilDict{tx@CoilDict begin }
\def\tx@CoilLoop  {\pst@CoilDict CoilLoop   end }
\def\tx@Coil      {\pst@CoilDict Coil       end }
\def\tx@AltCoil   {\pst@CoilDict AltCoil    end }
\def\tx@ZigZag    {\pst@CoilDict ZigZag     end }
\def\tx@ZigZagCirc{\pst@CoilDict ZigZagCirc end }
\def\tx@Sin       {\pst@CoilDict Sin        end }
\define@key[psset]{pst-coil}{coilwidth}[1cm]{\pst@getlength{#1}\psk@coilwidth}
\define@key[psset]{pst-coil}{coilheight}[1]{\pst@checknum{#1}\pscoilheight}
\define@key[psset]{pst-coil}{coilarmA}[0.5cm]{\pst@getlength{#1}\psk@coilarmA}
\define@key[psset]{pst-coil}{coilarmB}[0.5cm]{\pst@getlength{#1}\psk@coilarmB}
\define@key[psset]{pst-coil}{coilarm}[0.5cm]{%
  \pst@getlength{#1}\psk@coilarmA%
  \let\psk@coilarmB\psk@coilarmA}
\define@key[psset]{pst-coil}{coilaspect}[45]{\pst@getangle{#1}\psk@coilaspect}
\define@key[psset]{pst-coil}{coilinc}[10]{\pst@getangle{#1}\psk@coilinc}
\psset[pst-coil]{coilaspect=45,coilarm=.5cm,coilheight=1,coilwidth=1cm,coilinc=10}
\def\pscoil{\def\pst@par{}\pst@object{pscoil}}
\def\pscoil@i{\pst@getarrows\pscoil@ii}
\def\pscoil@ii(#1){\@ifnextchar({\pscoil@iii{1}(#1)}{\pscoil@iii{\z@}(0,0)(#1)}}
\def\pscoil@iii#1(#2)(#3){%
  \begin@OpenObj
  \pst@getcoor{#2}\pst@tempa
  \pst@getcoor{#3}\pst@tempb
  \pst@optcp{#1}\pst@tempa
  \addto@pscode{%
    \pst@tempa \pst@tempb
    \psk@coilwidth \pscoilheight
    \psk@coilarmA \psk@coilarmB
    \psk@coilaspect \psk@coilinc
    \tx@Coil }%
    \showpointsfalse
  \end@OpenObj}
\def\psCoil{\def\pst@par{}\pst@object{psCoil}}
\def\psCoil@i#1#2{%
  \begin@AltOpenObj
  \showpointsfalse
  \pst@getangle{#1}\pst@tempa
  \pst@getangle{#2}\pst@tempb
  \addto@pscode{%
    \pst@tempa
    \pst@tempb
    \psk@coilwidth
    \pscoilheight
    \psk@coilaspect
    \psk@coilinc
    \tx@AltCoil  
    \@nameuse{psls@\pslinestyle} }%
  \end@OpenObj}
\define@key[psset]{pst-coil}{bow}[0]{%
  \pst@getlength{#1}\psk@bow% 
  \pst@dimm=\psk@bow pt%
  \pst@absdim{\pst@dimm}{\pst@dimn}%
  \ifdim\pst@dimn<1pt \def\psk@bow{0}\fi}%
\psset[pst-coil]{bow=0}
\def\pszigzag{\def\pst@par{}\pst@object{pszigzag}}
\def\pszigzag@i{\pst@getarrows\pszigzag@ii}
\def\pszigzag@ii(#1){\@ifnextchar({\pszigzag@iii{1}(#1)}{\pszigzag@iii{\z@}(0,0)(#1)}}
\def\pszigzag@iii#1(#2)(#3){%
  \addbefore@par{bow=0}%
  \begin@OpenObj%
  \pst@getcoor{#2}\pst@tempA%
  \pst@getcoor{#3}\pst@tempB%
  \pst@optcp{#1}\pst@tempA%
  \addto@pscode{%
    \pst@tempA
    \pst@tempB
    \pscoilheight
    \psk@coilwidth
    \psk@coilarmA
    \psk@coilarmB 
    \ifdim\psk@bow pt=\z@ \tx@ZigZag \else \psk@bow\space \tx@ZigZagCirc \fi
    \psline@iii
    \tx@Line }%
  \end@OpenObj}
\def\nccoil{\pst@object{nccoil}}
\def\nccoil@i{\check@arrow{\nccoil@ii}}
\def\nccoil@ii#1#2{\nc@object{Open}{#1}{#2}{.5}{%
  \tx@NCCoor
  tx@Dict begin
  4 2 roll
  \psk@coilwidth \pscoilheight
  \psk@coilarmA \psk@coilarmB
  \psk@coilaspect \psk@coilinc
  \tx@Coil 
  end }}
\def\pccoil{\def\pst@par{}\pst@object{pccoil}}
\def\pccoil@i{\pc@object\nccoil@ii}
\def\nczigzag{\pst@object{nczigzag}}
\def\nczigzag@i{\check@arrow{\nczigzag@ii}}
\def\nczigzag@ii#1#2{\nc@object{Open}{#1}{#2}{.5}{%
  \tx@NCCoor
  tx@Dict begin
  4 2 roll
  \pscoilheight
  \psk@coilwidth
  \psk@coilarmA
  \psk@coilarmB
  \ifdim\psk@bow pt=\z@\tx@ZigZag\else\psk@bow\space\tx@ZigZagCirc\fi 
  \psline@iii
  \tx@Line
  end }}
\def\pczigzag{\def\pst@par{}\pst@object{pczigzag}}
\def\pczigzag@i{\pc@object\nczigzag@ii}
\def\pst@checkUnit#1#2{\expandafter\pst@checkUnit@i#1!!#2}
\def\pst@checkUnit@i{\@ifnextchar*%
  {\def\pst@roundValue{0 }\pst@checkUnit@ii}%
  {\def\pst@roundValue{-1 }\pst@checkUnit@iii**}}
\def\pst@checkUnit@ii*{\@ifnextchar*%
  {\def\pst@roundValue{1 }\pst@checkUnit@iii*}%
  {\pst@checkUnit@iii**}}
\def\pst@checkUnit@iii**#1!!#2{%
  \edef\ps@next{#1}%
  \ifx\ps@next\@empty\let\pst@num\z@%
  \else\expandafter\pst@@checknum\ps@next..\@nil%
  \fi%
  \ifnum\pst@num=\z@\pst@getlength{#1}{#2}\def\pst@relativePeriod{false }%
  \else%
    \def\pst@relativePeriod{true }%
    \edef#2{\ifnum\pst@num=\tw@-\fi\the\pst@cntg.%
    \expandafter\@gobble\the\pst@cnth\space}%
  \fi}
\define@key[psset]{pst-coil}{periods}[1]{\pst@checkUnit{#1}{\psk@periods}}
\define@key[psset]{pst-coil}{amplitude}[1]{\def\psk@amplitude{#1 }}
\define@key[psset]{pst-coil}{ppoints}[360]{\def\psk@ppoints{#1 }}
\define@key[psset]{pst-coil}{function}[sin]{\def\psk@function{#1 }}
\psset[pst-coil]{periods=1,amplitude=1,ppoints=360,function=sin}
\def\pssin{\pst@object{pssin}}
\def\pssin@i{\pst@getarrows\pssin@ii}
\def\pssin@ii(#1){\@ifnextchar({\pssin@iii{1}(#1)}{\pssin@iii{\z@}(0,0)(#1)}}
\def\pssin@iii#1(#2)(#3){%
  \begin@OpenObj
  \pst@getcoor{#2}\pst@tempa
  \pst@getcoor{#3}\pst@tempb
  \pst@optcp{#1}\pst@tempa
  \addto@pscode{%
    \pst@tempa \pst@tempb
    \psk@periods 
    \pst@relativePeriod 
    \pst@roundValue
    \psk@amplitude \pst@number\psyunit mul
    \psk@coilarmA \psk@coilarmB 
    \psk@ppoints
    { \psk@function }
    \tx@Sin
  }%
  \showpointsfalse%
  \end@OpenObj}
\def\ncsin{\pst@object{ncsin}}
\def\ncsin@i{\check@arrow{\ncsin@ii}}
\def\ncsin@ii#1#2{\nc@object{Open}{#1}{#2}{.5}{%
  \tx@NCCoor
  tx@Dict begin
  4 2 roll
  \psk@periods 
  \pst@relativePeriod 
  \pst@roundValue
  \psk@amplitude \pst@number\psyunit mul
  \psk@coilarmA \psk@coilarmB 
  \psk@ppoints
  { \psk@function }
  \tx@Sin 
  end }}
\def\pcsin{\def\pst@par{}\pst@object{pcsin}}
\def\pcsin@i{\pc@object\ncsin@ii}
\catcode`\@=\TheAtCode\relax

\usepackage{fancyhdr}
\usepackage{enumerate}
\usepackage{amsmath,mathtools}
\usepackage{color}
\usepackage{multirow, hhline}
\usepackage{hyperref}
\usepackage{amsmath,amssymb,amsfonts}
\usepackage{algorithmic}
\usepackage{graphicx}
\usepackage{textcomp}
\usepackage{multicol}

\def\a{{\bf a}} \def\b{{\bf b}}
\def\d{{\bf d}}
\def\e{{\bf e}} \def\f{{\bf f}} \def\g{{\bf g}} \def\h{{\bf h}}
\def\ib{{\bf i}} \def\j{{\bf j}} \def\k{{\bf k}} \def\l{{\bf l}}
\def\m{{\bf m}} \def\n{{\bf n}} \def\o{{\bf o}} \def\p{{\bf p}}
\def\q{{\bf q}} \def\r{{\bf r}} \def\s{{\bf s}} \def\t{{\bf t}}
\def\u{{\bf u}} \def\v{{\bf v}} \def\w{{\bf w}} \def\x{{\bf x}}
\def\y{{\bf y}} \def\z{{\bf z}}

\def\A{{\bf A}} \def\B{{\bf B}} \def\C{{\bf C}} \def\D{{\bf D}}
\def\E{{\bf E}} \def\F{{\bf F}} \def\G{{\bf G}} \def\H{{\bf H}}
\def\I{{\bf I}} \def\J{{\bf J}} \def\K{{\bf K}} \def\L{{\bf L}}
\def\M{{\bf M}} \def\N{{\bf N}} \def\O{{\bf O}} \def\P{{\bf P}}
\def\Q{{\bf Q}} \def\R{{\bf R}} \def\S{{\bf S}} \def\T{{\bf T}}
\def\U{{\bf U}} \def\V{{\bf V}} \def\W{{\bf W}} \def\X{{\bf X}}
\def\Y{{\bf Y}} \def\Z{{\bf Z}}

\newcommand{\mat}[2]{\left[\begin{array}{#1} #2 \end{array}\right]}
\newcommand{\tras}{^{\mbox{\tiny T}}}
\newcommand{\0}{\boldsymbol{0}}
\newcommand{\ds}{\displaystyle}
\newcommand{\und}{\underline}
\newcommand{\muno}{^{\mbox{\tiny -1}}}
\newcommand{\ts}{\textstyle}
\definecolor{mygreen}{RGB}{28,172,0}
\newcommand{\punto}{\pscircle*{0.15}}

\definecolor{myorangee}{rgb}{1.0, 0.31, 0.0}
\definecolor{mygreenn}{rgb}{0, 0.56, 0.0}
\definecolor{Green}{rgb}{0.0, 0.5, 0.0}

\definecolor{myoran}{rgb}{1.0, 0.36, 0.0}
\definecolor{mygre}{rgb}{0.0 0.5 0.0}
\definecolor{mygrey}{rgb}{0.4 0.4 0.4}

 \newcommand{\Vcm}{\overline{\V}_{\!c}}
 \newcommand{\Vcmi}{\overline{V}_{\!c1}}
 \newcommand{\Vcmii}{\overline{V}_{\!c2}}
 \newcommand{\Vcmif}{\overline{V}_{\!c1_f}}
 \newcommand{\Vcmiif}{\overline{V}_{\!c2_f}}
 \newcommand{\dotVcm}{\dot{\overline{\V}}_{\!c}}
 \newcommand{\dotVcmi}{\dot{\overline{V}}_{\!c1}}
 \newcommand{\dotVcmii}{\dot{\overline{V}}_{\!c2}}
 \newcommand{\Vcmref}{\overline{V}_{\!c12_{des}}}
 \newcommand{\Vcmmis}{\overline{V}_{\!c12_{mis}}}
 \newcommand{\Iaref}{\tilde{I}_{a}}
 \newcommand{\Idref}{\tilde{I}_{d}}
 \newcommand{\Vdref}{\tilde{V}_{d}}
 \newcommand{\dotIaref}{\dot{\tilde{I}}_a}
 \newcommand{\Vsref}{\tilde{V}_s}
 \newcommand{\Voneref}{\tilde{V}_1}
 \newcommand{\Vtworef}{\tilde{V}_2}
 \newcommand{\VoneMref}{\tilde{V}_{1M}}
 \newcommand{\VtwoMref}{\tilde{V}_{2M}}
 \newcommand{\VotMref}{\tilde{V}_{12M}}

\newcommand{\greentickcircle}{%
  \tikz[baseline=-0.6ex]\draw[green!60!black] (0,0) circle (1.2ex);\hspace{-3.86mm}\textcolor{green!60!black}{\footnotesize \checkmark}%
}
\newcommand{\redcrosscircle}{%
  \tikz[baseline=-0.6ex]\draw[red!90!black] (0,0) circle (1.2ex);\hspace{-3.86mm}\textcolor{red!90!black}{\times}%
}
\newcommand{\greylinecircle}{%
  \tikz[baseline=-0.6ex]\draw[mygrey] (0,0) circle (1.2ex);\hspace{-3.86mm}\textcolor{mygrey}{ii-}%
}

\date{}

\begin{document}

\makeatletter
\def\ps@pprintTitle{
  \let\@oddhead\@empty
  \let\@evenhead\@empty
  \let\@oddfoot\@empty
  \let\@evenfoot\@oddfoot
}
\makeatother

\begin{frontmatter}

\title{Model-Based Adaptive Control of Modular Multilevel Converters}

\author{Davide Tebaldi\corref{mycorrespondingauthor}}
\cortext[mycorrespondingauthor]{Corresponding author}
\ead{davide.tebaldi@unimore.it}

\author{Roberto Zanasi}
\ead{roberto.zanasi@unimore.it}

\affiliation{organization={University of Modena and Reggio Emilia},
            addressline={Via Pietro Vivarelli 10 - int. 1},
            city={Modena},
            postcode={41125},
            country={Italy}}

\begin{abstract}
Electrical power conversions are common in a large variety of
engineering applications. With reference to AC/DC and DC/AC power
conversions, a strong research interest resides in multilevel
converters, thanks to the many advantages they provide over standard
two-level converters. In this paper, we first provide a
power-oriented model of Modular Multilevel Converters (MMCs),
followed by a detailed harmonic analysis. The model is given in the
form of a block scheme that can be directly implemented in the
Matlab/Simulink environment. The performed harmonic analysis gives a
deep and exact understanding of the different terms affecting the
evolution of the voltage trajectories in the upper and lower arms of
the converter. Next, we propose a new model-based adaptive control
scheme for MMCs.
The proposed control allows to determine the optimal average capacitor
voltages reference in real-time,
thus allowing to properly track the desired load current while minimizing
the harmonic content in the generated load current itself.
\end{abstract}


\end{frontmatter}

%
%%%%%%%%%%%%%%%%%%%%%%%%%%%%%%%%%%%%%%%%%%%%%%%%%%%%%%
%
\section{Introduction}
\label{sec:introduction} Power conversions are performed in many
engineering applications, including power
grids~\cite{CEP_3}-\cite{CEP_4}, hybrid electric
vehicles~\cite{Nostro_3}-\cite{CEP_5} and many other applications
involving electric motor drives. The devices which are responsible
for performing such power conversions are power converters, and can be
mainly classified into DC/DC~\cite{CEP_NEW_2}-\cite{TCCT_6}, AC/DC and
DC/AC ~\cite{CEP_1}-\cite{Nostro_1} power converters.
Multilevel converter topologies offer many pros when compared to
two-level power converters~\cite{TCCT_4}, including distortion
reduction in the output voltage waveform and in the absorbed input
current as well as a reduced dv/dt effect.
Different multilevel converter topologies are available in the
literature, including cascaded H-bridges multilevel
converters~\cite{TCCT_0}-\cite{TCCT_3}, converter topologies with
flying capacitors~\cite{Nostro_1},~\cite{CEP_NEW_1}-\cite{TCCT_2},
 and  Modular Multilevel Converters
(MMCs)~\cite{Energies_1}-\cite{Altro_6}.
In this paper, MMCs in half-bridge configuration are subject of
study.

The first operation to be performed is a correct and accurate
modeling of the considered multilevel converter topology, for which
many different approaches can be found in the literature. Modular
Multilevel Matrix Converters are modeled in~\cite{Energies_1} in a
matrix form with the definition of a power-capacitor voltage model
and of a voltage-current model. A state-space model of Modular
Multilevel Converters is proposed in~\cite{Energies_2} instead.
In the present paper, Modular Multilevel Converters are modeled
using a new and effective approach which is based on the
Power-Oriented Graphs (POG) modeling technique~\cite{POG_Technique}.
In the literature, two additional main graphical techniques for
modeling physical systems can be found: Bond Graphs
(BG)~\cite{Bond_Graphs_1}-\cite{Bond_Graphs_2} and Energetic
Macroscopic Representation
(EMR)~\cite{Energetic_Macroscopic_Representation_2},~\cite{Energetic_Macroscopic_Representation_1}.
In the present paper, the POG technique is employed as a tool to
develop the proposed converter model, since it provides user-friendly
block schemes which are directly implementable in the Simulink
environment using simple blocks available from standard libraries, such as integrators and gains for instance.
Some examples where the POG technique can be
effectively applied can be found in~\cite{Nostro_1} for the modeling
of multilevel flying-capacitor converters, in~\cite{Nostro_2} for
the modeling of Permanent Magnet Synchronous Motors,
in~\cite{Nostro_3} for the modeling of planetary gear sets and
in~\cite{Nostro_4} for the modeling of multiphase diode bridge
rectifiers. In this paper, two MMC models are derived: a complete model and an average model. The complete model is very suitable for a detailed simulation of the converter dynamics and provides the full differential equations describing it, whereas the average model proves to be very suitable for deriving the proposed new model-based adaptive control of MMCs, thanks to the detailed harmonic analysis performed on it.

The next important step consists in properly controlling the
considered multilevel topology. The control of MMC can be divided
into two different parts~\cite{TCCT_1}: the control of the power and
current inside the MMC and the voltage balancing control. However, addressing all these control objectives is not an easy task. The
full-order nonlinear control of the MMC is addressed
in~\cite{TCCT_1} without considering
 the voltage balancing
issue. In~\cite{Altro_4}, a novel modulation scheme and a
closed-loop method for voltage balancing are proposed.
In~\cite{Altro_3}, the authors propose a new modulation technique
named Integral Modulation Technique to achieve load current control
and voltage balancing, whereas a control approach based on Weighted
Model Predictive Control (WMPC) for a half-bridge MMC is proposed
in~\cite{Altro_2}. However, this latter approach only relies on MPC
to minimize an objective function to achieve both voltage balancing
and load current control and does not fully analyze the complex
dynamics of MMC.
A control approach based on MPC is also proposed in~\cite{Altro_5},
named Model Predictive Direct Current Control, proposing a cost
function that also accounts for the number of switching transitions.
An interesting dynamic analysis of the modular multilevel converter
is proposed in \cite{Altro_6} where, assuming that the output
current tracks the desired profile, the nonlinear model of the MMC
is derived. The system equilibrium points are then computed
neglecting the oscillatory terms, based on which the control of the
circulating current is evinced.
In the present paper, we introduce the following new contributions which, to the best of
our knowledge, have not yet been fully addressed in the literature: 1) the proposal of a compact block scheme modeling MMCs. The block scheme is
directly implementable in the Matlab/Simulink environment using
simple blocks which are available in standard Simulink libraries, such as integrators and gains. The proposed model is also verified against the PLECS (Piecewise Linear Electrical Circuit Simulation) simulator \cite{Plecs_Ref}.
2) the development of an
harmonic analysis in which all the oscillatory terms are taken into account. This enables a deep and exact understanding of the MMC
dynamics, based on which we can determine the optimal tracking reference for the circulating current in order to make the average capacitor voltages follow the desired reference.
3) the exact computation of the optimal voltage reference for the average capacitor voltages in the
upper and lower arms of the converter. Such optimal voltage
reference is the minimum value which is strictly necessary to properly track the
desired load current while, at the same time, minimizing the harmonic content in the generated load current itself.
 4) the proposal of a new model-based adaptive control which allows to effectively achieve
all the control objectives at the same time: a) balancing of the capacitor voltages; b) tracking of the
optimal voltage reference for the average capacitor voltages in order to minimize the harmonic content in the load current; c) tracking of the desired load
current profile.

One of the new important concepts that we introduce in this paper is that of adaptively varying the average capacitor voltages in the upper and lower arms of the MMC, which is what makes the proposed model-based control adaptive to the operating conditions. The average capacitor voltages are
properly controlled to follow an optimal voltage reference using the
circulating current, where such optimal voltage reference is adapted in real-time as a function of the desired load current. This represents a crucial advantage with respect to maintaining the average capacitor voltages at a constant value, as it is typically done in literature, because it allows to reduce the harmonic content in the generated load current whenever the operating conditions allow it. This is done by minimizing the level-to-level distance in the commutating voltage signals, thus enhancing all the intrinsic main advantages of multilevel converters.

The remainder of this paper is structured as follows. The complete dynamic model
of MMCs is derived in Sec.~\ref{Complete_syst} and verified in Sec.~\ref{Model_Verification}. An average model of the converter is proposed in
Sec.~\ref{V_C1_V_C2}, which will be fundamental for the harmonic analysis and for the model-based adaptive control derivation. The Control Problem is addressed in Sec.~\ref{Control_of_MMC_system} based on the model transformation illustrated in Sec.~\ref{L_red_syst}. The capacitor voltages balancing problem in the upper and lower arms of the converter is addressed in Sec.~\ref{prob_11}, whereas the control of the average capacitor voltages and of the load current is described in Sec.~\ref{Idref_calc} and in Sec.~\ref{prob_12_2} using the proposed model-based adaptive control.
Simulation results
showing the effectiveness of the proposed control are
reported and discussed in Sec.~\ref{Simulations_sect}, whereas the
conclusions of this work are given in Sec.~\ref{Conclusion_sect}.
Finally, the detailed calculations employed when performing the
system harmonic analysis are reported in App.~A,
App.~B, App.~C, App.~D, App.~E and App.~F.

 \section{Dynamic Model of Modular Multilevel Converters}\label{Complete_syst_Model}

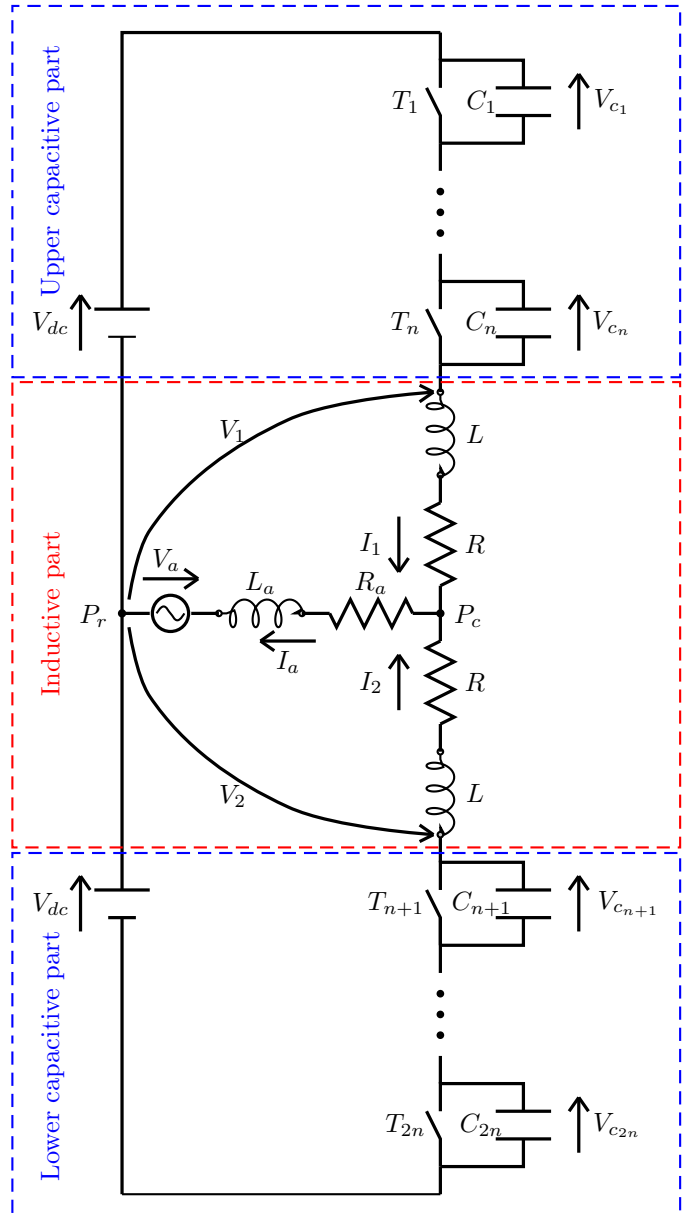
\begin{figure}[tp]
 \centering \footnotesize
 \setlength{\unitlength}{3.68mm} \psset{unit=\unitlength}
\SpecialCoor \hspace*{-10mm}
 \begin{pspicture}(-2.9,-11)(25,33)
 \newrgbcolor{dark_pastel_green}{0.01 0.75 0.24}
 \newrgbcolor{dark_pastel_red}{0.76 0.23 0.16}
 \newrgbcolor{dark_orange}{1.0 0.5 0.0}
 \newrgbcolor{dark_pastel_blue}{0.47 0.62 0.8}
 %%%%%%%%%%%%%%%%%%%%%%%%%%%%%%%%%%%%%%%%%%%%%%%%%%%%
% Dc ramo di sotto
\psline[linewidth=1.22pt](5,-10)(5,0)
\psline[linewidth=1.22pt](4.5,0)(5.5,0)
\psline[linewidth=1.22pt](4,1)(6,1)
\psline[linewidth=1.22pt](5,1)(5,11)
%%%%%%%%%
 \psline[linewidth=1.22pt](3.5,1.5)(3.5,-0.5)
 \psline[linewidth=1.22pt](3.5,1.5)(3.15,1)
 \psline[linewidth=1.22pt](3.5,1.5)(3.85,1)
 \rput(2.42,0.5){$V_{dc}$}
\rput(2.5,0){
 %%%%%%%%%%%%%%%%%%%%%%%%%%%%%%%%%%%%%%%%%%%%%%%%%%%%
% Ramo switches and capacitors giù
%%%%%%%%%%
\psline[linewidth=1.22pt](14,-10)(14,-9)
%%%%%%%%%%
\rput(15.5,-7.5){$C_{2n}$}
\psline[linewidth=1.22pt](19,-8.5)(19,-6.5)
\psline[linewidth=1.22pt](19,-6.5)(18.65,-7)\psline[linewidth=1.22pt](19,-6.5)(19.35,-7)
\rput(20.4,-7.5){$V_{c_{2n}}$} \rput(12.75,-7.5){$T_{2n}$}
\psline[linewidth=1.22pt](14,-9)(14,-8)(13.5,-7)
\psline[linewidth=1.22pt](14,-7)(14,-6)(17,-6)(17,-7)
\psline[linewidth=1.22pt](16,-7)(18,-7)
\psline[linewidth=1.22pt](16,-8)(18,-8)
\psline[linewidth=1.22pt](17,-8)(17,-9)(14,-9)
%%%%%%%%%%
\psline[linewidth=1.22pt](14,-6)(14,-5)
%%%%%%%%%%
 %%%%%%%%%%%%%%%%%%%%%%%%%%%%%%%%%%%%%%%%%%%%%%%%%%%%
% puntini ramo giu:
 \pscircle[fillstyle=solid,fillcolor=black](14,-4.25){0.1}
 \pscircle[fillstyle=solid,fillcolor=black](14,-3.5){0.1}
 \pscircle[fillstyle=solid,fillcolor=black](14,-2.75){0.1}
%%%%%%%%%%
\psline[linewidth=1.22pt](14,0)(14,-2)
%%%%%%%%%%
\rput(0,8){ \rput(15.5,-7.5){$C_{n+1}$}
\psline[linewidth=1.22pt](19,-8.5)(19,-6.5)
\psline[linewidth=1.22pt](19,-6.5)(18.65,-7)\psline[linewidth=1.22pt](19,-6.5)(19.35,-7)
\rput(20.8,-7.5){$V_{c_{n+1}}$} \rput(12.4,-7.5){$T_{n+1}$}
\psline[linewidth=1.22pt](14,-9)(14,-8)(13.5,-7)
\psline[linewidth=1.22pt](14,-7)(14,-6)(17,-6)(17,-7)
\psline[linewidth=1.22pt](16,-7)(18,-7)
\psline[linewidth=1.22pt](16,-8)(18,-8)
\psline[linewidth=1.22pt](17,-8)(17,-9)(14,-9)}
 %%%%%%%%%%%%%%%%%%%%%%%%%%%%%%%%%%%%%%%%%%%%%%%%%%%%
% Ramo R and L giù
\psline[linewidth=1.22pt](14,2)(14,3) \cnode(14,6){.1}{A}
\cnode[fillstyle=solid,fillcolor=lightgray](14,3){.1}{B}
\nccoil[coilwidth=1,coilarm=.065cm,coilaspect=35,linecolor=black]{A}{B}
\psline[linewidth=1.22pt](14,6)(14,7)
\psline[linewidth=1.22pt](14,7)(14.5,7.25)(13.5,7.75)(14.5,8.25)(13.5,8.75)(14.5,9.25)(13.5,9.75)(14,10)
\rput(15.25,4.5){$L$} \rput(15.25,8.5){$R$}
\psline[linewidth=1.22pt](12.5,7.5)(12.5,9.5)
\psline[linewidth=1.22pt](12.5,9.5)(12.85,9)
\psline[linewidth=1.22pt](12.5,9.5)(12.15,9)\rput(11.5,8.5){$I_2$}
\psline[linewidth=1.22pt](14,10)(14,11)}
 %%%%%%%%%%%%%%%%%%%%%%%%%%%%%%%%%%%%%%%%%%%%%%%%%%%%
% Load
\psline[linewidth=1.22pt](5,11)(6,11)
\pscircle[linewidth=1.22pt](6.75,11){0.75}
\pscurve(6.25,11)(6.5,11.25)(6.75,11)(7,10.75)(7.25,11)
\psline[linewidth=1.22pt](7.5,11)(8.5,11) \cnode(8.5,11){.1}{A}
\cnode[fillstyle=solid,fillcolor=lightgray](11.5,11){.1}{B}
\nccoil[coilwidth=1,coilarm=.065cm,coilaspect=35,linecolor=black]{A}{B}
\psline[linewidth=1.22pt](11.5,11)(12.5,11)
\psline[linewidth=1.22pt](12.5,11)(12.75,10.5)(13.25,11.5)(13.75,10.5)(14.25,11.5)(14.75,10.5)(15.25,11.5)(15.5,11)
\psline[linewidth=1.22pt](15.5,11)(16.5,11)
\psline[linewidth=1.22pt](5.75,12.25)(7.75,12.25)
\psline[linewidth=1.22pt](7.75,12.25)(7.25,12.6)
\psline[linewidth=1.22pt](7.75,12.25)(7.25,11.9)
\rput(6.6,12.92){$V_a$} \rput(10,12.10){$L_a$}
\rput(14,12.10){$R_a$}
\psline[linewidth=1.22pt](12,10)(10,10)
\psline[linewidth=1.22pt](10,10)(10.5,9.65)
\psline[linewidth=1.22pt](10,10)(10.5,10.35) \rput(11,9.25){$I_a$}
 %%%%%%%%%%%%%%%%%%%%%%%%%%%%%%%%%%%%%%%%%%%%%%%%%%%%
% Ramo chiusura giù
\psline(5,-10)(16.5,-10)
 %%%%%%%%%%%%%%%%%%%%%%%%%%%%%%%%%%%%%%%%%%%%%%%%%%%%
% Ramo R and L su
\psline[linewidth=1.22pt](16.5,11)(16.5,12) \cnode(16.5,19){.1}{A}
\cnode[fillstyle=solid,fillcolor=lightgray](16.5,16){.1}{B}
\nccoil[coilwidth=1,coilarm=.065cm,coilaspect=35,linecolor=black]{A}{B}
\psline[linewidth=1.22pt](16.5,16)(16.5,15) \rput(2.5,5){
\psline[linewidth=1.22pt](14,7)(14.5,7.25)(13.5,7.75)(14.5,8.25)(13.5,8.75)(14.5,9.25)(13.5,9.75)(14,10)}
\psline[linewidth=1.22pt](15,14.5)(15,12.5)
\psline[linewidth=1.22pt](15,12.5)(14.65,13)
\psline[linewidth=1.22pt](15,12.5)(15.35,13) \rput(14,13.5){$I_1$}
\rput(17.75,13.5){$R$} \rput(17.75,17.5){$L$}
 %%%%%%%%%%%%%%%%%%%%%%%%%%%%%%%%%%%%%%%%%%%%%%%%%%%%
% Ramo switches and capacitors su
\rput(2.5,29){
%%%%%%%%%%
\psline[linewidth=1.22pt](14,-10)(14,-9)
%%%%%%%%%%
\rput(15.5,-7.5){$C_n$} \psline[linewidth=1.22pt](19,-8.5)(19,-6.5)
\psline[linewidth=1.22pt](19,-6.5)(18.65,-7)\psline[linewidth=1.22pt](19,-6.5)(19.35,-7)
\rput(20.2,-7.5){$V_{c_n}$} \rput(12.75,-7.5){$T_n$}
\psline[linewidth=1.22pt](14,-9)(14,-8)(13.5,-7)
\psline[linewidth=1.22pt](14,-7)(14,-6)(17,-6)(17,-7)
\psline[linewidth=1.22pt](16,-7)(18,-7)
\psline[linewidth=1.22pt](16,-8)(18,-8)
\psline[linewidth=1.22pt](17,-8)(17,-9)(14,-9)}
%%%%%%%%%%
\rput(2.5,33){
%%%%%%%%%%%
\psline[linewidth=1.22pt](14,-10)(14,-9)
%%%%%%%%%%%
}
%%%%%%%%%%
\rput(2.5,37){
%%%%%%%%%%
\psline[linewidth=1.22pt](14,-10)(14,-9)
%%%%%%%%%%
\rput(15.5,-7.5){$C_1$} \psline[linewidth=1.22pt](19,-8.5)(19,-6.5)
\psline[linewidth=1.22pt](19,-6.5)(18.65,-7)\psline[linewidth=1.22pt](19,-6.5)(19.35,-7)
\rput(20.2,-7.5){$V_{c_1}$} \rput(12.75,-7.5){$T_1$}
\psline[linewidth=1.22pt](14,-9)(14,-8)(13.5,-7)
\psline[linewidth=1.22pt](14,-7)(14,-6)(17,-6)(17,-7)
\psline[linewidth=1.22pt](16,-7)(18,-7)
\psline[linewidth=1.22pt](16,-8)(18,-8)
\psline[linewidth=1.22pt](17,-8)(17,-9)(14,-9)}
\psline[linewidth=1.22pt](16.5,31)(16.5,32)
 %%%%%%%%%%%%%%%%%%%%%%%%%%%%%%%%%%%%%%%%%%%%%%%%%%%%
% Ramo chiusura su e Ramo Dc su
\psline[linewidth=1.22pt](16.5,32)(5,32)(5,22)
\psline[linewidth=1.22pt](5,11)(5,21) \psline(4.5,21)(5.5,21)
\psline[linewidth=1.22pt](4,22)(6,22)
 \rput(0,21){
%%%%%%%%%
 \psline[linewidth=1.22pt](3.5,-0.5)(3.5,1.5)
 \psline[linewidth=1.22pt](3.5,1.5)(3.15,1)
 \psline[linewidth=1.22pt](3.5,1.5)(3.85,1)
 \rput(2.42,0.5){$V_{dc}$}}
 %%%%%%%%%%%%%%%%%%%%%%%%%%%%%%%%%%%%%%%%%%%%%%%%%%%%
% puntini ramo su:
 \pscircle[fillstyle=solid,fillcolor=black](16.5,24.75){0.1}
 \pscircle[fillstyle=solid,fillcolor=black](16.5,25.5){0.1}
 \pscircle[fillstyle=solid,fillcolor=black](16.5,26.25){0.1}
 %%%%%%%%%%%%%%%%%%%%%%%%%%%%%%%%%%%%%%%%%%%%%%%%%%%%
 \rput[lb](16.5,11){\punto}
 \rput[lb](17.0,10.5){$P_c$}
 \rput[lb](5,11){\punto}
 \rput[rb](4.5,10.5){$P_r$}
% \rput[rb](16,19.3){\red $V_1$}
% \rput[rt](16,3.7){\red $V_2$}
 \psframe[linestyle=dashed,linecolor=blue,linewidth=0.8pt](1,19.5)(25.2,33)
 \rput{90}(2.5,27.0){\blue Upper capacitive part}
 \psframe[linestyle=dashed,linecolor=red,linewidth=0.8pt](1,2.5)(25.2,19.4)
 \rput{90}(2.5,11.0){\red Inductive part}
 \psframe[linestyle=dashed,linecolor=blue,linewidth=0.8pt](1,-11)(25.2,2.375)
 \rput{90}(2.5,-5.0){\blue Lower capacitive part}
%%%%%%%%%%%%%%%%%%%%%%%%%%%%%%%%%%%%%%%%%%%%%%%%%%%%
% Tensioni V1 e V2:
\pscurve[linewidth=1.22pt](5.25,11.5)(6,14)(11,18)(16.25,19)
\psline[linewidth=1.22pt](16.25,19)(15.75,19.25)
\psline[linewidth=1.22pt](16.25,19)(15.75,18.65)
\rput(9,17.6){$V_1$}
\pscurve[linewidth=1.22pt](5.25,10.5)(6,8)(11,4)(16.25,3)
\psline[linewidth=1.22pt](16.25,3)(15.75,3.35)
\psline[linewidth=1.22pt](16.25,3)(15.75,2.75) \rput(9,4.4){$V_2$}
%
%\psgrid[gridwidth=0.15pt,subgridwidth=0.1pt,subgriddiv=2,gridlabels=5pt](0,-11)(25,33)
\end{pspicture}
\caption{Circuit diagram of the considered Modular Multilevel
Converter.}\label{2n_Cond_MMC_figure}
\end{figure}

The circuit diagram of a MMC with $n$ of capacitors on each arm is shown in
Fig.~\ref{2n_Cond_MMC_figure}. The
switches reported in the
figure are supposed to be in half-bridge configuration.
The complete dynamic model of the MMC is derived in
Sec.~\ref{Complete_syst} and verified in Sec.~\ref{Model_Verification} against the PLECS circuit simulator. Next, an average model of the MMC
capacitive part which is suitable for the MMC harmonic analysis and for describing the model-based adaptive
control proposed in this paper is derived in Sec.~\ref{V_C1_V_C2}.

\subsection{Full Dynamic Model of Modular Multilevel Converters}\label{Complete_syst}

\begin{figure}[tp] \centering
 \includegraphics[clip,width=0.9\columnwidth]{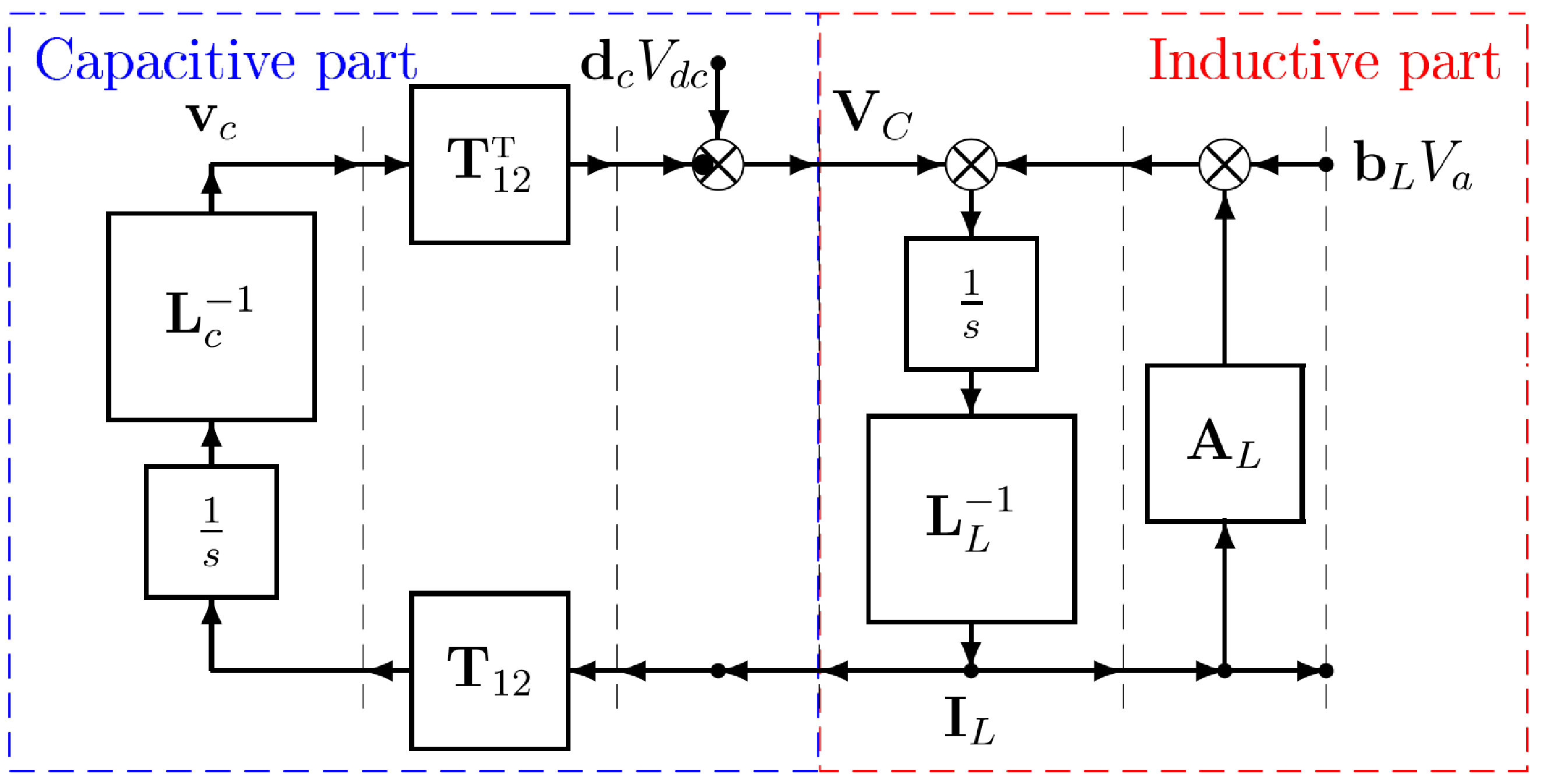}
 \caption{Block scheme of the complete MMC dynamic model.}\label{MMC_complete_POG}
\end{figure}

The dynamic model of the MMC inductive part can be expressed
as $\L_L\,\dot\I_L \!=\! \A_L\,\I_L \!+\! \V_C \!+\! \b_L\,V_a$:
\begin{equation}\label{POG_Trifase_Vin_SCH_LAB_red}
 \underbrace{
 \mat{@{\,}c@{\;}c@{\,}}{
  L+L_{a} & L_{a}\\
  L_{a} & L+L_{a}}
    }_{\L_L}
    \dot\I_L
 =
 \underbrace{
 -\!\mat{@{\,}c@{\;}c@{\,}}{
  R+R_{a} & R_{a}\\
  R_{a} & R+R_{a}}
     }_{\A_L}
 \underbrace{\mat{@{\,}c@{\,}}{ I_{1}\\ I_{2}}\!}_{\I_L}
 +
 \underbrace{\mat{@{\,}c@{\,}}{ V_{1}\\ V_{2} }}_{\V_C}
 +
 \underbrace{\mat{@{\,}c@{\,}}{-1\\ -1 }}_{\b_L}
 V_a,
\end{equation}
where $R,\,L$ are the resistance and the inductance in the lower and
upper arms of the converter; $R_a,\,L_a$ are the load resistance and the load
inductance; $I_2,\,I_1$ are the currents flowing in the lower and
upper arms of the converter; $V_2,\,V_1$ are the commutating voltages defined as in
Fig.~\ref{2n_Cond_MMC_figure}, and $V_a$ is a load sinusoidal
voltage source, see Fig.~\ref{2n_Cond_MMC_figure}.
 The dynamic model of the MMC capacitive part can be expressed as $\L_c\,\dot\v_c  = \T_{12}\,\I_L$, $\V_C
=  -\T_{12}\tras\,\v_c + \d_c\,V_{dc}$:
\vspace{-9.5mm}
\begin{multicols}{2}
\begin{equation}\label{MMC_input_syst}
 \underbrace{\mat{@{}c@{\,}c@{}}{\C_1 & \0 \\ \0  & \C_2}}_{\L_c}\!
 \dot\v_c
  \!=\!
 \underbrace{\mat{@{}c@{\,}c@{}}{\T_1 & \0  \\   \0 & -\T_2  }}_{\T_{12}}
 \underbrace{\mat{@{}c@{}}{ I_{1} \\ I_{2}}}_{\I_L},
 %%%%%%%
 \end{equation}\break
%\vspace{-3.5mm}
\begin{equation}\label{MMC_input_syst_bis}
 \underbrace{\mat{@{}c@{}}{ V_1 \\ V_2}}_{\V_C}
  \!=\!
 -\!\T_{12}\tras
 \underbrace{\mat{@{}c@{}}{ \v_{c1}\\ \v_{c2}}}_{\v_c}
 \!+\!
 \underbrace{\mat{@{}c@{}}{ 1 \\  -1 }}_{\d_c}
 V_{dc},
 \end{equation}
%\vspace{-2.2mm}
\end{multicols}
 where
\begin{equation}\label{MMC_input_syst_bis_matr_vect}
  \C_1 \!\!= \!\!\!\mat{@{\!}c@{}c@{}c@{\!}}{
    C_1 & \cdots & 0 \\
    \vdots & \ddots & \vdots \\
    0 & \dots & C_n }\!\!\!,
% \hspace{8mm}
  \C_2 \!\!=\!\! \!\mat{@{\!}c@{}c@{}c@{\!}}{
    C_{n+\!1} & \cdots & 0 \\
    \vdots & \ddots & \vdots \\
    0 & \dots & C_{2n} }\!\!\!,
 %\hspace{8mm}
 \v_{c1}\!\!=\!\! \!\mat{@{\!}c@{\!}}{ V_{c_1}\\ \vdots\\ V_{c_n}}\!\!\!,
%\]
%\[
 \v_{c2}\!\!=\!\!\! \mat{@{\!}c@{\!}}{ V_{c_{n+\!1}}\\ \vdots\\ V_{c_{2n}}}\!\!\!,
% \hspace{8mm}
 \T_1\!\!=\!\! \!\mat{@{\!}c@{\!}}{ T_1\\ \vdots\\ T_n}\!\!\!,
% \hspace{8mm}
 \T_2\!\!=\!\!\! \mat{@{\!}c@{\!}}{ T_{n+\!1}\\ \vdots\\ T_{2n}}\!\!\!.
\end{equation}
 The signals $T_i$, for $i\in\{1,\;2,\;\ldots,\;2n\}$, are the control variables that define the state of the switches (on/off)
 and the use (no/yes) of the capacitors $C_i$ in the definition of the voltages $V_1$ and $V_2$:
  a) if $T_i=0$, the $i$-th switch is on and the voltage $V_{c_i}$ of capacitor
  $C_i$ is not used in the Kirchhoff's Voltage Law in
  \eqref{MMC_input_syst_bis};
  b) if $T_i=1$, the $i$-th switch is off and the voltage $V_{c_i}$ of capacitor
  $C_i$ is used in the Kirchhoff's Voltage Law in
  \eqref{MMC_input_syst_bis}.
By combining together
systems~\eqref{POG_Trifase_Vin_SCH_LAB_red}-\eqref{MMC_input_syst_bis},
the following  $\L \dot \x = \A \, \x + \B \, \u$ complete dynamic
model of the MMC is obtained:
\begin{equation}\label{MMC_complt_syst}
  \underbrace{
  \mat{@{}c@{\;\;}c@{}}{\L_{c} & \boldsymbol{0} \\ \boldsymbol{0} & \L_{L} }
  }_{\L}
  \dot \x
  =
  \underbrace{
  \mat{@{}c@{\;\;}c@{}}{\0 & \T_{12} \\ -\T_{12}\tras & \A_{L} \;}
  }_{\A}
  \underbrace{
  \mat{@{}c@{}}{\v_c  \\ \I_L }
  }_{\x}
  +
  \underbrace{
  \mat{@{}c@{\;\;}c@{}}{ \0 & \0 \\ \d_c & \b_{L} \;}
  }_{\B}
  \underbrace{
  \mat{@{}c@{}}{V_{dc} \\ V_a }
  }_{\u}.
\end{equation}
%%%
 %
The order $m$ of the dynamic system (\ref{MMC_complt_syst}) is
$m=2n+2$.
A very
compact block scheme of system (\ref{MMC_complt_syst}) is shown in
Fig.~\ref{MMC_complete_POG}, having the interesting feature of being
directly implemented in the Simulink environment using simple blocks
which are available in standard Simulink
libraries~\cite{POG_Technique}.

%%%%%%%%%%%%%%%%%%%%%%%%%%%%%%%%%%%%%%%%%%%%%%%%%%%%%%%%%%%%%%%%%%%%%%%%%%%%%%%%%%%%%%%%%%%
%%%%%%%%%%%%%%%%%%%%%%%%%%%%%%%%%%%%%%%%%%%%%%%%%%%%%%%%%%%%%%%%%%%%%%%%%%%%%%%%%%%%%%%%%%%
%%%%%%%%%%%%%%%%%%%%%%%%%%%%%%%%%%%%%%%%%%%%%%%%%%%%%%%%%%%%%%%%%%%%%%%%%%%%%%%%%%%%%%%%%%%
%%%%%%%%%%%%%%%%%%%%%%%%%%%%%%%%%%%%%%%%%%%%%%%%%%%%%%%%%%%%%%%%%%%%%%%%%%%%%%%%%%%%%%%%%%%

 \subsection{Model Verification}\label{Model_Verification}

\begin{table}[t]
  \centering
  \caption{Parameters and initial conditions of the proposed model~\eqref{MMC_complt_syst} and of the PLECS model in Fig.~\ref{PLECS_implem}.}\label{MMC_for_Plecs_param_table}
\begin{tabular}{|c|c|c|c|c|c|}
\hline
% a & b & c & e & f & g \kill
 \multicolumn{2}{|c|}{$L = 10$ [mH]} & \multicolumn{2}{|c|}{$R = 0.1$ [$\Omega$]} & \multicolumn{2}{|c|}{$C_i = 1000$ [$\mu$ F]} \\ \hline
 \multicolumn{2}{|c|}{$L_a = 50$ [mH]} & \multicolumn{2}{|c|}{$R_a = 19$ [$\Omega$]} &  \multicolumn{2}{|c|}{$V_a\!=\!10\sin(2\pi 50\,t\!+\!\pi/6)$ [V]}  \\ \hline
 \multicolumn{2}{|c|}{$n = 3$} & \multicolumn{2}{|c|}{$V_{dc} = 250$ [V]} & \multicolumn{2}{|c|}{$T_s=10^{-4}$ [s]}  \\ \hline
 \multicolumn{6}{|c|}{$I_{1_0} = I_{2_0} = I_{a_0} = 0$ [H], $\;\;\;$ $V_{c_{1_0}},\,\cdots,\, V_{c_{6_0}} = 110$ [V]}
 \\ \hline
\end{tabular}
\end{table}

%%%%%%%%%%%%%%%%%%%%%%%%%%%%%%%%%%%%%%%%%%%%%%%%%%%%%%%%%%%%%%%%%%%%%%
%%%%%%%%%%%%%%%%%%%%%%%%%%%%%%%%%%%%%%%%%%%%%%%%%%%%%%%%%%%%%%%%%%%%%%

 \begin{figure}[t]
  \centering
\includegraphics[clip,width=0.4\columnwidth]{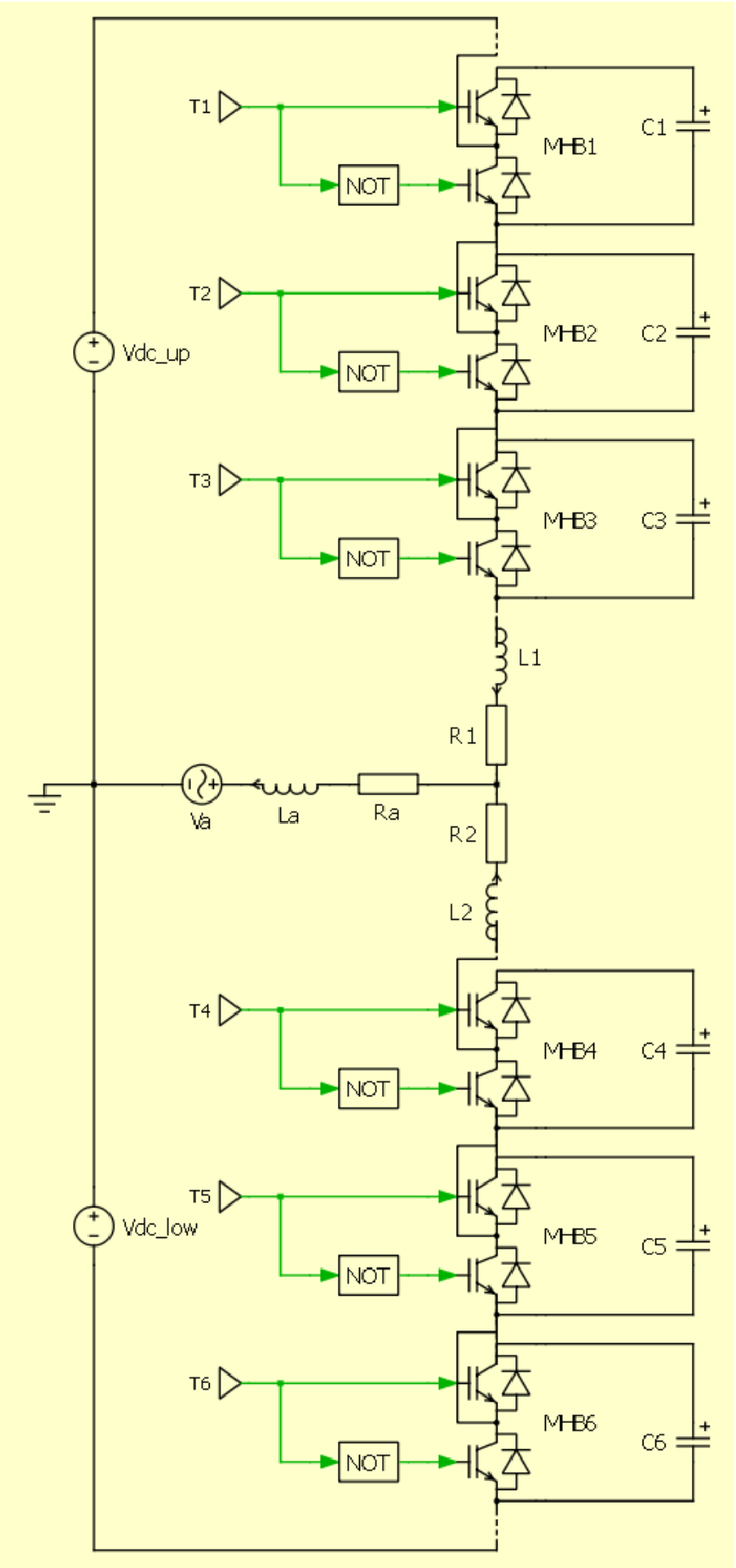}
%}
 \caption{PLECS implementation of the MMC circuit diagram in Fig.~\ref{2n_Cond_MMC_figure} for the case $n=3$.}\label{PLECS_implem}
\end{figure}

%%%%%%%%%%%%%%%%%%%%%%%%%%%%%%%%%%%%%%%%%%%%%%%%%%%%%%%%%%%%%%%%%%%%%%
%%%%%%%%%%%%%%%%%%%%%%%%%%%%%%%%%%%%%%%%%%%%%%%%%%%%%%%%%%%%%%%%%%%%%%

%%%%%%%%%%%%%%%%%%%%%%%%%%%%%%%%%%%%%%%%%%%%%%%%%%%%%%%%%%%%%%%%%%%%%%
%%%%%%%%%%%%%%%%%%%%%%%%%%%%%%%%%%%%%%%%%%%%%%%%%%%%%%%%%%%%%%%%%%%%%%

 \begin{figure}[htbp]
  \centering
\includegraphics[clip,width=13.6cm, height=8.28cm]{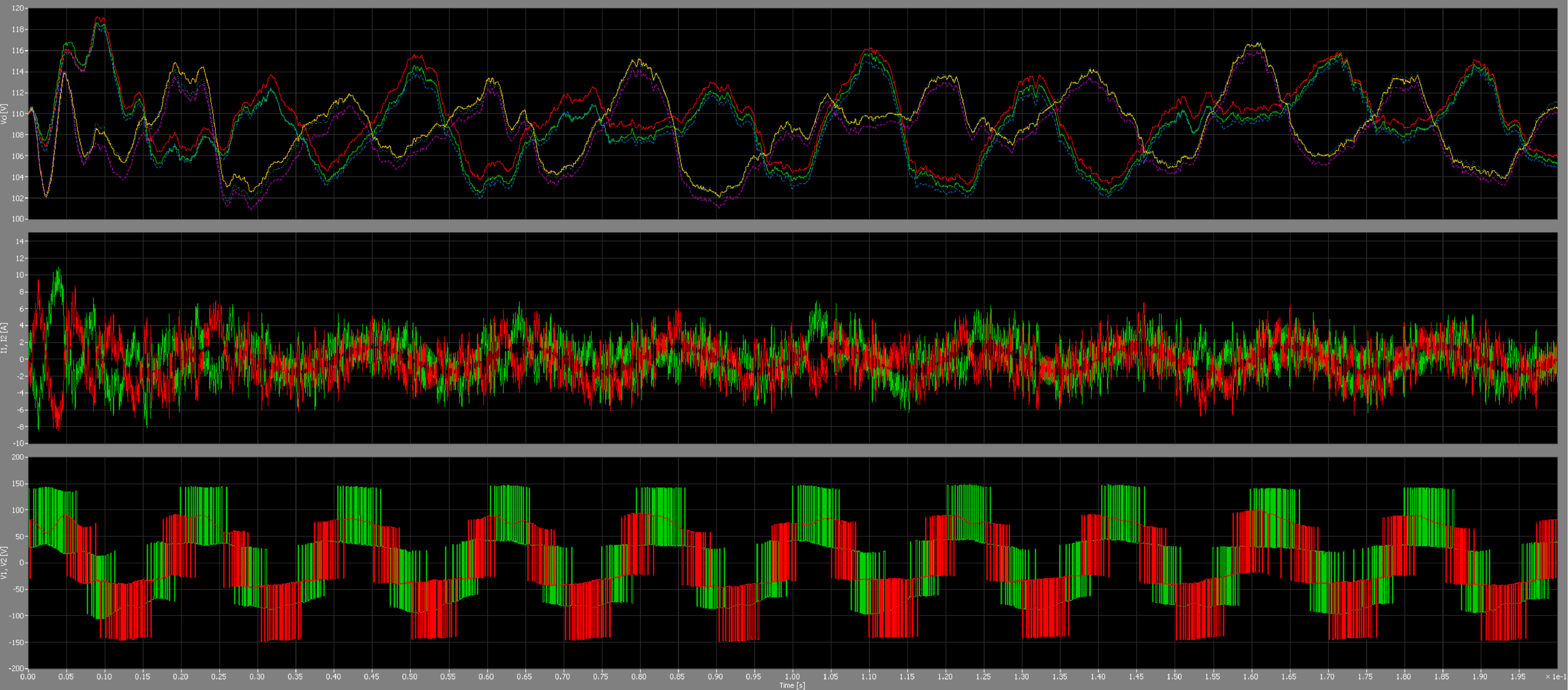}
%}
 \caption{Capacitor voltages $V_{c_{1}},\,\cdots,\, V_{c_{6}}$, currents $I_1,\,I_2$ and voltages $V_1,\,V_2$ given by the PLECS scope.}\label{Tutto_from_Plecs}
\end{figure}
%

%%%%%%%%%%%%%%%%%%%%%%%%%%%%%%%%%%%%%%%%%%%%%%%%%%%%%%%%%%%%%%%%%%%%%%
%%%%%%%%%%%%%%%%%%%%%%%%%%%%%%%%%%%%%%%%%%%%%%%%%%%%%%%%%%%%%%%%%%%%%%

%
 \begin{figure}[htbp]
 \psfrag{I1 Pl}[][][0.85]{$I_1$ Pl}
 \psfrag{I2 Pl}[][][0.85]{$I_2$ Pl}
 \psfrag{Mt}[][][0.85]{Mt}
 \psfrag{V1 Pl}[][][0.85]{$V_1$ Pl}
 \psfrag{V2 Pl}[][][0.85]{$V_2$ Pl}
 \psfrag{Mt}[][][0.85]{Mt}
 \psfrag{Vc1 Pl}[][][0.85]{$V_{c_1}$ Pl}
 \psfrag{Vc2 Pl}[][][0.85]{$V_{c_2}$ Pl}
 \psfrag{Vc3 Pl}[][][0.85]{$V_{c_3}$ Pl}
 \psfrag{Vc4 Pl}[][][0.85]{$V_{c_4}$ Pl}
 \psfrag{Vc5 Pl}[][][0.85]{$V_{c_5}$ Pl}
 \psfrag{Vc6 Pl}[][][0.85]{$V_{c_6}$ Pl}
 \psfrag{Mt}[][][0.85]{Mt}
  \psfrag{Time [s]}[t][t][0.85]{Time [s]}
 \psfrag{[A]}[b][b][0.85]{[A]}
 \psfrag{[V]}[b][b][0.85]{[V]}
 \psfrag{[V]}[b][b][0.85]{[V]}
  \psfrag{I1, I2}[b][b][0.75]{Currents $I_1,\,I_2$}
  \psfrag{V1, V2}[b][b][0.75]{Voltages $V_1,\,V_2$}
  \psfrag{Capacitor Voltages Vci}[b][b][0.75]{Capacitor Voltages $V_{c_i}$}
\includegraphics[clip,width=1\columnwidth]{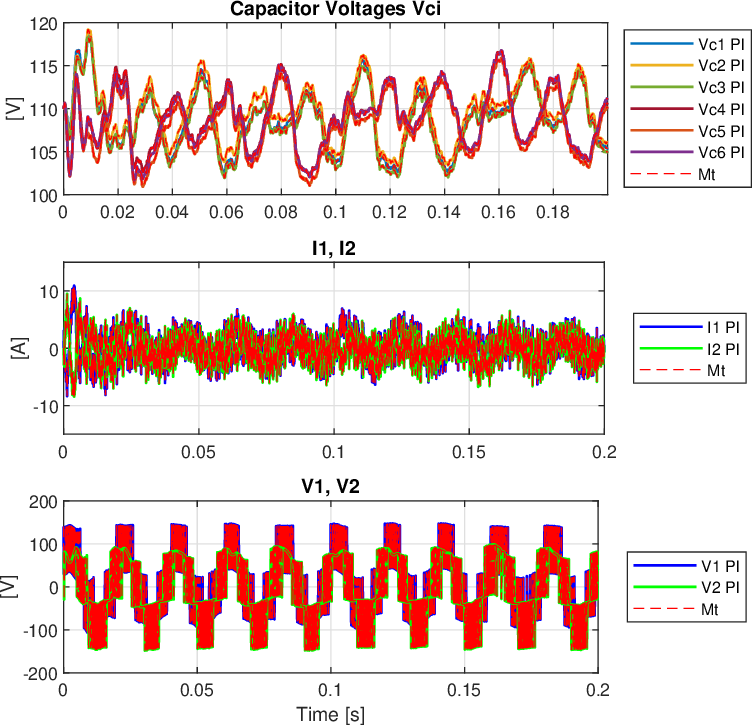}
 \caption{Comparison between Matlab/Simulink and PLECS results: capacitor
 voltages $V_{c_i}$, currents $I_1,\,I_2$ and voltages $V_1,\,V_2$.}\label{MMC_Matlab_Plecs_Comparison_Figure_10}
\end{figure}

The MMC model proposed in~\eqref{MMC_complt_syst} and %schematized in
in Fig.~\ref{MMC_complete_POG} has been tested
against one of the most widespread platforms for simulating power
electronics systems: PLECS (Piecewise Linear Electrical Circuit Simulation)~\cite{Plecs_Ref}. The PLECS implementation of the
considered MMC is reported in Fig.~\ref{PLECS_implem}, in which the case
$n=3$ has been taken as an example. Since the objective is the verification of the proposed
model~\eqref{MMC_complt_syst}, the simulation of the system
model~\eqref{MMC_complt_syst} and the simulation of the PLECS model
have been performed in open-loop by applying the same control
signals $T_1,\,\cdots,\,T_6$ in vectors $\T_1$ and $\T_2$ in~\eqref{MMC_input_syst_bis_matr_vect}. The considered model parameters and
initial conditions are reported in
Table~\ref{MMC_for_Plecs_param_table}. The results in terms of currents $I_1$ and $I_2$, voltages $V_1$ and $V_2$ and capacitor voltages $V_{c_{1}},\,\cdots,\, V_{c_{6}}$ given
by the PLECS model in Fig.~\ref{PLECS_implem} are shown in the
PLECS scopes in Fig.~\ref{Tutto_from_Plecs}.
 The comparison between
the results given by the model ~\eqref{MMC_complt_syst} and those
given by the PLECS model
 are reported in
Fig.~\ref{MMC_Matlab_Plecs_Comparison_Figure_10}. In this figure, the legend notation ``Pl'' stands for PLECS and
``Mt'' stands for Matlab/Simulink. From Fig.~\ref{MMC_Matlab_Plecs_Comparison_Figure_10}, it is possible to appreciate the very good superposition
between the results of model~\eqref{MMC_complt_syst} and those given
the PLECS model in Fig.~\ref{PLECS_implem}. In fact, the maximum absolute difference between the capacitor voltages $V_{c_{1}},\,\cdots,\, V_{c_{6}}$ given by the simulation of the PLECS model and those given by the simulation of model~\eqref{MMC_complt_syst} is $12.5$ mV, the maximum absolute difference between currents $I_1$, $I_2$ given by the simulation of the PLECS model and those given by the simulation of model~\eqref{MMC_complt_syst} is $14.2$ mA, and the maximum absolute difference between voltages $V_1$, $V_2$ given by the simulation of the PLECS model and those given by the simulation of model~\eqref{MMC_complt_syst} is $31.66$ mV. This verifies the
correctness of the MMC model proposed in~\eqref{MMC_complt_syst}.

%%%%%%%%%%%%%%%%%%%%%%%%%%%%%%%%%%%%%%%%%%%%%%%%%%%%%%%%%%%%%%%%%%%%%%%%%%%%%%%%%%%%%%%%%%%
%%%%%%%%%%%%%%%%%%%%%%%%%%%%%%%%%%%%%%%%%%%%%%%%%%%%%%%%%%%%%%%%%%%%%%%%%%%%%%%%%%%%%%%%%%%
%%%%%%%%%%%%%%%%%%%%%%%%%%%%%%%%%%%%%%%%%%%%%%%%%%%%%%%%%%%%%%%%%%%%%%%%%%%%%%%%%%%%%%%%%%%
%%%%%%%%%%%%%%%%%%%%%%%%%%%%%%%%%%%%%%%%%%%%%%%%%%%%%%%%%%%%%%%%%%%%%%%%%%%%%%%%%%%%%%%%%%%

 \subsection{Average dynamic model of the MMC capacitive part}\label{V_C1_V_C2}
 The order of the capacitive part \eqref{MMC_input_syst}-\eqref{MMC_input_syst_bis_matr_vect} in the MMC dynamic model is $2n$, that is the total number of capacitors
 in the upper and lower arms of the
 converter. Let $C$ be the capacitance value of capacitors
 $C_1,\,\ldots,\,C_n,\,C_{n+1},\,\ldots,\,C_{2n}$ in the MMC of
 Fig.~\ref{2n_Cond_MMC_figure}. In order to have equally spaced voltage levels for voltages $V_1$ and
 $V_2$, see Fig.~\ref{2n_Cond_MMC_figure}, the
 following balancing conditions must stand:
 \begin{equation}\label{average_cond}
 V_{c_1}\!\simeq \!\cdots \!\simeq\!
 V_{c_n} \!\simeq \!\Vcmi, \hspace{16mm} V_{c_{n+1}}\!\simeq \!\cdots \!\simeq \!V_{c_{2n}} \!\simeq
 \!\Vcmii,
 \end{equation}
 where $\Vcmi$ and $\Vcmii$ are the average values of the capacitors voltages
 in the upper and lower arms of the converter, respectively. Condition~\eqref{average_cond} can be effectively achieved using the algorithm described in Sec.~\ref{prob_11}.
If~\eqref{average_cond} holds true, the dynamic
 model of the MMC capacitive part can be approximated using the
 following average dynamic model.
 Let  $\Vcm$ denote the following vector:
\begin{equation}\label{MMC_V1_V2_first}
 \Vcm = \mat{c}{\Vcmi \\ \Vcmii}.
\end{equation}
The variables $V_1$ and $V_2$ of vector $\V_C$ in
(\ref{POG_Trifase_Vin_SCH_LAB_red}) can only take on the following
admissible voltage values:
\begin{equation}\label{MMC_V1_V2}
\V_C= \left[\begin{array}{c} V_1\\ V_2
\end{array}\right]=
 \left[\begin{array}{c}
  V_{dc}-n_1 \Vcmi\\
 -V_{dc}+n_2 \Vcmii
\end{array}\right],
\end{equation}
where the integers  $n_1\in\{0,\;1,\;2,\;\ldots,\;n\}$ and
$n_2\in\{0,\;1,\;2,\;\ldots,\;n\}$ denote the number of capacitors for which the corresponding switch is off in the upper and lower arms of the converter, i.e. for which $T_i=1$ in Fig.~\ref{2n_Cond_MMC_figure}.
The time derivative of the average voltage vector
$\Vcm$ in \eqref{MMC_V1_V2_first} can be expressed as follows:
\begin{equation}\label{MMC_VC1_VC2}
\dotVcm
 \!=\!\!
 \mat{@{}c@{}}{\ds \frac{d\Vcmi}{dt} \\[3mm] \ds \frac{d\Vcmii}{dt}}
 \!\!=\!\!
 \mat{@{}c@{}}{ \ds \frac{n_1 \,I_{1}  }{n\, C} \\[3mm] \ds -\frac{n_2 \,I_{2} }{n\, C} }
 \!\!=\!
 \frac{1}{C_T} \!
 \mat{@{}c@{}c@{}}{
  n_1\, I_{1} \\
  -n_2\, I_{2}},
\end{equation}
 where $C_T=n\,C$.
Using (\ref{MMC_V1_V2}) and \eqref{MMC_VC1_VC2}, the dynamic model of the MMC can be expressed as
  $\L_s \dot \x_s = \A_s \; \x_s + \B_s \; \u$:
\begin{equation}\label{MMC_complt_syst_simplifed}
  \underbrace{\!
  \mat{@{\!}c@{}c@{\!}}{\C_{T} & \boldsymbol{0} \\ \boldsymbol{0} & \L_{L} }
  \!}_{\L_s}\!
  \dot \x_s\!
  =\!
  \underbrace{\!
  \mat{@{\!}c@{}c@{\!}}{\0 & \A_{12} \\ -\A_{12}\tras & \A_{L} \;}
  \!}_{\A_s}
  \underbrace{\!
  \mat{@{\!}c@{\!}}{\Vcm  \\ \I_L }
  \!}_{\x_s}\!\!
  +\!
  \underbrace{\!
  \mat{@{}c@{}c@{}}{ \0 & \0 \\ \d_c & \b_{L} \;}
  \!}_{\B_s}
  \underbrace{\!
  \mat{@{\!}c@{\!}}{V_{dc} \\ V_a }
  \!}_{\u}\!,
 \C_T\!=\!\!\mat{@{\!}c@{}c@{\!}}{C_T & 0 \\ 0 & C_T}\!,
 \A_{12}\!=\!\!\mat{@{\!}c@{}c@{\!}}{n_1 & 0 \\ 0 & -n_2}\!.
\end{equation}

 \subsection{Transformed average model of the MMC}\label{L_red_syst}

 Applying the following transformations:
\begin{equation}  \label{Vs_Vd}
  \underbrace{\mat{@{}c@{}}{I_{1}\\I_{2}}}_{\I_L} =
  \underbrace{\frac{1}{2}\mat{@{}c@{\;}c@{}}{1 & 1\\ 1 & -1}}_{\T_w}
  \underbrace{\mat{@{}c@{}}{I_s\\I_d}}_{\I_w},
 \hspace{5mm}
  \underbrace{\mat{@{}c@{}}{V_{1}\\V_{2}}}_{\V_C} =
  \underbrace{\frac{1}{2}\mat{@{}c@{\;}c@{}}{1 & 1\\ 1 & -1}}_{\T_w}
  \underbrace{\mat{@{}c@{}}{V_s\\V_d}}_{\V_w}.
\end{equation}
to the inductive part
(\ref{POG_Trifase_Vin_SCH_LAB_red}) of the MMC complete model, one obtains:
\begin{equation}
\label{POG_transformed_W_L1_L2_equal}
\begin{array}{@{}r@{}c@{\,}l@{}}
\underbrace{\!
 \frac{1}{4}\!\!
 \mat{@{}c@{\,}c@{}}{
 2 L \!+\! 4 L_a & 0\\
 0 & 2 L}
 \!}_{\L_w}\!\dot{\I}_w
 &=&
 \underbrace{\!
 -\frac{1}{4}\!\!\mat{@{}c@{\,}c@{}}{
 2 R \!+\! 4 R_a & 0\\
 0 & 2 R}
 \!}_{\A_w}
 \underbrace{\!
 \mat{@{}c@{}}{ I_s\\ I_d }
 \!}_{\I_w}
 \!+
 \underbrace{\!
 \frac{1}{2}\!\!\mat{@{}c@{}}{ V_{s}\\ V_{d}}
 \!}_{\V_w}
 \!+\!
 \underbrace{\!
 \mat{@{}c@{}}{ -1\\ 0}
 \!}_{\b_w}
  V_{a},
%%%%%%%%%%%%%%%%%%%%%%%%%%%%%%%%%%%%%%%%%%%%%%%%%%%%
 \end{array}
 \end{equation}
where $\L_w=\T_w\tras\L_L\T_w$,  $\A_w=\T_w\tras\A_L\T_w$,
$\V_w=\T_w\tras\V_C$ and $\b_w=\T_w\tras\b_L$, see \cite{POG_Technique}. Note that variable $I_s$ is \eqref{Vs_Vd} is the load current $I_a$, by applying the Kirchhoff's Current Law to the circuit diagram in Fig.~\ref{2n_Cond_MMC_figure}. Hereinafter, variable $I_s$ will be used to denote the load current $I_a$.
 The two equations of system (\ref{POG_transformed_W_L1_L2_equal})
 can be expanded as follows:
\begin{equation}  \label{V_s_and_I_d}
 \left\{\begin{array}{@{\;}r@{\;\;}c@{\;\;}l@{\;}}
 L_T\,\dot{I}_{s} &=& -R_T\,I_{s}  -2\,V_a + V_s
 \\
 L\,\dot{I}_d
 &=&
 -R\,I_d
 +V_d
 \end{array}\right.,
\end{equation}
  where $L_T=L+2\,L_a$ and  $R_T=R+2\,R_a$.
 System (\ref{V_s_and_I_d}) can be controlled  by using the input voltages $V_s$ and $V_d$ introduced in \eqref{Vs_Vd}.
 An important observation has to be made on system (\ref{V_s_and_I_d}):
 the dynamics of currents $I_s$ and $I_d$ are \emph{decoupled}: current $I_s$ can only be controlled
 by using the input voltage
 $V_s$, see the first equation of system (\ref{V_s_and_I_d}),
 whereas current $I_d$ can only be controlled by using the input voltage $V_d$,
 see the second equation of system (\ref{V_s_and_I_d}).
 By replacing current $I_s$ in \eqref{V_s_and_I_d} with the desired one $\Iaref$, one obtains the desired value  $\Vsref$ of voltage $V_s$:
\begin{equation}\label{f_Ia_Va}
 \Vsref
 = f(t)=
 L_T\,\dotIaref +R_T\,\Iaref   +2\,V_a.
\end{equation}

 Inverting (\ref{MMC_V1_V2}), the control variables $n_1$ and $n_2$ can be expressed
 as follows:
\begin{equation}\label{MMC_n1_n2}
\left[\begin{array}{@{\;}c@{\;}} n_1\\ n_2
\end{array}\right]=
 \left[\begin{array}{@{\;}c@{\;}}
 \ds \frac{V_{dc}-V_1  }{\Vcmi} \\[3mm]
 \ds \frac{V_{dc}+V_2  }{\Vcmii}
\end{array}\right].
\end{equation}
Substituting (\ref{MMC_n1_n2}) in (\ref{MMC_VC1_VC2}), one obtains
the following nonlinear dynamic equations describing the capacitive
part of the MMC:
\begin{equation}\label{New_VC1_VC2}
\left\{
\begin{array}{@{\;}r@{\;\;}c@{\;\;}l@{\;}}
C_T\,\Vcmi\,\dotVcmi &=& \ds (V_{dc}-V_1) \, I_{1}
\\
C_T\,\Vcmii\,\dotVcmii &=& \ds-(V_{dc}+V_2 ) \,I_{2}
\end{array}
\right..
\end{equation}
Applying the transformations in (\ref{Vs_Vd})  to system
(\ref{New_VC1_VC2}), one obtains:
\begin{equation}\label{New_VC1_VC2_due}
\left\{
\begin{array}{@{\;}r@{\;\;}c@{\;\;}l@{\;}}
4\,C_T\,\Vcmi\,\dotVcmi &=& \ds (2\,V_{dc}-V_s-V_d ) \, (I_s+I_d)
\\
4\,C_T\,\Vcmii\,\dotVcmii &=& \ds -(2\,V_{dc}+ V_s-V_d ) \,(I_s-I_d)
\end{array}
\right..
\end{equation}

 The overall dynamic equations of the transformed
 average
 MMC
model
are given by combining together the transformed equations
(\ref{V_s_and_I_d}) and (\ref{New_VC1_VC2_due}) of the inductive and
capacitive parts of the system, respectively.

 \section{Model-Based Adaptive Control of Modular Multilevel Converters}\label{Control_of_MMC_system}

\noindent \und{\bf Control Problem.} {\sl  Let $\Iaref =
I_{aM}\sin(\omega t)$ be the desired load current, and let $V_{a} =
V_{aM}\sin(\omega t+\alpha_{V_a})$ be the load generator voltage source
in Fig.~\ref{2n_Cond_MMC_figure}.
The control Goals are: \\[1mm]
1) Capacitor voltages control: %\\[1mm]
1.1) Keeping $V_{c_1}\simeq\cdots\simeq V_{c_n} \simeq \Vcmi$ and $V_{c_{n+1}}\simeq\cdots\simeq V_{c_{2\,n}}\simeq \Vcmii$ (capacitor voltages balancing); %\\[1mm]
1.2) Keeping $\Vcmi\simeq \Vcmii\simeq \Vcmref$, where $\Vcmref$ is the optimal average capacitor voltages reference, which in turn translates into the definition of a desired profile $\Idref$ for current $I_d$. \\[1mm]
2) Load current control: the load current $I_s$ must track the
desired profile $\Iaref$; \\[1mm]
}
The solution of Goal 1.1 of the Control Problem, that is the capacitor voltages balancing problem, is addressed in Sec.~\ref{prob_11}. In Sec.~\ref{Idref_calc}, the transformed average MMC model in (\ref{V_s_and_I_d}) and (\ref{New_VC1_VC2_due}) is properly linearized.
The harmonic analysis performed in Sec.~\ref{Harm_analysis_sect} on the linearized model allows to determine the algorithm to compute the desired current profile $\Idref$ in Sec.~\ref{Control_Strategy_sect}, on the basis of the optimal average capacitor voltages reference $\Vcmref$ computed in Sec.~\ref{Desired_common_voltage_reference_sect}. Finally, the optimal control solving Goals 1.2 and 2 of the Control Problem is discussed in Sec.~\ref{prob_12_2}.

 \subsection{Solution of Goal 1.1 of the Control Problem}\label{prob_11}

\begin{figure*}[t]
  \centering \footnotesize
 \setlength{\unitlength}{2.65mm}
 \psset{unit=\unitlength}
 \vspace{26.3mm}
   \begin{pspicture}(-16.6,-22)(16,-9)
\rput(0.36,-19){\includegraphics[clip,width=2.06\textwidth]{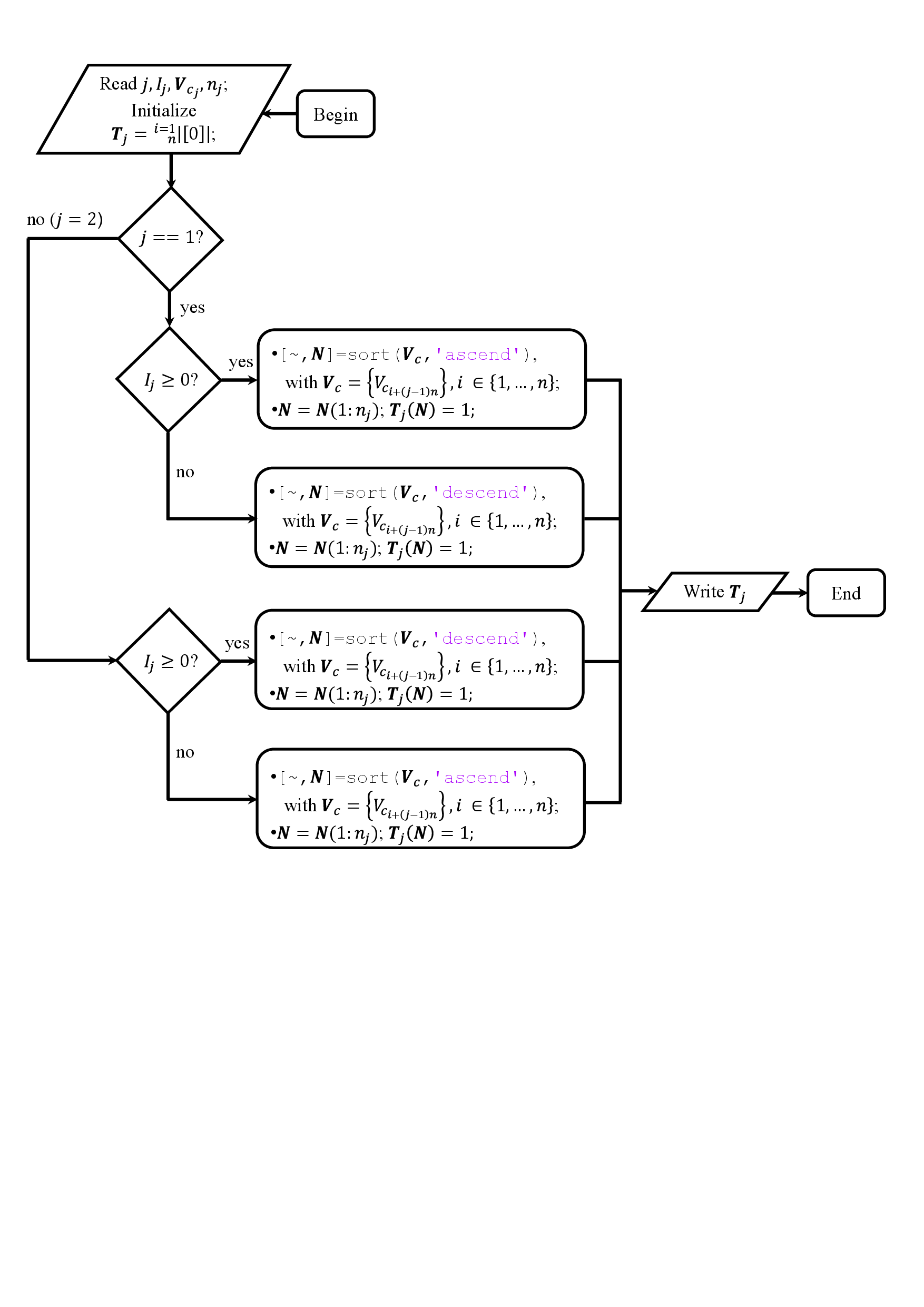}}
 %\psgrid[gridwidth=0.15pt,subgridwidth=0.1pt,subgriddiv=2,gridlabels=5pt](-16,-22)(16,-9)
\end{pspicture}\vspace{48.3mm}
 \caption{Algorithm solving Goal 1.1 of the Control Problem in Sec.~\ref{Control_of_MMC_system}.}\label{Control_of_Vci_algorithm}
\vspace{-3mm}\end{figure*}

Recalling that $n$ is the number of capacitors in the upper and lower arms of
the MMC, let $i \in \left\{1,\,\cdots\,n\right\}$ and $j \in
\left\{1,\,2\right\}$ be the indexes identifying the
$({i+(j-1)\times n})-th$ capacitor voltage $V_{c_{i+(j-1)\times
n}}$, where $j=1$ identifies the upper MMC arm and $j=2$ identifies
the lower MMC arm, see Fig.~\ref{2n_Cond_MMC_figure}.
The Goal 1.1 of the Control Problem in Sec.~\ref{Control_of_MMC_system}
is achieved by implementing the algorithm
schematized in Fig.~\ref{Control_of_Vci_algorithm}. The algorithm
takes as input: the index $j$ identifying the MMC arm under
consideration, the arm current $I_j$ (see~\eqref{POG_Trifase_Vin_SCH_LAB_red}), the capacitor voltages vector
$\V_{c_j}$
(see~\eqref{MMC_input_syst_bis}) and the index $n_j$ defining the
desired level for voltage $V_j$ in the considered arm $j \in
\left\{1,\,2\right\}$ (see~\eqref{MMC_V1_V2}). The indexes $n_j$ are the solution of the optimization problem~\eqref{min_e} described in Sec.~\ref{Ia_Id_prediction}. The algorithm in Fig.~\ref{Control_of_Vci_algorithm} generates as output the control
vector $\T_{j}$ in the considered arm $j$ %, that is for $j \in \left\{1,\,2\right\}$.
(see \eqref{MMC_input_syst_bis}), and works
as
follows.
If $j=1$, according to the
sign notations adopted in Fig.~\ref{2n_Cond_MMC_figure}, a
positive current $I_1 > 0$ recharges the capacitors $C_i$ in the upper arm for which $T_i=1$, making the corresponding voltages $V_{c_i}$ increase.
Therefore, if $j=1$
and $I_1>0$, $T_i$ is set to 1 for the $n_1$ capacitors having the
lowest voltages $V_{c_{i}}$, where $n_1$ identifies the required
number of capacitors to be connected in order to generate the
desired voltage level $V_1$ in the upper arm according
to~\eqref{MMC_V1_V2}. Conversely, if $j=1$ and $I_1<0$, $T_i$ is set
to 1 for the $n_1$ capacitors having the highest voltages
$V_{c_{i}}$. In doing so, the control vector $\T_1$ in \eqref{MMC_input_syst_bis} is determined.
If $j=2$, according to the sign notations adopted in
Fig.~\ref{2n_Cond_MMC_figure}, a positive current $I_2 > 0$  discharges the capacitors $C_i$ in the upper arm for which $T_i=1$, making the corresponding voltages $V_{c_i}$ increase. Therefore, the logic for sorting the capacitors to be activated is opposite with respect to the upper arm $j=1$, in order to determine the control vector $\T_2$ in \eqref{MMC_input_syst_bis}.

\subsection{Calculation of the desired current profile $\Idref$}\label{Idref_calc}

 Substituting $I_s=\Iaref$ and  $\Vsref$ given  in
 (\ref{f_Ia_Va}) in  the second equation of
 (\ref{V_s_and_I_d}) and in
 (\ref{New_VC1_VC2_due}), one obtains the following set of differential equations:
\begin{equation}\label{New_VC1_VC2_due_final}
\left\{
\begin{array}{@{\;}r@{\;\;}c@{\;\;}l@{\;}}
 L\,\dot{I}_d
 &=&
 -R\,I_d
 +V_d
\\
 4\,C_T\,\Vcmi\,\dotVcmi &=& \ds [  2\,V_{dc} - V_d -f(t)] \, (I_d+\Iaref)
\\
 4\,C_T\,\Vcmii\,\dotVcmii &=& \ds [ 2\,V_{dc} - V_d + f(t)] \,(I_d-\Iaref)
\end{array}
\right..
\end{equation}
Note: the last two equations of system (\ref{New_VC1_VC2_due_final})
are nonlinear with respect to variables $\Vcmi$ and $\Vcmii$.
 By introducing the substitutions $2\,\Vcmi\,\dotVcmi=\dotVcmi^2$ and
$2\,\Vcmii\,\dotVcmii=\dotVcmii^2$ in system
(\ref{New_VC1_VC2_due_final}), and expanding the right part, one can
rewrite the system in the following form:
\begin{equation}\label{New_VC1_VC2_due_final_3}
\left\{
\begin{array}{@{}r@{\;}c@{\;}l@{}}
 L\,\dot{I}_d
 &=&
 -R\,I_d
 +V_d,
\\
 2\,C_T\,\dotVcmi^2 &=& \ds   P_1(t) + P_2(t),
\\
 2\,C_T\,\dotVcmii^2 &=& \ds   P_1(t) - P_2(t),
\end{array} \hspace{4mm} \mbox{where} \hspace{4mm}  \begin{array}{@{}r@{\;}c@{\;}l@{}}
 P_1(t) &=& 2\,V_{dc} I_d - V_d I_d - f(t) \, \Iaref,
 \\[1mm]
 P_2(t) &=& 2\,V_{dc}\, \Iaref - V_d \Iaref - f(t) I_d.
  \end{array}
\right.
\end{equation}
The last two equations of \eqref{New_VC1_VC2_due_final_3} are now linear with respect to
variables $\Vcmi^2$ and $\Vcmii^2$.
From (\ref{New_VC1_VC2_due_final_3}), it is also evident that:
 % \\[1mm]
  a)  when  $P_1(t)$ is positive, both voltages $\Vcmi$ and $\Vcmii$
 increase; when $P_1(t)$ is negative, both  voltages $\Vcmi$ and $\Vcmii$
 decrease. This means that function $P_1(t)$ can be used to make
 the two voltages $\Vcmi$ and $\Vcmii$ follow the optimal average capacitor voltages reference $\Vcmref$.
 % \\[1mm]
  b)  when  $P_2(t)$ is positive, voltage  $\Vcmi$ increases and voltage $\Vcmii$
 decreases; when  $P_2(t)$ is negative, voltage  $\Vcmi$ decreases and voltage $\Vcmii$
 increases. This means that function $P_2(t)$ can be used to enforce the condition $\Vcmi\simeq \Vcmii$.
   %}

 \subsubsection{Harmonic analysis of dynamic system
 (\ref{New_VC1_VC2_due_final_3})}\label{Harm_analysis_sect}
Recalling that the desired load current $\Iaref$  and the input voltage
$V_a$ exhibit a sinusoidal behavior as in \eqref{Ia_Vd_1}, we are going to design voltage $V_d$ to exhibit a sinusoidal behavior as well, with the addition of an offset $V_{d0}$ as in \eqref{Ia_Vd_2}:
\vspace{-9.5mm}
\begin{multicols}{2}
\begin{equation}\label{Ia_Vd_1}
 \begin{array}{@{}r@{\;}c@{\;}l@{}}
 \Iaref &=& I_{aM}\sin(\omega t),
 \\[2mm]
 V_{a} &=& V_{aM}\sin(\omega t+\alpha_{V_a}),
  \end{array} \end{equation}\!\!\!\!\!\!\break
%\vspace{-4mm}
%\hspace{-4mm}
\begin{equation}\label{Ia_Vd_2}
 \begin{array}{@{}r@{\;}c@{\;}l@{}} \\[-10mm]
 V_{d} = V_{d0} + V_{dM}\sin(\omega t+\alpha_{V_d}).
  \end{array}
  \end{equation}%\vspace{0.2mm}
\end{multicols}
 The design of the parameters $V_{d0}$, $V_{dM}$ and
 $\alpha_{V_d}$ of voltage $V_d$ in \eqref{Ia_Vd_2} is addressed based on the following harmonic analysis.
 %

%\noindent \mbox{
%\pscircle[fillstyle=solid,fillcolor=black,linecolor=black](1,1){0.5}\hspace{2.1mm}}
\emph{Sinusoidal  behavior of function $I_d$:}
 at steady-state, when voltage $V_d$ in (\ref{Ia_Vd_2}) is applied, the solution of the first equation of system (\ref{New_VC1_VC2_due_final_3}) is:
\begin{equation}\label{Id}
 I_d = I_{d0} + I_{dM}\sin\left(\omega
 t+\alpha_{V_d}-\alpha_{LR}\right) \hspace{4mm} \mbox{where}\hspace{4mm}
\left\{\begin{array}{r@{\;}c@{\;}l}
 I_{d0} &=& \frac{V_{d0}}{R},
\\ I_{dM}&=&\frac{V_{dM}}{\sqrt{R^2\!+\!L^2\omega^2}},
 \\
 \alpha_{LR}&=& \arctan\left(\!\frac{L\omega}{R}\!\right).
 \end{array}\right.
\end{equation}

%\noindent \mbox{
%\pscircle[fillstyle=solid,fillcolor=black,linecolor=black](1,1){0.5}\hspace{2.1mm}}
\emph{Sinusoidal  behavior of function $P_1(t)$:}
 It can be proven, see App.~A, that function  $P_1(t)$ in~\eqref{New_VC1_VC2_due_final_3} can be written as $ P_1(t)  = P_{10} + P_{1}(\omega t)$, where:
\begin{equation}\label{T1t}
\!\!\left\{ \begin{array}{@{\!}r@{}c@{}l}
 \!P_{10}
 &=&
2 V_{dc} I_{d0} \!- \!V_{d0} I_{d0}
%  \\[1mm] & & \hspace{4mm}
 - \frac{V_{dM} I_{dM}}{2}\cos(\!\alpha_{LR}\!)
 \!-\! \frac{f_M I_{aM}}{2}\cos(\!\alpha_f\!), \\[4mm]
  P_{1}(\omega t)
 &=&
 2\,V_{dc} I_{dM}\sin\left(\!\omega
 t\!+\!\alpha_{V_d}\!-\!\alpha_{LR}\right)
%  \\[1mm] & & \hspace{4mm}
  \!-\! F_{V_dI_d}(\omega t)
  \!+ \!\frac{f_M\,I_{aM}}{2}\cos(2\omega t\!+\!\alpha_f).\!\!
\end{array}\right.
\end{equation}
Parameters $f_M$ and $\alpha_f$ are defined in \eqref{fM_alphaf}. Using (\ref{Id}), the constant term $P_{10}$ in \eqref{T1t} can be expressed as:
\begin{equation}\label{T10t}
\begin{array}{@{\!}c}
 P_{10}
 \!=\!
   \frac{2V_{dc} V_{d0}}{R} \!- \!\frac{V_{d0}^2}{R}
  \! -\!\frac{V_{dM}^2}{2\sqrt{R^2+L^2\omega^2}}\!\cos\!\left(\!\alpha_{LR}\!\right)
 \!-\! \frac{f_MI_{aM}}{2} \!\cos(\!\alpha_f\!)
\!=\!  \frac{2V_{dc} V_{d0} - V_{d0}^2-C_0}{R},
\end{array}
\end{equation}
 where parameter $C_0$ is  defined as follows:
\begin{equation}\label{C0_exp}
 C_0= \frac{R\,V_{dM}^2}{2\sqrt{R^2+L^2\omega^2}}\cos\left(\alpha_{LR}\right)+ \frac{R\,f_M\,I_{aM}}{2}
 \cos(\alpha_f).
\end{equation}
 From (\ref{T10t}), one can verify  that term $P_{10}$ is positive if:
\begin{equation}\label{VD0min_exp}
 V_{d0}^{-} = V_{dc} \!-\! \sqrt{V_{dc}^2 \!-\!C_0}<V_{d0} < V_{dc} \!+\! \sqrt{V_{dc}^2
 \!-\!C_0}.
\end{equation}

%\noindent \mbox{
%\pscircle[fillstyle=solid,fillcolor=black,linecolor=black](1,1){0.5}\hspace{2.1mm}}
\emph{Sinusoidal  behavior of function $P_2(t)$:}
It can be proven, see App.~C, that function $P_2(t)$
in~\eqref{New_VC1_VC2_due_final_3} can be written as $ P_2(t)
 = P_{20} + P_{2}(\omega t)$, where:
\begin{equation}\label{T2t}
\left\{ \begin{array}{@{\!}r@{\,}c@{\,}l@{}}
 P_{20}
 &=&
  -
  \frac{V_{dM}\, I_{aM}}{2}\cos(\alpha_{V_d})
  \!-\!
 \frac{f_M\,I_{dM}}{2} \cos(\alpha_f\!-\!\alpha_{V_d}\!+\!\alpha_{LR}), \\[4mm]
  P_{2}(\omega t)
 &=&
 2V_{dc} I_{aM}\sin(\omega t)
  %\\[1mm] & & \hspace{4mm}
  \!-\!
  F_2(\omega t)
 \!-\!
  F_3(\omega t).
\end{array}\right.
\end{equation}
%%%%%%%%%%%%%%%
The sinusoidal functions $F_2(\omega t)$ and $F_3(\omega t)$ are defined in App.~C. After some elaboration, see App.~F, the constant term $P_{20}$ can be written as:
\begin{equation}\label{T20t}
 P_{20}
   \!=\!
 -\frac{V_{dM}\sqrt{a^2\!+\!b^2}}{2} \cos(\alpha_{V_d}\!+\!\gamma), \;
 %\; \mbox{where}\;
   a
  \!=\!I_{aM}  \!+ \!\frac{f_M\cos(\beta)}{\sqrt{R^2\!+\!L^2\omega^2}},
%  \hspace{5mm}
  b
  \!=\! \frac{f_M\sin(\beta)}{\sqrt{R^2\!+\!L^2\omega^2}},
 \end{equation}
where $\beta  =-\alpha_{f}-\alpha_{LR}$ and $ \gamma = \arctan\!2(b,a)$.
The values of parameters $f_M$, $\alpha_{f}$ and $\alpha_{LR}$ are
given  in (\ref{fM_alphaf}) and (\ref{Id}).

 Using (\ref{T1t}) and   (\ref{T2t}),
  system
(\ref{New_VC1_VC2_due_final_3}) can be rewritten as follows:
\begin{equation}\label{Transformed_system_final}
\left\{
\begin{array}{@{\;}r@{\;\;}c@{\;\;}l@{\;}}
 L\,\dot{I}_d
 &=&
 -R\,I_d
 +V_d
\\
 2\,C_T\,\dotVcmi^2 &=& \ds   P_{10} + P_{20} + P_1(\omega t)+ P_2(\omega t)
\\
 2\,C_T\,\dotVcmii^2 &=& \ds   P_{10} - P_{20} + P_1(\omega t)- P_2(\omega t)
\end{array}
\right..
\end{equation}
 The terms $P_1(\omega t)$ and $P_2(\omega t)$ in \eqref{Transformed_system_final} are the sum of
 sinusoidal functions at frequency $\omega$ and $2\omega$ with zero average value. Therefore, they produce a periodic oscillation on variables $\Vcmi^2$ and
  $\Vcmii^2$ at steady-state, but they do not modify their average values. It follows that the average values of the two variables $\Vcmi^2$ and
  $\Vcmii^2$ can only be modified by  the two constant terms $P_{10}$ and  $P_{20}$.

\subsubsection{Algorithm computing the desired current profile $\Idref$}\label{Control_Strategy_sect}

Observation 1): choosing the value  $\alpha_{V_d}=-\gamma$ for the design
parameter $\alpha_{V_d}$, the expression of constant term $P_{20}$
in~\eqref{T20t} simplifies as follows:
\begin{equation}\label{T20t_simp}
 P_{20}
 =
 -\frac{V_{dM}\,\sqrt{a^2+b^2}}{2}.
 \end{equation}
 From (\ref{T20t_simp}), it is clear that parameter $V_{dM}$ directly affects the value of the constant term
 $P_{20}$.  Specifically, the design parameter $V_{dM}$ has the following impact on term
 $P_{20}$ and, consequently, on variables $\Vcmi$ and $\Vcmii$:
\begin{equation}\label{obs_3_eqs}
\left\{\begin{array}{l}
 V_{dM}>0\;\leftrightarrow \;P_{20}<0\;\leftrightarrow \; (\Vcmi\;{\setlength{\unitlength}{3.4mm}  \psset{unit=\unitlength} \SpecialCoor \pcline[linecolor=myorangee,linewidth=1.25pt]{->}(0,1)(0,0)}\;)
 \wedge(\Vcmii\;{\setlength{\unitlength}{3.4mm}  \psset{unit=\unitlength} \SpecialCoor
 \pcline[linecolor=mygreen,linewidth=1.25pt]{->}(0,0)(0,1)}\;) \\
 V_{dM}<0\;\leftrightarrow \;P_{20}>0\;\leftrightarrow \; (\Vcmi\;{\setlength{\unitlength}{3.4mm}  \psset{unit=\unitlength} \SpecialCoor \pcline[linecolor=mygreen,linewidth=1.25pt]{->}(0,0)(0,1)}\;)\wedge(\Vcmii\;{\setlength{\unitlength}{3.4mm}
 \psset{unit=\unitlength} \SpecialCoor \pcline[linecolor=myorangee,linewidth=1.25pt]{->}(0,1)(0,0)}\;)
\end{array}\right.,
\end{equation}
where $\;{\setlength{\unitlength}{3.4mm} \psset{unit=\unitlength}
\SpecialCoor
\pcline[linecolor=myorangee,linewidth=1.25pt]{->}(0,1)(0,0)}\;$ and
$\;{\setlength{\unitlength}{3.4mm}  \psset{unit=\unitlength}
\SpecialCoor
\pcline[linecolor=mygreen,linewidth=1.25pt]{->}(0,0)(0,1)}\;$
 denote a decreasing and an increasing variation of the variable next to the arrow.
 Parameter $V_{dM}$ can therefore be effectively exploited to enforce the condition $\Vcmi \simeq \Vcmii$.

Observation 2):  the design parameter $V_{d0}$ in \eqref{Ia_Vd_2} can be effectively exploited in order
to make the two voltages $\Vcmi$ and $\Vcmii$ track the optimal average capacitor voltages reference $\Vcmref$.
Once the parameters $V_{dM}$ and $\alpha_{V_d}$ have been chosen
as described  in 1)  in order to enforce $\Vcmi\simeq\Vcmii$, and using $P_{10}$ in~\eqref{T10t} and $C_0$
in~\eqref{C0_exp}, the design parameter $V_{d0}$ has the following impact on term
 $P_{10}$ and on variables $\Vcmi$ and $\Vcmii$, see \eqref{T10t}, \eqref{VD0min_exp} and \eqref{Transformed_system_final}:
\begin{equation}\label{obs_5_eqs}
 \left\{\begin{array}{l}
 V_{d0}>V_{d0}^{-}\;
 \leftrightarrow \;
 P_{10}>0\;
 \leftrightarrow \;
 (\Vcmi\;{\setlength{\unitlength}{3.4mm}  \psset{unit=\unitlength} \SpecialCoor \pcline[linecolor=mygreen,linewidth=1.25pt]{->}(0,0)(0,1)}\;)
 \wedge
 (\Vcmii\;{\setlength{\unitlength}{3.4mm}  \psset{unit=\unitlength} \SpecialCoor \pcline[linecolor=mygreen,linewidth=1.25pt]{->}(0,0)(0,1)}\;)
 \\
 V_{d0}<V_{d0}^{-}\;
 \leftrightarrow \;
 P_{10}<0\;
 \leftrightarrow \;
 (\Vcmi\;{\setlength{\unitlength}{3.4mm}  \psset{unit=\unitlength} \SpecialCoor \pcline[linecolor=myorangee,linewidth=1.25pt]{->}(0,1)(0,0)}\;)
 \wedge
 (\Vcmii\;{\setlength{\unitlength}{3.4mm}  \psset{unit=\unitlength} \SpecialCoor \pcline[linecolor=myorangee,linewidth=1.25pt]{->}(0,1)(0,0)}\;)
\end{array}\right.,
\end{equation}
In fact, the constant term $P_{10}$ is positive if parameter
$V_{d0}$ belongs to the range defined in (\ref{VD0min_exp}).
Parameter $V_{d0}$ can therefore be effectively exploited to
increase or decrease the value of both voltages $\Vcmi$ and
$\Vcmii$, in order to make track follow the optimal average capacitor voltages reference $\Vcmref$.

\begin{figure}[tp]
 \includegraphics[clip,width=1\columnwidth]{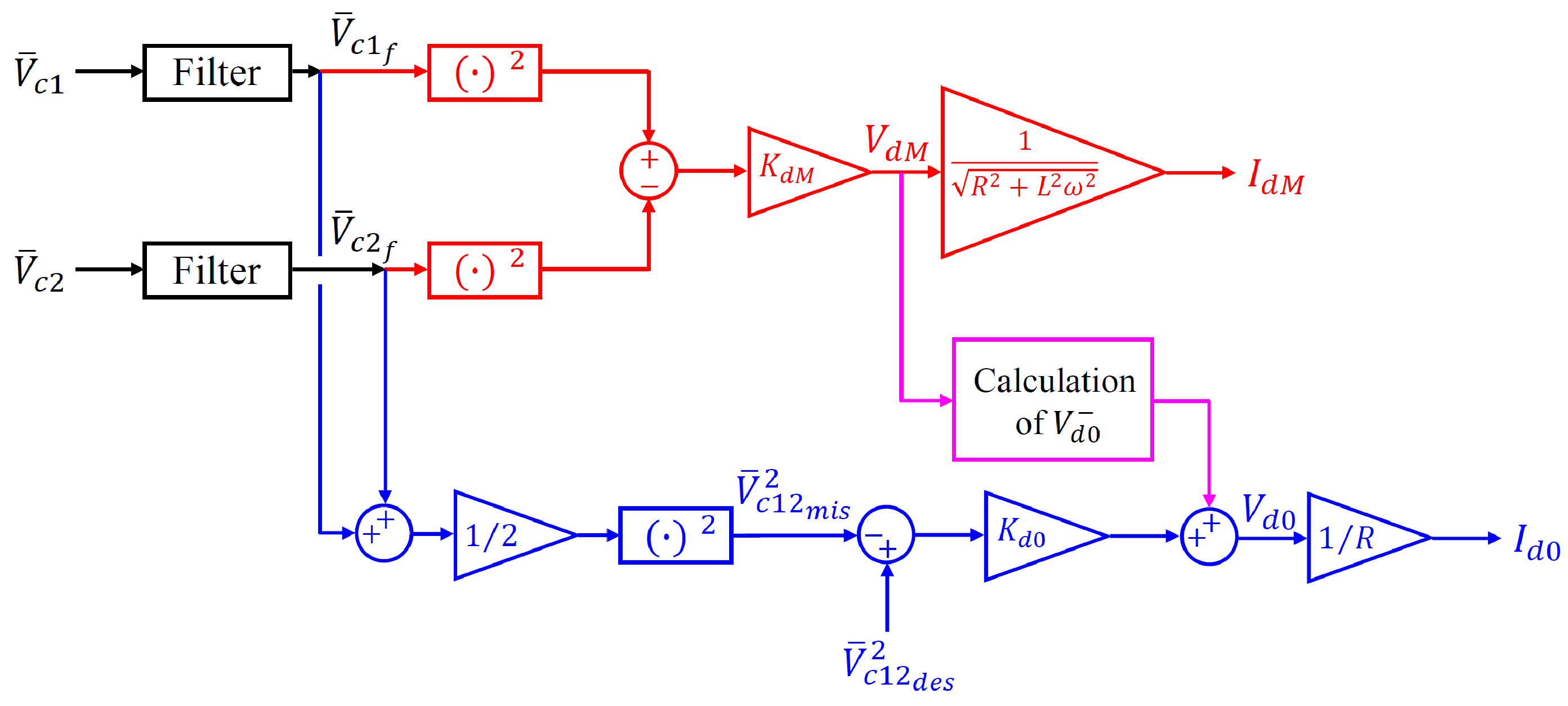}
\caption{Algorithm for the generation of parameters $I_{d0}$ and $I_{dM}$ of the desired current $\Idref$.}\label{simpl_syst_ctrl_sch}
\end{figure}

 Based on the observations 1) and 2) reported above, the parameters $I_{d0}$ and $I_{dM}$ of the desired current profile $\Idref$ having the form described in \eqref{Id}
are computed using the algorithm
reported in Fig.~\ref{simpl_syst_ctrl_sch} consisting in two control loops, where the first loop affects the
second one.

%%%

\vspace{2mm} \noindent \textbf{\red Loop
 1:} The first loop,
highlighted in red in Fig.~\ref{simpl_syst_ctrl_sch}, computes:
\begin{equation}\label{Control_law_V_dM}
 I_{dM}=\dfrac{V_{dM}}{\sqrt{R^2+L^2\,\omega^2}}= \dfrac{K_{dM}}{\sqrt{R^2+L^2\,\omega^2}} (\Vcmif^2-\Vcmiif^2),
\end{equation}
 where $\Vcmif$ and $\Vcmiif$ are a filtered version of the
 voltages $\Vcmi$ and $\Vcmii$.
 This control law, based on the choice $\alpha_{V_d}=-\gamma$,  provides the value of the design parameter $I_{dM}$
 in order  to enforce $\Vcmi \simeq \Vcmii$, and is based on relations (\ref{obs_3_eqs}) and \eqref{Id}.

\vspace{2mm} \noindent \textbf{\blue Loop 2:}
The second loop, highlighted in blue in
Fig.~\ref{simpl_syst_ctrl_sch},computes:
\begin{equation}\label{Control_law_V_d0}
I_{d0}= \dfrac{V_{d0}}{R}=\dfrac{V_{d0}^{-}\! +\! K_{d0} (\Vcmref^2\!-\!\Vcmmis^2)}{R},
 \hspace{1.4mm} \ts
 \Vcmmis \!=\! \frac{\Vcmif\!+\!\Vcmiif}{2},
\end{equation}
 where  $V_{d0}^{-}$ is the threshold  given  in~\eqref{VD0min_exp},
  $\Vcmmis$ is the measured mean value  of voltages $\Vcmif$ and
  $\Vcmiif$, and  $\Vcmref$ is the optimal reference of the average capacitor voltages $\Vcmi$ and
  $\Vcmii$. Note that the
error considered in \eqref{Control_law_V_d0} is
$\Vcmref^2-\Vcmmis^2$ instead of $\Vcmref-\Vcmmis$; this is because
system \eqref{New_VC1_VC2_due_final_3} is linear with respect to
variables $\Vcmif^2$ and $\Vcmiif^2$. The control law in
\eqref{Control_law_V_d0} provides the value of the design parameter
$I_{d0}$ in order to maintain the mean value $\Vcmmis$ track the optimal reference $\Vcmref$ exploiting relations (\ref{obs_5_eqs}).

 The part highlighted in magenta in Fig.~\ref{simpl_syst_ctrl_sch} represents
 the calculation of
 parameter $V_{d0}^{-}=V_{d0}^{-}(V_{dM})$ according
 to~\eqref{C0_exp},~\eqref{VD0min_exp} and using the control parameter $V_{dM}$ generated by the first control loop
 (red one in Fig.~\ref{simpl_syst_ctrl_sch}). The quantity $V_{d0}^{-}=V_{d0}^{-}(V_{dM})$ is then injected into the second
 control loop (blue one in Fig.~\ref{simpl_syst_ctrl_sch}).

 \vspace{2mm} \noindent

 \subsubsection{Optimal average capacitor voltages reference $\Vcmref$}\label{Desired_common_voltage_reference_sect}
  On one hand, \eqref{MMC_V1_V2} shows that smaller values for voltages $\Vcmi$ and $\Vcmii$ allow
to obtain discrete voltage levels for $V_1$ and $V_2$ which are
closer to each other, thus reducing their harmonic content as well as the harmonic content of the generated load current $I_s$ through \eqref{MMC_input_syst}, \eqref{MMC_input_syst_bis} and recalling that $I_s=I_a=I_1+I_2$, see Fig.~\ref{2n_Cond_MMC_figure}. This significantly  enhances the main intrinsic advantages of multilevel converters.  On the other hand, from~\eqref{MMC_V1_V2} it is clear that small
values for voltages $\Vcmi$ and $\Vcmii$ limit the lower boundary of
voltage $V_1$ and the upper boundary of voltage $V_2$. This may
compromise the generation of currents $I_1$ and $I_2$ in the complete
MMC system since the condition \eqref{f_Ia_Va} may not be satisfied, given that $V_s=V_1+V_2$.
The amplitude $I_{aM}$ and the frequency $\omega$ of the desired
load current $\Iaref=\Iaref(t)$ given in~\eqref{Ia_Vd_1} can be,  in general,
time-variant due to  variations of the operating conditions, such as a fault for example, leading to variations of the desired voltage $\Vsref$ through~\eqref{f_Ia_Va}.
 Using the desired $\Idref$ from the algorithm in Fig.~\ref{simpl_syst_ctrl_sch} and the first equation in \eqref{Transformed_system_final}, one obtains the desired reference $\Vdref$ for voltage $V_d$. From $\Vdref$, $\Vsref$ in~\eqref{f_Ia_Va} and using \eqref{Vs_Vd}, one obtains the desired references $\Voneref$ and $\Vtworef$ for voltages $V_1$ and $V_2$:
% %
\begin{equation}  \label{des_V1_V2}
 \Voneref=\dfrac{\Vsref+V_d}{2} \hspace{6.2mm} \mbox{and} \hspace{6.2mm} \Vtworef=\dfrac{\Vsref-V_d}{2}.
\end{equation}
 Let
$\VoneMref$ and $\VtwoMref$ denote the maximum values of the two
desired voltages $\Voneref$ and $\Vtworef$, and let
$\VotMref=(\VoneMref+\VtwoMref)/2$. Since $\VoneMref\simeq
\VtwoMref$ is desired, then $\VotMref\simeq \VoneMref\simeq
\VtwoMref$. The optimal voltage reference $\Vcmref$ used in the control scheme of Fig.~\ref{simpl_syst_ctrl_sch} can be
obtained, for example, from the second equation in \eqref{MMC_V1_V2} when the
desired voltage $\Vtworef$ takes on its maximum value $\VotMref$
(i.e. when $n_2=n$):
\begin{equation}  \label{VC12_des}
\Vcmref = \dfrac{\VotMref+V_{dc}}{n}.
\end{equation}

%%%%%%%%%%%%%%%%%%%%%%%%%%%%%%%%%%%%%%%%%%%%%%%%%%%%%%%%%%%%%%%%%%%
%%%%%%%%%%%%%%%%%%%%%%%%%%%%%%%%%%%%%%%%%%%%%%%%%%%%%%%%%%%%%%%%%%%
%%%%%%%%%%%%%%%%%%%%%%%%%%%%%%%%%%%%%%%%%%%%%%%%%%%%%%%%%%%%%%%%%%%
%%%%%%%%%%%%%%%%%%%%%%%%%%%%%%%%%%%%%%%%%%%%%%%%%%%%%%%%%%%%%%%%%%%

\subsection{Solution of Goals 1.2 and 2 of the Control Problem}\label{prob_12_2}
The inductive part of the MMC model in \eqref{POG_Trifase_Vin_SCH_LAB_red} can be rewritten as:
\begin{equation}\label{POG_Trifase_Vin_SCH_LAB_red_BIS}
 \mat{@{\,}c@{\,}}{ \dot I_{1}\\ \dot I_{2}}={\L}^{-1}_L \A_L \mat{@{\,}c@{\,}}{ I_{1}\\ I_{2}}+{\L}^{-1}_L\; \mat{@{\,}c@{\,}}{  V_1-V_a\\ V_2-V_a}.
\end{equation}
Let $I_1(k)$, $I_2(k)$ and $I_1(k+1)$,  $I_2(k+1)$ be the values of currents $I_1$ and $I_2$ at the discrete time instants $t=k\,T_s$ and $t=(k+1)\,T_s$, where $T_s$ is the sampling time. Recalling the definition of incremental ratio, one can write:
\begin{equation}\label{Incr_Ratio}
\dot I_{1} \approx \dfrac{I_1(k+1)-I_1(k)}{T_s}, \hspace{10mm}
\dot I_{2} \approx \dfrac{I_2(k+1)-I_2(k)}{T_s}.
\end{equation}
Replacing \eqref{Incr_Ratio} in \eqref{POG_Trifase_Vin_SCH_LAB_red_BIS}, one can discretize system \eqref{POG_Trifase_Vin_SCH_LAB_red_BIS} using the Euler's forward method as follows:
\begin{equation}\label{POG_Trifase_Vin_SCH_LAB_red_BIS_BIS}
 \mat{@{\,}c@{\,}}{ I_{1}(k+1)\\ I_{2}(k+1)}\!=\! \mat{@{\,}c@{\,}}{ I_{1}(k)\\ I_{2}(k)}\! +\!{\L}^{-1}_L \A_L \!\mat{@{\,}c@{\,}}{ I_{1}(k)\\ I_{2}(k)}\!T_s+{\L}^{-1}_L\!\mat{@{\,}c@{\,}}{ V_1(k)-V_a(k)\\ V_2(k)-V_a(k)}\!T_s,
\end{equation}
where $V_1(k)$, $V_2(k)$ and $V_a(k)$ are the values of voltages $V_1$, $V_2$ and $V_a$ at $t=k\,T_s$.

\subsubsection{Optimal Control Problem on Currents $I_a$ and $I_d$}\label{Ia_Id_prediction}
The output of the optimal control problem shown in Fig.~\ref{var_ctrl_alg_figure} are the optimal indexes $n_1=n_{1_{opt}}(k)$ and $n_2=n_{2_{opt}}(k)$ to be fed to the algorithm in Fig.~\ref{Control_of_Vci_algorithm} described in Sec.~\ref{prob_11}.
Let $n_{1}(k-1)$ and $n_{2}(k-1)$ be the optimal indexes given by the optimal control
at the previous time step, in order to generate the voltage levels $V_1(k-1)$
and $V_2(k-1)$ according to~\eqref{MMC_V1_V2}, and let $w_n$ be a window
parameter that can take on discrete values from the set
$\{0,\,\cdots,\,n\}$. The indexes $n_1(k)$ and $n_2(k)$ can take on values such that $n_1 \in
\{n_{1}(k-1)-w_n,\,\cdots,\, n_{1}(k-1)+w_n\}$ and $n_2 \in
\{n_{2}(k-1)-w_n,\,\cdots,\, n_{2}(k-1)+w_n\}$. Note that decreasing
the window $w_n$ has the following consequences: 1) the maximum step
between two consecutive voltage levels is decreased, thus
emphasizing the intrinsic advantages of multilevel converters; 2)
there are fewer values of $n_1(k)$ and $n_2(k)$ to choose from for the
current control step, thus reducing the computational burden. Two
nested ``for'' cycles are implemented over the possible values of $n_1(k)$ and $n_2(k)$, predicting the
future errors $e_{I_a}(k+1)$ and $e_{I_d}(k+1)$ between the
desired and actual load currents $\Iaref(k+1)$ and $I_s(k+1)$
and between the desired and actual circulating
currents $\Idref(k+1)$ and $I_d(k+1)$. These predictions are
made for each admissible value of $n_1(k)$ and $n_2(k)$, as depicted by the pseudo-code reported in Fig.~\ref{var_ctrl_alg_figure}.

\begin{figure}[h] \centering
%\hspace{-26mm}
\begin{tabular}{l@{\;}l@{\;\;}l}\hline
 %%%%%%%%%%%%%%%%%%%%%%%%%%%%%%%%%%%%%%%%%%%%%%%%%%%%%%%%%%%%%%
 1. & {\bf for} $n_{1}(k)=n_{1}(k-1)-w_n:n_{1}(k-1)+w_n$                                                                               & \%  \emph{cycle $n_1$} \\
  %\in \{n_{1_{o}}\!-\!w_n,\cdots, n_{1_{o}}\!+\!w_n\}$} \\
 2. & \hspace{3mm}     {\bf for} $n_{2}(k)=n_{2}(k-1)-w_n:n_{2}(k-1)+w_n$                                                              & \%  \emph{cycle $n_2$} \\
  %\in \{n_{2_{o}}\!-\!w_n,\cdots, n_{2_{o}}\!+\!w_n\}$} \\
 3. & \hspace{6mm}         $\left[\begin{array}{@{\!}c@{\!}}V_{1}(k)\\[2mm]V_{2}(k)\end{array}\right]\!\!=\!\!
                             \left[\begin{array}{@{\!}c@{\!}} V_{dc}\!-n_{1}(k) V_{C1}(k)\\[2mm]
                              -\!V_{dc}\!+\!n_{2}(k) V_{C2}(k)\end{array}\right]\!\!;$                                    & \%  \emph{Eq.~\eqref{MMC_V1_V2}} \\[8mm]
 4. & \hspace{6mm}          $\mat{@{\!}c@{\!}}{ I_{1}(k\!+\!1)\\ I_{2}(k\!+\!1)}\!\!=\!\! \mat{@{\!}c@{\!}}{ I_{1}(k)\\ I_{2}(k)} \!\!+\!{\L}^{-1}_L \! \A_L \!\!\mat{@{\!}c@{\!}}{ I_{1}(k)\\ I_{2}(k)}\!\!T_s\!+\!{\L}^{-1}_L\!\!\mat{@{\!}c@{\!}}{V_1(k)\!\!-\!\!V_a(k)\\ V_2(k)\!\!-\!\!V_a(k)}\!\!T_s$
                             & \%  \emph{Eq.~\eqref{POG_Trifase_Vin_SCH_LAB_red_BIS_BIS}}  \\[8mm]
 5. & \hspace{6mm}         $I_{s}(k\!+\!1)\!=\!I_{1}(k\!+\!1)\!+\!I_{2}(k\!+\!1)$;                                           & \%  \emph{Eq.~\eqref{Vs_Vd}} \\
 6. & \hspace{6mm}         $I_{d}(k\!+\!1)\!=\!I_{1}(k\!+\!1)\!-\!I_{2}(k\!+\!1)$;                                           & \%  \emph{Eq.~\eqref{Vs_Vd}} \\
 7. & \hspace{6mm}         $e_{I_{a}}(k\!+\!1)\!=\!\Iaref(k\!+\!1)\!-\!I_{s}(k\!+\!1)$;                                   & \\ %\%  \emph{Error $e_{I_{a}}(k+1)$} \\
 8. & \hspace{6mm}         $e_{I_{d}}(k\!+\!1)\!=\!I_{d_{des}}(k\!+\!1)\!-\!I_{d}(k\!+\!1)$;                                   & \\ %\%  \emph{Error $e_{I_{d}}(k+1)$} \\
 9. & \hspace{3mm}     {\bf end}     \\
 10. & {\bf end}     \\ \hline
\end{tabular}
 \vspace{-3  mm}
 \caption{Pseudo-code for the prediction of current errors $e_{I_{a}}(k\!+\!1)$ and $e_{I_{d}}(k\!+\!1)$.}\label{var_ctrl_alg_figure}
 \end{figure}

The optimal indexes $n_{1_{opt}}(k)$ and $n_{2_{opt}}(k)$ are determined by solving the following optimal control problem:
\begin{equation}\label{min_e}
\ds \{n_{1_{opt}}(k),\,n_{2_{opt}}(k)\}= \min_{n_{1}(k),\,n_{1}(k)}{(\alpha_1 e_{I_a}(k+1)+\alpha_2 e_{I_d}(k+1))},
\end{equation}
where the two weights $\alpha_1$ and $\alpha_2$ composing the objective function in~\eqref{min_e} satisfy the constraint $\alpha_1+\alpha_2=1$. The optimal control problem \eqref{min_e} can be solved exactly thanks to the current errors prediction described in Fig.~\ref{var_ctrl_alg_figure}.

\section{Simulations}\label{Simulations_sect}

\begin{figure*}[htbp]
\psfrag{Iades (r--) and Is (b)}[b][b][0.86]{Currents $I_s$ and $\Iaref$}
\psfrag{Iat}[][][0.55]{$\;\;\;\;I_s$}
\psfrag{Iades}[][][0.55]{$\Iaref$}
\psfrag{Vc1 (b), Vc2 (or), Vc12des (g--)}[b][b][0.86]{Voltages $\Vcmi$, $\Vcmii$ and  $\Vcmref$}
\psfrag{Vc1m}[][][0.55]{$\Vcmi$}
\psfrag{Vc2m}[][][0.55]{$\Vcmii$}
\psfrag{Vcreft}[][][0.55]{$\Vcmref$}
\psfrag{V1des (b), V1Mdes (g), V1Mmin (m)}[b][b][0.86]{Voltage $V_1$ actual its minimum allowed boundary $-V_{12M}$}
\psfrag{V1}[][][0.5]{$\;\;\;\;\;\;V_1$}
\psfrag{V1lb}[][][0.5]{$\;\;-\!V_{12M}$}
\psfrag{mMaxV1}[][][0.5]{$-V_{12M}$}
\psfrag{V2des (b), V2Mdes (g), V2Mmax (m)}[b][b][0.86]{Voltage $V_2$ and its maximum allowed boundary $V_{12M}$}
\psfrag{V2}[][][0.5]{$\;\;\;\;\;V_2$}
\psfrag{V2lb}[][][0.5]{$\;\;\VotMref$}
\psfrag{mMaxV2}[][][0.5]{$V_{12M}$}
\psfrag{[A]}[][t][0.8]{[A]}
\psfrag{[V]}[][t][0.8]{[V]}
\psfrag{Time [s]}[b][b][0.8]{Time [s]}
\includegraphics[clip,width=1\textwidth]{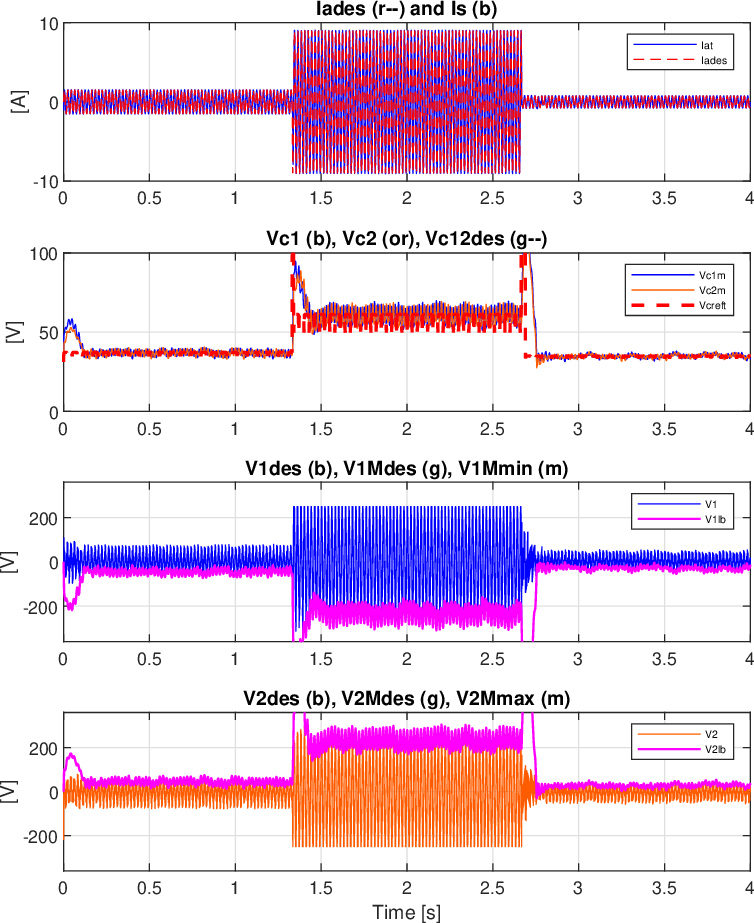}
 \caption{Results of the first simulation.}\label{Res_Figure_1}
\end{figure*}

\begin{figure*}[htbp]
\psfrag{Iades (r--) and Is (b)}[b][b][0.86]{Currents $I_s$ and $\Iaref$}
\psfrag{Iat}[][][0.55]{$\;\;\;\;\;I_s$}
\psfrag{Iades}[][][0.55]{$\Iaref$}
\psfrag{Vc1 (b), Vc2 (or), Vc1f and Vc2f (m--), Vc12des (g--)}[b][b][0.86]{Voltages $\Vcmi$, $\Vcmii$ and  $\Vcmref$}
\psfrag{Vc1m}[][][0.55]{$\Vcmi$}
%\psfrag{Vc1mf}[][][0.55]{$\Vcmif$}
\psfrag{Vc2m}[][][0.55]{$\Vcmii$}
%\psfrag{Vc2mf}[][][0.55]{$\Vcmiif$}
\psfrag{Vcreft}[][][0.55]{$\Vcmref$}
\psfrag{V1 (c)}[b][b][0.86]{Voltage $V_1$ and its filtering ($V_{1_{f}}$), $V_1$ levels}
\psfrag{V1}[][][0.55]{$\;\;\;\;V_1$}
\psfrag{V1f}[][][0.55]{$\;\;V_{1_{f}}$}
\psfrag{V1Lv}[][][0.55]{$V_1$ lv}
\psfrag{V2 (c)}[b][b][0.86]{Voltage $V_2$ and its filtering ($V_{2_{f}}$), $V_2$ levels}
\psfrag{V2}[][][0.55]{$\;\;\;\;V_2$}
\psfrag{V2f}[][][0.55]{$\;\;V_{2_{f}}$}
\psfrag{V2Lv}[][][0.55]{$V_2$ lv}
\psfrag{[A]}[][t][0.8]{[A]}
\psfrag{[V]}[][t][0.8]{[V]}
\psfrag{Time [s]}[b][b][0.8]{Time [s]}
\includegraphics[clip,width=1\textwidth]{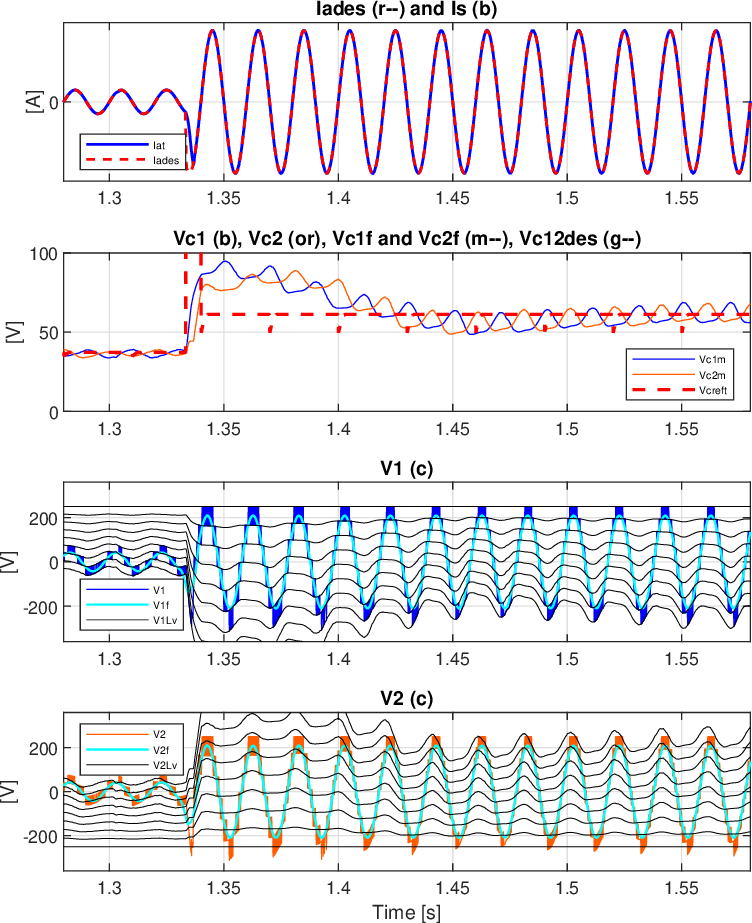}
 \caption{Results of the first simulation: zoom-in.}\label{Res_Figure_2}
\end{figure*}

\begin{figure*}[htbp]
\psfrag{Iades (r--) and Is (b)}[b][b][0.86]{Currents $I_s$ and $\Iaref$}
\psfrag{Iat}[][][0.55]{$\;\;\;\;I_s$}
\psfrag{Iades}[][][0.55]{$\Iaref$}
\psfrag{Vc1 (b), Vc2 (or), Vc12des (g--)}[b][b][0.86]{Voltages $\Vcmi$, $\Vcmii$ and  $\Vcmref$}
\psfrag{Vc1m}[][][0.55]{$\Vcmi$}
\psfrag{Vc2m}[][][0.55]{$\Vcmii$}
\psfrag{Vcreft}[][][0.55]{$\Vcmref$}
\psfrag{V1des (b), V1Mdes (g), V1Mmin (m)}[b][b][0.86]{Voltage $V_1$ actual its minimum allowed boundary $-V_{12M}$}
\psfrag{V1}[][][0.5]{$\;\;\;\;\;\;V_1$}
\psfrag{V1lb}[][][0.5]{$\;\;-\!V_{12M}$}
\psfrag{mMaxV1}[][][0.5]{$-V_{12M}$}
\psfrag{V2des (b), V2Mdes (g), V2Mmax (m)}[b][b][0.86]{Voltage $V_2$ and its maximum allowed boundary $V_{12M}$}
\psfrag{V2}[][][0.5]{$\;\;\;\;\;V_2$}
\psfrag{V2lb}[][][0.5]{$\;\;\VotMref$}
\psfrag{mMaxV2}[][][0.5]{$V_{12M}$}
\psfrag{[A]}[][t][0.8]{[A]}
\psfrag{[V]}[][t][0.8]{[V]}
\psfrag{Time [s]}[b][b][0.8]{Time [s]}
\includegraphics[clip,width=1\textwidth]{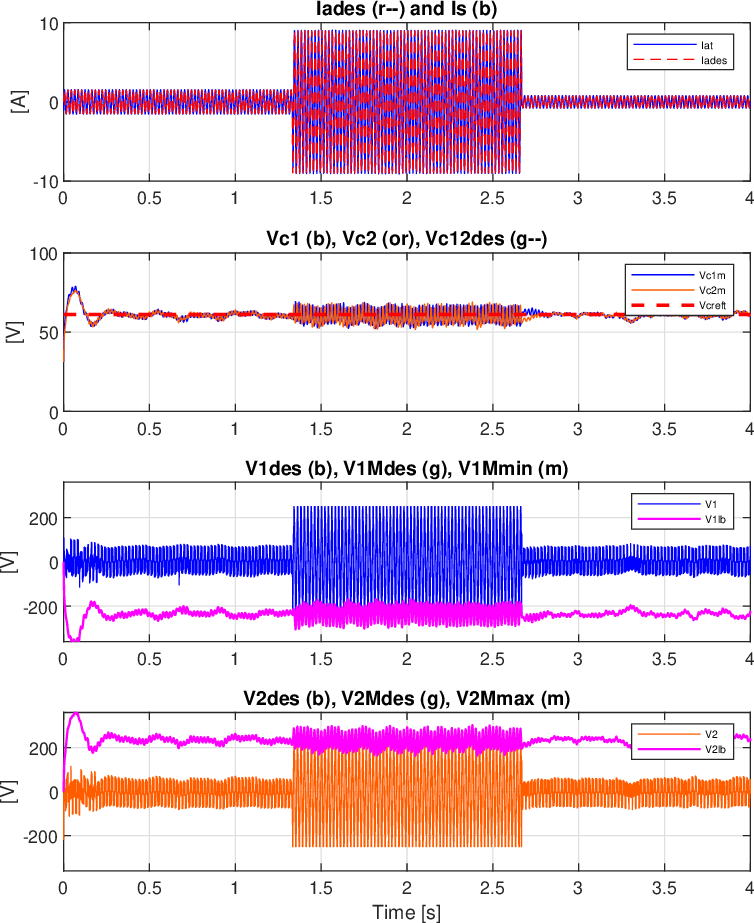}
 \caption{Results of the second simulation.}\label{Res_Figure_4}
\end{figure*}

\begin{figure*}[htbp]
\psfrag{Iades (r--) and Is (b)}[b][b][0.86]{Currents $I_s$ and $\Iaref$}
\psfrag{Iat}[][][0.55]{$\;\;\;\;\;I_s$}
\psfrag{Iades}[][][0.55]{$\Iaref$}
\psfrag{Vc1 (b), Vc2 (or), Vc1f and Vc2f (m--), Vc12des (g--)}[b][b][0.86]{Voltages $\Vcmi$, $\Vcmii$ and  $\Vcmref$}
\psfrag{Vc1m}[][][0.55]{$\Vcmi$}
%\psfrag{Vc1mf}[][][0.55]{$\Vcmif$}
\psfrag{Vc2m}[][][0.55]{$\Vcmii$}
%\psfrag{Vc2mf}[][][0.55]{$\Vcmiif$}
\psfrag{Vcreft}[][][0.55]{$\Vcmref$}
\psfrag{V1 (c)}[b][b][0.86]{Voltage $V_1$ and its filtering ($V_{1_{f}}$), $V_1$ levels}
\psfrag{V1}[][][0.55]{$\;\;\;\;V_1$}
\psfrag{V1f}[][][0.55]{$\;\;V_{1_{f}}$}
\psfrag{V1Lv}[][][0.55]{$V_1$ lv}
\psfrag{V2 (c)}[b][b][0.86]{Voltage $V_2$ and its filtering ($V_{2_{f}}$), $V_2$ levels}
\psfrag{V2}[][][0.55]{$\;\;\;\;V_2$}
\psfrag{V2f}[][][0.55]{$\;\;V_{2_{f}}$}
\psfrag{V2Lv}[][][0.55]{$V_2$ lv}
\psfrag{[A]}[][t][0.8]{[A]}
\psfrag{[V]}[][t][0.8]{[V]}
\psfrag{Time [s]}[b][b][0.8]{Time [s]}
\includegraphics[clip,width=1\textwidth]{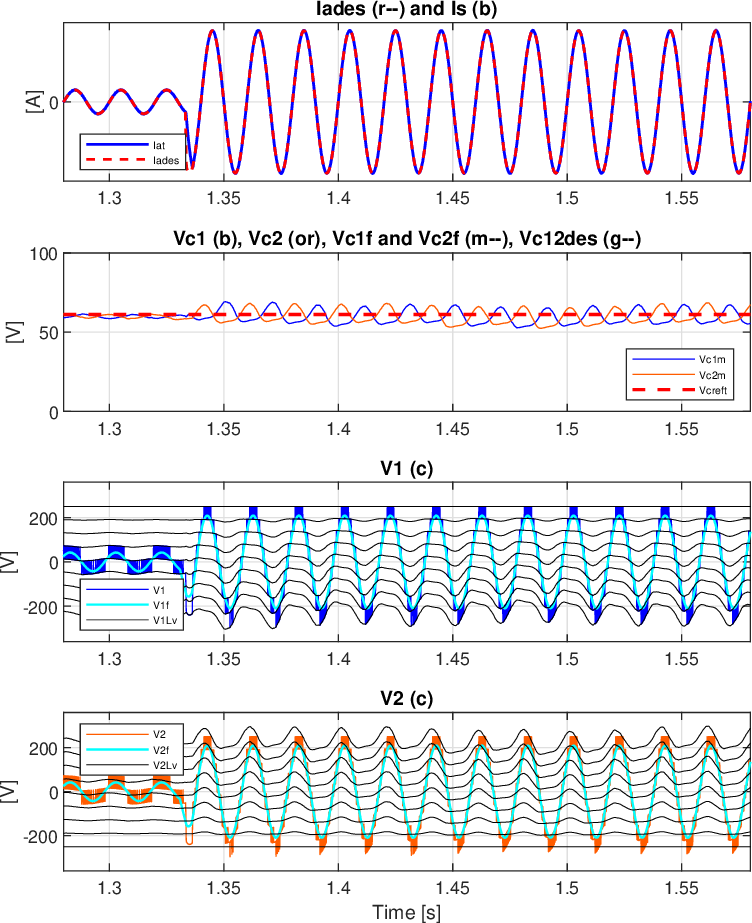}
 \caption{Results of the second simulation: zoom-in.}\label{Res_Figure_5}
\end{figure*}

Two simulations have been performed on the complete MMC system~\eqref{MMC_complt_syst} using the simulink implementation in Fig.~\ref{MMC_complete_POG}.
The capacitor voltages balancing conditions $V_{c_1}\simeq\cdots\simeq V_{c_n} \simeq \Vcmi$ and $V_{c_{n+1}}\simeq\cdots\simeq V_{c_{2\,n}}\simeq \Vcmii$ on the capacitor voltages in the upper and lower arms of the converter are enforced using the algorithm reported in Fig.~\ref{Control_of_Vci_algorithm} and described in Sec.~\ref{prob_11}, which solved the Goal 1.1 of the Control Problem in Sec.~\ref{Control_of_MMC_system}. The desired profile $\Idref$ for current $I_d$ is generated using the scheme reported
in Fig.~\ref{simpl_syst_ctrl_sch}. The conditions $\Vcmi\simeq \Vcmii\simeq \Vcmref$ (that is $I_d \simeq \Idref$) and $I_a \simeq \Iaref$ are achieved by solving the optimal control problem~\eqref{min_e} in Sec.~\ref{prob_12_2} and implementing the prediction for the future values of current errors $e_{I_d(k+1)}$ and $e_{I_a(k+1)}$ reported in Fig.~\ref{var_ctrl_alg_figure}.  The MMC system and control parameters are reported in Table~\ref{syst_and_ctrl_param_varying}.
In order to simulate changes in the system operating conditions, such as for example faults, the amplitude $I_{aM}$ of the desired load current $\Iaref$ has been
chosen to vary as shown in the top subplot of Fig.~\ref{Res_Figure_1}, Fig.~\ref{Res_Figure_4} and in Table~\ref{syst_and_ctrl_param_varying}. In the first simulation, see Fig.~\ref{Res_Figure_1} and Fig.~\ref{Res_Figure_2}, the reference $\Vcmref$ for the average capacitor voltages $\Vcmi$ and $\Vcmii$ is the optimal one computed as in Sec.~\ref{Desired_common_voltage_reference_sect}. In the second simulation, see Fig.~\ref{Res_Figure_4} and Fig.~\ref{Res_Figure_5}, the reference $\Vcmref$ for the average capacitor voltages $\Vcmi$ and $\Vcmii$ is kept constant and equal to the minimum value which is strictly needed to follow the desired load current $\Iaref$ during the whole simulation, which is given by the case $I_{aM}=9$ A representing the most demanding situation.
The frequency $\omega$ of
the desired current $\Iaref$ and of the input voltage $V_a$ has been
chosen to be constant and equal to $2\,\pi\,50$ rad/s. The initial capacitor voltages are equal to $31.25$ V, whereas the initial inductor currents are equal to $0$ A.

%\noindent \mbox{
%\pscircle[fillstyle=solid,fillcolor=black,linecolor=black](1,1){0.5}\hspace{2.1mm}}
\emph{First Simulation:}
 The results of the simulation using the optimal average capacitor voltages reference $\Vcmref$ are shown in
Fig.~\ref{Res_Figure_1} and in Fig.~\ref{Res_Figure_2}.
\begin{table}[t]
  \centering
  \caption{MMC system and control parameters.}\label{syst_and_ctrl_param_varying}
\begin{tabular}{|c|c|c|c|c|c|}
\hline
 \multicolumn{2}{|c|}{$L = 10$ [mH]} & \multicolumn{2}{|c|}{$R = 0.1$ [$\Omega$]} & \multicolumn{2}{|c|}{$C_i = 1000$ [$\mu$ F]} \\ \hline
 \multicolumn{2}{|c|}{$L_a = 50$ [mH]} & \multicolumn{2}{|c|}{$R_a = 19$ [$\Omega$]} &  \multicolumn{2}{|c|}{$V_a\!=\!10\sin(2\pi 50\,t\!+\!\pi/6)$ [V]}  \\ \hline
 \multicolumn{2}{|c|}{$n = 8$} & \multicolumn{2}{|c|}{$V_{dc} = 250$ [V]} & \multicolumn{2}{|c|}{$T_s=10^{-4}$ [s]}  \\ \hline
 \multicolumn{2}{|c|}{$\!\!I_{aM}\!=\!1.5, 9 ,  0.75\!\!$ [A]$\!\!$} & \multicolumn{2}{|c|}{$\!\!1^{\!st}\!\!$ order $\!\!\!$ filt. $\!\tau\!=\!0.0318\!\!$} & \multicolumn{2}{|c|}{$\!\!K_{d0} \!=\! 0.10\!\cdot\! 10^{-3}$}   \\ \hline
 \multicolumn{2}{|c|}{$\!K_{dM} \!=\!\ 1.5\!\cdot\! 10^{-3}\!\!$} & \multicolumn{2}{|c|}{$\alpha_1=0.99$, $\alpha_2=0.01$} &
 \multicolumn{2}{|c|}{$w_n=1$}
 \\ \hline
\end{tabular}
\end{table}
%
%%%%%%%%%%%%%%%%%%%%%%%%%%%%%%%%%%%%%%%%%%%%%%%%%%%%%%%%%%%%%%%%%%%%%%%%%%%%%%%%%%
%%%%%%%%%%%%%%%%%%%%%%%%%%%%%%%%%%%%%%%%%%%%%%%%%%%%%%%%%%%%%%%%%%%%%%%%%%%%%%%%%%
%%%%%%%%%%%%%%%%%%%%%%%%%%%%%%%%%%%%%%%%%%%%%%%%%%%%%%%%%%%%%%%%%%%%%%%%%%%%%%%%%%
%%%%%%%%%%%%%%%%%%%%%%%%%%%%%%%%%%%%%%%%%%%%%%%%%%%%%%%%%%%%%%%%%%%%%%%%%%%%%%%%%%
%%%%%%%%%%%%%%%%%%%%%%%%%%%%%%%%%%%%%%%%%%%%%%%%%%%%%%%%%%%%%%%%%%%%%%%%%%%%%%%%%%
%%%%%%%%%%%%%%%%%%%%%%%%%%%%%%%%%%%%%%%%%%%%%%%%%%%%%%%%%%%%%%%%%%%%%%%%%%%%%%%%%%
%%%%%%%%%%%%%%%%%%%%%%%%%%%%%%%%%%%%%%%%%%%%%%%%%%%%%%%%%%%%%%%%%%%%%%%%%%%%%%%%%%
%%%%%%%%%%%%%%%%%%%%%%%%%%%%%%%%%%%%%%%%%%%%%%%%%%%%%%%%%%%%%%%%%%%%%%%%%%%%%%%%%%
%%%%%%%%%%%%%%%%%%%%%%%%%%%%%%%%%%%%%%%%%%%%%%%%%%%%%%%%%%%%%%%%%%%%%%%%%%%%%%%%%%
%%%%%%%%%%%%%%%%%%%%%%%%%%%%%%%%%%%%%%%%%%%%%%%%%%%%%%%%%%%%%%%%%%%%%%%%%%%%%%%%%%
%%%%%%%%%%%%%%%%%%%%%%%%%%%%%%%%%%%%%%%%%%%%%%%%%%%%%%%%%%%%%%%%%%%%%%%%%%%%%%%%%%
%%%%%%%%%%%%%%%%%%%%%%%%%%%%%%%%%%%%%%%%%%%%%%%%%%%%%%%%%%%%%%%%%%%%%%%%%%%%%%%%%%
%%%%%%%%%%%%%%%%%%%%%%%%%%%%%%%%%%%%%%%%%%%%%%%%%%%%%%%%%%%%%%%%%%%%%%%%%%%%%%%%%%
%%%%%%%%%%%%%%%%%%%%%%%%%%%%%%%%%%%%%%%%%%%%%%%%%%%%%%%%%%%%%%%%%%%%%%%%%%%%%%%%%%
%%%%%%%%%%%%%%%%%%%%%%%%%%%%%%%%%%%%%%%%%%%%%%%%%%%%%%%%%%%%%%%%%%%%%%%%%%%%%%%%%%
%%%%%%%%%%%%%%%%%%%%%%%%%%%%%%%%%%%%%%%%%%%%%%%%%%%%%%%%%%%%%%%%%%%%%%%%%%%%%%%%%%
%%%%%%%%%%%%%%%%%%%%%%%%%%%%%%%%%%%%%%%%%%%%%%%%%%%%%%%%%%%%%%%%%%%%%%%%%%%%%%%%%%
%%%%%%%%%%%%%%%%%%%%%%%%%%%%%%%%%%%%%%%%%%%%%%%%%%%%%%%%%%%%%%%%%%%%%%%%%%%%%%%%%%
%%%%%%%%%%%%%%%%%%%%%%%%%%%%%%%%%%%%%%%%%%%%%%%%%%%%%%%%%%%%%%%%%%%%%%%%%%%%%%%%%%
%%%%%%%%%%%%%%%%%%%%%%%%%%%%%%%%%%%%%%%%%%%%%%%%%%%%%%%%%%%%%%%%%%%%%%%%%%%%%%%%%%
%%%%%%%%%%%%%%%%%%%%%%%%%%%%%%%%%%%%%%%%%%%%%%%%%%%%%%%%%%%%%%%%%%%%%%%%%%%%%%%%%%
%%%%%%%%%%%%%%%%%%%%%%%%%%%%%%%%%%%%%%%%%%%%%%%%%%%%%%%%%%%%%%%%%%%%%%%%%%%%%%%%%%
%%%%%%%%%%%%%%%%%%%%%%%%%%%%%%%%%%%%%%%%%%%%%%%%%%%%%%%%%%%%%%%%%%%%%%%%%%%%%%%%%%
%%%%%%%%%%%%%%%%%%%%%%%%%%%%%%%%%%%%%%%%%%%%%%%%%%%%%%%%%%%%%%%%%%%%%%%%%%%%%%%%%%
%%%%%%%%%%%%%%%%%%%%%%%%%%%%%%%%%%%%%%%%%%%%%%%%%%%%%%%%%%%%%%%%%%%%%%%%%%%%%%%%%%
%%%%%%%%%%%%%%%%%%%%%%%%%%%%%%%%%%%%%%%%%%%%%%%%%%%%%%%%%%%%%%%%%%%%%%%%%%%%%%%%%%
From Fig.~\ref{Res_Figure_1}, one can observe that the reference $\Vcmref$ for the average capacitor voltages $\Vcmi$ and $\Vcmii$ computed as in Sec.~\ref{Desired_common_voltage_reference_sect} is indeed the optimal one: 1) the load current $I_s$ always follows the desired profile $\Iaref$ as shown in the first subplot, except for very short transients when the amplitude $I_{aM}$ changes; 2) the lower boundary $-V_{12M}$ of voltage $V_1$ and the upper boundary $V_{12M}$ of voltage $V_2$, computed by replacing $n_1=n_2=n$ in \eqref{MMC_V1_V2}, always coincide with the peak value of voltages $V_1$ and $V_2$ except for the very short transients when the amplitude $I_{aM}$ changes, as shown in the third and fourth subplots of Fig.~\ref{Res_Figure_1}. This proves the effectiveness of the approach described in Sec.~\ref{Desired_common_voltage_reference_sect} for the computation of the optimal voltage reference $\Vcmref$ for voltages $\Vcmi$ and $\Vcmii$. The capacitor voltages $V_{c_{i+(j-1)\times
n}}$ for $i \in \left\{1,\,\cdots\, n\right\}$, in the lower and upper converter arms  $j \in \left\{1,\,2\right\}$, are kept balanced and equal to the corresponding mean values $\Vcmi$ and $\Vcmii$ by the algorithm in Fig.~\ref{Control_of_Vci_algorithm}.
%, see \cite{Suppl_Mat_Ref}.
The good tracking of the optimal reference $\Vcmref$ by voltages $\Vcmi$ and $\Vcmii$ can be observed from the second subplot of Fig.~\ref{Res_Figure_1}, and shows: 1) The effectiveness of the algorithm proposed in Fig.~\ref{simpl_syst_ctrl_sch} for the generation of the desired current $\Idref$; 2) the effectiveness of the optimal control problem in~\eqref{min_e}, which could be solved exactly thanks to the prediction described by the code in Fig.~\ref{var_ctrl_alg_figure}. Furthermore, Fig.~\ref{Res_Figure_1} also shows the very good superposition between the desired load current $\Iaref$ and the load current $I_s$.

The optimal voltage reference $\Vcmref$ is computed using \eqref{des_V1_V2}-\eqref{VC12_des} as described in Sec.~\ref{Desired_common_voltage_reference_sect},
and represents the minimum value which is strictly needed
 for voltage $V_s$ to track the desired value $\Vsref$ in \eqref{f_Ia_Va},
 and for the load current $I_a$  to track the desired value $\Iaref$. Note that it is very convenient to make voltages $\Vcmi$ and $\Vcmii$ follow $\Vcmref$; this is true because larger $\Vcmi$ and $\Vcmii$ would cause the discrete voltage levels generating voltages $V_1$ and $V_2$ to be more distant from each other.
As an example, Fig.~\ref{Res_Figure_2} shows a zoom-in of Fig.~\ref{Res_Figure_1} for $t\in[1.28,\;1.58]$ s. Note that the transient on voltages $\Vcmi$ and $\Vcmii$ in correspondence of $I_{aM}= 1.5 \rightarrow 9$ A is very short. Furthermore, one can observe that the available voltage levels (black characteristics in the figure) generating voltages $V_1$ and $V_2$ are much closer to each other whenever $I_{aM}$ is lower. This brings two important advantages: 1) a lower harmonic content in the resulting load current $I_s$; 2) a better tracking of current $I_s$ itself, leading to a cleaner sinusoid, as proved in the comparison with the second simulation.

\begin{table}[t]
  \centering
  \caption{Metrics evaluating the resulting load current $I_s$ in the first and second simulation.}\label{MMC_metrics_table}
\begin{tabular}{|p{3cm}|p{2.2cm}|p{2.2cm}|p{2.2cm}|p{1.86cm}c}  \cline{1-4}
\centering $t$ [s]&
\centering $\!\!\in \![0.38,0.4] \!\!$ &
\centering $\!\! \in\! [2.14,2.16] \!\! $ &
\centering $\!\! \in \![3.08,3.1] \!\! $  & &\\ \cline{1-4}
\end{tabular} \\[1mm]
\begin{tabular}{|p{3cm}|p{2.2cm}|p{2.2cm}|p{2.2cm}|p{1.86cm}|c}  \cline{1-5}
\centering RMS$(e_{I_{a}})$ [mA] & \centering 25.0 & \centering 28.19 & \centering 24.43 & \multirow{4}{1.86cm}{\centering Constant $\Vcmref$} & \\ \cline{1-4}
\centering max$(\left|e_{I_{a}}\right|)$ [mA] & \centering 56.4 & \centering 81.4 & \centering 54.8 & \\ \cline{1-4}
\centering \multirow{2}{3cm}{avg(fft($I_s$)) [A]} & \centering 1.82 \\ (Ideal: 1.5) & \centering 9.33 \\ (Ideal: 9) & \centering 1.04 \\ (Ideal: \!0.75) & \\ \cline{1-5}
\end{tabular} \\[1mm]
\begin{tabular}{|p{3cm}|p{2.2cm}|p{2.2cm}|p{2.2cm}|p{1.86cm}|c}  \cline{1-5}
%\hline
\centering \centering \multirow{2}{3cm}{RMS$(e_{I_{a}})$ [mA]} & \centering 14.97 (\textbf{-40.12 \%}) & \centering 28.15 (\textbf{-0.14 \%}) & \centering 13.44  (\textbf{-44.99 \%}) & \multirow{6.28}{1.86cm}{\centering Optimized $\Vcmref$} \\ \cline{1-4}
\centering \centering \multirow{2}{3cm}{max$(\left|e_{I_{a}}\right|)$ [mA]} & \centering 39.06 (\textbf{-30.74 \%})& \centering 81.34 (\textbf{-0.07 \%})& \centering 30.95 (\textbf{-43.52 \%})& \\ \cline{1-4}
\centering \centering \multirow{2}{3cm}{avg(fft($I_s$)) [A]} & \centering 1.68 \\ (Ideal: 1.5) & \centering 9.34 \\ (Ideal: 9) & \centering 0.92 \\ \!\!(Ideal: \! 0.75) & \\ \cline{1-5}
\end{tabular}
\end{table}

%\noindent \mbox{
%\pscircle[fillstyle=solid,fillcolor=black,linecolor=black](1,1){0.5}\hspace{2.1mm}}
\emph{Second Simulation:}
 The results of the simulation using the constant reference $\Vcmref$ are shown in
Fig.~\ref{Res_Figure_4} and Fig.~\ref{Res_Figure_5}.
In this case, $\Vcmi$ and $\Vcmii$ have to track the constant reference $\Vcmref$ representing the minimum value which is strictly needed to always track the desired $\Iaref$, that is the value of the optimal reference $\Vcmref$ in the first simulation in the most demanding situation represented by the case $I_{aM}=9$ A.
Since $\Vcmref$ is constant, from the zoomed-in results of Fig.~\ref{Res_Figure_5}, one can observe that the level-to-level distance when generating voltages $V_1$ and $V_2$ does not change throughout the simulation even when there would be room to reduce it, that is when $I_{aM}$ decreases.
This causes two main disadvantages: 1) a higher harmonic content in the resulting load current $I_s$; 2) a worse tracking of the load current $I_s$ itself. This can be verified by the results reported in Table~\ref{MMC_metrics_table}, which contains the following three metrics for comparison: the root mean square value of the tracking error $e_{I_a}$ on the load current $I_s$: RMS$(e_{I_a})$, the maximum absolute value of the tracking error $e_{I_a}$ on the load current $I_s$: max$(\left|e_{I_a}\right|)$, and the average value of the amplitude spectrum resulting after applying the Fast-Fourier Transform (FFT) on the load current $I_s$: avg(fft($I_s$)). The FFT on the load current $I_s$ has been computed using the ``fft'' Matlab function. These three metrics have been computed on the load current $I_s$ for three different time intervals, corresponding to one period of the load current taken when $I_{aM}=1.5$ A, $I_{aM}=9$ A and $I_{aM}=0.75$ A, respectively. Let us first consider the time interval $t \in [2.14,2.16]$ s, corresponding to one period of the load current when $I_{aM}=9$ A. In this case, Table~\ref{MMC_metrics_table} shows that RMS$(e_{I_a})$ is approximately the same both in the first simulation, in which the optimal reference $\Vcmref$ is used for voltages $\Vcmi$ and $\Vcmii$, and in the second simulation. This happens because the constant reference $\Vcmref$ in the second simulation has indeed been set to the exact minimum value which is strictly needed to track a load current having an amplitude $I_{aM}=9$ A. A similar observation applies to the metrics max$(\left|e_{I_a}\right|)$
and avg(fft($I_s$)). Let us now consider the time intervals $t \in [0.38,0.4]$ s and $t \in [3.08,3.1]$ s, corresponding to one period of the load current when $I_{aM}=1.5$ A and $I_{aM}=0.75$ A, respectively. In this case, one can observe that RMS$(e_{I_a})$ and max$(\left|e_{I_a}\right|)$ are significantly larger in the second simulation (``Constant $\Vcmref$'' case) than in the first simulation (``Optimized $\Vcmref$'' case), highlighting a tracking of the desired load current profile $\Iaref$ which is much worse.
This is due to the fact that voltages $V_1$ and $V_2$ are generated using larger discrete voltage levels, as shown in Fig.~\ref{Res_Figure_5}. This also implies a higher harmonic content in the generated load current $I_s$, as quantified by the quantity avg(fft($I_s$)) in the second simulation with a constant $\Vcmref$, which is more distant from the theoretical value that it should exhibit in the case of a pure sinusoid (1.5 A and 0.75 A, respectively) with respect to the first simulation. This proves the effectiveness of the approach for computing the optimal average capacitor voltages reference $\Vcmref$ and of the model-based adaptive control proposed in this paper.

\section{Conclusion}\label{Conclusion_sect}

In this paper, we addressed the modeling, the harmonic analysis and the model-based adaptive control of MMCs. As far as the modeling part is concerned, we proposed a new compact block scheme to model MMCs, which is
directly implementable in the Matlab/Simulink environment using
simple blocks which are available in standard Simulink libraries.
The proposed model has then been verified against the PLECS simulator.
We then introduced two congruent state-space transformations
in order to decouple the MMC dynamics, thus enabling the proposed harmonic analysis of modular multilevel converters. The performed harmonic analysis has given a deep and exact understanding of the MMC
dynamics. Thanks to this, we could determine the optimal tracking reference for the circulating current in order to make the average capacitor voltages follow the desired reference, and we could compute exactly the optimal voltage reference for the average capacitor voltages in the
upper and lower arms of the converter. Such optimal voltage
reference represents the minimum value which is strictly necessary to properly track the
desired load current while, at the same time, minimizing the harmonic content in the generated load current itself.
The simulation results show how the new proposed model-based adaptive control allows to effectively achieve
all the goals of the Control Problem at the same time: a) balancing of the capacitor voltages; b) tracking of the
optimal voltage reference for the average capacitor voltages in order to minimize the harmonic content in the load current; c) tracking of the desired load
current profile. The simulation results also show the effectiveness of one of the new important concepts that we have introduced in this paper, that is to adaptively vary the average capacitor voltages in the upper and lower arms of the MMC thus making the proposed model-based control adaptive to the operating conditions. The circulating current is exploited in order to make the average capacitor voltages follow the optimal voltage reference which is adapted in real-time as a function of the desired load current. This represents a crucial advantage with respect to the classical approach of maintaining the average capacitor voltages at a constant value, because it enables the reduction of the harmonic content in the generated load current by minimizing the level-to-level distance in the commutating voltage signals, thus enhancing all the intrinsic main advantages of multilevel converters.

\setcounter{secnumdepth}{0}
\section{Funding}
This research did not receive any specific grant from funding
agencies in the public, commercial, or not-for-profit sectors.

\setcounter{secnumdepth}{0}
\section{Declaration of Generative AI and AI-assisted technologies in the writing process}
 During the preparation of this work the authors did not use any Generative AI and AI-assisted technologies in the writing process.

%%%%%%%%%%%%%%%%%%%%%%%%%%%%%%%%%%%%%%%%%%%%%%%%%%%%%%%%%%%%%%%%%%%%
\renewcommand{\theequation}{A.\arabic{equation}}
\setcounter{equation}{0} \setcounter{secnumdepth}{0}
\section{Appendix A: Sinusoidal behavior of function $P_1(t)$}\label{App_P1t}
Function  $P_1(t)$ in~\eqref{New_VC1_VC2_due_final_3} is composed of the three terms $2\,V_{dc}
I_d$, $V_d I_d$ and $f(t)\Iaref$. The first term $2\,V_{dc} I_d$ is
known, see (\ref{Id}). The second term $V_d I_d$ can be expressed as
follows, see App.~B:
\begin{equation}\label{VId}
 V_d\,I_d
 =
  V_{d0}\,I_{d0} + \frac{V_{dM}\, I_{dM}}{2}\cos(\alpha_{LR})
  + F_{V_dI_d}(\omega t),
\end{equation}
where $F_{V_dI_d}(\omega t)$  is the sum of two sinusoidal
terms at frequency $\omega$ and $2\omega$, see \eqref{new_B1}.
 Function $f(t)= L_T\, \dotIaref \!+\!R_T\,\Iaref \!+\!2 V_a$ in \eqref{f_Ia_Va} and \eqref{New_VC1_VC2_due_final_3} can be written as $f(t) =   f_M\,\sin(\omega t+\alpha_f)$, see App.~D,
where:
\begin{equation}\label{fM_alphaf}
  f_M =  \sqrt{S_f^2 + C_f^2},
 \hspace{16mm}
 \alpha_f =
 \ds\arctan\left(\frac{S_f}{C_f}\right),
\end{equation}
$ S_f = L_T\, I_{aM}\,\omega\, \!+\! 2 V_{aM}\sin(\alpha_{V_a})$ and $C_f = R_T\,I_{aM} \!+\! 2  V_{aM}\cos(\alpha_{V_a})$.
Using (\ref{F1_F2}), the third term $f(t)\Iaref$ of function
$P_1(t)$ can be expressed as follows:
\begin{equation}\label{fIa}
\begin{array}{@{\!\!}c}
f(t) \Iaref
  \!=\!\!
  f_M\!\sin(\omega t\!+\!\alpha_f)I_{aM}\!\sin(\omega t)
  \!=\!
  \frac{f_MI_{aM}}{2}\! \cos(\!\alpha_f\!) \!- \!\frac{f_MI_{aM}}{2}\!\cos(2\omega
  t\!+\!\alpha_f).
\end{array}
\end{equation}
 From (\ref{f_Ia_Va}), (\ref{Id}), (\ref{VId}) and  (\ref{fIa}), it follows that function $P_1(t)
 $ can be rewritten as in \eqref{T1t}.

%%%%%%%%%%%%%%%%%%%%%%%%%%%%%%%%%%%%%%%%%%%%%%%%%%%%%%%%%%%%%%%%%%%%
\renewcommand{\theequation}{B.\arabic{equation}}
\setcounter{equation}{0} \setcounter{secnumdepth}{0}
\section{Appendix B: Sinusoidal behavior of function $V_d I_d$}\label{App_Vd_Id}

Using (\ref{Ia_Vd_2}), (\ref{Id}) and (\ref{F1_F2}), function
$V_d\,I_d$ can be expressed as follows:
\[
\begin{array}{@{\!}r@{}c@{}l@{\!}}
 V_dI_d
 & = &
 [V_{d0} \!+ \!V_{dM} \sin(\omega t\!+\!\alpha_{V_d})]
 \cdot
  [ I_{d0}\!+\!I_{dM} \sin(\omega  t\!+\!\alpha_{V_d}-\alpha_{LR})]
 \\[1mm]
 & = &
 V_{d0} I_{d0} \! +\! V_{dM}I_{dM} \sin(\omega t\!+\!\alpha_{V_d})
 \cdot
  \sin(\omega t\!+\!\alpha_{V_d}\!-\!\alpha_{LR}) \!+\! F_{1}(\omega t)
 \\[1mm]
 & = &
  V_{d0} I_{d0} \!+\! \frac{V_{dM} I_{dM}}{2}\cos(\alpha_{LR})
 \!- \frac{V_{dM} I_{dM}}{2} \cos(2\omega
 t\!+\!2\alpha_{V_d}\!-\!\alpha_{LR}) \!+\! F_{1}(\omega t)
 \\[1mm]
 & = &
  V_{d0}I_{d0} \!+\! \frac{V_{dM}
  I_{dM}}{2}\cos(\alpha_{LR}) \!+\!F_{V_dI_d}(\omega t),
\end{array}
\]
 where $ F_{1}(\omega t) = I_{d0}\;V_{dM}\,\sin(\omega t+\alpha_{V_d}) +  V_{d0}\;I_{dM}\,\sin(\omega  t+\alpha_{V_d}-\alpha_{LR})$ and:
\begin{equation}\label{new_B1}
\begin{array}{r@{\;}c@{\;}l}
  F_{V_dI_d}(\omega t)
 &=&
  F_{1}(\omega t)  - \frac{V_{dM} I_{dM}}{2} \cos(2\omega
 t\!+\!2\alpha_{V_d}\!-\alpha_{LR})
 \\[1mm]
 &=&
   I_{d0}V_{dM} \sin(\omega t\!+\!\alpha_{V_d})
% \\ && \hspace{4mm}
   +  V_{d0}I_{dM}\sin(\omega  t\!+\!\alpha_{V_d}-\alpha_{LR})
 \\ && \hspace{8mm}
   - \frac{V_{dM} I_{dM}}{2} \cos(2\omega  t\!+\!2\alpha_{V_d}\!-\alpha_{LR}).
\end{array}
\end{equation}

%%%%%%%%%%%%%%%%%%%%%%%%%%%%%%%%%%%%%%%%%%%%%%%%%%%%%%%%%%%%%%%%%%%%
\renewcommand{\theequation}{C.\arabic{equation}}
\setcounter{equation}{0} \setcounter{secnumdepth}{0}
\section{Appendix C: Sinusoidal behavior of function $P_2(t)$}\label{App_P2t}
 Function $P_2(t)$
in~\eqref{New_VC1_VC2_due_final_3} is composed of the three terms $2\,V_{dc}\,
\Iaref$, $V_d \Iaref$ and $f(t) I_d$.
  The first term $2\,V_{dc}\, \Iaref$ is known, see (\ref{Ia_Vd_1}).
Using (\ref{F1_F2}), the second term $V_d\,\Iaref$ can be expressed
as follows:
\begin{equation}\label{VdIa}
\begin{array}{@{}r@{}c@{}l@{}}
 V_d\Iaref
  \!=\!
 [V_{d0}\!+\!V_{dM}\sin(\omega t \!+\!\alpha_{V_d})] I_{aM} \sin(\omega t)
  \!=\! %\ds
 \frac{V_{dM}\, I_{aM}}{2}\cos(\alpha_{V_d}) \!+\! F_2(\omega t),
\end{array}
\end{equation}
where
%\[
 $F_2(\omega t)
 =
 V_{d0}\,I_{aM}\, \sin(\omega t) - \frac{V_{dM}\, I_{aM}}{2} \cos(2\omega
 t+\alpha_{V_d})$.
 %\]
Using (\ref{F1_F2}), the third term $f(t) \, I_d$ of function
$P_2(t)$ can be expressed as follows:
\begin{equation}\label{FIaId}
\begin{array}{@{}r@{\,}c@{\,}l@{}}
\!\!\!f(t) I_d
  &=&
  f_M\!\sin(\omega t\!+\!\alpha_f\!) [I_{d0}\!+\!I_{dM}\!\sin(\omega  t\!+\!\alpha_{V_d}\!-\!\alpha_{LR})]\!\!
 \\[1mm]
  &=& \ds
   \frac{f_M\,I_{dM}}{2} \cos(\alpha_f-\alpha_{V_d}+\alpha_{LR})
  +
  F_3(\omega t),
\end{array}
\end{equation}
where
%\[
 $F_3(\omega t)
 =
 I_{d0}\,f_M\,\sin(\omega t+\alpha_f)
 - \frac{f_M\,I_{dM}}{2}\cos(2\omega t+\alpha_f+\alpha_{V_d}-\alpha_{LR})$.
 %\]
From (\ref{f_Ia_Va}), (\ref{Ia_Vd_1}), (\ref{Ia_Vd_2}), (\ref{VdIa}) and (\ref{FIaId}),
it follows that function $P_2(t)$  can be expressed as in \eqref{T2t}.

%%%%%%%%%%%%%%%%%%%%%%%%%%%%%%%%%%%%%%%%%%%%%%%%%%%%%%%%%%%%%%%%%%%%
\renewcommand{\theequation}{D.\arabic{equation}}
\setcounter{equation}{0} \setcounter{secnumdepth}{0}
\section{Appendix D: Sinusoidal behavior of function $f(t)$}\label{App_ft}

Function $f(t)$ is the sum of three sinusoidal terms characterized
by the same frequency $\omega$, and can therefore be expressed as
follows:
\[
\begin{array}{@{\!}r@{}c@{}l@{\!}}
f(t)
  &=&
 L_T \dotIaref \!+\!R_T \Iaref \!+\!2 V_a
\!=\!  L_T I_{aM} \omega \cos(\omega t) \!+\!R_T I_{aM}\sin(\omega t)
  \!+\! 2  V_{aM}\sin(\omega t\!+\!\alpha_{V_a}\!)
 \\[1mm]
  &=&
  L_T I_{aM} \omega\! \cos(\omega t) \!\!+\!\!R_T \!I_{aM}\!\sin(\omega t)
   % \\ && \hspace{4mm}
 \!\!+\!\! 2  V_{aM}\!\sin(\omega t)\!\cos(\!\alpha_{V_a}\!)
  %\\ && \hspace{4mm}
 \!\!+\!\! 2  V_{aM}\!\cos(\omega t)\!\sin(\!\alpha_{V_a}\!)\!
 \\[1mm]
  &=&
  [R_T I_{aM} \!+\! 2  V_{aM}\cos(\alpha_{V_a})]\sin(\omega t)
  %\\ && \hspace{4mm}
  +
  [L_T I_{aM} \omega \!+\! 2  V_{aM}\!\sin(\!\alpha_{V_a}\!)]\!\cos(\omega t)
 \\[1mm]
  &=&
  f_M\,\sin(\omega t+\alpha_f),
\end{array}
\]
where $  f_M =  \sqrt{S_f^2 + C_f^2}$, $ \alpha_f =
 \ds\arctan\left(\frac{S_f}{C_f}\right)$, $S_f = L_T  I_{aM} \omega \!+\! 2 V_{aM}\sin(\alpha_{V_a})$ and $C_f = R_T I_{aM} \!+\! 2  V_{aM}\cos(\alpha_{V_a})$.

%%%%%%%%%%%%%%%%%%%%%%%%%%%%%%%%%%%%%%%%%%%%%%%%%%%%%%%%%%%%%%%%%%%%
\renewcommand{\theequation}{E.\arabic{equation}}
\setcounter{equation}{0} \setcounter{secnumdepth}{0}
\section{Appendix E: Product of two sinusoidal functions}\label{App_B}

The product $F_1(t)F_2(t)$ of two sinusoidal signals
$F_1(t)=a_1\sin(\omega t+\alpha_1)$ and  $F_2(t)=a_2\sin(\omega
t+\alpha_2)$  can always be written as follows:

\begin{equation}\label{F1_F2}
\begin{array}{@{}r@{\;}c@{\;}l@{}}
F_1(t)F_2(t)
 & = &
 a_1\, a_2\, \sin(\omega t+\alpha_1) \sin(\omega t+\alpha_2)
 \\[1mm]
% & = &
& = &   \frac{a_1\, a_2}{2}\cos(\alpha_1-\alpha_2)
 % \\ && \hspace{4mm}
 - \frac{a_1\, a_2}{2} \cos(2\omega  t+\alpha_1+\alpha_2).
% \\
\end{array}
\end{equation}

%%%%%%%%%%%%%%%%%%%%%%%%%%%%%%%%%%%%%%%%%%%%%%%%%%%%%%%%%%%%%%%%%%%%
\renewcommand{\theequation}{F.\arabic{equation}}
\setcounter{equation}{0} \setcounter{secnumdepth}{0}
\section{Appendix F: The constant term $P_{20}$}\label{App_P_20}

 Using (\ref{Id}), the constant term $P_{20}$ can also be expressed as
 follows:
\[
\begin{array}{r@{}c@{}l}
 P_{20}
  & = &
 -\frac{V_{dM}\, I_{aM}}{2}\cos(\alpha_{V_d}) - \frac{f_M\,I_{dM}}{2} \cos(\alpha_{V_d}-\alpha_{f}-\alpha_{LR})
 \\[1mm]
  & = &
 -\frac{V_{dM}\, I_{aM}}{2}\cos(\alpha_{V_d}) - \frac{V_{dM} f_M}{2\sqrt{R^2+L^2\omega^2}} \cos(\alpha_{V_d}+\beta)
 \\[1mm]
      & = &
 -\frac{V_{dM} I_{aM}}{2}\! \cos(\alpha_{V_d})
  %\\ && \hspace{4mm}
   \!+\! \frac{V_{dM} f_M}{2 \sqrt{R^2+L^2\omega^2}}
   \left[
   \cos(\alpha_{V_d})\!\cos(\beta)
   \!-\! \sin(\alpha_{V_d})\!\sin(\beta)
   \right]
 \\[2mm]
  & = &
 -\frac{V_{dM}}{2}\!
   \underbrace{\left( I_{aM}  \!+ \! \frac{f_M\cos(\beta)}{\sqrt{R^2+L^2\omega^2}}\right)}_{\ds a} \! \cos(\alpha_{V_d})
 %\\[7mm] && \hspace{8mm}
   \!+\!\frac{V_{dM}}{2}\!
   \underbrace{\frac{f_M\sin(\beta)}{\sqrt{R^2+L^2\omega^2}}}_{\ds b} \sin(\alpha_{V_d})
 \\[9mm]
  & = &
 -\frac{V_{dM}\sqrt{a^2\!+\!b^2}}{2}\!\left[\cos(\gamma)\!\cos(\alpha_{V_d})\!-\!
  \sin(\gamma)\!\sin(\alpha_{V_d})\right]
% \\[2mm]
%  & = &
\!\!=\!\! -\frac{V_{dM}\sqrt{a^2\!+\!b^2}}{2} \!\cos(\alpha_{V_d}\!+\!\gamma\!),
 \end{array}
 \]
where $\beta  =-\alpha_{f}-\alpha_{LR}$, $a=\cos(\gamma)$,
$b=\sin(\gamma)$ and $ \gamma = \arctan\!2(b,a)$. The values of
parameters $f_M$ and $\alpha_{f}$ have been defined in
(\ref{fM_alphaf}), and the value of parameter $\alpha_{LR}$ has been
defined in (\ref{Id}).


\begin{thebibliography}{00}
%
\bibitem{CEP_3} M.~Romero-Rodr\'iguez, R.~Delpoux, L.~Pi\'etrac, J.~Dai, A.~Benchaib, E.~Niel, ``An implementation method for the supervisory control of
time-driven systems applied to high-voltage direct current
transmission grids'', \emph{Control Engineering Practice}, vol.~82,
pp.~97-107, Jan.~2019, DOI: 10.1016/j.conengprac.2018.10.002.
%
\bibitem{CEP_4} M.~Emin Meral, D.~\c Celik, ``Proportional complex integral based control of distributed energy converters
connected to unbalanced grid system'', \emph{Control Engineering
Practice}, vol.~103, Oct.~2020, DOI:
10.1016/j.conengprac.2020.104574.

%
\bibitem{Nostro_3} R.~Zanasi, D.~Tebaldi, ``Modeling of complex planetary gear sets using power-oriented graphs'', \emph{IEEE
Trans. Veh. Technol.}, vol.~69, no.~12, pp.~14470-14483, Dec.~2020, DOI:
10.1109/TVT.2020.3040899.
%
\bibitem{CEP_5} D.~Tebaldi, R.~Zanasi, ``Systematic modeling of complex time-variant gear systems
using a Power-Oriented approach'', \emph{Control Engineering
Practice}, vol.~132, Mar.~2023, DOI:
10.1016/j.conengprac.2022.105420.
%
\bibitem{CEP_NEW_2} W.~He, Y.~Shang, M.~Masoud Namazi, R.~Ortega, ``Adaptive sensorless control for buck converter with constant power load'', \emph{Control Engineering Practice}, vol.~126, Sep.~2022, DOI: 10.1016/j.conengprac.2022.105237.
%
\bibitem{CEP_NEW_3} Q.~Guo, I.~Bahri, D.~Diallo, E.~Berthelot, ``Model predictive control and linear control of DC-DC boost converter in low voltage DC microgrid: An experimental comparative study'', vol.~131, Feb.~2023, DOI: 10.1016/j.conengprac.2022.105387.
%
\bibitem{TCCT_6} J.~Saeed, L.~Wang, N.~Fernando, ``Model predictive control of phase shift full-bridge DC-DC converter using
laguerre functions'', \emph{IEEE Trans. Control Syst. Technol.},
vol.~30, no.~2, pp.~819-826, Mar.~2022, DOI:
10.1109/TCST.2021.3069148.


%
\bibitem{CEP_1} R.~Cisneros, M.~Pirro, G.Bergna, R.~Ortega, G.~Ippoliti, M.~Molinas, ``Global tracking passivity-based PI
control of bilinear systems: Application to the interleaved boost
and modular multilevel converters'', \emph{Control Engineering
Practice}, vol.~43, pp.~109-119, Oct.~2015, DOI:
10.1016/j.conengprac.2015.07.002.
%
\bibitem{CEP_2} M.~Kamarzarrin, M.~Hossein Refan, P.~Amiri, ``Open-circuit faults diagnosis and Fault-Tolerant Control
scheme based on Sliding-Mode Observer for DFIG back-to-back
converters: wind turbine applications'', \emph{Control Engineering
Practice}, vol.~126, Sep.~2022, DOI:
10.1016/j.conengprac.2022.105235.



%
\bibitem{Nostro_1} R.~Zanasi, D.~Tebaldi, ``Modeling control and robustness assessment of multilevel flying-capacitor converters'',
\emph{Energies}, 2021, 14(7), 1903, DOI: 10.3390/en14071903.
%
\bibitem{TCCT_4} A.~Bouarfa, M.~Bodson, M.~Fadel, ``An optimization formulation of converter control and its general
solution for the four-leg two-level inverter'', \emph{IEEE Trans.
Control Syst. Technol.}, vol.~26, no.~5, pp.~1901-1908, Sep.~2018,
DOI: 10.1109/TCST.2017.2738608.
%
\bibitem{TCCT_0} J.-N.~Chiasson, L.-M.~Tolbert, K.-J.~McKenzie, Z.~Du, ``Control of a multilevel converter
using resultant theory'', \emph{IEEE Trans. Control Syst. Technol.},
vol.~11, no.~3, pp.~345-354, May.~2003, DOI:
10.1109/TCST.2003.810382.
%
\bibitem{TCCT_3} J.-N.~Chiasson, L.-M.~Tolbert, K.-J.~McKenzie, Z.~Du, ``Elimination of harmonics in a multilevel converter
using the theory of symmetric polynomials and resultants'',
\emph{IEEE Trans. Control Syst. Technol.}, vol.~13, no.~2,
pp.~216-223, Mar.~2005, DOI: 10.1109/TCST.2004.839556.
%
\bibitem{CEP_NEW_1} S.~Laamiri, M.~Ghanes, G.~Santomenna, ``Observer based direct control strategy for a multi-level three phase
flying-capacitor inverter'', \emph{Control Engineering Practice}, vol.~86,
pp.~155-165, May~2019, DOI: 10.1016/j.conengprac.2019.03.011.






%
\bibitem{TCCT_2} L.~Hetel, M.~Defoort, M.~Djema$\ddot{\i}$, ``Binary control design for a class of bilinear systems:
application to a multilevel power converter'', \emph{IEEE Trans.
Control Syst. Technol.}, vol.~24, no.~2, pp.~719-726, Mar.~2016,
DOI: 10.1109/TCST.2015.2460696.


%
\bibitem{Energies_1}
M.~Diaz, R.~Cardenas, E.~Ibaceta, A.~Mora, M.~Urrutia, M.~Espinoza,
F.~Rojas, P.~Wheeler, ``An overview of modelling techniques and
control strategies for modular multilevel matrix converters'',
\emph{Energies}, 2020, 13(18), 4678, DOI: 10.3390/en13184678.
%
\bibitem{Energies_2}
M.~Liu, Z.~Li, X.~Yang, ``A universal mathematical model of modular
multilevel converter with half-bridge'', \emph{Energies}, 2020,
13(17), 4464, DOI: 10.3390/en13174464.
%
\bibitem{Energetic_Macroscopic_Representation_2} P.~Delarue, F.~Gruson, X.~Guillaud, ``Energetic macroscopic representation and inversion based control
of a modular multilevel converter'', \emph{15th European Conference
on Power Electronics and Applications (EPE)}, Lille, France,
Sep.~2-6, 2013, DOI: 10.1109/EPE.2013.6631859.


%
\bibitem{TCCT_1} P.-B.~Steckler, J.-Y.~Gauthier, X.~Lin-Shi, F.~Wallart, ``Differential flatness-based, full-order nonlinear
control of a modular multilevel converter (MMC)'', \emph{IEEE Trans.
Control Syst. Technol.}, vol.~30, no.~2, pp.~547-557, Mar.~2022,
DOI: 10.1109/TCST.2021.3067887.



%
\bibitem{Altro_4} S.~Du, J.~Liu, T.~Liu, ``Modulation and closed-loop-based DC capacitor voltage control for MMC with fundamental switching frequency'',
\emph{IEEE Trans. Power Electron.}, vol.~30, no.~1, pp.~327-338,
Jan.~2015, DOI: 10.1109/TPEL.2014.2301836.

%
\bibitem{Altro_3} A.~Abdayem, J.~Sawma, F.~Khatounian, E.~Monmasson, ``Control of a single phase modular
multilevel converter based on a new modulation technique'',
\emph{22nd IEEE International Conference on Industrial Technology
(ICIT)}, Valencia, Spain, Mar.~10-12, 2021, DOI:
10.1109/ICIT46573.2021.9453516.
%
\bibitem{Altro_2} L.~Ben-Brahim, A.~Gastli, M.~Trabelsi, K.-A.~Ghazi, M.~Houchati, H. Abu-Rub, ``Modular multilevel converter
circulating current reduction using model predictive control'',
\emph{IEEE Trans. Ind. Electron.}, vol.~63, no.~6, pp.~3857-3866,
Jun.~2016, DOI: 10.1109/TIE.2016.2519320.

%
\bibitem{Altro_5} B.~S.~Riar, T.~Geyer, U.~K.~Madawala, ``Model predictive direct current control of modular multilevel converters: modeling, analysis, and experimental evaluation'', \emph{IEEE Trans. Power Electron.}, vol.~30, no.~1, pp.~431-439,
Jan.~2015, DOI: 10.1109/TPEL.2014.2301438.

%
\bibitem{Altro_6} L.~Harnefors, A.~Antonopoulos, S.~Norrga, L.~\mbox{$\ddot{A}$}ngquist, H.-P.~Nee, ``Dynamic analysis of modular multilevel converters'',
\emph{IEEE Trans. Ind. Electron.}, vol.~60, no.~7, pp.~2526-2537,
Jul.~2013, DOI: 10.1109/TIE.2012.2194974.


%
\bibitem{POG_Technique} R.~Zanasi, ``The power-oriented graphs technique: system modeling and
basic properties'', \emph{Proceedings of IEEE Vehicle Power and
Propulsion Conference (VPPC)}, Lille, France, Sep.~1-3, 2010, DOI:
10.1109/VPPC.2010.5729018.

%%%%%%%%%%%%%%%%%%%%%%%%%%%%%%%%%%%%%%%%%%%%%%%%%%%%%%%%%%%%%%%%%%%%%%%%%%%%%%%%%%%%%
%%%%%%%%%%%%%%%%%%%%%%%%%%%%%%%%%%%%%%%%%%%%%%%%%%%%%%%%%%%%%%%%%%%%%%%%%%%%%%%%%%%%%

%
\bibitem{Bond_Graphs_1} M.-S.~Jha, G.~Dauphin-Tanguy, B.~Ould-Bouamama, ``Robust fault detection with interval valued uncertainties in bond graph
framework'', \emph{Control Engineering Practice}, vol.~71, pp.
61-78, Feb.~2018, DOI: 10.1016/j.conengprac.2017.10.009.
%
\bibitem{Bond_Graphs_2} A.~Badoud, M.~Khemliche, S.~Bacha, B.~Raison, ``Modeling and performance analysis of multilevel inverter for single-phase grid
connected photovoltaic modules'', \emph{International Renewable and
Sustainable Energy Conference (IRSEC)}, Ouarzazate, Morocco,
Mar.~7-9, 2013, DOI: 10.1109/IRSEC.2013.6529727.

%
\bibitem{Energetic_Macroscopic_Representation_1} I.~Garc\'ia-Herreros, X.~Kestelyn, J.~Gomand, R.~Coleman, P.-J.~Barre,
``Model-based decoupling control method for dual-drive gantry
stages: a case study with experimental validations'', \emph{Control
Engineering Practice}, vol.~21, no. 3, pp. 298-307, Mar.~2013, DOI:
10.1016/j.conengprac.2012.10.010.


%
\bibitem{Nostro_2} D.~Tebaldi, ``Efficiency map-based PMSM parameters estimation using
power-oriented modeling'', \emph{IEEE Access}, vol.~10,
pp.~45954-45961, Apr.~2022, DOI:
10.1109/ACCESS.2022.3169149.

%
\bibitem{Nostro_4} D.~Tebaldi, R.~Zanasi, ``Modeling and simulation of a multiphase
diode bridge rectifier'', \emph{IEEE European Control Conference
(ECC)}, St.~Petersburg, Russia, May~12-15,~2020, DOI:
10.23919/ECC51009.2020.9143693.


\bibitem{Plecs_Ref}
PLECS Documentation:
\url{https://www.plexim.com/products/plecs}.

%
\end{thebibliography}
\end{document}